\documentclass[apjl,iop]{emulateapj}
\usepackage{comment}
\usepackage{ifthen}

\newcommand{\forloop}[5][1]%
{%
\setcounter{#2}{#3}%
\ifthenelse{#4}%
    {%
    #5%
    \addtocounter{#2}{#1}%
    \forloop[#1]{#2}{\value{#2}}{#4}{#5}%
    }%
    {%
    }%
}%

\newcommand{\ctbd}[1]{}

\newcommand{\lc}{light curve}
\newcommand{\lcs}{light curves}
\newcommand{\Lc}{Light curve}

\newcommand{\masy}{\ensuremath{\rm mas\,yr^{-1}}}
\newcommand{\kms}{\ensuremath{\rm km\,s^{-1}}}
\newcommand{\ms}{\ensuremath{\rm m\,s^{-1}}}

\newcommand{\gcmc}{\ensuremath{\rm g\,cm^{-3}}}

\newcommand{\vsini}{\ensuremath{v \sin{i}}}
\newcommand{\feh}{\ensuremath{\rm [Fe/H]}}

\newcommand{\vmac}{\ensuremath{v_{\rm mac}}}
\newcommand{\vmic}{\ensuremath{v_{\rm mic}}}

\newcommand{\rsun}{\ensuremath{R_\sun}}
\newcommand{\msun}{\ensuremath{M_\sun}}
\newcommand{\lsun}{\ensuremath{L_\sun}}

\newcommand{\rstar}{\ensuremath{R_\star}}
\newcommand{\mstar}{\ensuremath{M_\star}}
\newcommand{\lstar}{\ensuremath{L_\star}}

\newcommand{\teffstar}{\ensuremath{T_{\rm eff\star}}}
\newcommand{\rhostar}{\ensuremath{\rho_\star}}
\newcommand{\loggstar}{\ensuremath{\log{g_{\star}}}}

\newcommand{\rpl}{\ensuremath{R_{p}}}
\newcommand{\mpl}{\ensuremath{M_{p}}}

\newcommand{\rhopl}{\ensuremath{\rho_{p}}}

\newcommand{\arstar}{\ensuremath{a/\rstar}}
\newcommand{\zrstar}{\ensuremath{\zeta/\rstar}}

\newcommand{\rjup}{\ensuremath{R_{\rm J}}}
\newcommand{\mjup}{\ensuremath{M_{\rm J}}}

\newcommand{\reffigl}[1]{Figure~\ref{fig:#1}}
\newcommand{\refsecl}[1]{\mbox{Section \ref{sec:#1}}}

\newcommand{\reftabl}[1]{Table~\ref{tab:#1}}

\newcommand{\loopand}{\ifnum\value{planetcounter}=2 and \else\fi}
\newcommand{\loopcomma}{\ifnum\value{planetcounter}<2 ,\else. \fi}
\newcommand{\loopcommanoperiod}{\ifnum\value{planetcounter}<2 ,\else \space\fi}
\newcommand{\loopcommanospace}{\ifnum\value{planetcounter}<2 ,\else \fi}

\newcommand{\hatcurhtrxxxxxA}{HATS568-001}                                   
\newcommand{\hatcurfieldxxxxxA}{\ensuremath{string}}                         
\newcommand{\hatcurCCraxxxxxA}{\ensuremath{13^{\mathrm h}51^{\mathrm m}37.80{\mathrm s}}}                                  
\newcommand{\hatcurCCdecxxxxxA}{\ensuremath{-23{\arcdeg}46{\arcmin}52.2{\arcsec}}}                                 
\newcommand{\hatcurCCmagxxxxxA}{13.097}                                      
\newcommand{\hatcurCCtwomassxxxxxA}{2MASS~13513786-2346522}                  
\newcommand{\hatcurCCgscxxxxxA}{GSC~6716-01190}                              
\newcommand{\hatcurCCtassmvxxxxxA}{\ensuremath{13.097\pm0.030}}              
\newcommand{\hatcurCCtassmvshortxxxxxA}{\ensuremath{13.1}}                   
\newcommand{\hatcurCCtassmBxxxxxA}{\ensuremath{13.812\pm0.030}}              
\newcommand{\hatcurCCtassmBshortxxxxxA}{\ensuremath{13.8}}                   
\newcommand{\hatcurCCtassmIxxxxxA}{\ensuremath{nff\pmnff}}                   
\newcommand{\hatcurCCtassmIshortxxxxxA}{\ensuremath{0.0}}                    
\newcommand{\hatcurCCtassmgxxxxxA}{\ensuremath{13.380\pm0.020}}              
\newcommand{\hatcurCCtassmgshortxxxxxA}{\ensuremath{13.4}}                   
\newcommand{\hatcurCCtassmrxxxxxA}{\ensuremath{12.909\pm0.040}}              
\newcommand{\hatcurCCtassmrshortxxxxxA}{\ensuremath{12.9}}                   
\newcommand{\hatcurCCtassmixxxxxA}{\ensuremath{12.687\pm0.050}}              
\newcommand{\hatcurCCtassmishortxxxxxA}{\ensuremath{12.7}}                   
\newcommand{\hatcurCCtwomassJmagxxxxxA}{\ensuremath{11.788\pm0.022}}         
\newcommand{\hatcurCCtwomassHmagxxxxxA}{\ensuremath{11.487\pm0.024}}         
\newcommand{\hatcurCCtwomassKmagxxxxxA}{\ensuremath{11.416\pm0.021}}         
\newcommand{\hatcurCCcitJmagxxxxxA}{\ensuremath{11.805\pm0.022}}             
\newcommand{\hatcurCCcitHmagxxxxxA}{\ensuremath{11.482\pm0.024}}             
\newcommand{\hatcurCCcitKmagxxxxxA}{\ensuremath{11.440\pm0.021}}             
\newcommand{\hatcurCCbbJmagxxxxxA}{\ensuremath{11.854\pm0.024}}              
\newcommand{\hatcurCCbbHmagxxxxxA}{\ensuremath{11.503\pm0.025}}              
\newcommand{\hatcurCCbbKmagxxxxxA}{\ensuremath{11.460\pm0.021}}              
\newcommand{\hatcurCCesoJmagxxxxxA}{\ensuremath{11.857\pm0.025}}             
\newcommand{\hatcurCCesoHmagxxxxxA}{\ensuremath{11.498\pm0.028}}             
\newcommand{\hatcurCCesoKmagxxxxxA}{\ensuremath{11.459\pm0.022}}             
\newcommand{\hatcurCCesoJHmagxxxxxA}{\ensuremath{0.358\pm0.036}}             
\newcommand{\hatcurCCesoJKmagxxxxxA}{\ensuremath{0.398\pm0.033}}             
\newcommand{\hatcurCCesoHKmagxxxxxA}{\ensuremath{0.040\pm0.036}}             
\newcommand{\hatcurLCdipxxxxxA}{\ensuremath{16.8}}                           
\newcommand{\hatcurLCrprstarxxxxxA}{\ensuremath{0.1171\pm0.0026}}            
\newcommand{\hatcurLCbsqxxxxxA}{\ensuremath{0.290_{-0.102}^{+0.087}}}        
\newcommand{\hatcurLCimpxxxxxA}{\ensuremath{0.538_{-0.105}^{+0.075}}}        
\newcommand{\hatcurLCzetaxxxxxA}{\ensuremath{17.39\pm0.17}}                  
\newcommand{\hatcurLCdurxxxxxA}{\ensuremath{0.1335\pm0.0025}}                
\newcommand{\hatcurLCdurshortxxxxxA}{\ensuremath{0.1335}}                    
\newcommand{\hatcurLCdurhrxxxxxA}{\ensuremath{3.205\pm0.059}}                
\newcommand{\hatcurLCdurhrshortxxxxxA}{\ensuremath{3.205}}                   
\newcommand{\hatcurLCqxxxxxA}{\ensuremath{0.03110\pm0.00057}}                
\newcommand{\hatcurLCqshortxxxxxA}{\ensuremath{0.031}}                       
\newcommand{\hatcurLCingdurxxxxxA}{\ensuremath{0.0190\pm0.0027}}             
\newcommand{\hatcurLCPxxxxxA}{\ensuremath{4.2986432\pm0.0000045}}            
\newcommand{\hatcurLCPprecxxxxxA}{\ensuremath{4.2986432}}                    
\newcommand{\hatcurLCPshortxxxxxA}{\ensuremath{4.2986}}                      
\newcommand{\hatcurLCTxxxxxA}{\ensuremath{2456870.36872\pm0.00051}}          
\newcommand{\hatcurLCTAxxxxxA}{\ensuremath{2455645.2555\pm0.0013}}           
\newcommand{\hatcurLCTBxxxxxA}{\ensuremath{2457098.19685\pm0.00058}}         
\newcommand{\hatcurLChatnetmxxxxxA}{\ensuremath{12.864680\pm0.000093}}       
\newcommand{\hatcurLCiblendxxxxxA}{\ensuremath{0.959\pm0.046}}               
\newcommand{\hatcurLCrhoxxxxxA}{\ensuremath{1.03\pm0.20}}                    
\newcommand{\hatcurSMEiteffxxxxxA}{\ensuremath{5860\pm180}}                  
\newcommand{\hatcurSMEizfehxxxxxA}{\ensuremath{0.100\pm0.080}}               
\newcommand{\hatcurSMEizfehshortxxxxxA}{\ensuremath{0.10}}                   
\newcommand{\hatcurSMEiloggxxxxxA}{\ensuremath{4.55\pm0.17}}                 
\newcommand{\hatcurSMEivsinxxxxxA}{\ensuremath{3.14\pm0.34}}                 
\newcommand{\hatcurSMEivmacxxxxxA}{\ensuremath{0.0}}                         
\newcommand{\hatcurSMEivmicxxxxxA}{\ensuremath{0.0}}                         
\newcommand{\hatcurSMEiiteffxxxxxA}{\ensuremath{5715\pm73}}                  
\newcommand{\hatcurSMEiizfehxxxxxA}{\ensuremath{0.020\pm0.050}}              
\newcommand{\hatcurSMEiizfehshortxxxxxA}{\ensuremath{0.02}}                  
\newcommand{\hatcurSMEiiloggxxxxxA}{\ensuremath{4\pm0}}                      
\newcommand{\hatcurSMEiivsinxxxxxA}{\ensuremath{3.88\pm0.50}}                
\newcommand{\hatcurLBizxxxxxA}{\ensuremath{0.2160}}                          
\newcommand{\hatcurLBiizxxxxxA}{\ensuremath{0.3244}}                         
\newcommand{\hatcurLBiixxxxxA}{\ensuremath{0.2774}}                          
\newcommand{\hatcurLBiiixxxxxA}{\ensuremath{0.3246}}                         
\newcommand{\hatcurLBiIxxxxxA}{\ensuremath{0.2567}}                          
\newcommand{\hatcurLBiiIxxxxxA}{\ensuremath{0.3254}}                         
\newcommand{\hatcurLBigxxxxxA}{\ensuremath{0.5644}}                          
\newcommand{\hatcurLBiigxxxxxA}{\ensuremath{0.2277}}                         
\newcommand{\hatcurLBirxxxxxA}{\ensuremath{0.3674}}                          
\newcommand{\hatcurLBiirxxxxxA}{\ensuremath{0.3192}}                         
\newcommand{\hatcurLBiRxxxxxA}{\ensuremath{0.3425}}                          
\newcommand{\hatcurLBiiRxxxxxA}{\ensuremath{0.3217}}                         
\newcommand{\hatcurLBikepxxxxxA}{\ensuremath{0.1000}}                        
\newcommand{\hatcurLBiikepxxxxxA}{\ensuremath{0.1000}}                       
\newcommand{\hatcurISOmxxxxxA}{\ensuremath{0.994\pm0.035}}                   
\newcommand{\hatcurISOmshortxxxxxA}{\ensuremath{0.99}}                       
\newcommand{\hatcurISOmlongxxxxxA}{\ensuremath{0.994\pm0.035}}               
\newcommand{\hatcurISOrxxxxxA}{\ensuremath{1.107\pm0.069}}                   
\newcommand{\hatcurISOrshortxxxxxA}{\ensuremath{1.11}}                       
\newcommand{\hatcurISOrlongxxxxxA}{\ensuremath{1.107\pm0.069}}               
\newcommand{\hatcurISOrhoxxxxxA}{\ensuremath{1.03\pm0.20}}                   
\newcommand{\hatcurISOrholongxxxxxA}{\ensuremath{1.03\pm0.20}}               
\newcommand{\hatcurISOloggxxxxxA}{\ensuremath{4.347\pm0.053}}                
\newcommand{\hatcurISOlumxxxxxA}{\ensuremath{1.17\pm0.17}}                   
\newcommand{\hatcurISOlumshortxxxxxA}{\ensuremath{1.17}}                     
\newcommand{\hatcurISOmvxxxxxA}{\ensuremath{4.67\pm0.16}}                    
\newcommand{\hatcurISOvixxxxxA}{\ensuremath{0.715\pm0.023}}                  
\newcommand{\hatcurISOagexxxxxA}{\ensuremath{7.5\pm1.9}}                     
\newcommand{\hatcurISOsigmaxxxxxA}{\ensuremath{0.00070\pm0.00011}}           
\newcommand{\hatcurISOMJxxxxxA}{\ensuremath{3.50\pm0.14}}                    
\newcommand{\hatcurISOMHxxxxxA}{\ensuremath{3.15\pm0.14}}                    
\newcommand{\hatcurISOMKxxxxxA}{\ensuremath{3.10\pm0.14}}                    
\newcommand{\hatcurISOJKxxxxxA}{\ensuremath{0.410\pm0.020}}                  
\newcommand{\hatcurISOspecxxxxxA}{G}                                         
\newcommand{\hatcurRVKxxxxxA}{\ensuremath{76.8\pm5.0}}                       
\newcommand{\hatcurRVrkxxxxxA}{\ensuremath{0\pm0}}                           
\newcommand{\hatcurRVrhxxxxxA}{\ensuremath{0\pm0}}                           
\newcommand{\hatcurRVkxxxxxA}{\ensuremath{0\pm0}}                            
\newcommand{\hatcurRVhxxxxxA}{\ensuremath{0\pm0}}                            
\newcommand{\hatcurRVtronexxxxxA}{\ensuremath{0\pm0}}                        
\newcommand{\hatcurRVtrtwoxxxxxA}{\ensuremath{0\pm0}}                        
\newcommand{\hatcurRVgammaxxxxxA}{\ensuremath{31663.2\pm3.6}}                
\newcommand{\hatcurRVjitterxxxxxA}{\ensuremath{0.000\pm0.051}}               
\newcommand{\hatcurRVjittertwosiglimxxxxxA}{\ensuremath{<0.12}}               
\newcommand{\hatcurRVfitrmsxxxxxA}{\ensuremath{.1fym}}                       %
\newcommand{\hatcurRVeccenxxxxxA}{\ensuremath{0\pm0}}                        
\newcommand{\hatcurRVeccentwosiglimxxxxxA}{\ensuremath{<0.000}}              
\newcommand{\hatcurRVomegaxxxxxA}{\ensuremath{0\pm0}}                        
\newcommand{\hatcurPPixxxxxA}{\ensuremath{86.93\pm0.71}}                     
\newcommand{\hatcurPPgxxxxxA}{\ensuremath{9.5\pm1.7}}                        
\newcommand{\hatcurPPloggxxxxxA}{\ensuremath{2.976\pm0.075}}                 
\newcommand{\hatcurPParxxxxxA}{\ensuremath{10.03\pm0.62}}                    
\newcommand{\hatcurPParelxxxxxA}{\ensuremath{0.05163\pm0.00060}}             
\newcommand{\hatcurPPrhoxxxxxA}{\ensuremath{0.38\pm0.10}}                    
\newcommand{\hatcurPPmxxxxxA}{\ensuremath{0.613\pm0.042}}                    
\newcommand{\hatcurPPmshortxxxxxA}{\ensuremath{0.61}}                        
\newcommand{\hatcurPPmlongxxxxxA}{\ensuremath{0.613\pm0.042}}                
\newcommand{\hatcurPPmexxxxxA}{\ensuremath{195\pm13}}                        
\newcommand{\hatcurPPmeshortxxxxxA}{\ensuremath{194.7}}                      
\newcommand{\hatcurPPmelongxxxxxA}{\ensuremath{195\pm13}}                    
\newcommand{\hatcurPPrxxxxxA}{\ensuremath{1.26\pm0.10}}                      
\newcommand{\hatcurPPrshortxxxxxA}{\ensuremath{1.26}}                        
\newcommand{\hatcurPPrlongxxxxxA}{\ensuremath{1.26\pm0.10}}                  
\newcommand{\hatcurPPrexxxxxA}{\ensuremath{14.1\pm1.1}}                      
\newcommand{\hatcurPPreshortxxxxxA}{\ensuremath{14.1}}                       
\newcommand{\hatcurPPrelongxxxxxA}{\ensuremath{14.1\pm1.1}}                  
\newcommand{\hatcurPPmrcorrxxxxxA}{\ensuremath{0.01}}                        
\newcommand{\hatcurPPteffxxxxxA}{\ensuremath{1277\pm42}}                     
\newcommand{\hatcurPPthetaxxxxxA}{\ensuremath{0.0500\pm0.0054}}              
\newcommand{\hatcurPPfluxperixxxxxA}{\ensuremath{6.00\pm0.79}}               
\newcommand{\hatcurPPfluxperidimxxxxxA}{\ensuremath{8}}                      
\newcommand{\hatcurPPfluxapxxxxxA}{\ensuremath{6.00\pm0.79}}                 
\newcommand{\hatcurPPfluxapdimxxxxxA}{\ensuremath{8}}                        
\newcommand{\hatcurPPfluxavgxxxxxA}{\ensuremath{6.00\pm0.79}}                
\newcommand{\hatcurPPfluxavgdimxxxxxA}{\ensuremath{8}}                       
\newcommand{\hatcurPPfluxavglogxxxxxA}{\ensuremath{8.778\pm0.057}}           
\newcommand{\hatcurXsecphasexxxxxA}{\ensuremath{0\pm0}}                      
\newcommand{\hatcurXsecondaryxxxxxA}{\ensuremath{2456872.51805\pm0.00051}}   
\newcommand{\hatcurXsecdurxxxxxA}{\ensuremath{0.1335\pm0.0025}}              
\newcommand{\hatcurXsecingdurxxxxxA}{\ensuremath{0.0190\pm0.0027}}           
\newcommand{\hatcurPPphiconjxxxxxA}{\ensuremath{0\pm0}}                      
\newcommand{\hatcurPPperixxxxxA}{\ensuremath{2456869.29406\pm0.00051}}       
\newcommand{\hatcurPPaequivxxxxxA}{\ensuremath{0.0477\pm0.0032}}             
\newcommand{\hatcurPPtcircxxxxxA}{\ensuremath{320_{-110}^{+160}}}            
\newcommand{\hatcurPPtinfallxxxxxA}{\ensuremath{6700_{-1800}^{+2700}}}       
\newcommand{\hatcurXdistxxxxxA}{\ensuremath{471\pm30}}                       
\newcommand{\hatcurXAvxxxxxA}{\ensuremath{0.083\pm0.061}}                    
\newcommand{\hatcurXdistredxxxxxA}{\ensuremath{466\pm30}}                    
\newcommand{\hatcurXEBVxxxxxA}{\ensuremath{0.027\pm0.020}}                   
\newcommand{\hatcurXmvisoredxxxxxA}{\ensuremath{13.096\pm0.029}}             
\newcommand{\hatcurXmiisoredxxxxxA}{\ensuremath{12.336\pm0.016}}             
\newcommand{\hatcurXmjisoredxxxxxA}{\ensuremath{11.868\pm0.014}}             
\newcommand{\hatcurXmhisoredxxxxxA}{\ensuremath{11.508\pm0.015}}             
\newcommand{\hatcurXmkisoredxxxxxA}{\ensuremath{11.445\pm0.016}}             
\newcommand{\hatcurXviisoredxxxxxA}{\ensuremath{0.759\pm0.020}}              
\newcommand{\hatcurXvkisoredxxxxxA}{\ensuremath{1.651\pm0.035}}              
\newcommand{\hatcurXjhisoredxxxxxA}{\ensuremath{0.3590\pm0.0097}}            
\newcommand{\hatcurXjkisoredxxxxxA}{\ensuremath{0.422\pm0.010}}              
\newcommand{\hatcurCCpmraxxxxxA}{\ensuremath{-20.5\pm1.0}}                   
\newcommand{\hatcurCCpmdecxxxxxA}{\ensuremath{-11.9\pm1.1}}                  
\newcommand{\hatcurCCpmxxxxxA}{\ensuremath{23.7\pm1.5}}                      

\newcommand{\hatcurhtrxxxxxB}{HATS606-007}                                   
\newcommand{\hatcurfieldxxxxxB}{\ensuremath{string}}                         
\newcommand{\hatcurCCraxxxxxB}{\ensuremath{09^{\mathrm h}39^{\mathrm m}42.44{\mathrm s}}}                                  
\newcommand{\hatcurCCdecxxxxxB}{\ensuremath{-28{\arcdeg}35{\arcmin}08.1{\arcsec}}}                                 
\newcommand{\hatcurCCmagxxxxxB}{12.955}                                      
\newcommand{\hatcurCCtwomassxxxxxB}{2MASS~09394244-2835081}                  
\newcommand{\hatcurCCgscxxxxxB}{GSC~6614-01083}                              
\newcommand{\hatcurCCtassmvxxxxxB}{\ensuremath{12.955\pm0.030}}              
\newcommand{\hatcurCCtassmvshortxxxxxB}{\ensuremath{13.0}}                   
\newcommand{\hatcurCCtassmBxxxxxB}{\ensuremath{13.553\pm0.030}}              
\newcommand{\hatcurCCtassmBshortxxxxxB}{\ensuremath{13.6}}                   
\newcommand{\hatcurCCtassmIxxxxxB}{\ensuremath{100\pm1000}}                  
\newcommand{\hatcurCCtassmIshortxxxxxB}{\ensuremath{100.0}}                  
\newcommand{\hatcurCCtassmgxxxxxB}{\ensuremath{13.229\pm0.010}}              
\newcommand{\hatcurCCtassmgshortxxxxxB}{\ensuremath{13.2}}                   
\newcommand{\hatcurCCtassmrxxxxxB}{\ensuremath{12.822\pm0.010}}              
\newcommand{\hatcurCCtassmrshortxxxxxB}{\ensuremath{12.8}}                   
\newcommand{\hatcurCCtassmixxxxxB}{\ensuremath{12.695\pm0.030}}              
\newcommand{\hatcurCCtassmishortxxxxxB}{\ensuremath{12.7}}                   
\newcommand{\hatcurCCtwomassJmagxxxxxB}{\ensuremath{11.839\pm0.024}}         
\newcommand{\hatcurCCtwomassHmagxxxxxB}{\ensuremath{11.510\pm0.024}}         
\newcommand{\hatcurCCtwomassKmagxxxxxB}{\ensuremath{11.435\pm0.021}}         
\newcommand{\hatcurCCcitJmagxxxxxB}{\ensuremath{11.854\pm0.024}}             
\newcommand{\hatcurCCcitHmagxxxxxB}{\ensuremath{11.505\pm0.025}}             
\newcommand{\hatcurCCcitKmagxxxxxB}{\ensuremath{11.459\pm0.021}}             
\newcommand{\hatcurCCbbJmagxxxxxB}{\ensuremath{11.906\pm0.026}}              
\newcommand{\hatcurCCbbHmagxxxxxB}{\ensuremath{11.526\pm0.025}}              
\newcommand{\hatcurCCbbKmagxxxxxB}{\ensuremath{11.479\pm0.021}}              
\newcommand{\hatcurCCesoJmagxxxxxB}{\ensuremath{11.909\pm0.027}}             
\newcommand{\hatcurCCesoHmagxxxxxB}{\ensuremath{11.521\pm0.028}}             
\newcommand{\hatcurCCesoKmagxxxxxB}{\ensuremath{11.478\pm0.022}}             
\newcommand{\hatcurCCesoJHmagxxxxxB}{\ensuremath{0.388\pm0.011}}             
\newcommand{\hatcurCCesoJKmagxxxxxB}{\ensuremath{0.431\pm0.034}}             
\newcommand{\hatcurCCesoHKmagxxxxxB}{\ensuremath{0.043\pm0.035}}             
\newcommand{\hatcurLCdipxxxxxB}{\ensuremath{8.3}}                            
\newcommand{\hatcurLCrprstarxxxxxB}{\ensuremath{0.0879\pm0.0055}}            
\newcommand{\hatcurLCbsqxxxxxB}{\ensuremath{0.109_{-0.084}^{+0.098}}}        
\newcommand{\hatcurLCimpxxxxxB}{\ensuremath{0.33_{-0.17}^{+0.12}}}           
\newcommand{\hatcurLCzetaxxxxxB}{\ensuremath{10.14\pm0.13}}                  
\newcommand{\hatcurLCdurxxxxxB}{\ensuremath{0.2173\pm0.0041}}                
\newcommand{\hatcurLCdurshortxxxxxB}{\ensuremath{0.2173}}                    
\newcommand{\hatcurLCdurhrxxxxxB}{\ensuremath{5.215\pm0.099}}                
\newcommand{\hatcurLCdurhrshortxxxxxB}{\ensuremath{5.215}}                   
\newcommand{\hatcurLCqxxxxxB}{\ensuremath{0.0658\pm0.0013}}                  
\newcommand{\hatcurLCqshortxxxxxB}{\ensuremath{0.066}}                       
\newcommand{\hatcurLCingdurxxxxxB}{\ensuremath{0.0196\pm0.0035}}             
\newcommand{\hatcurLCPxxxxxB}{\ensuremath{3.3023881\pm0.0000076}}            
\newcommand{\hatcurLCPprecxxxxxB}{\ensuremath{3.3023881}}                    
\newcommand{\hatcurLCPshortxxxxxB}{\ensuremath{3.3024}}                      
\newcommand{\hatcurLCTxxxxxB}{\ensuremath{2456867.4232\pm0.0012}}            
\newcommand{\hatcurLCTAxxxxxB}{\ensuremath{2455972.4757\pm0.0024}}           
\newcommand{\hatcurLCTBxxxxxB}{\ensuremath{2457177.8478\pm0.0014}}           
\newcommand{\hatcurLChatnetmxxxxxB}{\ensuremath{12.975610\pm0.000081}}       
\newcommand{\hatcurLCiblendxxxxxB}{\ensuremath{0.775\pm0.077}}               
\newcommand{\hatcurLCrhoxxxxxB}{\ensuremath{0.219\pm0.033}}                    
\newcommand{\hatcurSMEiteffxxxxxB}{\ensuremath{6062\pm98}}                   
\newcommand{\hatcurSMEizfehxxxxxB}{\ensuremath{-0.020\pm0.050}}              
\newcommand{\hatcurSMEizfehshortxxxxxB}{\ensuremath{-0.02}}                  
\newcommand{\hatcurSMEiloggxxxxxB}{\ensuremath{3.97\pm0.19}}                 
\newcommand{\hatcurSMEivsinxxxxxB}{\ensuremath{7.43\pm0.50}}                 
\newcommand{\hatcurSMEivmacxxxxxB}{\ensuremath{0.0}}                         
\newcommand{\hatcurSMEivmicxxxxxB}{\ensuremath{0.0}}                         
\newcommand{\hatcurSMEiiteffxxxxxB}{\ensuremath{6071\pm81}}                  
\newcommand{\hatcurSMEiizfehxxxxxB}{\ensuremath{-0.020\pm0.050}}             
\newcommand{\hatcurSMEiizfehshortxxxxxB}{\ensuremath{-0.02}}                 
\newcommand{\hatcurSMEiiloggxxxxxB}{\ensuremath{3.951\pm0.042}}              
\newcommand{\hatcurSMEiivsinxxxxxB}{\ensuremath{7.48\pm0.50}}                
\newcommand{\hatcurLBizxxxxxB}{\ensuremath{0.1616}}                          
\newcommand{\hatcurLBiizxxxxxB}{\ensuremath{0.3511}}                         
\newcommand{\hatcurLBiixxxxxB}{\ensuremath{0.2145}}                          
\newcommand{\hatcurLBiiixxxxxB}{\ensuremath{0.3580}}                         
\newcommand{\hatcurLBiIxxxxxB}{\ensuremath{0.1954}}                          
\newcommand{\hatcurLBiiIxxxxxB}{\ensuremath{0.3572}}                         
\newcommand{\hatcurLBigxxxxxB}{\ensuremath{0.4752}}                          
\newcommand{\hatcurLBiigxxxxxB}{\ensuremath{0.2907}}                         
\newcommand{\hatcurLBirxxxxxB}{\ensuremath{0.2947}}                          
\newcommand{\hatcurLBiirxxxxxB}{\ensuremath{0.3611}}                         
\newcommand{\hatcurLBiRxxxxxB}{\ensuremath{0.2721}}                          
\newcommand{\hatcurLBiiRxxxxxB}{\ensuremath{0.3616}}                         
\newcommand{\hatcurLBikepxxxxxB}{\ensuremath{0.1000}}                        
\newcommand{\hatcurLBiikepxxxxxB}{\ensuremath{0.1000}}                       
\newcommand{\hatcurISOmxxxxxB}{\ensuremath{1.299_{-0.056}^{+0.113}}}         
\newcommand{\hatcurISOmshortxxxxxB}{\ensuremath{1.30}}                       
\newcommand{\hatcurISOmlongxxxxxB}{\ensuremath{1.299_{-0.056}^{+0.113}}}     
\newcommand{\hatcurISOrxxxxxB}{\ensuremath{2.04_{-0.11}^{+0.15}}}            
\newcommand{\hatcurISOrshortxxxxxB}{\ensuremath{2.04}}                       
\newcommand{\hatcurISOrlongxxxxxB}{\ensuremath{2.04_{-0.11}^{+0.15}}}        
\newcommand{\hatcurISOrhoxxxxxB}{\ensuremath{0.218\pm0.034}}                 
\newcommand{\hatcurISOrholongxxxxxB}{\ensuremath{0.218\pm0.034}}             
\newcommand{\hatcurISOloggxxxxxB}{\ensuremath{3.936\pm0.046}}                
\newcommand{\hatcurISOlumxxxxxB}{\ensuremath{5.06_{-0.64}^{+0.90}}}          
\newcommand{\hatcurISOlumshortxxxxxB}{\ensuremath{5.06}}                     
\newcommand{\hatcurISOmvxxxxxB}{\ensuremath{3.03\pm0.17}}                    
\newcommand{\hatcurISOvixxxxxB}{\ensuremath{0.608\pm0.023}}                  
\newcommand{\hatcurISOagexxxxxB}{\ensuremath{4.04_{-0.94}^{+0.62}}}          
\newcommand{\hatcurISOsigmaxxxxxB}{\ensuremath{0.000200\pm0.000029}}         
\newcommand{\hatcurISOMJxxxxxB}{\ensuremath{2.03\pm0.16}}                    
\newcommand{\hatcurISOMHxxxxxB}{\ensuremath{1.74\pm0.15}}                    
\newcommand{\hatcurISOMKxxxxxB}{\ensuremath{1.68\pm0.15}}                    
\newcommand{\hatcurISOJKxxxxxB}{\ensuremath{0.340\pm0.010}}                  
\newcommand{\hatcurISOspecxxxxxB}{F}                                         
\newcommand{\hatcurRVKxxxxxB}{\ensuremath{73.3\pm8.0}}                       
\newcommand{\hatcurRVrkxxxxxB}{\ensuremath{0\pm0}}                           
\newcommand{\hatcurRVrhxxxxxB}{\ensuremath{0\pm0}}                           
\newcommand{\hatcurRVkxxxxxB}{\ensuremath{0\pm0}}                            
\newcommand{\hatcurRVhxxxxxB}{\ensuremath{0\pm0}}                            
\newcommand{\hatcurRVtronexxxxxB}{\ensuremath{0\pm0}}                        
\newcommand{\hatcurRVtrtwoxxxxxB}{\ensuremath{0\pm0}}                        
\newcommand{\hatcurRVgammaAxxxxxB}{\ensuremath{-12515.9\pm6.7}}              
\newcommand{\hatcurRVjitterAxxxxxB}{\ensuremath{1\pm11}}                     
\newcommand{\hatcurRVjittertwosiglimAxxxxxB}{\ensuremath{<28}}               
\newcommand{\hatcurRVfitrmsAxxxxxB}{\ensuremath{0.0}}                        
\newcommand{\hatcurRVgammaBxxxxxB}{\ensuremath{-12489\pm20}}                 
\newcommand{\hatcurRVjitterBxxxxxB}{\ensuremath{0.1\pm5.0}}                  
\newcommand{\hatcurRVjittertwosiglimBxxxxxB}{\ensuremath{<8.1}}               
\newcommand{\hatcurRVfitrmsBxxxxxB}{\ensuremath{0.0}}                        
\newcommand{\hatcurRVgammaCxxxxxB}{\ensuremath{-12560.8\pm9.4}}              
\newcommand{\hatcurRVjitterCxxxxxB}{\ensuremath{0.4\pm4.3}}                  
\newcommand{\hatcurRVjittertwosiglimCxxxxxB}{\ensuremath{<9.1}}               
\newcommand{\hatcurRVfitrmsCxxxxxB}{\ensuremath{0.0}}                        
\newcommand{\hatcurRVeccenxxxxxB}{\ensuremath{0\pm0}}                        
\newcommand{\hatcurRVeccentwosiglimxxxxxB}{\ensuremath{<0.000}}              
\newcommand{\hatcurRVomegaxxxxxB}{\ensuremath{0\pm0}}                        
\newcommand{\hatcurPPixxxxxB}{\ensuremath{86.2\pm1.9}}                       
\newcommand{\hatcurPPgxxxxxB}{\ensuremath{5.3\pm1.0}}                        
\newcommand{\hatcurPPloggxxxxxB}{\ensuremath{2.724_{-0.103}^{+0.074}}}       
\newcommand{\hatcurPParxxxxxB}{\ensuremath{5.01\pm0.27}}                     
\newcommand{\hatcurPParelxxxxxB}{\ensuremath{0.04735_{-0.00068}^{+0.00133}}} 
\newcommand{\hatcurPPrhoxxxxxB}{\ensuremath{0.153\pm0.042}}                  
\newcommand{\hatcurPPmxxxxxB}{\ensuremath{0.650\pm0.076}}                    
\newcommand{\hatcurPPmshortxxxxxB}{\ensuremath{0.65}}                        
\newcommand{\hatcurPPmlongxxxxxB}{\ensuremath{0.650\pm0.076}}                
\newcommand{\hatcurPPmexxxxxB}{\ensuremath{207\pm24}}                        
\newcommand{\hatcurPPmeshortxxxxxB}{\ensuremath{206.6}}                      
\newcommand{\hatcurPPmelongxxxxxB}{\ensuremath{207\pm24}}                    
\newcommand{\hatcurPPrxxxxxB}{\ensuremath{1.75\pm0.21}}                      
\newcommand{\hatcurPPrshortxxxxxB}{\ensuremath{1.75}}                        
\newcommand{\hatcurPPrlongxxxxxB}{\ensuremath{1.75\pm0.21}}                  
\newcommand{\hatcurPPrexxxxxB}{\ensuremath{19.6\pm2.4}}                      
\newcommand{\hatcurPPreshortxxxxxB}{\ensuremath{19.6}}                       
\newcommand{\hatcurPPrelongxxxxxB}{\ensuremath{19.6\pm2.4}}                  
\newcommand{\hatcurPPmrcorrxxxxxB}{\ensuremath{0.27}}                        
\newcommand{\hatcurPPteffxxxxxB}{\ensuremath{1918\pm61}}                     
\newcommand{\hatcurPPthetaxxxxxB}{\ensuremath{0.0264\pm0.0038}}              
\newcommand{\hatcurPPfluxperixxxxxB}{\ensuremath{3.06_{-0.30}^{+0.43}}}      
\newcommand{\hatcurPPfluxperidimxxxxxB}{\ensuremath{9}}                      
\newcommand{\hatcurPPfluxapxxxxxB}{\ensuremath{3.06_{-0.30}^{+0.43}}}        
\newcommand{\hatcurPPfluxapdimxxxxxB}{\ensuremath{9}}                        
\newcommand{\hatcurPPfluxavgxxxxxB}{\ensuremath{3.06_{-0.30}^{+0.43}}}       
\newcommand{\hatcurPPfluxavgdimxxxxxB}{\ensuremath{9}}                       
\newcommand{\hatcurPPfluxavglogxxxxxB}{\ensuremath{9.485\pm0.054}}           
\newcommand{\hatcurXsecphasexxxxxB}{\ensuremath{0\pm0}}                      
\newcommand{\hatcurXsecondaryxxxxxB}{\ensuremath{2456869.0744\pm0.0012}}     
\newcommand{\hatcurXsecdurxxxxxB}{\ensuremath{0.2173\pm0.0041}}              
\newcommand{\hatcurXsecingdurxxxxxB}{\ensuremath{0.0196\pm0.0035}}           
\newcommand{\hatcurPPphiconjxxxxxB}{\ensuremath{0\pm0}}                      
\newcommand{\hatcurPPperixxxxxB}{\ensuremath{2456866.5976\pm0.0012}}         
\newcommand{\hatcurPPaequivxxxxxB}{\ensuremath{0.0211\pm0.0013}}             
\newcommand{\hatcurPPtcircxxxxxB}{\ensuremath{25\pm11}}                      
\newcommand{\hatcurPPtinfallxxxxxB}{\ensuremath{199\pm61}}                   
\newcommand{\hatcurXdistxxxxxB}{\ensuremath{910_{-51}^{+70}}}                
\newcommand{\hatcurXAvxxxxxB}{\ensuremath{0.140\pm0.070}}                    
\newcommand{\hatcurXdistredxxxxxB}{\ensuremath{907_{-49}^{+69}}}             
\newcommand{\hatcurXEBVxxxxxB}{\ensuremath{0.045\pm0.023}}                   
\newcommand{\hatcurXmvisoredxxxxxB}{\ensuremath{12.965\pm0.029}}             
\newcommand{\hatcurXmiisoredxxxxxB}{\ensuremath{12.282\pm0.016}}             
\newcommand{\hatcurXmjisoredxxxxxB}{\ensuremath{11.856\pm0.014}}             
\newcommand{\hatcurXmhisoredxxxxxB}{\ensuremath{11.552\pm0.015}}             
\newcommand{\hatcurXmkisoredxxxxxB}{\ensuremath{11.489\pm0.016}}             
\newcommand{\hatcurXviisoredxxxxxB}{\ensuremath{0.682\pm0.022}}              
\newcommand{\hatcurXvkisoredxxxxxB}{\ensuremath{1.475\pm0.035}}              
\newcommand{\hatcurXjhisoredxxxxxB}{\ensuremath{0.3040\pm0.0086}}            
\newcommand{\hatcurXjkisoredxxxxxB}{\ensuremath{0.3670\pm0.0083}}            
\newcommand{\hatcurCCpmraxxxxxB}{\ensuremath{-1.3\pm1.4}}                    
\newcommand{\hatcurCCpmdecxxxxxB}{\ensuremath{-6.1\pm1.3}}                   
\newcommand{\hatcurCCpmxxxxxB}{\ensuremath{6.2\pm1.9}}                       

\newcommand{\hatcurhtrxxxxxxC}{HATS700-004}                            
\newcommand{\hatcurfieldxxxxxxC}{\ensuremath{string}}                  
\newcommand{\hatcurCCraxxxxxxC}{\ensuremath{12^{\mathrm h}54^{\mathrm m}12.60{\mathrm s}}}                           
\newcommand{\hatcurCCdecxxxxxxC}{\ensuremath{-46{\arcdeg}35{\arcmin}15.8{\arcsec}}}                          
\newcommand{\hatcurCCmagxxxxxxC}{12.766}                               
\newcommand{\hatcurCCtwomassxxxxxxC}{2MASS~12541261-4635157}           
\newcommand{\hatcurCCgscxxxxxxC}{GSC~8245-02236}                       
\newcommand{\hatcurCCtassmvxxxxxxC}{\ensuremath{12.766\pm0.040}}       
\newcommand{\hatcurCCtassmvshortxxxxxxC}{\ensuremath{12.8}}            
\newcommand{\hatcurCCtassmBxxxxxxC}{\ensuremath{13.239\pm0.050}}       
\newcommand{\hatcurCCtassmBshortxxxxxxC}{\ensuremath{13.2}}            
\newcommand{\hatcurCCtassmIxxxxxxC}{\ensuremath{100\pm100}}            
\newcommand{\hatcurCCtassmIshortxxxxxxC}{\ensuremath{100.0}}           
\newcommand{\hatcurCCtassmgxxxxxxC}{\ensuremath{12.927\pm0.040}}       
\newcommand{\hatcurCCtassmgshortxxxxxxC}{\ensuremath{12.9}}            
\newcommand{\hatcurCCtassmrxxxxxxC}{\ensuremath{12.665\pm0.040}}       
\newcommand{\hatcurCCtassmrshortxxxxxxC}{\ensuremath{12.7}}            
\newcommand{\hatcurCCtassmixxxxxxC}{\ensuremath{12.515\pm0.080}}       
\newcommand{\hatcurCCtassmishortxxxxxxC}{\ensuremath{12.5}}            
\newcommand{\hatcurCCtwomassJmagxxxxxxC}{\ensuremath{11.831\pm0.022}}  
\newcommand{\hatcurCCtwomassHmagxxxxxxC}{\ensuremath{11.651\pm0.023}}  
\newcommand{\hatcurCCtwomassKmagxxxxxxC}{\ensuremath{11.550\pm0.023}}  
\newcommand{\hatcurCCcitJmagxxxxxxC}{\ensuremath{11.853\pm0.022}}      
\newcommand{\hatcurCCcitHmagxxxxxxC}{\ensuremath{11.645\pm0.024}}      
\newcommand{\hatcurCCcitKmagxxxxxxC}{\ensuremath{11.574\pm0.023}}      
\newcommand{\hatcurCCbbJmagxxxxxxC}{\ensuremath{11.895\pm0.024}}       
\newcommand{\hatcurCCbbHmagxxxxxxC}{\ensuremath{11.667\pm0.024}}       
\newcommand{\hatcurCCbbKmagxxxxxxC}{\ensuremath{11.594\pm0.023}}       
\newcommand{\hatcurCCesoJmagxxxxxxC}{\ensuremath{11.895\pm0.025}}      
\newcommand{\hatcurCCesoHmagxxxxxxC}{\ensuremath{11.665\pm0.026}}      
\newcommand{\hatcurCCesoKmagxxxxxxC}{\ensuremath{11.594\pm0.024}}      
\newcommand{\hatcurCCesoJHmagxxxxxxC}{\ensuremath{0.231\pm0.035}}      
\newcommand{\hatcurCCesoJKmagxxxxxxC}{\ensuremath{0.302\pm0.034}}      
\newcommand{\hatcurCCesoHKmagxxxxxxC}{\ensuremath{0.070\pm0.012}}      
\newcommand{\hatcurLCdipxxxxxxC}{\ensuremath{8.2}}                     
\newcommand{\hatcurLCrprstarxxxxxxC}{\ensuremath{0.0895\pm0.0043}}     
\newcommand{\hatcurLCbsqxxxxxxC}{\ensuremath{0.13_{-0.10}^{+0.13}}}    
\newcommand{\hatcurLCimpxxxxxxC}{\ensuremath{0.36_{-0.19}^{+0.15}}}    
\newcommand{\hatcurLCzetaxxxxxxC}{\ensuremath{10.98_{-0.14}^{+0.11}}}  
\newcommand{\hatcurLCdurxxxxxxC}{\ensuremath{0.2013\pm0.0033}}         
\newcommand{\hatcurLCdurshortxxxxxxC}{\ensuremath{0.2013}}             
\newcommand{\hatcurLCdurhrxxxxxxC}{\ensuremath{4.831\pm0.079}}         
\newcommand{\hatcurLCdurhrshortxxxxxxC}{\ensuremath{4.831}}            
\newcommand{\hatcurLCqxxxxxxC}{\ensuremath{0.04340\pm0.00071}}         
\newcommand{\hatcurLCqshortxxxxxxC}{\ensuremath{0.043}}                
\newcommand{\hatcurLCingdurxxxxxxC}{\ensuremath{0.0186\pm0.0030}}      
\newcommand{\hatcurLCPxxxxxxC}{\ensuremath{4.637038\pm0.000014}}       
\newcommand{\hatcurLCPprecxxxxxxC}{\ensuremath{4.6370382}}             
\newcommand{\hatcurLCPshortxxxxxxC}{\ensuremath{4.6370}}               
\newcommand{\hatcurLCTxxxxxxC}{\ensuremath{2457029.3374\pm0.0011}}     
\newcommand{\hatcurLCTAxxxxxxC}{\ensuremath{2455679.9595\pm0.0042}}    
\newcommand{\hatcurLCTBxxxxxxC}{\ensuremath{2457122.0782\pm0.0011}}    
\newcommand{\hatcurLChatnetmxxxxxxC}{\ensuremath{12.698730\pm0.000073}} 
\newcommand{\hatcurLCiblendxxxxxxC}{\ensuremath{0.778\pm0.086}}        
\newcommand{\hatcurLCrhoxxxxxxC}{\ensuremath{0.380\pm0.063}}           
\newcommand{\hatcurSMEiteffxxxxxxC}{\ensuremath{6438\pm64}}            
\newcommand{\hatcurSMEizfehxxxxxxC}{\ensuremath{0.090\pm0.040}}        
\newcommand{\hatcurSMEizfehshortxxxxxxC}{\ensuremath{0.09}}            
\newcommand{\hatcurSMEiloggxxxxxxC}{\ensuremath{4.07\pm0.10}}          
\newcommand{\hatcurSMEivsinxxxxxxC}{\ensuremath{9.32\pm0.50}}          
\newcommand{\hatcurSMEivmacxxxxxxC}{\ensuremath{0.0}}                  
\newcommand{\hatcurSMEivmicxxxxxxC}{\ensuremath{0.0}}                  
\newcommand{\hatcurLBizxxxxxxC}{\ensuremath{0.1252}}                   
\newcommand{\hatcurLBiizxxxxxxC}{\ensuremath{0.3688}}                  
\newcommand{\hatcurLBiixxxxxxC}{\ensuremath{0.1754}}                   
\newcommand{\hatcurLBiiixxxxxxC}{\ensuremath{0.3788}}                  
\newcommand{\hatcurLBiIxxxxxxC}{\ensuremath{0.1568}}                   
\newcommand{\hatcurLBiiIxxxxxxC}{\ensuremath{0.3769}}                  
\newcommand{\hatcurLBigxxxxxxC}{\ensuremath{0.4175}}                   
\newcommand{\hatcurLBiigxxxxxxC}{\ensuremath{0.3314}}                  
\newcommand{\hatcurLBirxxxxxxC}{\ensuremath{0.2511}}                   
\newcommand{\hatcurLBiirxxxxxxC}{\ensuremath{0.3857}}                  
\newcommand{\hatcurLBiRxxxxxxC}{\ensuremath{0.2295}}                   
\newcommand{\hatcurLBiiRxxxxxxC}{\ensuremath{0.3855}}                  
\newcommand{\hatcurLBikepxxxxxxC}{\ensuremath{0.1000}}                 
\newcommand{\hatcurLBiikepxxxxxxC}{\ensuremath{0.1000}}                
\newcommand{\hatcurISOmxxxxxxC}{\ensuremath{1.415\pm0.048}}            
\newcommand{\hatcurISOmshortxxxxxxC}{\ensuremath{1.41}}                
\newcommand{\hatcurISOmlongxxxxxxC}{\ensuremath{1.415\pm0.048}}        
\newcommand{\hatcurISOrxxxxxxC}{\ensuremath{1.74_{-0.10}^{+0.17}}}     
\newcommand{\hatcurISOrshortxxxxxxC}{\ensuremath{1.74}}                
\newcommand{\hatcurISOrlongxxxxxxC}{\ensuremath{1.74_{-0.10}^{+0.17}}} 
\newcommand{\hatcurISOrhoxxxxxxC}{\ensuremath{0.379\pm0.064}}          
\newcommand{\hatcurISOrholongxxxxxxC}{\ensuremath{0.379\pm0.064}}      
\newcommand{\hatcurISOloggxxxxxxC}{\ensuremath{4.107\pm0.049}}         
\newcommand{\hatcurISOlumxxxxxxC}{\ensuremath{4.67_{-0.58}^{+0.92}}}   
\newcommand{\hatcurISOlumshortxxxxxxC}{\ensuremath{4.67}}              
\newcommand{\hatcurISOmvxxxxxxC}{\ensuremath{3.06\pm0.17}}             
\newcommand{\hatcurISOvixxxxxxC}{\ensuremath{0.513\pm0.016}}           
\newcommand{\hatcurISOagexxxxxxC}{\ensuremath{2.30\pm0.22}}            
\newcommand{\hatcurISOsigmaxxxxxxC}{\ensuremath{0.000200\pm0.000072}}  
\newcommand{\hatcurISOMJxxxxxxC}{\ensuremath{2.24\pm0.16}}             
\newcommand{\hatcurISOMHxxxxxxC}{\ensuremath{2.02\pm0.15}}             
\newcommand{\hatcurISOMKxxxxxxC}{\ensuremath{1.98\pm0.15}}             
\newcommand{\hatcurISOJKxxxxxxC}{\ensuremath{0.270\pm0.010}}           
\newcommand{\hatcurISOspecxxxxxxC}{F}                                  
\newcommand{\hatcurRVKxxxxxxC}{\ensuremath{51\pm13}}                   
\newcommand{\hatcurRVrkxxxxxxC}{\ensuremath{0\pm0}}                    
\newcommand{\hatcurRVrhxxxxxxC}{\ensuremath{0\pm0}}                    
\newcommand{\hatcurRVkxxxxxxC}{\ensuremath{0\pm0}}                     
\newcommand{\hatcurRVhxxxxxxC}{\ensuremath{0\pm0}}                     
\newcommand{\hatcurRVtronexxxxxxC}{\ensuremath{0\pm0}}                 
\newcommand{\hatcurRVtrtwoxxxxxxC}{\ensuremath{0\pm0}}                 
\newcommand{\hatcurRVgammaAxxxxxxC}{\ensuremath{-3527\pm22}}           
\newcommand{\hatcurRVjitterAxxxxxxC}{\ensuremath{72\pm17}}             
\newcommand{\hatcurRVjittertwosiglimAxxxxxxC}{\ensuremath{<109.6}}     
\newcommand{\hatcurRVfitrmsAxxxxxxC}{\ensuremath{0.0}}                 
\newcommand{\hatcurRVgammaBxxxxxxC}{\ensuremath{-3526\pm52}}           
\newcommand{\hatcurRVjitterBxxxxxxC}{\ensuremath{0\pm74}}              
\newcommand{\hatcurRVjittertwosiglimBxxxxxxC}{\ensuremath{<142.2}}     
\newcommand{\hatcurRVfitrmsBxxxxxxC}{\ensuremath{0.0}}                 
\newcommand{\hatcurRVgammaCxxxxxxC}{\ensuremath{-3582\pm12}}           
\newcommand{\hatcurRVjitterCxxxxxxC}{\ensuremath{13\pm15}}             
\newcommand{\hatcurRVjittertwosiglimCxxxxxxC}{\ensuremath{<38.0}}      
\newcommand{\hatcurRVfitrmsCxxxxxxC}{\ensuremath{0.0}}                 
\newcommand{\hatcurRVeccenxxxxxxC}{\ensuremath{0\pm0}}                 
\newcommand{\hatcurRVeccentwosiglimxxxxxxC}{\ensuremath{<0.000}}       
\newcommand{\hatcurRVomegaxxxxxxC}{\ensuremath{0\pm0}}                 
\newcommand{\hatcurPPixxxxxxC}{\ensuremath{87.3\pm1.3}}                
\newcommand{\hatcurPPgxxxxxxC}{\ensuremath{5.6\pm1.8}}                 
\newcommand{\hatcurPPloggxxxxxxC}{\ensuremath{2.75\pm0.15}}            
\newcommand{\hatcurPParxxxxxxC}{\ensuremath{7.55_{-0.59}^{+0.42}}}     
\newcommand{\hatcurPParelxxxxxxC}{\ensuremath{0.06110\pm0.00068}}      
\newcommand{\hatcurPPrhoxxxxxxC}{\ensuremath{0.180_{-0.057}^{+0.083}}} 
\newcommand{\hatcurPPmxxxxxxC}{\ensuremath{0.53\pm0.13}}               
\newcommand{\hatcurPPmshortxxxxxxC}{\ensuremath{0.53}}                 
\newcommand{\hatcurPPmlongxxxxxxC}{\ensuremath{0.53\pm0.13}}           
\newcommand{\hatcurPPmexxxxxxC}{\ensuremath{168\pm42}}                 
\newcommand{\hatcurPPmeshortxxxxxxC}{\ensuremath{168.2}}               
\newcommand{\hatcurPPmelongxxxxxxC}{\ensuremath{168\pm42}}             
\newcommand{\hatcurPPrxxxxxxC}{\ensuremath{1.50_{-0.11}^{+0.20}}}      
\newcommand{\hatcurPPrshortxxxxxxC}{\ensuremath{1.50}}                 
\newcommand{\hatcurPPrlongxxxxxxC}{\ensuremath{1.50_{-0.11}^{+0.20}}}  
\newcommand{\hatcurPPrexxxxxxC}{\ensuremath{16.8_{-1.2}^{+2.2}}}       
\newcommand{\hatcurPPreshortxxxxxxC}{\ensuremath{16.8}}                
\newcommand{\hatcurPPrelongxxxxxxC}{\ensuremath{16.8_{-1.2}^{+2.2}}}   
\newcommand{\hatcurPPmrcorrxxxxxxC}{\ensuremath{-0.00}}                
\newcommand{\hatcurPPteffxxxxxxC}{\ensuremath{1659_{-46}^{+66}}}       
\newcommand{\hatcurPPthetaxxxxxxC}{\ensuremath{0.0292\pm0.0081}}       
\newcommand{\hatcurPPfluxperixxxxxxC}{\ensuremath{1.71_{-0.18}^{+0.29}}} 
\newcommand{\hatcurPPfluxperidimxxxxxxC}{\ensuremath{9}}               
\newcommand{\hatcurPPfluxapxxxxxxC}{\ensuremath{1.71_{-0.18}^{+0.29}}} 
\newcommand{\hatcurPPfluxapdimxxxxxxC}{\ensuremath{9}}                 
\newcommand{\hatcurPPfluxavgxxxxxxC}{\ensuremath{1.71_{-0.18}^{+0.29}}} 
\newcommand{\hatcurPPfluxavgdimxxxxxxC}{\ensuremath{9}}                
\newcommand{\hatcurPPfluxavglogxxxxxxC}{\ensuremath{9.232_{-0.050}^{+0.067}}} 
\newcommand{\hatcurXsecphasexxxxxxC}{\ensuremath{0\pm0}}               
\newcommand{\hatcurXsecondaryxxxxxxC}{\ensuremath{2457031.6559\pm0.0011}} 
\newcommand{\hatcurXsecdurxxxxxxC}{\ensuremath{0.2013\pm0.0033}}       
\newcommand{\hatcurXsecingdurxxxxxxC}{\ensuremath{0.0186\pm0.0030}}    
\newcommand{\hatcurPPphiconjxxxxxxC}{\ensuremath{0\pm0}}               
\newcommand{\hatcurPPperixxxxxxC}{\ensuremath{2457028.1782\pm0.0011}}  
\newcommand{\hatcurPPaequivxxxxxxC}{\ensuremath{0.0283_{-0.0021}^{+0.0016}}} 
\newcommand{\hatcurPPtcircxxxxxxC}{\ensuremath{190_{-82}^{+113}}}      
\newcommand{\hatcurPPtinfallxxxxxxC}{\ensuremath{2820_{-830}^{+1600}}} 
\newcommand{\hatcurXdistxxxxxxC}{\ensuremath{840_{-51}^{+79}}}         
\newcommand{\hatcurXAvxxxxxxC}{\ensuremath{0.084\pm0.066}}             
\newcommand{\hatcurXdistredxxxxxxC}{\ensuremath{840_{-51}^{+80}}}      
\newcommand{\hatcurXEBVxxxxxxC}{\ensuremath{0.027\pm0.021}}            
\newcommand{\hatcurXmvisoredxxxxxxC}{\ensuremath{12.774\pm0.036}}      
\newcommand{\hatcurXmiisoredxxxxxxC}{\ensuremath{12.215\pm0.020}}      
\newcommand{\hatcurXmjisoredxxxxxxC}{\ensuremath{11.889\pm0.014}}      
\newcommand{\hatcurXmhisoredxxxxxxC}{\ensuremath{11.655\pm0.015}}      
\newcommand{\hatcurXmkisoredxxxxxxC}{\ensuremath{11.606\pm0.016}}      
\newcommand{\hatcurXviisoredxxxxxxC}{\ensuremath{0.557\pm0.026}}       
\newcommand{\hatcurXvkisoredxxxxxxC}{\ensuremath{1.168\pm0.042}}       
\newcommand{\hatcurXjhisoredxxxxxxC}{\ensuremath{0.2340\pm0.0091}}     
\newcommand{\hatcurXjkisoredxxxxxxC}{\ensuremath{0.2830\pm0.0099}}     
\newcommand{\hatcurCCpmraxxxxxxC}{\ensuremath{-10.2\pm1.1}}            
\newcommand{\hatcurCCpmdecxxxxxxC}{\ensuremath{4.7\pm1.1}}             
\newcommand{\hatcurCCpmxxxxxxC}{\ensuremath{11.2\pm1.6}}               

\newcommand{\hatcurhtrxxxxxD}{HATS746-005}                      
\newcommand{\hatcurfieldxxxxxD}{\ensuremath{string}}            
\newcommand{\hatcurCCraxxxxxD}{\ensuremath{18^{\mathrm h}57^{\mathrm m}36.00{\mathrm s}}}                     
\newcommand{\hatcurCCdecxxxxxD}{\ensuremath{-49{\arcdeg}08{\arcmin}18.5{\arcsec}}}                    
\newcommand{\hatcurCCmagxxxxxD}{13.934}                         
\newcommand{\hatcurCCtwomassxxxxxD}{2MASS~18573592-4908184}     
\newcommand{\hatcurCCgscxxxxxD}{GSC~8382-00661}                 
\newcommand{\hatcurCCtassmvxxxxxD}{\ensuremath{13.934\pm0.080}} 
\newcommand{\hatcurCCtassmvshortxxxxxD}{\ensuremath{13.9}}      
\newcommand{\hatcurCCtassmBxxxxxD}{\ensuremath{14.697\pm0.020}} 
\newcommand{\hatcurCCtassmBshortxxxxxD}{\ensuremath{14.7}}      
\newcommand{\hatcurCCtassmIxxxxxD}{\ensuremath{nff\pmnff}}      
\newcommand{\hatcurCCtassmIshortxxxxxD}{\ensuremath{0.0}}       
\newcommand{\hatcurCCtassmgxxxxxD}{\ensuremath{14.274\pm0.030}} 
\newcommand{\hatcurCCtassmgshortxxxxxD}{\ensuremath{14.3}}      
\newcommand{\hatcurCCtassmrxxxxxD}{\ensuremath{13.717\pm0.010}} 
\newcommand{\hatcurCCtassmrshortxxxxxD}{\ensuremath{13.7}}      
\newcommand{\hatcurCCtassmixxxxxD}{\ensuremath{13.615\pm0.010}} 
\newcommand{\hatcurCCtassmishortxxxxxD}{\ensuremath{13.6}}      
\newcommand{\hatcurCCtwomassJmagxxxxxD}{\ensuremath{12.522\pm0.026}} 
\newcommand{\hatcurCCtwomassHmagxxxxxD}{\ensuremath{12.188\pm0.025}} 
\newcommand{\hatcurCCtwomassKmagxxxxxD}{\ensuremath{12.086\pm0.029}} 
\newcommand{\hatcurCCcitJmagxxxxxD}{\ensuremath{12.536\pm0.026}} 
\newcommand{\hatcurCCcitHmagxxxxxD}{\ensuremath{12.182\pm0.025}} 
\newcommand{\hatcurCCcitKmagxxxxxD}{\ensuremath{12.110\pm0.029}} 
\newcommand{\hatcurCCbbJmagxxxxxD}{\ensuremath{12.590\pm0.028}} 
\newcommand{\hatcurCCbbHmagxxxxxD}{\ensuremath{12.204\pm0.026}} 
\newcommand{\hatcurCCbbKmagxxxxxD}{\ensuremath{12.130\pm0.029}} 
\newcommand{\hatcurCCesoJmagxxxxxD}{\ensuremath{12.593\pm0.030}} 
\newcommand{\hatcurCCesoHmagxxxxxD}{\ensuremath{12.200\pm0.031}} 
\newcommand{\hatcurCCesoKmagxxxxxD}{\ensuremath{12.129\pm0.030}} 
\newcommand{\hatcurCCesoJHmagxxxxxD}{\ensuremath{0.392\pm0.040}} 
\newcommand{\hatcurCCesoJKmagxxxxxD}{\ensuremath{0.465\pm0.042}} 
\newcommand{\hatcurCCesoHKmagxxxxxD}{\ensuremath{0.073\pm0.043}} 
\newcommand{\hatcurLCdipxxxxxD}{\ensuremath{17.6}}              
\newcommand{\hatcurLCrprstarxxxxxD}{\ensuremath{0.1331\pm0.0029}} 
\newcommand{\hatcurLCbsqxxxxxD}{\ensuremath{0.414_{-0.059}^{+0.052}}} 
\newcommand{\hatcurLCimpxxxxxD}{\ensuremath{0.643_{-0.047}^{+0.039}}} 
\newcommand{\hatcurLCzetaxxxxxD}{\ensuremath{24.84\pm0.28}}     
\newcommand{\hatcurLCdurxxxxxD}{\ensuremath{0.0981\pm0.0018}}   
\newcommand{\hatcurLCdurshortxxxxxD}{\ensuremath{0.0981}}       
\newcommand{\hatcurLCdurhrxxxxxD}{\ensuremath{2.354\pm0.042}}   
\newcommand{\hatcurLCdurhrshortxxxxxD}{\ensuremath{2.354}}      
\newcommand{\hatcurLCqxxxxxD}{\ensuremath{0.03080\pm0.00056}}   
\newcommand{\hatcurLCqshortxxxxxD}{\ensuremath{0.031}}          
\newcommand{\hatcurLCingdurxxxxxD}{\ensuremath{0.0185\pm0.0020}} 
\newcommand{\hatcurLCPxxxxxD}{\ensuremath{3.1810781\pm0.0000039}} 
\newcommand{\hatcurLCPprecxxxxxD}{\ensuremath{3.1810781}}       
\newcommand{\hatcurLCPshortxxxxxD}{\ensuremath{3.1811}}         
\newcommand{\hatcurLCTxxxxxD}{\ensuremath{2457034.28300\pm0.00046}} 
\newcommand{\hatcurLCTAxxxxxD}{\ensuremath{2456366.25661\pm0.00093}} 
\newcommand{\hatcurLCTBxxxxxD}{\ensuremath{2457269.68278\pm0.00055}} 
\newcommand{\hatcurLChatnetmxxxxxD}{\ensuremath{13.681490\pm0.000077}} 
\newcommand{\hatcurLCiblendxxxxxD}{\ensuremath{0.893\pm0.042}}  
\newcommand{\hatcurLCrhoxxxxxD}{\ensuremath{1.68\pm0.27}}                    
\newcommand{\hatcurSMEiteffxxxxxD}{\ensuremath{5600\pm110}}     
\newcommand{\hatcurSMEizfehxxxxxD}{\ensuremath{0.080\pm0.080}}  
\newcommand{\hatcurSMEizfehshortxxxxxD}{\ensuremath{0.08}}      
\newcommand{\hatcurSMEiloggxxxxxD}{\ensuremath{4.60\pm0.14}}    
\newcommand{\hatcurSMEivsinxxxxxD}{\ensuremath{1.8\pm1.2}}      
\newcommand{\hatcurSMEivmacxxxxxD}{\ensuremath{0.0}}            
\newcommand{\hatcurSMEivmicxxxxxD}{\ensuremath{0.0}}            
\newcommand{\hatcurSMEiiteffxxxxxD}{\ensuremath{5498\pm84}}     
\newcommand{\hatcurSMEiizfehxxxxxD}{\ensuremath{0.010\pm0.060}} 
\newcommand{\hatcurSMEiizfehshortxxxxxD}{\ensuremath{0.01}}     
\newcommand{\hatcurSMEiiloggxxxxxD}{\ensuremath{4.477\pm0.035}} 
\newcommand{\hatcurSMEiivsinxxxxxD}{\ensuremath{2.6\pm1.0}}     
\newcommand{\hatcurLBizxxxxxD}{\ensuremath{0.2476}}             
\newcommand{\hatcurLBiizxxxxxD}{\ensuremath{0.3079}}            
\newcommand{\hatcurLBiixxxxxD}{\ensuremath{0.3148}}             
\newcommand{\hatcurLBiiixxxxxD}{\ensuremath{0.3030}}            
\newcommand{\hatcurLBiIxxxxxD}{\ensuremath{0.2926}}             
\newcommand{\hatcurLBiiIxxxxxD}{\ensuremath{0.3051}}            
\newcommand{\hatcurLBigxxxxxD}{\ensuremath{0.6241}}             
\newcommand{\hatcurLBiigxxxxxD}{\ensuremath{0.1819}}            
\newcommand{\hatcurLBirxxxxxD}{\ensuremath{0.4137}}             
\newcommand{\hatcurLBiirxxxxxD}{\ensuremath{0.2900}}            
\newcommand{\hatcurLBiRxxxxxD}{\ensuremath{0.3864}}             
\newcommand{\hatcurLBiiRxxxxxD}{\ensuremath{0.2944}}            
\newcommand{\hatcurLBikepxxxxxD}{\ensuremath{0.1000}}           
\newcommand{\hatcurLBiikepxxxxxD}{\ensuremath{0.1000}}          
\newcommand{\hatcurISOmxxxxxD}{\ensuremath{0.929\pm0.036}}      
\newcommand{\hatcurISOmshortxxxxxD}{\ensuremath{0.93}}          
\newcommand{\hatcurISOmlongxxxxxD}{\ensuremath{0.929\pm0.036}}  
\newcommand{\hatcurISOrxxxxxD}{\ensuremath{0.922\pm0.040}}      
\newcommand{\hatcurISOrshortxxxxxD}{\ensuremath{0.92}}          
\newcommand{\hatcurISOrlongxxxxxD}{\ensuremath{0.922\pm0.040}}  
\newcommand{\hatcurISOrhoxxxxxD}{\ensuremath{1.67\pm0.22}}      
\newcommand{\hatcurISOrholongxxxxxD}{\ensuremath{1.67\pm0.22}}  
\newcommand{\hatcurISOloggxxxxxD}{\ensuremath{4.476\pm0.039}}   
\newcommand{\hatcurISOlumxxxxxD}{\ensuremath{0.696\pm0.084}}    
\newcommand{\hatcurISOlumshortxxxxxD}{\ensuremath{0.70}}        
\newcommand{\hatcurISOmvxxxxxD}{\ensuremath{5.28\pm0.14}}       
\newcommand{\hatcurISOvixxxxxD}{\ensuremath{0.782\pm0.023}}     
\newcommand{\hatcurISOagexxxxxD}{\ensuremath{6.2\pm2.8}}        
\newcommand{\hatcurISOsigmaxxxxxD}{\ensuremath{0.00110\pm0.00018}} 
\newcommand{\hatcurISOMJxxxxxD}{\ensuremath{4.00\pm0.11}}       
\newcommand{\hatcurISOMHxxxxxD}{\ensuremath{3.60\pm0.10}}       
\newcommand{\hatcurISOMKxxxxxD}{\ensuremath{3.53\pm0.10}}       
\newcommand{\hatcurISOJKxxxxxD}{\ensuremath{0.460\pm0.020}}     
\newcommand{\hatcurISOspecxxxxxD}{G}                            
\newcommand{\hatcurRVKxxxxxD}{\ensuremath{97\pm12}}             
\newcommand{\hatcurRVrkxxxxxD}{\ensuremath{0\pm0}}              
\newcommand{\hatcurRVrhxxxxxD}{\ensuremath{0\pm0}}              
\newcommand{\hatcurRVkxxxxxD}{\ensuremath{0\pm0}}               
\newcommand{\hatcurRVhxxxxxD}{\ensuremath{0\pm0}}               
\newcommand{\hatcurRVtronexxxxxD}{\ensuremath{0\pm0}}           
\newcommand{\hatcurRVtrtwoxxxxxD}{\ensuremath{0\pm0}}           
\newcommand{\hatcurRVgammaxxxxxD}{\ensuremath{-8650.5\pm9.1}}   
\newcommand{\hatcurRVjitterxxxxxD}{\ensuremath{32.9\pm8.7}}     
\newcommand{\hatcurRVjittertwosiglimxxxxxD}{\ensuremath{<49}}               
\newcommand{\hatcurRVfitrmsxxxxxD}{\ensuremath{.1fym}}          %
\newcommand{\hatcurRVeccenxxxxxD}{\ensuremath{0\pm0}}           
\newcommand{\hatcurRVeccentwosiglimxxxxxD}{\ensuremath{<0.000}} 
\newcommand{\hatcurRVomegaxxxxxD}{\ensuremath{0\pm0}}           
\newcommand{\hatcurPPixxxxxD}{\ensuremath{86.17\pm0.42}}        
\newcommand{\hatcurPPgxxxxxD}{\ensuremath{11.6\pm2.0}}          
\newcommand{\hatcurPPloggxxxxxD}{\ensuremath{3.065\pm0.076}}    
\newcommand{\hatcurPParxxxxxD}{\ensuremath{9.63\pm0.42}}        
\newcommand{\hatcurPParelxxxxxD}{\ensuremath{0.04131\pm0.00053}} 
\newcommand{\hatcurPPrhoxxxxxD}{\ensuremath{0.48\pm0.11}}       
\newcommand{\hatcurPPmxxxxxD}{\ensuremath{0.672\pm0.087}}       
\newcommand{\hatcurPPmshortxxxxxD}{\ensuremath{0.67}}           
\newcommand{\hatcurPPmlongxxxxxD}{\ensuremath{0.672\pm0.087}}   
\newcommand{\hatcurPPmexxxxxD}{\ensuremath{213\pm28}}           
\newcommand{\hatcurPPmeshortxxxxxD}{\ensuremath{213.4}}         
\newcommand{\hatcurPPmelongxxxxxD}{\ensuremath{213\pm28}}       
\newcommand{\hatcurPPrxxxxxD}{\ensuremath{1.194\pm0.070}}       
\newcommand{\hatcurPPrshortxxxxxD}{\ensuremath{1.19}}           
\newcommand{\hatcurPPrlongxxxxxD}{\ensuremath{1.194\pm0.070}}   
\newcommand{\hatcurPPrexxxxxD}{\ensuremath{13.38\pm0.78}}       
\newcommand{\hatcurPPreshortxxxxxD}{\ensuremath{13.4}}          
\newcommand{\hatcurPPrelongxxxxxD}{\ensuremath{13.38\pm0.78}}   
\newcommand{\hatcurPPmrcorrxxxxxD}{\ensuremath{0.01}}           
\newcommand{\hatcurPPteffxxxxxD}{\ensuremath{1253\pm35}}        
\newcommand{\hatcurPPthetaxxxxxD}{\ensuremath{0.0498\pm0.0070}} 
\newcommand{\hatcurPPfluxperixxxxxD}{\ensuremath{5.57\pm0.62}}  
\newcommand{\hatcurPPfluxperidimxxxxxD}{\ensuremath{8}}         
\newcommand{\hatcurPPfluxapxxxxxD}{\ensuremath{5.57\pm0.62}}    
\newcommand{\hatcurPPfluxapdimxxxxxD}{\ensuremath{8}}           
\newcommand{\hatcurPPfluxavgxxxxxD}{\ensuremath{5.57\pm0.62}}   
\newcommand{\hatcurPPfluxavgdimxxxxxD}{\ensuremath{8}}          
\newcommand{\hatcurPPfluxavglogxxxxxD}{\ensuremath{8.746\pm0.048}} 
\newcommand{\hatcurXsecphasexxxxxD}{\ensuremath{0\pm0}}         
\newcommand{\hatcurXsecondaryxxxxxD}{\ensuremath{2457035.87353\pm0.00046}} 
\newcommand{\hatcurXsecdurxxxxxD}{\ensuremath{0.0981\pm0.0018}} 
\newcommand{\hatcurXsecingdurxxxxxD}{\ensuremath{0.0185\pm0.0020}} 
\newcommand{\hatcurPPphiconjxxxxxD}{\ensuremath{0\pm0}}         
\newcommand{\hatcurPPperixxxxxD}{\ensuremath{2457033.48773\pm0.00046}} 
\newcommand{\hatcurPPaequivxxxxxD}{\ensuremath{0.0495\pm0.0028}} 
\newcommand{\hatcurPPtcircxxxxxD}{\ensuremath{120_{-33}^{+46}}} 
\newcommand{\hatcurPPtinfallxxxxxD}{\ensuremath{3490_{-800}^{+1050}}} 
\newcommand{\hatcurXdistxxxxxD}{\ensuremath{525\pm25}}          
\newcommand{\hatcurXAvxxxxxD}{\ensuremath{0.055_{-0.055}^{+0.124}}} 
\newcommand{\hatcurXdistredxxxxxD}{\ensuremath{521\pm25}}       
\newcommand{\hatcurXEBVxxxxxD}{\ensuremath{0.018_{-0.018}^{+0.040}}} 
\newcommand{\hatcurXmvisoredxxxxxD}{\ensuremath{13.944\pm0.060}} 
\newcommand{\hatcurXmiisoredxxxxxD}{\ensuremath{13.123\pm0.032}} 
\newcommand{\hatcurXmjisoredxxxxxD}{\ensuremath{12.602\pm0.019}} 
\newcommand{\hatcurXmhisoredxxxxxD}{\ensuremath{12.195\pm0.018}} 
\newcommand{\hatcurXmkisoredxxxxxD}{\ensuremath{12.123\pm0.020}} 
\newcommand{\hatcurXviisoredxxxxxD}{\ensuremath{0.816_{-0.027}^{+0.046}}} 
\newcommand{\hatcurXvkisoredxxxxxD}{\ensuremath{1.820\pm0.067}} 
\newcommand{\hatcurXjhisoredxxxxxD}{\ensuremath{0.407\pm0.015}} 
\newcommand{\hatcurXjkisoredxxxxxD}{\ensuremath{0.479\pm0.017}} 
\newcommand{\hatcurCCpmraxxxxxD}{\ensuremath{10.3\pm1.6}}       
\newcommand{\hatcurCCpmdecxxxxxD}{\ensuremath{-2.4\pm1.4}}      
\newcommand{\hatcurCCpmxxxxxD}{\ensuremath{10.6\pm2.1}}         

\newcommand{\hatcurhtrxxxxxE}{HATS747-002}                      
\newcommand{\hatcurfieldxxxxxE}{\ensuremath{string}}            
\newcommand{\hatcurCCraxxxxxE}{\ensuremath{19^{\mathrm h}00^{\mathrm m}23.04{\mathrm s}}}                     
\newcommand{\hatcurCCdecxxxxxE}{\ensuremath{-54{\arcdeg}53{\arcmin}35.5{\arcsec}}}                    
\newcommand{\hatcurCCmagxxxxxE}{12.612}                         
\newcommand{\hatcurCCtwomassxxxxxE}{2MASS~19002314-5453354}     
\newcommand{\hatcurCCgscxxxxxE}{GSC~8763-00475}                 
\newcommand{\hatcurCCtassmvxxxxxE}{\ensuremath{12.612\pm0.010}} 
\newcommand{\hatcurCCtassmvshortxxxxxE}{\ensuremath{12.6}}      
\newcommand{\hatcurCCtassmBxxxxxE}{\ensuremath{13.361\pm0.010}} 
\newcommand{\hatcurCCtassmBshortxxxxxE}{\ensuremath{13.4}}      
\newcommand{\hatcurCCtassmIxxxxxE}{\ensuremath{100\pm1000}}     
\newcommand{\hatcurCCtassmIshortxxxxxE}{\ensuremath{100.0}}     
\newcommand{\hatcurCCtassmgxxxxxE}{\ensuremath{12.950\pm0.010}} 
\newcommand{\hatcurCCtassmgshortxxxxxE}{\ensuremath{12.9}}      
\newcommand{\hatcurCCtassmrxxxxxE}{\ensuremath{12.430\pm0.010}} 
\newcommand{\hatcurCCtassmrshortxxxxxE}{\ensuremath{12.4}}      
\newcommand{\hatcurCCtassmixxxxxE}{\ensuremath{12.154\pm0.010}} 
\newcommand{\hatcurCCtassmishortxxxxxE}{\ensuremath{12.2}}      
\newcommand{\hatcurCCtwomassJmagxxxxxE}{\ensuremath{11.286\pm0.026}} 
\newcommand{\hatcurCCtwomassHmagxxxxxE}{\ensuremath{10.933\pm0.021}} 
\newcommand{\hatcurCCtwomassKmagxxxxxE}{\ensuremath{10.877\pm0.019}} 
\newcommand{\hatcurCCcitJmagxxxxxE}{\ensuremath{11.301\pm0.026}} 
\newcommand{\hatcurCCcitHmagxxxxxE}{\ensuremath{10.928\pm0.022}} 
\newcommand{\hatcurCCcitKmagxxxxxE}{\ensuremath{10.901\pm0.019}} 
\newcommand{\hatcurCCbbJmagxxxxxE}{\ensuremath{11.353\pm0.028}} 
\newcommand{\hatcurCCbbHmagxxxxxE}{\ensuremath{10.950\pm0.022}} 
\newcommand{\hatcurCCbbKmagxxxxxE}{\ensuremath{10.921\pm0.019}} 
\newcommand{\hatcurCCesoJmagxxxxxE}{\ensuremath{11.356\pm0.030}} 
\newcommand{\hatcurCCesoHmagxxxxxE}{\ensuremath{10.943\pm0.025}} 
\newcommand{\hatcurCCesoKmagxxxxxE}{\ensuremath{10.920\pm0.020}} 
\newcommand{\hatcurCCesoJHmagxxxxxE}{\ensuremath{0.412\pm0.036}} 
\newcommand{\hatcurCCesoJKmagxxxxxE}{\ensuremath{0.437\pm0.035}} 
\newcommand{\hatcurCCesoHKmagxxxxxE}{\ensuremath{0.024\pm0.032}} 
\newcommand{\hatcurLCdipxxxxxE}{\ensuremath{15.6}}              
\newcommand{\hatcurLCrprstarxxxxxE}{\ensuremath{0.1201\pm0.0027}} 
\newcommand{\hatcurLCbsqxxxxxE}{\ensuremath{0.252_{-0.040}^{+0.049}}} 
\newcommand{\hatcurLCimpxxxxxE}{\ensuremath{0.502_{-0.041}^{+0.046}}} 
\newcommand{\hatcurLCzetaxxxxxE}{\ensuremath{17.30\pm0.12}}     
\newcommand{\hatcurLCdurxxxxxE}{\ensuremath{0.1338\pm0.0014}}   
\newcommand{\hatcurLCdurshortxxxxxE}{\ensuremath{0.1338}}       
\newcommand{\hatcurLCdurhrxxxxxE}{\ensuremath{3.212\pm0.034}}   
\newcommand{\hatcurLCdurhrshortxxxxxE}{\ensuremath{3.212}}      
\newcommand{\hatcurLCqxxxxxE}{\ensuremath{0.02910\pm0.00031}}   
\newcommand{\hatcurLCqshortxxxxxE}{\ensuremath{0.029}}          
\newcommand{\hatcurLCingdurxxxxxE}{\ensuremath{0.0186\pm0.0014}} 
\newcommand{\hatcurLCPxxxxxE}{\ensuremath{4.6058749\pm0.0000063}} 
\newcommand{\hatcurLCPprecxxxxxE}{\ensuremath{4.6058749}}       
\newcommand{\hatcurLCPshortxxxxxE}{\ensuremath{4.6059}}         
\newcommand{\hatcurLCTxxxxxE}{\ensuremath{2457031.95618\pm0.00038}} 
\newcommand{\hatcurLCTAxxxxxE}{\ensuremath{2456368.71012\pm0.00098}} 
\newcommand{\hatcurLCTBxxxxxE}{\ensuremath{2457174.73825\pm0.00044}} 
\newcommand{\hatcurLChatnetmxxxxxE}{\ensuremath{12.463790\pm0.000049}} 
\newcommand{\hatcurLCiblendxxxxxE}{\ensuremath{0.859\pm0.039}}  
\newcommand{\hatcurLCrhoxxxxxE}{\ensuremath{1.17\pm0.11}}                    
\newcommand{\hatcurSMEiteffxxxxxE}{\ensuremath{5680\pm120}}     
\newcommand{\hatcurSMEizfehxxxxxE}{\ensuremath{0.190\pm0.080}}  
\newcommand{\hatcurSMEizfehshortxxxxxE}{\ensuremath{0.19}}      
\newcommand{\hatcurSMEiloggxxxxxE}{\ensuremath{4.39\pm0.20}}    
\newcommand{\hatcurSMEivsinxxxxxE}{\ensuremath{2.50\pm0.94}}    
\newcommand{\hatcurSMEivmacxxxxxE}{\ensuremath{0.0}}            
\newcommand{\hatcurSMEivmicxxxxxE}{\ensuremath{0.0}}            
\newcommand{\hatcurSMEiiteffxxxxxE}{\ensuremath{5670\pm110}}    
\newcommand{\hatcurSMEiizfehxxxxxE}{\ensuremath{0.160\pm0.080}} 
\newcommand{\hatcurSMEiizfehshortxxxxxE}{\ensuremath{0.16}}     
\newcommand{\hatcurSMEiiloggxxxxxE}{\ensuremath{4.382\pm0.046}} 
\newcommand{\hatcurSMEiivsinxxxxxE}{\ensuremath{2.35\pm0.80}}   
\newcommand{\hatcurLBizxxxxxE}{\ensuremath{0.2248}}             
\newcommand{\hatcurLBiizxxxxxE}{\ensuremath{0.3233}}            
\newcommand{\hatcurLBiixxxxxE}{\ensuremath{0.2914}}             
\newcommand{\hatcurLBiiixxxxxE}{\ensuremath{0.3213}}            
\newcommand{\hatcurLBiIxxxxxE}{\ensuremath{0.2690}}             
\newcommand{\hatcurLBiiIxxxxxE}{\ensuremath{0.3226}}            
\newcommand{\hatcurLBigxxxxxE}{\ensuremath{0.5937}}             
\newcommand{\hatcurLBiigxxxxxE}{\ensuremath{0.2066}}            
\newcommand{\hatcurLBirxxxxxE}{\ensuremath{0.3875}}             
\newcommand{\hatcurLBiirxxxxxE}{\ensuremath{0.3096}}            
\newcommand{\hatcurLBiRxxxxxE}{\ensuremath{0.3607}}             
\newcommand{\hatcurLBiiRxxxxxE}{\ensuremath{0.3138}}            
\newcommand{\hatcurLBikepxxxxxE}{\ensuremath{0.1000}}           
\newcommand{\hatcurLBiikepxxxxxE}{\ensuremath{0.1000}}          
\newcommand{\hatcurISOmxxxxxE}{\ensuremath{1.032\pm0.049}}      
\newcommand{\hatcurISOmshortxxxxxE}{\ensuremath{1.03}}          
\newcommand{\hatcurISOmlongxxxxxE}{\ensuremath{1.032\pm0.049}}  
\newcommand{\hatcurISOrxxxxxE}{\ensuremath{1.073\pm0.038}}      
\newcommand{\hatcurISOrshortxxxxxE}{\ensuremath{1.07}}          
\newcommand{\hatcurISOrlongxxxxxE}{\ensuremath{1.073\pm0.038}}  
\newcommand{\hatcurISOrhoxxxxxE}{\ensuremath{1.17\pm0.11}}      
\newcommand{\hatcurISOrholongxxxxxE}{\ensuremath{1.17\pm0.11}}  
\newcommand{\hatcurISOloggxxxxxE}{\ensuremath{4.389\pm0.027}}   
\newcommand{\hatcurISOlumxxxxxE}{\ensuremath{1.07\pm0.13}}      
\newcommand{\hatcurISOlumshortxxxxxE}{\ensuremath{1.07}}        
\newcommand{\hatcurISOmvxxxxxE}{\ensuremath{4.77\pm0.15}}       
\newcommand{\hatcurISOvixxxxxE}{\ensuremath{0.734\pm0.035}}     
\newcommand{\hatcurISOagexxxxxE}{\ensuremath{5.5_{-1.7}^{+2.6}}} 
\newcommand{\hatcurISOsigmaxxxxxE}{\ensuremath{0.000800\pm0.000093}} 
\newcommand{\hatcurISOMJxxxxxE}{\ensuremath{3.58\pm0.11}}       
\newcommand{\hatcurISOMHxxxxxE}{\ensuremath{3.224\pm0.091}}     
\newcommand{\hatcurISOMKxxxxxE}{\ensuremath{3.166\pm0.088}}     
\newcommand{\hatcurISOJKxxxxxE}{\ensuremath{0.420\pm0.030}}     
\newcommand{\hatcurISOspecxxxxxE}{G}                            
\newcommand{\hatcurRVKxxxxxE}{\ensuremath{78.4\pm7.1}}          
\newcommand{\hatcurRVrkxxxxxE}{\ensuremath{0\pm0}}              
\newcommand{\hatcurRVrhxxxxxE}{\ensuremath{0\pm0}}              
\newcommand{\hatcurRVkxxxxxE}{\ensuremath{0\pm0}}               
\newcommand{\hatcurRVhxxxxxE}{\ensuremath{0\pm0}}               
\newcommand{\hatcurRVtronexxxxxE}{\ensuremath{0\pm0}}           
\newcommand{\hatcurRVtrtwoxxxxxE}{\ensuremath{0\pm0}}           
\newcommand{\hatcurRVgammaAxxxxxE}{\ensuremath{-19698.8\pm6.9}} 
\newcommand{\hatcurRVjitterAxxxxxE}{\ensuremath{0.00\pm0.39}}   
\newcommand{\hatcurRVjittertwosiglimAxxxxxE}{\ensuremath{<0.68}}               
\newcommand{\hatcurRVfitrmsAxxxxxE}{\ensuremath{0.0}}           
\newcommand{\hatcurRVgammaBxxxxxE}{\ensuremath{-19719.3\pm6.9}} 
\newcommand{\hatcurRVjitterBxxxxxE}{\ensuremath{0.0\pm3.1}}     
\newcommand{\hatcurRVjittertwosiglimBxxxxxE}{\ensuremath{<4.5}}               
\newcommand{\hatcurRVfitrmsBxxxxxE}{\ensuremath{0.0}}           
\newcommand{\hatcurRVgammaCxxxxxE}{\ensuremath{-19722\pm14}}    
\newcommand{\hatcurRVjitterCxxxxxE}{\ensuremath{36\pm11}}       
\newcommand{\hatcurRVjittertwosiglimCxxxxxE}{\ensuremath{<59}}               
\newcommand{\hatcurRVfitrmsCxxxxxE}{\ensuremath{0.0}}           
\newcommand{\hatcurRVeccenxxxxxE}{\ensuremath{0\pm0}}           
\newcommand{\hatcurRVeccentwosiglimxxxxxE}{\ensuremath{<0.000}} 
\newcommand{\hatcurRVomegaxxxxxE}{\ensuremath{0\pm0}}           
\newcommand{\hatcurPPixxxxxE}{\ensuremath{87.37\pm0.34}}        
\newcommand{\hatcurPPgxxxxxE}{\ensuremath{10.2\pm1.2}}          
\newcommand{\hatcurPPloggxxxxxE}{\ensuremath{3.010\pm0.049}}    
\newcommand{\hatcurPParxxxxxE}{\ensuremath{10.96\pm0.34}}       
\newcommand{\hatcurPParelxxxxxE}{\ensuremath{0.05475\pm0.00088}} 
\newcommand{\hatcurPPrhoxxxxxE}{\ensuremath{0.411\pm0.060}}     
\newcommand{\hatcurPPmxxxxxE}{\ensuremath{0.653\pm0.063}}       
\newcommand{\hatcurPPmshortxxxxxE}{\ensuremath{0.65}}           
\newcommand{\hatcurPPmlongxxxxxE}{\ensuremath{0.653\pm0.063}}   
\newcommand{\hatcurPPmexxxxxE}{\ensuremath{208\pm20}}           
\newcommand{\hatcurPPmeshortxxxxxE}{\ensuremath{207.7}}         
\newcommand{\hatcurPPmelongxxxxxE}{\ensuremath{208\pm20}}       
\newcommand{\hatcurPPrxxxxxE}{\ensuremath{1.251\pm0.061}}       
\newcommand{\hatcurPPrshortxxxxxE}{\ensuremath{1.25}}           
\newcommand{\hatcurPPrlongxxxxxE}{\ensuremath{1.251\pm0.061}}   
\newcommand{\hatcurPPrexxxxxE}{\ensuremath{14.03\pm0.68}}       
\newcommand{\hatcurPPreshortxxxxxE}{\ensuremath{14.0}}          
\newcommand{\hatcurPPrelongxxxxxE}{\ensuremath{14.03\pm0.68}}   
\newcommand{\hatcurPPmrcorrxxxxxE}{\ensuremath{0.31}}           
\newcommand{\hatcurPPteffxxxxxE}{\ensuremath{1212\pm30}}        
\newcommand{\hatcurPPthetaxxxxxE}{\ensuremath{0.0557\pm0.0051}} 
\newcommand{\hatcurPPfluxperixxxxxE}{\ensuremath{4.87\pm0.49}}  
\newcommand{\hatcurPPfluxperidimxxxxxE}{\ensuremath{8}}         
\newcommand{\hatcurPPfluxapxxxxxE}{\ensuremath{4.87\pm0.49}}    
\newcommand{\hatcurPPfluxapdimxxxxxE}{\ensuremath{8}}           
\newcommand{\hatcurPPfluxavgxxxxxE}{\ensuremath{4.87\pm0.49}}   
\newcommand{\hatcurPPfluxavgdimxxxxxE}{\ensuremath{8}}          
\newcommand{\hatcurPPfluxavglogxxxxxE}{\ensuremath{8.687\pm0.044}} 
\newcommand{\hatcurXsecphasexxxxxE}{\ensuremath{0\pm0}}         
\newcommand{\hatcurXsecondaryxxxxxE}{\ensuremath{2457034.25912\pm0.00038}} 
\newcommand{\hatcurXsecdurxxxxxE}{\ensuremath{0.1338\pm0.0014}} 
\newcommand{\hatcurXsecingdurxxxxxE}{\ensuremath{0.0186\pm0.0014}} 
\newcommand{\hatcurPPphiconjxxxxxE}{\ensuremath{0\pm0}}         
\newcommand{\hatcurPPperixxxxxE}{\ensuremath{2457030.80471\pm0.00038}} 
\newcommand{\hatcurPPaequivxxxxxE}{\ensuremath{0.0530\pm0.0027}} 
\newcommand{\hatcurPPtcircxxxxxE}{\ensuremath{490\pm110}}       
\newcommand{\hatcurPPtinfallxxxxxE}{\ensuremath{10800_{-1800}^{+2400}}} 
\newcommand{\hatcurXdistxxxxxE}{\ensuremath{355\pm15}}          
\newcommand{\hatcurXAvxxxxxE}{\ensuremath{0.111\pm0.082}}       
\newcommand{\hatcurXdistredxxxxxE}{\ensuremath{351\pm15}}       
\newcommand{\hatcurXEBVxxxxxE}{\ensuremath{0.036\pm0.026}}      
\newcommand{\hatcurXmvisoredxxxxxE}{\ensuremath{12.613\pm0.012}} 
\newcommand{\hatcurXmiisoredxxxxxE}{\ensuremath{11.821\pm0.014}} 
\newcommand{\hatcurXmjisoredxxxxxE}{\ensuremath{11.338\pm0.014}} 
\newcommand{\hatcurXmhisoredxxxxxE}{\ensuremath{10.972\pm0.020}} 
\newcommand{\hatcurXmkisoredxxxxxE}{\ensuremath{10.905\pm0.021}} 
\newcommand{\hatcurXviisoredxxxxxE}{\ensuremath{0.792\pm0.014}} 
\newcommand{\hatcurXvkisoredxxxxxE}{\ensuremath{1.707\pm0.028}} 
\newcommand{\hatcurXjhisoredxxxxxE}{\ensuremath{0.366\pm0.015}} 
\newcommand{\hatcurXjkisoredxxxxxE}{\ensuremath{0.432_{-0.010}^{+0.014}}} 
\newcommand{\hatcurCCpmraxxxxxE}{\ensuremath{2.8\pm1.3}}        
\newcommand{\hatcurCCpmdecxxxxxE}{\ensuremath{-37.1\pm3.7}}     
\newcommand{\hatcurCCpmxxxxxE}{\ensuremath{37.2\pm3.9}}         

\newcommand{\hatcurhtrxxxxxF}{HATS754-005}                                   
\newcommand{\hatcurfieldxxxxxF}{\ensuremath{string}}                         
\newcommand{\hatcurCCraxxxxxF}{\ensuremath{00^{\mathrm h}22^{\mathrm m}28.49{\mathrm s}}}                                  
\newcommand{\hatcurCCdecxxxxxF}{\ensuremath{-59{\arcdeg}56{\arcmin}33.2{\arcsec}}}                                 
\newcommand{\hatcurCCmagxxxxxF}{12.192}                                      
\newcommand{\hatcurCCtwomassxxxxxF}{2MASS~00222848-5956331}                  
\newcommand{\hatcurCCgscxxxxxF}{GSC~8471-00231}                              
\newcommand{\hatcurCCtassmvxxxxxF}{\ensuremath{12.192\pm0.010}}              
\newcommand{\hatcurCCtassmvshortxxxxxF}{\ensuremath{12.2}}                   
\newcommand{\hatcurCCtassmBxxxxxF}{\ensuremath{12.790\pm0.010}}              
\newcommand{\hatcurCCtassmBshortxxxxxF}{\ensuremath{12.8}}                   
\newcommand{\hatcurCCtassmIxxxxxF}{\ensuremath{100\pm1000}}                  
\newcommand{\hatcurCCtassmIshortxxxxxF}{\ensuremath{100.0}}                  
\newcommand{\hatcurCCtassmgxxxxxF}{\ensuremath{12.439\pm0.010}}              
\newcommand{\hatcurCCtassmgshortxxxxxF}{\ensuremath{12.4}}                   
\newcommand{\hatcurCCtassmrxxxxxF}{\ensuremath{12.046\pm0.010}}              
\newcommand{\hatcurCCtassmrshortxxxxxF}{\ensuremath{12.0}}                   
\newcommand{\hatcurCCtassmixxxxxF}{\ensuremath{11.935\pm0.010}}              
\newcommand{\hatcurCCtassmishortxxxxxF}{\ensuremath{11.9}}                   
\newcommand{\hatcurCCtwomassJmagxxxxxF}{\ensuremath{11.129\pm0.024}}         
\newcommand{\hatcurCCtwomassHmagxxxxxF}{\ensuremath{10.826\pm0.024}}         
\newcommand{\hatcurCCtwomassKmagxxxxxF}{\ensuremath{10.793\pm0.019}}         
\newcommand{\hatcurCCcitJmagxxxxxF}{\ensuremath{11.147\pm0.024}}             
\newcommand{\hatcurCCcitHmagxxxxxF}{\ensuremath{10.822\pm0.024}}             
\newcommand{\hatcurCCcitKmagxxxxxF}{\ensuremath{10.817\pm0.019}}             
\newcommand{\hatcurCCbbJmagxxxxxF}{\ensuremath{11.193\pm0.026}}              
\newcommand{\hatcurCCbbHmagxxxxxF}{\ensuremath{10.842\pm0.025}}              
\newcommand{\hatcurCCbbKmagxxxxxF}{\ensuremath{10.837\pm0.019}}              
\newcommand{\hatcurCCesoJmagxxxxxF}{\ensuremath{11.196\pm0.027}}             
\newcommand{\hatcurCCesoHmagxxxxxF}{\ensuremath{10.835\pm0.027}}             
\newcommand{\hatcurCCesoKmagxxxxxF}{\ensuremath{10.836\pm0.020}}             
\newcommand{\hatcurCCesoJHmagxxxxxF}{\ensuremath{0.3610\pm0.0080}}           
\newcommand{\hatcurCCesoJKmagxxxxxF}{\ensuremath{0.360\pm0.033}}             
\newcommand{\hatcurCCesoHKmagxxxxxF}{\ensuremath{-0.001\pm0.033}}            
\newcommand{\hatcurLCdipxxxxxF}{\ensuremath{14.0}}                           
\newcommand{\hatcurLCrprstarxxxxxF}{\ensuremath{0.1137\pm0.0017}}            
\newcommand{\hatcurLCbsqxxxxxF}{\ensuremath{0.237_{-0.056}^{+0.053}}}        
\newcommand{\hatcurLCimpxxxxxF}{\ensuremath{0.487_{-0.061}^{+0.052}}}        
\newcommand{\hatcurLCzetaxxxxxF}{\ensuremath{19.99\pm0.13}}                  
\newcommand{\hatcurLCdurxxxxxF}{\ensuremath{0.1146\pm0.0012}}                
\newcommand{\hatcurLCdurshortxxxxxF}{\ensuremath{0.1146}}                    
\newcommand{\hatcurLCdurhrxxxxxF}{\ensuremath{2.751\pm0.029}}                
\newcommand{\hatcurLCdurhrshortxxxxxF}{\ensuremath{2.751}}                   
\newcommand{\hatcurLCqxxxxxF}{\ensuremath{0.03610\pm0.00039}}                
\newcommand{\hatcurLCqshortxxxxxF}{\ensuremath{0.036}}                       
\newcommand{\hatcurLCingdurxxxxxF}{\ensuremath{0.0150\pm0.0012}}             
\newcommand{\hatcurLCPxxxxxF}{\ensuremath{3.1743516\pm0.0000026}}            
\newcommand{\hatcurLCPprecxxxxxF}{\ensuremath{3.1743516}}                    
\newcommand{\hatcurLCPshortxxxxxF}{\ensuremath{3.1744}}                      
\newcommand{\hatcurLCTxxxxxF}{\ensuremath{2456629.76156\pm0.00036}}          
\newcommand{\hatcurLCTAxxxxxF}{\ensuremath{2455759.98923\pm0.00077}}         
\newcommand{\hatcurLCTBxxxxxF}{\ensuremath{2456953.54540\pm0.00047}}         
\newcommand{\hatcurLChatnetmAxxxxxF}{\ensuremath{12.184490\pm0.000073}}      
\newcommand{\hatcurLCiblendAxxxxxF}{\ensuremath{0.963\pm0.027}}              
\newcommand{\hatcurLChatnetmBxxxxxF}{\ensuremath{12.184740\pm0.000061}}      
\newcommand{\hatcurLCiblendBxxxxxF}{\ensuremath{0.823\pm0.031}}              
\newcommand{\hatcurLChatnetmCxxxxxF}{\ensuremath{12.184520\pm0.000065}}      
\newcommand{\hatcurLCiblendCxxxxxF}{\ensuremath{0.967\pm0.027}}              
\newcommand{\hatcurLCrhoxxxxxF}{\ensuremath{1.34\pm0.19}}                    
\newcommand{\hatcurSMEiteffxxxxxF}{\ensuremath{5947\pm70}}                   
\newcommand{\hatcurSMEizfehxxxxxF}{\ensuremath{0.040\pm0.040}}               
\newcommand{\hatcurSMEizfehshortxxxxxF}{\ensuremath{0.04}}                   
\newcommand{\hatcurSMEiloggxxxxxF}{\ensuremath{4.330\pm0.070}}               
\newcommand{\hatcurSMEivsinxxxxxF}{\ensuremath{4.24\pm0.50}}                 
\newcommand{\hatcurSMEivmacxxxxxF}{\ensuremath{0.0}}                         
\newcommand{\hatcurSMEivmicxxxxxF}{\ensuremath{0.0}}                         
\newcommand{\hatcurSMEiiteffxxxxxF}{\ensuremath{5943\pm70}}                  
\newcommand{\hatcurSMEiizfehxxxxxF}{\ensuremath{0.060\pm0.050}}              
\newcommand{\hatcurSMEiizfehshortxxxxxF}{\ensuremath{0.06}}                  
\newcommand{\hatcurSMEiiloggxxxxxF}{\ensuremath{4.426\pm0.033}}              
\newcommand{\hatcurSMEiivsinxxxxxF}{\ensuremath{4.11\pm0.50}}                
\newcommand{\hatcurLBizxxxxxF}{\ensuremath{0.1876}}                          
\newcommand{\hatcurLBiizxxxxxF}{\ensuremath{0.3386}}                         
\newcommand{\hatcurLBiixxxxxF}{\ensuremath{0.2450}}                          
\newcommand{\hatcurLBiiixxxxxF}{\ensuremath{0.3430}}                         
\newcommand{\hatcurLBiIxxxxxF}{\ensuremath{0.2254}}                          
\newcommand{\hatcurLBiiIxxxxxF}{\ensuremath{0.3425}}                         
\newcommand{\hatcurLBigxxxxxF}{\ensuremath{0.5117}}                          
\newcommand{\hatcurLBiigxxxxxF}{\ensuremath{0.2670}}                         
\newcommand{\hatcurLBirxxxxxF}{\ensuremath{0.3275}}                          
\newcommand{\hatcurLBiirxxxxxF}{\ensuremath{0.3438}}                         
\newcommand{\hatcurLBiRxxxxxF}{\ensuremath{0.3045}}                          
\newcommand{\hatcurLBiiRxxxxxF}{\ensuremath{0.3447}}                         
\newcommand{\hatcurLBikepxxxxxF}{\ensuremath{0.1000}}                        
\newcommand{\hatcurLBiikepxxxxxF}{\ensuremath{0.1000}}                       
\newcommand{\hatcurISOmxxxxxF}{\ensuremath{1.093\pm0.031}}                   
\newcommand{\hatcurISOmshortxxxxxF}{\ensuremath{1.09}}                       
\newcommand{\hatcurISOmlongxxxxxF}{\ensuremath{1.093\pm0.031}}               
\newcommand{\hatcurISOrxxxxxF}{\ensuremath{1.061\pm0.039}}                   
\newcommand{\hatcurISOrshortxxxxxF}{\ensuremath{1.06}}                       
\newcommand{\hatcurISOrlongxxxxxF}{\ensuremath{1.061\pm0.039}}               
\newcommand{\hatcurISOrhoxxxxxF}{\ensuremath{1.29\pm0.14}}                   
\newcommand{\hatcurISOrholongxxxxxF}{\ensuremath{1.29\pm0.14}}               
\newcommand{\hatcurISOloggxxxxxF}{\ensuremath{4.425\pm0.030}}                
\newcommand{\hatcurISOlumxxxxxF}{\ensuremath{1.25\pm0.12}}                   
\newcommand{\hatcurISOlumshortxxxxxF}{\ensuremath{1.25}}                     
\newcommand{\hatcurISOmvxxxxxF}{\ensuremath{4.57\pm0.12}}                    
\newcommand{\hatcurISOvixxxxxF}{\ensuremath{0.651\pm0.020}}                  
\newcommand{\hatcurISOagexxxxxF}{\ensuremath{2.3\pm1.2}}                     
\newcommand{\hatcurISOsigmaxxxxxF}{\ensuremath{0.000900\pm0.000096}}         
\newcommand{\hatcurISOMJxxxxxF}{\ensuremath{3.491\pm0.095}}                  
\newcommand{\hatcurISOMHxxxxxF}{\ensuremath{3.174\pm0.089}}                  
\newcommand{\hatcurISOMKxxxxxF}{\ensuremath{3.123\pm0.088}}                  
\newcommand{\hatcurISOJKxxxxxF}{\ensuremath{0.370\pm0.010}}                  
\newcommand{\hatcurISOspecxxxxxF}{G}                                         
\newcommand{\hatcurRVKxxxxxF}{\ensuremath{91.8\pm4.7}}                       
\newcommand{\hatcurRVrkxxxxxF}{\ensuremath{0\pm0}}                           
\newcommand{\hatcurRVrhxxxxxF}{\ensuremath{0\pm0}}                           
\newcommand{\hatcurRVkxxxxxF}{\ensuremath{0\pm0}}                            
\newcommand{\hatcurRVhxxxxxF}{\ensuremath{0\pm0}}                            
\newcommand{\hatcurRVtronexxxxxF}{\ensuremath{0\pm0}}                        
\newcommand{\hatcurRVtrtwoxxxxxF}{\ensuremath{0\pm0}}                        
\newcommand{\hatcurRVgammaAxxxxxF}{\ensuremath{-78.6\pm4.2}}                 
\newcommand{\hatcurRVjitterAxxxxxF}{\ensuremath{0.0\pm2.2}}                  
\newcommand{\hatcurRVjittertwosiglimAxxxxxF}{\ensuremath{<4.7}}               
\newcommand{\hatcurRVfitrmsAxxxxxF}{\ensuremath{0.0}}                        
\newcommand{\hatcurRVgammaBxxxxxF}{\ensuremath{-111.9\pm6.5}}                
\newcommand{\hatcurRVjitterBxxxxxF}{\ensuremath{2\pm12}}                     
\newcommand{\hatcurRVjittertwosiglimBxxxxxF}{\ensuremath{<25}}               
\newcommand{\hatcurRVfitrmsBxxxxxF}{\ensuremath{0.0}}                        
\newcommand{\hatcurRVeccenxxxxxF}{\ensuremath{0\pm0}}                        
\newcommand{\hatcurRVeccentwosiglimxxxxxF}{\ensuremath{<0.000}}              
\newcommand{\hatcurRVomegaxxxxxF}{\ensuremath{0\pm0}}                        
\newcommand{\hatcurPPixxxxxF}{\ensuremath{86.84\pm0.48}}                     
\newcommand{\hatcurPPgxxxxxF}{\ensuremath{12.7\pm1.3}}                       
\newcommand{\hatcurPPloggxxxxxF}{\ensuremath{3.105\pm0.044}}                 
\newcommand{\hatcurPParxxxxxF}{\ensuremath{8.82\pm0.31}}                     
\newcommand{\hatcurPParelxxxxxF}{\ensuremath{0.04354\pm0.00042}}             
\newcommand{\hatcurPPrhoxxxxxF}{\ensuremath{0.543\pm0.076}}                  
\newcommand{\hatcurPPmxxxxxF}{\ensuremath{0.706\pm0.039}}                    
\newcommand{\hatcurPPmshortxxxxxF}{\ensuremath{0.71}}                        
\newcommand{\hatcurPPmlongxxxxxF}{\ensuremath{0.706\pm0.039}}                
\newcommand{\hatcurPPmexxxxxF}{\ensuremath{224\pm12}}                        
\newcommand{\hatcurPPmeshortxxxxxF}{\ensuremath{224.3}}                      
\newcommand{\hatcurPPmelongxxxxxF}{\ensuremath{224\pm12}}                    
\newcommand{\hatcurPPrxxxxxF}{\ensuremath{1.175\pm0.052}}                    
\newcommand{\hatcurPPrshortxxxxxF}{\ensuremath{1.18}}                        
\newcommand{\hatcurPPrlongxxxxxF}{\ensuremath{1.175\pm0.052}}                
\newcommand{\hatcurPPrexxxxxF}{\ensuremath{13.18\pm0.59}}                    
\newcommand{\hatcurPPreshortxxxxxF}{\ensuremath{13.2}}                       
\newcommand{\hatcurPPrelongxxxxxF}{\ensuremath{13.18\pm0.59}}                
\newcommand{\hatcurPPmrcorrxxxxxF}{\ensuremath{0.10}}                        
\newcommand{\hatcurPPteffxxxxxF}{\ensuremath{1414\pm32}}                     
\newcommand{\hatcurPPthetaxxxxxF}{\ensuremath{0.0478\pm0.0033}}              
\newcommand{\hatcurPPfluxperixxxxxF}{\ensuremath{9.01\pm0.80}}               
\newcommand{\hatcurPPfluxperidimxxxxxF}{\ensuremath{8}}                      
\newcommand{\hatcurPPfluxapxxxxxF}{\ensuremath{9.01\pm0.80}}                 
\newcommand{\hatcurPPfluxapdimxxxxxF}{\ensuremath{8}}                        
\newcommand{\hatcurPPfluxavgxxxxxF}{\ensuremath{9.01\pm0.80}}                
\newcommand{\hatcurPPfluxavgdimxxxxxF}{\ensuremath{8}}                       
\newcommand{\hatcurPPfluxavglogxxxxxF}{\ensuremath{8.955\pm0.039}}           
\newcommand{\hatcurXsecphasexxxxxF}{\ensuremath{0\pm0}}                      
\newcommand{\hatcurXsecondaryxxxxxF}{\ensuremath{2456631.34874\pm0.00036}}   
\newcommand{\hatcurXsecdurxxxxxF}{\ensuremath{0.1146\pm0.0012}}              
\newcommand{\hatcurXsecingdurxxxxxF}{\ensuremath{0.0150\pm0.0012}}           
\newcommand{\hatcurPPphiconjxxxxxF}{\ensuremath{0\pm0}}                      
\newcommand{\hatcurPPperixxxxxF}{\ensuremath{2456628.96798\pm0.00036}}       
\newcommand{\hatcurPPaequivxxxxxF}{\ensuremath{0.0389\pm0.0018}}             
\newcommand{\hatcurPPtcircxxxxxF}{\ensuremath{152_{-31}^{+40}}}              
\newcommand{\hatcurPPtinfallxxxxxF}{\ensuremath{2510\pm460}}                 
\newcommand{\hatcurXdistxxxxxF}{\ensuremath{349\pm14}}                       
\newcommand{\hatcurXAvxxxxxF}{\ensuremath{0.0000\pm0.0066}}                  
\newcommand{\hatcurXdistredxxxxxF}{\ensuremath{339\pm16}}                    
\newcommand{\hatcurXEBVxxxxxF}{\ensuremath{0.0000\pm0.0021}}                 
\newcommand{\hatcurXmvisoredxxxxxF}{\ensuremath{12.217\pm0.015}}             
\newcommand{\hatcurXmiisoredxxxxxF}{\ensuremath{11.565\pm0.011}}             
\newcommand{\hatcurXmjisoredxxxxxF}{\ensuremath{11.142\pm0.018}}             
\newcommand{\hatcurXmhisoredxxxxxF}{\ensuremath{10.824\pm0.027}}             
\newcommand{\hatcurXmkisoredxxxxxF}{\ensuremath{10.773\pm0.028}}             
\newcommand{\hatcurXviisoredxxxxxF}{\ensuremath{0.651\pm0.019}}              
\newcommand{\hatcurXvkisoredxxxxxF}{\ensuremath{1.444\pm0.039}}              
\newcommand{\hatcurXjhisoredxxxxxF}{\ensuremath{0.318\pm0.010}}              
\newcommand{\hatcurXjkisoredxxxxxF}{\ensuremath{0.369\pm0.011}}              
\newcommand{\hatcurCCpmraxxxxxF}{\ensuremath{-25.3\pm1.0}}                   
\newcommand{\hatcurCCpmdecxxxxxF}{\ensuremath{-8.2\pm1.0}}                   
\newcommand{\hatcurCCpmxxxxxF}{\ensuremath{26.6\pm1.4}}                      

\newcommand{\hatcurCCbbHmag}[1]{\ifnum#1=25 %
\hatcurCCbbHmagxxxxxA
\else
\ifnum#1=26 %
\hatcurCCbbHmagxxxxxB
\else
\ifnum#1=27 %
\hatcurCCbbHmagxxxxxxC
\else
\ifnum#1=28 %
\hatcurCCbbHmagxxxxxD
\else
\ifnum#1=29 %
\hatcurCCbbHmagxxxxxE
\else
\ifnum#1=30 %
\hatcurCCbbHmagxxxxxF
\else
??????\fi
\fi
\fi
\fi
\fi
\fi
}
\newcommand{\hatcurCCbbJmag}[1]{\ifnum#1=25 %
\hatcurCCbbJmagxxxxxA
\else
\ifnum#1=26 %
\hatcurCCbbJmagxxxxxB
\else
\ifnum#1=27 %
\hatcurCCbbJmagxxxxxxC
\else
\ifnum#1=28 %
\hatcurCCbbJmagxxxxxD
\else
\ifnum#1=29 %
\hatcurCCbbJmagxxxxxE
\else
\ifnum#1=30 %
\hatcurCCbbJmagxxxxxF
\else
??????\fi
\fi
\fi
\fi
\fi
\fi
}
\newcommand{\hatcurCCbbKmag}[1]{\ifnum#1=25 %
\hatcurCCbbKmagxxxxxA
\else
\ifnum#1=26 %
\hatcurCCbbKmagxxxxxB
\else
\ifnum#1=27 %
\hatcurCCbbKmagxxxxxxC
\else
\ifnum#1=28 %
\hatcurCCbbKmagxxxxxD
\else
\ifnum#1=29 %
\hatcurCCbbKmagxxxxxE
\else
\ifnum#1=30 %
\hatcurCCbbKmagxxxxxF
\else
??????\fi
\fi
\fi
\fi
\fi
\fi
}
\newcommand{\hatcurCCcitHmag}[1]{\ifnum#1=25 %
\hatcurCCcitHmagxxxxxA
\else
\ifnum#1=26 %
\hatcurCCcitHmagxxxxxB
\else
\ifnum#1=27 %
\hatcurCCcitHmagxxxxxxC
\else
\ifnum#1=28 %
\hatcurCCcitHmagxxxxxD
\else
\ifnum#1=29 %
\hatcurCCcitHmagxxxxxE
\else
\ifnum#1=30 %
\hatcurCCcitHmagxxxxxF
\else
??????\fi
\fi
\fi
\fi
\fi
\fi
}
\newcommand{\hatcurCCcitJmag}[1]{\ifnum#1=25 %
\hatcurCCcitJmagxxxxxA
\else
\ifnum#1=26 %
\hatcurCCcitJmagxxxxxB
\else
\ifnum#1=27 %
\hatcurCCcitJmagxxxxxxC
\else
\ifnum#1=28 %
\hatcurCCcitJmagxxxxxD
\else
\ifnum#1=29 %
\hatcurCCcitJmagxxxxxE
\else
\ifnum#1=30 %
\hatcurCCcitJmagxxxxxF
\else
??????\fi
\fi
\fi
\fi
\fi
\fi
}
\newcommand{\hatcurCCcitKmag}[1]{\ifnum#1=25 %
\hatcurCCcitKmagxxxxxA
\else
\ifnum#1=26 %
\hatcurCCcitKmagxxxxxB
\else
\ifnum#1=27 %
\hatcurCCcitKmagxxxxxxC
\else
\ifnum#1=28 %
\hatcurCCcitKmagxxxxxD
\else
\ifnum#1=29 %
\hatcurCCcitKmagxxxxxE
\else
\ifnum#1=30 %
\hatcurCCcitKmagxxxxxF
\else
??????\fi
\fi
\fi
\fi
\fi
\fi
}
\newcommand{\hatcurCCdec}[1]{\ifnum#1=25 %
\hatcurCCdecxxxxxA
\else
\ifnum#1=26 %
\hatcurCCdecxxxxxB
\else
\ifnum#1=27 %
\hatcurCCdecxxxxxxC
\else
\ifnum#1=28 %
\hatcurCCdecxxxxxD
\else
\ifnum#1=29 %
\hatcurCCdecxxxxxE
\else
\ifnum#1=30 %
\hatcurCCdecxxxxxF
\else
??????\fi
\fi
\fi
\fi
\fi
\fi
}
\newcommand{\hatcurCCesoHKmag}[1]{\ifnum#1=25 %
\hatcurCCesoHKmagxxxxxA
\else
\ifnum#1=26 %
\hatcurCCesoHKmagxxxxxB
\else
\ifnum#1=27 %
\hatcurCCesoHKmagxxxxxxC
\else
\ifnum#1=28 %
\hatcurCCesoHKmagxxxxxD
\else
\ifnum#1=29 %
\hatcurCCesoHKmagxxxxxE
\else
\ifnum#1=30 %
\hatcurCCesoHKmagxxxxxF
\else
??????\fi
\fi
\fi
\fi
\fi
\fi
}
\newcommand{\hatcurCCesoHmag}[1]{\ifnum#1=25 %
\hatcurCCesoHmagxxxxxA
\else
\ifnum#1=26 %
\hatcurCCesoHmagxxxxxB
\else
\ifnum#1=27 %
\hatcurCCesoHmagxxxxxxC
\else
\ifnum#1=28 %
\hatcurCCesoHmagxxxxxD
\else
\ifnum#1=29 %
\hatcurCCesoHmagxxxxxE
\else
\ifnum#1=30 %
\hatcurCCesoHmagxxxxxF
\else
??????\fi
\fi
\fi
\fi
\fi
\fi
}
\newcommand{\hatcurCCesoJHmag}[1]{\ifnum#1=25 %
\hatcurCCesoJHmagxxxxxA
\else
\ifnum#1=26 %
\hatcurCCesoJHmagxxxxxB
\else
\ifnum#1=27 %
\hatcurCCesoJHmagxxxxxxC
\else
\ifnum#1=28 %
\hatcurCCesoJHmagxxxxxD
\else
\ifnum#1=29 %
\hatcurCCesoJHmagxxxxxE
\else
\ifnum#1=30 %
\hatcurCCesoJHmagxxxxxF
\else
??????\fi
\fi
\fi
\fi
\fi
\fi
}
\newcommand{\hatcurCCesoJKmag}[1]{\ifnum#1=25 %
\hatcurCCesoJKmagxxxxxA
\else
\ifnum#1=26 %
\hatcurCCesoJKmagxxxxxB
\else
\ifnum#1=27 %
\hatcurCCesoJKmagxxxxxxC
\else
\ifnum#1=28 %
\hatcurCCesoJKmagxxxxxD
\else
\ifnum#1=29 %
\hatcurCCesoJKmagxxxxxE
\else
\ifnum#1=30 %
\hatcurCCesoJKmagxxxxxF
\else
??????\fi
\fi
\fi
\fi
\fi
\fi
}
\newcommand{\hatcurCCesoJmag}[1]{\ifnum#1=25 %
\hatcurCCesoJmagxxxxxA
\else
\ifnum#1=26 %
\hatcurCCesoJmagxxxxxB
\else
\ifnum#1=27 %
\hatcurCCesoJmagxxxxxxC
\else
\ifnum#1=28 %
\hatcurCCesoJmagxxxxxD
\else
\ifnum#1=29 %
\hatcurCCesoJmagxxxxxE
\else
\ifnum#1=30 %
\hatcurCCesoJmagxxxxxF
\else
??????\fi
\fi
\fi
\fi
\fi
\fi
}
\newcommand{\hatcurCCesoKmag}[1]{\ifnum#1=25 %
\hatcurCCesoKmagxxxxxA
\else
\ifnum#1=26 %
\hatcurCCesoKmagxxxxxB
\else
\ifnum#1=27 %
\hatcurCCesoKmagxxxxxxC
\else
\ifnum#1=28 %
\hatcurCCesoKmagxxxxxD
\else
\ifnum#1=29 %
\hatcurCCesoKmagxxxxxE
\else
\ifnum#1=30 %
\hatcurCCesoKmagxxxxxF
\else
??????\fi
\fi
\fi
\fi
\fi
\fi
}
\newcommand{\hatcurCCgsc}[1]{\ifnum#1=25 %
\hatcurCCgscxxxxxA
\else
\ifnum#1=26 %
\hatcurCCgscxxxxxB
\else
\ifnum#1=27 %
\hatcurCCgscxxxxxxC
\else
\ifnum#1=28 %
\hatcurCCgscxxxxxD
\else
\ifnum#1=29 %
\hatcurCCgscxxxxxE
\else
\ifnum#1=30 %
\hatcurCCgscxxxxxF
\else
??????\fi
\fi
\fi
\fi
\fi
\fi
}
\newcommand{\hatcurCCmag}[1]{\ifnum#1=25 %
\hatcurCCmagxxxxxA
\else
\ifnum#1=26 %
\hatcurCCmagxxxxxB
\else
\ifnum#1=27 %
\hatcurCCmagxxxxxxC
\else
\ifnum#1=28 %
\hatcurCCmagxxxxxD
\else
\ifnum#1=29 %
\hatcurCCmagxxxxxE
\else
\ifnum#1=30 %
\hatcurCCmagxxxxxF
\else
??????\fi
\fi
\fi
\fi
\fi
\fi
}
\newcommand{\hatcurCCpm}[1]{\ifnum#1=25 %
\hatcurCCpmxxxxxA
\else
\ifnum#1=26 %
\hatcurCCpmxxxxxB
\else
\ifnum#1=27 %
\hatcurCCpmxxxxxxC
\else
\ifnum#1=28 %
\hatcurCCpmxxxxxD
\else
\ifnum#1=29 %
\hatcurCCpmxxxxxE
\else
\ifnum#1=30 %
\hatcurCCpmxxxxxF
\else
??????\fi
\fi
\fi
\fi
\fi
\fi
}
\newcommand{\hatcurCCpmdec}[1]{\ifnum#1=25 %
\hatcurCCpmdecxxxxxA
\else
\ifnum#1=26 %
\hatcurCCpmdecxxxxxB
\else
\ifnum#1=27 %
\hatcurCCpmdecxxxxxxC
\else
\ifnum#1=28 %
\hatcurCCpmdecxxxxxD
\else
\ifnum#1=29 %
\hatcurCCpmdecxxxxxE
\else
\ifnum#1=30 %
\hatcurCCpmdecxxxxxF
\else
??????\fi
\fi
\fi
\fi
\fi
\fi
}
\newcommand{\hatcurCCpmra}[1]{\ifnum#1=25 %
\hatcurCCpmraxxxxxA
\else
\ifnum#1=26 %
\hatcurCCpmraxxxxxB
\else
\ifnum#1=27 %
\hatcurCCpmraxxxxxxC
\else
\ifnum#1=28 %
\hatcurCCpmraxxxxxD
\else
\ifnum#1=29 %
\hatcurCCpmraxxxxxE
\else
\ifnum#1=30 %
\hatcurCCpmraxxxxxF
\else
??????\fi
\fi
\fi
\fi
\fi
\fi
}
\newcommand{\hatcurCCra}[1]{\ifnum#1=25 %
\hatcurCCraxxxxxA
\else
\ifnum#1=26 %
\hatcurCCraxxxxxB
\else
\ifnum#1=27 %
\hatcurCCraxxxxxxC
\else
\ifnum#1=28 %
\hatcurCCraxxxxxD
\else
\ifnum#1=29 %
\hatcurCCraxxxxxE
\else
\ifnum#1=30 %
\hatcurCCraxxxxxF
\else
??????\fi
\fi
\fi
\fi
\fi
\fi
}
\newcommand{\hatcurCCtassmB}[1]{\ifnum#1=25 %
\hatcurCCtassmBxxxxxA
\else
\ifnum#1=26 %
\hatcurCCtassmBxxxxxB
\else
\ifnum#1=27 %
\hatcurCCtassmBxxxxxxC
\else
\ifnum#1=28 %
\hatcurCCtassmBxxxxxD
\else
\ifnum#1=29 %
\hatcurCCtassmBxxxxxE
\else
\ifnum#1=30 %
\hatcurCCtassmBxxxxxF
\else
??????\fi
\fi
\fi
\fi
\fi
\fi
}
\newcommand{\hatcurCCtassmBshort}[1]{\ifnum#1=25 %
\hatcurCCtassmBshortxxxxxA
\else
\ifnum#1=26 %
\hatcurCCtassmBshortxxxxxB
\else
\ifnum#1=27 %
\hatcurCCtassmBshortxxxxxxC
\else
\ifnum#1=28 %
\hatcurCCtassmBshortxxxxxD
\else
\ifnum#1=29 %
\hatcurCCtassmBshortxxxxxE
\else
\ifnum#1=30 %
\hatcurCCtassmBshortxxxxxF
\else
??????\fi
\fi
\fi
\fi
\fi
\fi
}
\newcommand{\hatcurCCtassmg}[1]{\ifnum#1=25 %
\hatcurCCtassmgxxxxxA
\else
\ifnum#1=26 %
\hatcurCCtassmgxxxxxB
\else
\ifnum#1=27 %
\hatcurCCtassmgxxxxxxC
\else
\ifnum#1=28 %
\hatcurCCtassmgxxxxxD
\else
\ifnum#1=29 %
\hatcurCCtassmgxxxxxE
\else
\ifnum#1=30 %
\hatcurCCtassmgxxxxxF
\else
??????\fi
\fi
\fi
\fi
\fi
\fi
}
\newcommand{\hatcurCCtassmgshort}[1]{\ifnum#1=25 %
\hatcurCCtassmgshortxxxxxA
\else
\ifnum#1=26 %
\hatcurCCtassmgshortxxxxxB
\else
\ifnum#1=27 %
\hatcurCCtassmgshortxxxxxxC
\else
\ifnum#1=28 %
\hatcurCCtassmgshortxxxxxD
\else
\ifnum#1=29 %
\hatcurCCtassmgshortxxxxxE
\else
\ifnum#1=30 %
\hatcurCCtassmgshortxxxxxF
\else
??????\fi
\fi
\fi
\fi
\fi
\fi
}
\newcommand{\hatcurCCtassmi}[1]{\ifnum#1=25 %
\hatcurCCtassmixxxxxA
\else
\ifnum#1=26 %
\hatcurCCtassmixxxxxB
\else
\ifnum#1=27 %
\hatcurCCtassmixxxxxxC
\else
\ifnum#1=28 %
\hatcurCCtassmixxxxxD
\else
\ifnum#1=29 %
\hatcurCCtassmixxxxxE
\else
\ifnum#1=30 %
\hatcurCCtassmixxxxxF
\else
??????\fi
\fi
\fi
\fi
\fi
\fi
}
\newcommand{\hatcurCCtassmI}[1]{\ifnum#1=25 %
\hatcurCCtassmIxxxxxA
\else
\ifnum#1=26 %
\hatcurCCtassmIxxxxxB
\else
\ifnum#1=27 %
\hatcurCCtassmIxxxxxxC
\else
\ifnum#1=28 %
\hatcurCCtassmIxxxxxD
\else
\ifnum#1=29 %
\hatcurCCtassmIxxxxxE
\else
\ifnum#1=30 %
\hatcurCCtassmIxxxxxF
\else
??????\fi
\fi
\fi
\fi
\fi
\fi
}
\newcommand{\hatcurCCtassmishort}[1]{\ifnum#1=25 %
\hatcurCCtassmishortxxxxxA
\else
\ifnum#1=26 %
\hatcurCCtassmishortxxxxxB
\else
\ifnum#1=27 %
\hatcurCCtassmishortxxxxxxC
\else
\ifnum#1=28 %
\hatcurCCtassmishortxxxxxD
\else
\ifnum#1=29 %
\hatcurCCtassmishortxxxxxE
\else
\ifnum#1=30 %
\hatcurCCtassmishortxxxxxF
\else
??????\fi
\fi
\fi
\fi
\fi
\fi
}
\newcommand{\hatcurCCtassmIshort}[1]{\ifnum#1=25 %
\hatcurCCtassmIshortxxxxxA
\else
\ifnum#1=26 %
\hatcurCCtassmIshortxxxxxB
\else
\ifnum#1=27 %
\hatcurCCtassmIshortxxxxxxC
\else
\ifnum#1=28 %
\hatcurCCtassmIshortxxxxxD
\else
\ifnum#1=29 %
\hatcurCCtassmIshortxxxxxE
\else
\ifnum#1=30 %
\hatcurCCtassmIshortxxxxxF
\else
??????\fi
\fi
\fi
\fi
\fi
\fi
}
\newcommand{\hatcurCCtassmr}[1]{\ifnum#1=25 %
\hatcurCCtassmrxxxxxA
\else
\ifnum#1=26 %
\hatcurCCtassmrxxxxxB
\else
\ifnum#1=27 %
\hatcurCCtassmrxxxxxxC
\else
\ifnum#1=28 %
\hatcurCCtassmrxxxxxD
\else
\ifnum#1=29 %
\hatcurCCtassmrxxxxxE
\else
\ifnum#1=30 %
\hatcurCCtassmrxxxxxF
\else
??????\fi
\fi
\fi
\fi
\fi
\fi
}
\newcommand{\hatcurCCtassmrshort}[1]{\ifnum#1=25 %
\hatcurCCtassmrshortxxxxxA
\else
\ifnum#1=26 %
\hatcurCCtassmrshortxxxxxB
\else
\ifnum#1=27 %
\hatcurCCtassmrshortxxxxxxC
\else
\ifnum#1=28 %
\hatcurCCtassmrshortxxxxxD
\else
\ifnum#1=29 %
\hatcurCCtassmrshortxxxxxE
\else
\ifnum#1=30 %
\hatcurCCtassmrshortxxxxxF
\else
??????\fi
\fi
\fi
\fi
\fi
\fi
}
\newcommand{\hatcurCCtassmv}[1]{\ifnum#1=25 %
\hatcurCCtassmvxxxxxA
\else
\ifnum#1=26 %
\hatcurCCtassmvxxxxxB
\else
\ifnum#1=27 %
\hatcurCCtassmvxxxxxxC
\else
\ifnum#1=28 %
\hatcurCCtassmvxxxxxD
\else
\ifnum#1=29 %
\hatcurCCtassmvxxxxxE
\else
\ifnum#1=30 %
\hatcurCCtassmvxxxxxF
\else
??????\fi
\fi
\fi
\fi
\fi
\fi
}
\newcommand{\hatcurCCtassmvshort}[1]{\ifnum#1=25 %
\hatcurCCtassmvshortxxxxxA
\else
\ifnum#1=26 %
\hatcurCCtassmvshortxxxxxB
\else
\ifnum#1=27 %
\hatcurCCtassmvshortxxxxxxC
\else
\ifnum#1=28 %
\hatcurCCtassmvshortxxxxxD
\else
\ifnum#1=29 %
\hatcurCCtassmvshortxxxxxE
\else
\ifnum#1=30 %
\hatcurCCtassmvshortxxxxxF
\else
??????\fi
\fi
\fi
\fi
\fi
\fi
}
\newcommand{\hatcurCCtwomass}[1]{\ifnum#1=25 %
\hatcurCCtwomassxxxxxA
\else
\ifnum#1=26 %
\hatcurCCtwomassxxxxxB
\else
\ifnum#1=27 %
\hatcurCCtwomassxxxxxxC
\else
\ifnum#1=28 %
\hatcurCCtwomassxxxxxD
\else
\ifnum#1=29 %
\hatcurCCtwomassxxxxxE
\else
\ifnum#1=30 %
\hatcurCCtwomassxxxxxF
\else
??????\fi
\fi
\fi
\fi
\fi
\fi
}
\newcommand{\hatcurCCtwomassHmag}[1]{\ifnum#1=25 %
\hatcurCCtwomassHmagxxxxxA
\else
\ifnum#1=26 %
\hatcurCCtwomassHmagxxxxxB
\else
\ifnum#1=27 %
\hatcurCCtwomassHmagxxxxxxC
\else
\ifnum#1=28 %
\hatcurCCtwomassHmagxxxxxD
\else
\ifnum#1=29 %
\hatcurCCtwomassHmagxxxxxE
\else
\ifnum#1=30 %
\hatcurCCtwomassHmagxxxxxF
\else
??????\fi
\fi
\fi
\fi
\fi
\fi
}
\newcommand{\hatcurCCtwomassJmag}[1]{\ifnum#1=25 %
\hatcurCCtwomassJmagxxxxxA
\else
\ifnum#1=26 %
\hatcurCCtwomassJmagxxxxxB
\else
\ifnum#1=27 %
\hatcurCCtwomassJmagxxxxxxC
\else
\ifnum#1=28 %
\hatcurCCtwomassJmagxxxxxD
\else
\ifnum#1=29 %
\hatcurCCtwomassJmagxxxxxE
\else
\ifnum#1=30 %
\hatcurCCtwomassJmagxxxxxF
\else
??????\fi
\fi
\fi
\fi
\fi
\fi
}
\newcommand{\hatcurCCtwomassKmag}[1]{\ifnum#1=25 %
\hatcurCCtwomassKmagxxxxxA
\else
\ifnum#1=26 %
\hatcurCCtwomassKmagxxxxxB
\else
\ifnum#1=27 %
\hatcurCCtwomassKmagxxxxxxC
\else
\ifnum#1=28 %
\hatcurCCtwomassKmagxxxxxD
\else
\ifnum#1=29 %
\hatcurCCtwomassKmagxxxxxE
\else
\ifnum#1=30 %
\hatcurCCtwomassKmagxxxxxF
\else
??????\fi
\fi
\fi
\fi
\fi
\fi
}
\newcommand{\hatcurfield}[1]{\ifnum#1=25 %
\hatcurfieldxxxxxA
\else
\ifnum#1=26 %
\hatcurfieldxxxxxB
\else
\ifnum#1=27 %
\hatcurfieldxxxxxxC
\else
\ifnum#1=28 %
\hatcurfieldxxxxxD
\else
\ifnum#1=29 %
\hatcurfieldxxxxxE
\else
\ifnum#1=30 %
\hatcurfieldxxxxxF
\else
??????\fi
\fi
\fi
\fi
\fi
\fi
}
\newcommand{\hatcurhtr}[1]{\ifnum#1=25 %
\hatcurhtrxxxxxA
\else
\ifnum#1=26 %
\hatcurhtrxxxxxB
\else
\ifnum#1=27 %
\hatcurhtrxxxxxxC
\else
\ifnum#1=28 %
\hatcurhtrxxxxxD
\else
\ifnum#1=29 %
\hatcurhtrxxxxxE
\else
\ifnum#1=30 %
\hatcurhtrxxxxxF
\else
??????\fi
\fi
\fi
\fi
\fi
\fi
}
\newcommand{\hatcurISOage}[1]{\ifnum#1=25 %
\hatcurISOagexxxxxA
\else
\ifnum#1=26 %
\hatcurISOagexxxxxB
\else
\ifnum#1=27 %
\hatcurISOagexxxxxxC
\else
\ifnum#1=28 %
\hatcurISOagexxxxxD
\else
\ifnum#1=29 %
\hatcurISOagexxxxxE
\else
\ifnum#1=30 %
\hatcurISOagexxxxxF
\else
??????\fi
\fi
\fi
\fi
\fi
\fi
}
\newcommand{\hatcurISOJK}[1]{\ifnum#1=25 %
\hatcurISOJKxxxxxA
\else
\ifnum#1=26 %
\hatcurISOJKxxxxxB
\else
\ifnum#1=27 %
\hatcurISOJKxxxxxxC
\else
\ifnum#1=28 %
\hatcurISOJKxxxxxD
\else
\ifnum#1=29 %
\hatcurISOJKxxxxxE
\else
\ifnum#1=30 %
\hatcurISOJKxxxxxF
\else
??????\fi
\fi
\fi
\fi
\fi
\fi
}
\newcommand{\hatcurISOlogg}[1]{\ifnum#1=25 %
\hatcurISOloggxxxxxA
\else
\ifnum#1=26 %
\hatcurISOloggxxxxxB
\else
\ifnum#1=27 %
\hatcurISOloggxxxxxxC
\else
\ifnum#1=28 %
\hatcurISOloggxxxxxD
\else
\ifnum#1=29 %
\hatcurISOloggxxxxxE
\else
\ifnum#1=30 %
\hatcurISOloggxxxxxF
\else
??????\fi
\fi
\fi
\fi
\fi
\fi
}
\newcommand{\hatcurISOlum}[1]{\ifnum#1=25 %
\hatcurISOlumxxxxxA
\else
\ifnum#1=26 %
\hatcurISOlumxxxxxB
\else
\ifnum#1=27 %
\hatcurISOlumxxxxxxC
\else
\ifnum#1=28 %
\hatcurISOlumxxxxxD
\else
\ifnum#1=29 %
\hatcurISOlumxxxxxE
\else
\ifnum#1=30 %
\hatcurISOlumxxxxxF
\else
??????\fi
\fi
\fi
\fi
\fi
\fi
}
\newcommand{\hatcurISOlumshort}[1]{\ifnum#1=25 %
\hatcurISOlumshortxxxxxA
\else
\ifnum#1=26 %
\hatcurISOlumshortxxxxxB
\else
\ifnum#1=27 %
\hatcurISOlumshortxxxxxxC
\else
\ifnum#1=28 %
\hatcurISOlumshortxxxxxD
\else
\ifnum#1=29 %
\hatcurISOlumshortxxxxxE
\else
\ifnum#1=30 %
\hatcurISOlumshortxxxxxF
\else
??????\fi
\fi
\fi
\fi
\fi
\fi
}
\newcommand{\hatcurISOm}[1]{\ifnum#1=25 %
\hatcurISOmxxxxxA
\else
\ifnum#1=26 %
\hatcurISOmxxxxxB
\else
\ifnum#1=27 %
\hatcurISOmxxxxxxC
\else
\ifnum#1=28 %
\hatcurISOmxxxxxD
\else
\ifnum#1=29 %
\hatcurISOmxxxxxE
\else
\ifnum#1=30 %
\hatcurISOmxxxxxF
\else
??????\fi
\fi
\fi
\fi
\fi
\fi
}
\newcommand{\hatcurISOMH}[1]{\ifnum#1=25 %
\hatcurISOMHxxxxxA
\else
\ifnum#1=26 %
\hatcurISOMHxxxxxB
\else
\ifnum#1=27 %
\hatcurISOMHxxxxxxC
\else
\ifnum#1=28 %
\hatcurISOMHxxxxxD
\else
\ifnum#1=29 %
\hatcurISOMHxxxxxE
\else
\ifnum#1=30 %
\hatcurISOMHxxxxxF
\else
??????\fi
\fi
\fi
\fi
\fi
\fi
}
\newcommand{\hatcurISOMJ}[1]{\ifnum#1=25 %
\hatcurISOMJxxxxxA
\else
\ifnum#1=26 %
\hatcurISOMJxxxxxB
\else
\ifnum#1=27 %
\hatcurISOMJxxxxxxC
\else
\ifnum#1=28 %
\hatcurISOMJxxxxxD
\else
\ifnum#1=29 %
\hatcurISOMJxxxxxE
\else
\ifnum#1=30 %
\hatcurISOMJxxxxxF
\else
??????\fi
\fi
\fi
\fi
\fi
\fi
}
\newcommand{\hatcurISOMK}[1]{\ifnum#1=25 %
\hatcurISOMKxxxxxA
\else
\ifnum#1=26 %
\hatcurISOMKxxxxxB
\else
\ifnum#1=27 %
\hatcurISOMKxxxxxxC
\else
\ifnum#1=28 %
\hatcurISOMKxxxxxD
\else
\ifnum#1=29 %
\hatcurISOMKxxxxxE
\else
\ifnum#1=30 %
\hatcurISOMKxxxxxF
\else
??????\fi
\fi
\fi
\fi
\fi
\fi
}
\newcommand{\hatcurISOmlong}[1]{\ifnum#1=25 %
\hatcurISOmlongxxxxxA
\else
\ifnum#1=26 %
\hatcurISOmlongxxxxxB
\else
\ifnum#1=27 %
\hatcurISOmlongxxxxxxC
\else
\ifnum#1=28 %
\hatcurISOmlongxxxxxD
\else
\ifnum#1=29 %
\hatcurISOmlongxxxxxE
\else
\ifnum#1=30 %
\hatcurISOmlongxxxxxF
\else
??????\fi
\fi
\fi
\fi
\fi
\fi
}
\newcommand{\hatcurISOmshort}[1]{\ifnum#1=25 %
\hatcurISOmshortxxxxxA
\else
\ifnum#1=26 %
\hatcurISOmshortxxxxxB
\else
\ifnum#1=27 %
\hatcurISOmshortxxxxxxC
\else
\ifnum#1=28 %
\hatcurISOmshortxxxxxD
\else
\ifnum#1=29 %
\hatcurISOmshortxxxxxE
\else
\ifnum#1=30 %
\hatcurISOmshortxxxxxF
\else
??????\fi
\fi
\fi
\fi
\fi
\fi
}
\newcommand{\hatcurISOmv}[1]{\ifnum#1=25 %
\hatcurISOmvxxxxxA
\else
\ifnum#1=26 %
\hatcurISOmvxxxxxB
\else
\ifnum#1=27 %
\hatcurISOmvxxxxxxC
\else
\ifnum#1=28 %
\hatcurISOmvxxxxxD
\else
\ifnum#1=29 %
\hatcurISOmvxxxxxE
\else
\ifnum#1=30 %
\hatcurISOmvxxxxxF
\else
??????\fi
\fi
\fi
\fi
\fi
\fi
}
\newcommand{\hatcurISOr}[1]{\ifnum#1=25 %
\hatcurISOrxxxxxA
\else
\ifnum#1=26 %
\hatcurISOrxxxxxB
\else
\ifnum#1=27 %
\hatcurISOrxxxxxxC
\else
\ifnum#1=28 %
\hatcurISOrxxxxxD
\else
\ifnum#1=29 %
\hatcurISOrxxxxxE
\else
\ifnum#1=30 %
\hatcurISOrxxxxxF
\else
??????\fi
\fi
\fi
\fi
\fi
\fi
}
\newcommand{\hatcurISOrho}[1]{\ifnum#1=25 %
\hatcurISOrhoxxxxxA
\else
\ifnum#1=26 %
\hatcurISOrhoxxxxxB
\else
\ifnum#1=27 %
\hatcurISOrhoxxxxxxC
\else
\ifnum#1=28 %
\hatcurISOrhoxxxxxD
\else
\ifnum#1=29 %
\hatcurISOrhoxxxxxE
\else
\ifnum#1=30 %
\hatcurISOrhoxxxxxF
\else
??????\fi
\fi
\fi
\fi
\fi
\fi
}
\newcommand{\hatcurISOrholong}[1]{\ifnum#1=25 %
\hatcurISOrholongxxxxxA
\else
\ifnum#1=26 %
\hatcurISOrholongxxxxxB
\else
\ifnum#1=27 %
\hatcurISOrholongxxxxxxC
\else
\ifnum#1=28 %
\hatcurISOrholongxxxxxD
\else
\ifnum#1=29 %
\hatcurISOrholongxxxxxE
\else
\ifnum#1=30 %
\hatcurISOrholongxxxxxF
\else
??????\fi
\fi
\fi
\fi
\fi
\fi
}
\newcommand{\hatcurISOrlong}[1]{\ifnum#1=25 %
\hatcurISOrlongxxxxxA
\else
\ifnum#1=26 %
\hatcurISOrlongxxxxxB
\else
\ifnum#1=27 %
\hatcurISOrlongxxxxxxC
\else
\ifnum#1=28 %
\hatcurISOrlongxxxxxD
\else
\ifnum#1=29 %
\hatcurISOrlongxxxxxE
\else
\ifnum#1=30 %
\hatcurISOrlongxxxxxF
\else
??????\fi
\fi
\fi
\fi
\fi
\fi
}
\newcommand{\hatcurISOrshort}[1]{\ifnum#1=25 %
\hatcurISOrshortxxxxxA
\else
\ifnum#1=26 %
\hatcurISOrshortxxxxxB
\else
\ifnum#1=27 %
\hatcurISOrshortxxxxxxC
\else
\ifnum#1=28 %
\hatcurISOrshortxxxxxD
\else
\ifnum#1=29 %
\hatcurISOrshortxxxxxE
\else
\ifnum#1=30 %
\hatcurISOrshortxxxxxF
\else
??????\fi
\fi
\fi
\fi
\fi
\fi
}
\newcommand{\hatcurISOsigma}[1]{\ifnum#1=25 %
\hatcurISOsigmaxxxxxA
\else
\ifnum#1=26 %
\hatcurISOsigmaxxxxxB
\else
\ifnum#1=27 %
\hatcurISOsigmaxxxxxxC
\else
\ifnum#1=28 %
\hatcurISOsigmaxxxxxD
\else
\ifnum#1=29 %
\hatcurISOsigmaxxxxxE
\else
\ifnum#1=30 %
\hatcurISOsigmaxxxxxF
\else
??????\fi
\fi
\fi
\fi
\fi
\fi
}
\newcommand{\hatcurISOspec}[1]{\ifnum#1=25 %
\hatcurISOspecxxxxxA
\else
\ifnum#1=26 %
\hatcurISOspecxxxxxB
\else
\ifnum#1=27 %
\hatcurISOspecxxxxxxC
\else
\ifnum#1=28 %
\hatcurISOspecxxxxxD
\else
\ifnum#1=29 %
\hatcurISOspecxxxxxE
\else
\ifnum#1=30 %
\hatcurISOspecxxxxxF
\else
??????\fi
\fi
\fi
\fi
\fi
\fi
}
\newcommand{\hatcurISOvi}[1]{\ifnum#1=25 %
\hatcurISOvixxxxxA
\else
\ifnum#1=26 %
\hatcurISOvixxxxxB
\else
\ifnum#1=27 %
\hatcurISOvixxxxxxC
\else
\ifnum#1=28 %
\hatcurISOvixxxxxD
\else
\ifnum#1=29 %
\hatcurISOvixxxxxE
\else
\ifnum#1=30 %
\hatcurISOvixxxxxF
\else
??????\fi
\fi
\fi
\fi
\fi
\fi
}
\newcommand{\hatcurLBig}[1]{\ifnum#1=25 %
\hatcurLBigxxxxxA
\else
\ifnum#1=26 %
\hatcurLBigxxxxxB
\else
\ifnum#1=27 %
\hatcurLBigxxxxxxC
\else
\ifnum#1=28 %
\hatcurLBigxxxxxD
\else
\ifnum#1=29 %
\hatcurLBigxxxxxE
\else
\ifnum#1=30 %
\hatcurLBigxxxxxF
\else
??????\fi
\fi
\fi
\fi
\fi
\fi
}
\newcommand{\hatcurLBii}[1]{\ifnum#1=25 %
\hatcurLBiixxxxxA
\else
\ifnum#1=26 %
\hatcurLBiixxxxxB
\else
\ifnum#1=27 %
\hatcurLBiixxxxxxC
\else
\ifnum#1=28 %
\hatcurLBiixxxxxD
\else
\ifnum#1=29 %
\hatcurLBiixxxxxE
\else
\ifnum#1=30 %
\hatcurLBiixxxxxF
\else
??????\fi
\fi
\fi
\fi
\fi
\fi
}
\newcommand{\hatcurLBiI}[1]{\ifnum#1=25 %
\hatcurLBiIxxxxxA
\else
\ifnum#1=26 %
\hatcurLBiIxxxxxB
\else
\ifnum#1=27 %
\hatcurLBiIxxxxxxC
\else
\ifnum#1=28 %
\hatcurLBiIxxxxxD
\else
\ifnum#1=29 %
\hatcurLBiIxxxxxE
\else
\ifnum#1=30 %
\hatcurLBiIxxxxxF
\else
??????\fi
\fi
\fi
\fi
\fi
\fi
}
\newcommand{\hatcurLBiig}[1]{\ifnum#1=25 %
\hatcurLBiigxxxxxA
\else
\ifnum#1=26 %
\hatcurLBiigxxxxxB
\else
\ifnum#1=27 %
\hatcurLBiigxxxxxxC
\else
\ifnum#1=28 %
\hatcurLBiigxxxxxD
\else
\ifnum#1=29 %
\hatcurLBiigxxxxxE
\else
\ifnum#1=30 %
\hatcurLBiigxxxxxF
\else
??????\fi
\fi
\fi
\fi
\fi
\fi
}
\newcommand{\hatcurLBiii}[1]{\ifnum#1=25 %
\hatcurLBiiixxxxxA
\else
\ifnum#1=26 %
\hatcurLBiiixxxxxB
\else
\ifnum#1=27 %
\hatcurLBiiixxxxxxC
\else
\ifnum#1=28 %
\hatcurLBiiixxxxxD
\else
\ifnum#1=29 %
\hatcurLBiiixxxxxE
\else
\ifnum#1=30 %
\hatcurLBiiixxxxxF
\else
??????\fi
\fi
\fi
\fi
\fi
\fi
}
\newcommand{\hatcurLBiiI}[1]{\ifnum#1=25 %
\hatcurLBiiIxxxxxA
\else
\ifnum#1=26 %
\hatcurLBiiIxxxxxB
\else
\ifnum#1=27 %
\hatcurLBiiIxxxxxxC
\else
\ifnum#1=28 %
\hatcurLBiiIxxxxxD
\else
\ifnum#1=29 %
\hatcurLBiiIxxxxxE
\else
\ifnum#1=30 %
\hatcurLBiiIxxxxxF
\else
??????\fi
\fi
\fi
\fi
\fi
\fi
}
\newcommand{\hatcurLBiikep}[1]{\ifnum#1=25 %
\hatcurLBiikepxxxxxA
\else
\ifnum#1=26 %
\hatcurLBiikepxxxxxB
\else
\ifnum#1=27 %
\hatcurLBiikepxxxxxxC
\else
\ifnum#1=28 %
\hatcurLBiikepxxxxxD
\else
\ifnum#1=29 %
\hatcurLBiikepxxxxxE
\else
\ifnum#1=30 %
\hatcurLBiikepxxxxxF
\else
??????\fi
\fi
\fi
\fi
\fi
\fi
}
\newcommand{\hatcurLBiir}[1]{\ifnum#1=25 %
\hatcurLBiirxxxxxA
\else
\ifnum#1=26 %
\hatcurLBiirxxxxxB
\else
\ifnum#1=27 %
\hatcurLBiirxxxxxxC
\else
\ifnum#1=28 %
\hatcurLBiirxxxxxD
\else
\ifnum#1=29 %
\hatcurLBiirxxxxxE
\else
\ifnum#1=30 %
\hatcurLBiirxxxxxF
\else
??????\fi
\fi
\fi
\fi
\fi
\fi
}
\newcommand{\hatcurLBiiR}[1]{\ifnum#1=25 %
\hatcurLBiiRxxxxxA
\else
\ifnum#1=26 %
\hatcurLBiiRxxxxxB
\else
\ifnum#1=27 %
\hatcurLBiiRxxxxxxC
\else
\ifnum#1=28 %
\hatcurLBiiRxxxxxD
\else
\ifnum#1=29 %
\hatcurLBiiRxxxxxE
\else
\ifnum#1=30 %
\hatcurLBiiRxxxxxF
\else
??????\fi
\fi
\fi
\fi
\fi
\fi
}
\newcommand{\hatcurLBiiz}[1]{\ifnum#1=25 %
\hatcurLBiizxxxxxA
\else
\ifnum#1=26 %
\hatcurLBiizxxxxxB
\else
\ifnum#1=27 %
\hatcurLBiizxxxxxxC
\else
\ifnum#1=28 %
\hatcurLBiizxxxxxD
\else
\ifnum#1=29 %
\hatcurLBiizxxxxxE
\else
\ifnum#1=30 %
\hatcurLBiizxxxxxF
\else
??????\fi
\fi
\fi
\fi
\fi
\fi
}
\newcommand{\hatcurLBikep}[1]{\ifnum#1=25 %
\hatcurLBikepxxxxxA
\else
\ifnum#1=26 %
\hatcurLBikepxxxxxB
\else
\ifnum#1=27 %
\hatcurLBikepxxxxxxC
\else
\ifnum#1=28 %
\hatcurLBikepxxxxxD
\else
\ifnum#1=29 %
\hatcurLBikepxxxxxE
\else
\ifnum#1=30 %
\hatcurLBikepxxxxxF
\else
??????\fi
\fi
\fi
\fi
\fi
\fi
}
\newcommand{\hatcurLBir}[1]{\ifnum#1=25 %
\hatcurLBirxxxxxA
\else
\ifnum#1=26 %
\hatcurLBirxxxxxB
\else
\ifnum#1=27 %
\hatcurLBirxxxxxxC
\else
\ifnum#1=28 %
\hatcurLBirxxxxxD
\else
\ifnum#1=29 %
\hatcurLBirxxxxxE
\else
\ifnum#1=30 %
\hatcurLBirxxxxxF
\else
??????\fi
\fi
\fi
\fi
\fi
\fi
}
\newcommand{\hatcurLBiR}[1]{\ifnum#1=25 %
\hatcurLBiRxxxxxA
\else
\ifnum#1=26 %
\hatcurLBiRxxxxxB
\else
\ifnum#1=27 %
\hatcurLBiRxxxxxxC
\else
\ifnum#1=28 %
\hatcurLBiRxxxxxD
\else
\ifnum#1=29 %
\hatcurLBiRxxxxxE
\else
\ifnum#1=30 %
\hatcurLBiRxxxxxF
\else
??????\fi
\fi
\fi
\fi
\fi
\fi
}
\newcommand{\hatcurLBiz}[1]{\ifnum#1=25 %
\hatcurLBizxxxxxA
\else
\ifnum#1=26 %
\hatcurLBizxxxxxB
\else
\ifnum#1=27 %
\hatcurLBizxxxxxxC
\else
\ifnum#1=28 %
\hatcurLBizxxxxxD
\else
\ifnum#1=29 %
\hatcurLBizxxxxxE
\else
\ifnum#1=30 %
\hatcurLBizxxxxxF
\else
??????\fi
\fi
\fi
\fi
\fi
\fi
}
\newcommand{\hatcurLCbsq}[1]{\ifnum#1=25 %
\hatcurLCbsqxxxxxA
\else
\ifnum#1=26 %
\hatcurLCbsqxxxxxB
\else
\ifnum#1=27 %
\hatcurLCbsqxxxxxxC
\else
\ifnum#1=28 %
\hatcurLCbsqxxxxxD
\else
\ifnum#1=29 %
\hatcurLCbsqxxxxxE
\else
\ifnum#1=30 %
\hatcurLCbsqxxxxxF
\else
??????\fi
\fi
\fi
\fi
\fi
\fi
}
\newcommand{\hatcurLCdip}[1]{\ifnum#1=25 %
\hatcurLCdipxxxxxA
\else
\ifnum#1=26 %
\hatcurLCdipxxxxxB
\else
\ifnum#1=27 %
\hatcurLCdipxxxxxxC
\else
\ifnum#1=28 %
\hatcurLCdipxxxxxD
\else
\ifnum#1=29 %
\hatcurLCdipxxxxxE
\else
\ifnum#1=30 %
\hatcurLCdipxxxxxF
\else
??????\fi
\fi
\fi
\fi
\fi
\fi
}
\newcommand{\hatcurLCdur}[1]{\ifnum#1=25 %
\hatcurLCdurxxxxxA
\else
\ifnum#1=26 %
\hatcurLCdurxxxxxB
\else
\ifnum#1=27 %
\hatcurLCdurxxxxxxC
\else
\ifnum#1=28 %
\hatcurLCdurxxxxxD
\else
\ifnum#1=29 %
\hatcurLCdurxxxxxE
\else
\ifnum#1=30 %
\hatcurLCdurxxxxxF
\else
??????\fi
\fi
\fi
\fi
\fi
\fi
}
\newcommand{\hatcurLCdurhr}[1]{\ifnum#1=25 %
\hatcurLCdurhrxxxxxA
\else
\ifnum#1=26 %
\hatcurLCdurhrxxxxxB
\else
\ifnum#1=27 %
\hatcurLCdurhrxxxxxxC
\else
\ifnum#1=28 %
\hatcurLCdurhrxxxxxD
\else
\ifnum#1=29 %
\hatcurLCdurhrxxxxxE
\else
\ifnum#1=30 %
\hatcurLCdurhrxxxxxF
\else
??????\fi
\fi
\fi
\fi
\fi
\fi
}
\newcommand{\hatcurLCdurhrshort}[1]{\ifnum#1=25 %
\hatcurLCdurhrshortxxxxxA
\else
\ifnum#1=26 %
\hatcurLCdurhrshortxxxxxB
\else
\ifnum#1=27 %
\hatcurLCdurhrshortxxxxxxC
\else
\ifnum#1=28 %
\hatcurLCdurhrshortxxxxxD
\else
\ifnum#1=29 %
\hatcurLCdurhrshortxxxxxE
\else
\ifnum#1=30 %
\hatcurLCdurhrshortxxxxxF
\else
??????\fi
\fi
\fi
\fi
\fi
\fi
}
\newcommand{\hatcurLCdurshort}[1]{\ifnum#1=25 %
\hatcurLCdurshortxxxxxA
\else
\ifnum#1=26 %
\hatcurLCdurshortxxxxxB
\else
\ifnum#1=27 %
\hatcurLCdurshortxxxxxxC
\else
\ifnum#1=28 %
\hatcurLCdurshortxxxxxD
\else
\ifnum#1=29 %
\hatcurLCdurshortxxxxxE
\else
\ifnum#1=30 %
\hatcurLCdurshortxxxxxF
\else
??????\fi
\fi
\fi
\fi
\fi
\fi
}
\newcommand{\hatcurLChatnetm}[1]{\ifnum#1=25 %
\hatcurLChatnetmxxxxxA
\else
\ifnum#1=26 %
\hatcurLChatnetmxxxxxB
\else
\ifnum#1=27 %
\hatcurLChatnetmxxxxxxC
\else
\ifnum#1=28 %
\hatcurLChatnetmxxxxxD
\else
\ifnum#1=29 %
\hatcurLChatnetmxxxxxE
\else
??????\fi
\fi
\fi
\fi
\fi
}
\newcommand{\hatcurLChatnetmA}[1]{\ifnum#1=30 %
\hatcurLChatnetmAxxxxxF
\else
??????\fi
}
\newcommand{\hatcurLChatnetmB}[1]{\ifnum#1=30 %
\hatcurLChatnetmBxxxxxF
\else
??????\fi
}
\newcommand{\hatcurLChatnetmC}[1]{\ifnum#1=30 %
\hatcurLChatnetmCxxxxxF
\else
??????\fi
}
\newcommand{\hatcurLCiblend}[1]{\ifnum#1=25 %
\hatcurLCiblendxxxxxA
\else
\ifnum#1=26 %
\hatcurLCiblendxxxxxB
\else
\ifnum#1=27 %
\hatcurLCiblendxxxxxxC
\else
\ifnum#1=28 %
\hatcurLCiblendxxxxxD
\else
\ifnum#1=29 %
\hatcurLCiblendxxxxxE
\else
??????\fi
\fi
\fi
\fi
\fi
}
\newcommand{\hatcurLCiblendA}[1]{\ifnum#1=30 %
\hatcurLCiblendAxxxxxF
\else
??????\fi
}
\newcommand{\hatcurLCiblendB}[1]{\ifnum#1=30 %
\hatcurLCiblendBxxxxxF
\else
??????\fi
}
\newcommand{\hatcurLCiblendC}[1]{\ifnum#1=30 %
\hatcurLCiblendCxxxxxF
\else
??????\fi
}
\newcommand{\hatcurLCimp}[1]{\ifnum#1=25 %
\hatcurLCimpxxxxxA
\else
\ifnum#1=26 %
\hatcurLCimpxxxxxB
\else
\ifnum#1=27 %
\hatcurLCimpxxxxxxC
\else
\ifnum#1=28 %
\hatcurLCimpxxxxxD
\else
\ifnum#1=29 %
\hatcurLCimpxxxxxE
\else
\ifnum#1=30 %
\hatcurLCimpxxxxxF
\else
??????\fi
\fi
\fi
\fi
\fi
\fi
}
\newcommand{\hatcurLCingdur}[1]{\ifnum#1=25 %
\hatcurLCingdurxxxxxA
\else
\ifnum#1=26 %
\hatcurLCingdurxxxxxB
\else
\ifnum#1=27 %
\hatcurLCingdurxxxxxxC
\else
\ifnum#1=28 %
\hatcurLCingdurxxxxxD
\else
\ifnum#1=29 %
\hatcurLCingdurxxxxxE
\else
\ifnum#1=30 %
\hatcurLCingdurxxxxxF
\else
??????\fi
\fi
\fi
\fi
\fi
\fi
}
\newcommand{\hatcurLCP}[1]{\ifnum#1=25 %
\hatcurLCPxxxxxA
\else
\ifnum#1=26 %
\hatcurLCPxxxxxB
\else
\ifnum#1=27 %
\hatcurLCPxxxxxxC
\else
\ifnum#1=28 %
\hatcurLCPxxxxxD
\else
\ifnum#1=29 %
\hatcurLCPxxxxxE
\else
\ifnum#1=30 %
\hatcurLCPxxxxxF
\else
??????\fi
\fi
\fi
\fi
\fi
\fi
}
\newcommand{\hatcurLCPprec}[1]{\ifnum#1=25 %
\hatcurLCPprecxxxxxA
\else
\ifnum#1=26 %
\hatcurLCPprecxxxxxB
\else
\ifnum#1=27 %
\hatcurLCPprecxxxxxxC
\else
\ifnum#1=28 %
\hatcurLCPprecxxxxxD
\else
\ifnum#1=29 %
\hatcurLCPprecxxxxxE
\else
\ifnum#1=30 %
\hatcurLCPprecxxxxxF
\else
??????\fi
\fi
\fi
\fi
\fi
\fi
}
\newcommand{\hatcurLCPshort}[1]{\ifnum#1=25 %
\hatcurLCPshortxxxxxA
\else
\ifnum#1=26 %
\hatcurLCPshortxxxxxB
\else
\ifnum#1=27 %
\hatcurLCPshortxxxxxxC
\else
\ifnum#1=28 %
\hatcurLCPshortxxxxxD
\else
\ifnum#1=29 %
\hatcurLCPshortxxxxxE
\else
\ifnum#1=30 %
\hatcurLCPshortxxxxxF
\else
??????\fi
\fi
\fi
\fi
\fi
\fi
}
\newcommand{\hatcurLCq}[1]{\ifnum#1=25 %
\hatcurLCqxxxxxA
\else
\ifnum#1=26 %
\hatcurLCqxxxxxB
\else
\ifnum#1=27 %
\hatcurLCqxxxxxxC
\else
\ifnum#1=28 %
\hatcurLCqxxxxxD
\else
\ifnum#1=29 %
\hatcurLCqxxxxxE
\else
\ifnum#1=30 %
\hatcurLCqxxxxxF
\else
??????\fi
\fi
\fi
\fi
\fi
\fi
}
\newcommand{\hatcurLCqshort}[1]{\ifnum#1=25 %
\hatcurLCqshortxxxxxA
\else
\ifnum#1=26 %
\hatcurLCqshortxxxxxB
\else
\ifnum#1=27 %
\hatcurLCqshortxxxxxxC
\else
\ifnum#1=28 %
\hatcurLCqshortxxxxxD
\else
\ifnum#1=29 %
\hatcurLCqshortxxxxxE
\else
\ifnum#1=30 %
\hatcurLCqshortxxxxxF
\else
??????\fi
\fi
\fi
\fi
\fi
\fi
}
\newcommand{\hatcurLCrho}[1]{\ifnum#1=25 %
\hatcurLCrhoxxxxxA
\else
\ifnum#1=26 %
\hatcurLCrhoxxxxxB
\else
\ifnum#1=27 %
\hatcurLCrhoxxxxxxC
\else
\ifnum#1=28 %
\hatcurLCrhoxxxxxD
\else
\ifnum#1=29 %
\hatcurLCrhoxxxxxE
\else
\ifnum#1=30 %
\hatcurLCrhoxxxxxF
\else
??????\fi
\fi
\fi
\fi
\fi
\fi
}
\newcommand{\hatcurLCrprstar}[1]{\ifnum#1=25 %
\hatcurLCrprstarxxxxxA
\else
\ifnum#1=26 %
\hatcurLCrprstarxxxxxB
\else
\ifnum#1=27 %
\hatcurLCrprstarxxxxxxC
\else
\ifnum#1=28 %
\hatcurLCrprstarxxxxxD
\else
\ifnum#1=29 %
\hatcurLCrprstarxxxxxE
\else
\ifnum#1=30 %
\hatcurLCrprstarxxxxxF
\else
??????\fi
\fi
\fi
\fi
\fi
\fi
}
\newcommand{\hatcurLCT}[1]{\ifnum#1=25 %
\hatcurLCTxxxxxA
\else
\ifnum#1=26 %
\hatcurLCTxxxxxB
\else
\ifnum#1=27 %
\hatcurLCTxxxxxxC
\else
\ifnum#1=28 %
\hatcurLCTxxxxxD
\else
\ifnum#1=29 %
\hatcurLCTxxxxxE
\else
\ifnum#1=30 %
\hatcurLCTxxxxxF
\else
??????\fi
\fi
\fi
\fi
\fi
\fi
}
\newcommand{\hatcurLCTA}[1]{\ifnum#1=25 %
\hatcurLCTAxxxxxA
\else
\ifnum#1=26 %
\hatcurLCTAxxxxxB
\else
\ifnum#1=27 %
\hatcurLCTAxxxxxxC
\else
\ifnum#1=28 %
\hatcurLCTAxxxxxD
\else
\ifnum#1=29 %
\hatcurLCTAxxxxxE
\else
\ifnum#1=30 %
\hatcurLCTAxxxxxF
\else
??????\fi
\fi
\fi
\fi
\fi
\fi
}
\newcommand{\hatcurLCTB}[1]{\ifnum#1=25 %
\hatcurLCTBxxxxxA
\else
\ifnum#1=26 %
\hatcurLCTBxxxxxB
\else
\ifnum#1=27 %
\hatcurLCTBxxxxxxC
\else
\ifnum#1=28 %
\hatcurLCTBxxxxxD
\else
\ifnum#1=29 %
\hatcurLCTBxxxxxE
\else
\ifnum#1=30 %
\hatcurLCTBxxxxxF
\else
??????\fi
\fi
\fi
\fi
\fi
\fi
}
\newcommand{\hatcurLCzeta}[1]{\ifnum#1=25 %
\hatcurLCzetaxxxxxA
\else
\ifnum#1=26 %
\hatcurLCzetaxxxxxB
\else
\ifnum#1=27 %
\hatcurLCzetaxxxxxxC
\else
\ifnum#1=28 %
\hatcurLCzetaxxxxxD
\else
\ifnum#1=29 %
\hatcurLCzetaxxxxxE
\else
\ifnum#1=30 %
\hatcurLCzetaxxxxxF
\else
??????\fi
\fi
\fi
\fi
\fi
\fi
}
\newcommand{\hatcurPPaequiv}[1]{\ifnum#1=25 %
\hatcurPPaequivxxxxxA
\else
\ifnum#1=26 %
\hatcurPPaequivxxxxxB
\else
\ifnum#1=27 %
\hatcurPPaequivxxxxxxC
\else
\ifnum#1=28 %
\hatcurPPaequivxxxxxD
\else
\ifnum#1=29 %
\hatcurPPaequivxxxxxE
\else
\ifnum#1=30 %
\hatcurPPaequivxxxxxF
\else
??????\fi
\fi
\fi
\fi
\fi
\fi
}
\newcommand{\hatcurPPar}[1]{\ifnum#1=25 %
\hatcurPParxxxxxA
\else
\ifnum#1=26 %
\hatcurPParxxxxxB
\else
\ifnum#1=27 %
\hatcurPParxxxxxxC
\else
\ifnum#1=28 %
\hatcurPParxxxxxD
\else
\ifnum#1=29 %
\hatcurPParxxxxxE
\else
\ifnum#1=30 %
\hatcurPParxxxxxF
\else
??????\fi
\fi
\fi
\fi
\fi
\fi
}
\newcommand{\hatcurPParel}[1]{\ifnum#1=25 %
\hatcurPParelxxxxxA
\else
\ifnum#1=26 %
\hatcurPParelxxxxxB
\else
\ifnum#1=27 %
\hatcurPParelxxxxxxC
\else
\ifnum#1=28 %
\hatcurPParelxxxxxD
\else
\ifnum#1=29 %
\hatcurPParelxxxxxE
\else
\ifnum#1=30 %
\hatcurPParelxxxxxF
\else
??????\fi
\fi
\fi
\fi
\fi
\fi
}
\newcommand{\hatcurPPfluxap}[1]{\ifnum#1=25 %
\hatcurPPfluxapxxxxxA
\else
\ifnum#1=26 %
\hatcurPPfluxapxxxxxB
\else
\ifnum#1=27 %
\hatcurPPfluxapxxxxxxC
\else
\ifnum#1=28 %
\hatcurPPfluxapxxxxxD
\else
\ifnum#1=29 %
\hatcurPPfluxapxxxxxE
\else
\ifnum#1=30 %
\hatcurPPfluxapxxxxxF
\else
??????\fi
\fi
\fi
\fi
\fi
\fi
}
\newcommand{\hatcurPPfluxapdim}[1]{\ifnum#1=25 %
\hatcurPPfluxapdimxxxxxA
\else
\ifnum#1=26 %
\hatcurPPfluxapdimxxxxxB
\else
\ifnum#1=27 %
\hatcurPPfluxapdimxxxxxxC
\else
\ifnum#1=28 %
\hatcurPPfluxapdimxxxxxD
\else
\ifnum#1=29 %
\hatcurPPfluxapdimxxxxxE
\else
\ifnum#1=30 %
\hatcurPPfluxapdimxxxxxF
\else
??????\fi
\fi
\fi
\fi
\fi
\fi
}
\newcommand{\hatcurPPfluxavg}[1]{\ifnum#1=25 %
\hatcurPPfluxavgxxxxxA
\else
\ifnum#1=26 %
\hatcurPPfluxavgxxxxxB
\else
\ifnum#1=27 %
\hatcurPPfluxavgxxxxxxC
\else
\ifnum#1=28 %
\hatcurPPfluxavgxxxxxD
\else
\ifnum#1=29 %
\hatcurPPfluxavgxxxxxE
\else
\ifnum#1=30 %
\hatcurPPfluxavgxxxxxF
\else
??????\fi
\fi
\fi
\fi
\fi
\fi
}
\newcommand{\hatcurPPfluxavgdim}[1]{\ifnum#1=25 %
\hatcurPPfluxavgdimxxxxxA
\else
\ifnum#1=26 %
\hatcurPPfluxavgdimxxxxxB
\else
\ifnum#1=27 %
\hatcurPPfluxavgdimxxxxxxC
\else
\ifnum#1=28 %
\hatcurPPfluxavgdimxxxxxD
\else
\ifnum#1=29 %
\hatcurPPfluxavgdimxxxxxE
\else
\ifnum#1=30 %
\hatcurPPfluxavgdimxxxxxF
\else
??????\fi
\fi
\fi
\fi
\fi
\fi
}
\newcommand{\hatcurPPfluxavglog}[1]{\ifnum#1=25 %
\hatcurPPfluxavglogxxxxxA
\else
\ifnum#1=26 %
\hatcurPPfluxavglogxxxxxB
\else
\ifnum#1=27 %
\hatcurPPfluxavglogxxxxxxC
\else
\ifnum#1=28 %
\hatcurPPfluxavglogxxxxxD
\else
\ifnum#1=29 %
\hatcurPPfluxavglogxxxxxE
\else
\ifnum#1=30 %
\hatcurPPfluxavglogxxxxxF
\else
??????\fi
\fi
\fi
\fi
\fi
\fi
}
\newcommand{\hatcurPPfluxperi}[1]{\ifnum#1=25 %
\hatcurPPfluxperixxxxxA
\else
\ifnum#1=26 %
\hatcurPPfluxperixxxxxB
\else
\ifnum#1=27 %
\hatcurPPfluxperixxxxxxC
\else
\ifnum#1=28 %
\hatcurPPfluxperixxxxxD
\else
\ifnum#1=29 %
\hatcurPPfluxperixxxxxE
\else
\ifnum#1=30 %
\hatcurPPfluxperixxxxxF
\else
??????\fi
\fi
\fi
\fi
\fi
\fi
}
\newcommand{\hatcurPPfluxperidim}[1]{\ifnum#1=25 %
\hatcurPPfluxperidimxxxxxA
\else
\ifnum#1=26 %
\hatcurPPfluxperidimxxxxxB
\else
\ifnum#1=27 %
\hatcurPPfluxperidimxxxxxxC
\else
\ifnum#1=28 %
\hatcurPPfluxperidimxxxxxD
\else
\ifnum#1=29 %
\hatcurPPfluxperidimxxxxxE
\else
\ifnum#1=30 %
\hatcurPPfluxperidimxxxxxF
\else
??????\fi
\fi
\fi
\fi
\fi
\fi
}
\newcommand{\hatcurPPg}[1]{\ifnum#1=25 %
\hatcurPPgxxxxxA
\else
\ifnum#1=26 %
\hatcurPPgxxxxxB
\else
\ifnum#1=27 %
\hatcurPPgxxxxxxC
\else
\ifnum#1=28 %
\hatcurPPgxxxxxD
\else
\ifnum#1=29 %
\hatcurPPgxxxxxE
\else
\ifnum#1=30 %
\hatcurPPgxxxxxF
\else
??????\fi
\fi
\fi
\fi
\fi
\fi
}
\newcommand{\hatcurPPi}[1]{\ifnum#1=25 %
\hatcurPPixxxxxA
\else
\ifnum#1=26 %
\hatcurPPixxxxxB
\else
\ifnum#1=27 %
\hatcurPPixxxxxxC
\else
\ifnum#1=28 %
\hatcurPPixxxxxD
\else
\ifnum#1=29 %
\hatcurPPixxxxxE
\else
\ifnum#1=30 %
\hatcurPPixxxxxF
\else
??????\fi
\fi
\fi
\fi
\fi
\fi
}
\newcommand{\hatcurPPlogg}[1]{\ifnum#1=25 %
\hatcurPPloggxxxxxA
\else
\ifnum#1=26 %
\hatcurPPloggxxxxxB
\else
\ifnum#1=27 %
\hatcurPPloggxxxxxxC
\else
\ifnum#1=28 %
\hatcurPPloggxxxxxD
\else
\ifnum#1=29 %
\hatcurPPloggxxxxxE
\else
\ifnum#1=30 %
\hatcurPPloggxxxxxF
\else
??????\fi
\fi
\fi
\fi
\fi
\fi
}
\newcommand{\hatcurPPm}[1]{\ifnum#1=25 %
\hatcurPPmxxxxxA
\else
\ifnum#1=26 %
\hatcurPPmxxxxxB
\else
\ifnum#1=27 %
\hatcurPPmxxxxxxC
\else
\ifnum#1=28 %
\hatcurPPmxxxxxD
\else
\ifnum#1=29 %
\hatcurPPmxxxxxE
\else
\ifnum#1=30 %
\hatcurPPmxxxxxF
\else
??????\fi
\fi
\fi
\fi
\fi
\fi
}
\newcommand{\hatcurPPme}[1]{\ifnum#1=25 %
\hatcurPPmexxxxxA
\else
\ifnum#1=26 %
\hatcurPPmexxxxxB
\else
\ifnum#1=27 %
\hatcurPPmexxxxxxC
\else
\ifnum#1=28 %
\hatcurPPmexxxxxD
\else
\ifnum#1=29 %
\hatcurPPmexxxxxE
\else
\ifnum#1=30 %
\hatcurPPmexxxxxF
\else
??????\fi
\fi
\fi
\fi
\fi
\fi
}
\newcommand{\hatcurPPmelong}[1]{\ifnum#1=25 %
\hatcurPPmelongxxxxxA
\else
\ifnum#1=26 %
\hatcurPPmelongxxxxxB
\else
\ifnum#1=27 %
\hatcurPPmelongxxxxxxC
\else
\ifnum#1=28 %
\hatcurPPmelongxxxxxD
\else
\ifnum#1=29 %
\hatcurPPmelongxxxxxE
\else
\ifnum#1=30 %
\hatcurPPmelongxxxxxF
\else
??????\fi
\fi
\fi
\fi
\fi
\fi
}
\newcommand{\hatcurPPmeshort}[1]{\ifnum#1=25 %
\hatcurPPmeshortxxxxxA
\else
\ifnum#1=26 %
\hatcurPPmeshortxxxxxB
\else
\ifnum#1=27 %
\hatcurPPmeshortxxxxxxC
\else
\ifnum#1=28 %
\hatcurPPmeshortxxxxxD
\else
\ifnum#1=29 %
\hatcurPPmeshortxxxxxE
\else
\ifnum#1=30 %
\hatcurPPmeshortxxxxxF
\else
??????\fi
\fi
\fi
\fi
\fi
\fi
}
\newcommand{\hatcurPPmlong}[1]{\ifnum#1=25 %
\hatcurPPmlongxxxxxA
\else
\ifnum#1=26 %
\hatcurPPmlongxxxxxB
\else
\ifnum#1=27 %
\hatcurPPmlongxxxxxxC
\else
\ifnum#1=28 %
\hatcurPPmlongxxxxxD
\else
\ifnum#1=29 %
\hatcurPPmlongxxxxxE
\else
\ifnum#1=30 %
\hatcurPPmlongxxxxxF
\else
??????\fi
\fi
\fi
\fi
\fi
\fi
}
\newcommand{\hatcurPPmrcorr}[1]{\ifnum#1=25 %
\hatcurPPmrcorrxxxxxA
\else
\ifnum#1=26 %
\hatcurPPmrcorrxxxxxB
\else
\ifnum#1=27 %
\hatcurPPmrcorrxxxxxxC
\else
\ifnum#1=28 %
\hatcurPPmrcorrxxxxxD
\else
\ifnum#1=29 %
\hatcurPPmrcorrxxxxxE
\else
\ifnum#1=30 %
\hatcurPPmrcorrxxxxxF
\else
??????\fi
\fi
\fi
\fi
\fi
\fi
}
\newcommand{\hatcurPPmshort}[1]{\ifnum#1=25 %
\hatcurPPmshortxxxxxA
\else
\ifnum#1=26 %
\hatcurPPmshortxxxxxB
\else
\ifnum#1=27 %
\hatcurPPmshortxxxxxxC
\else
\ifnum#1=28 %
\hatcurPPmshortxxxxxD
\else
\ifnum#1=29 %
\hatcurPPmshortxxxxxE
\else
\ifnum#1=30 %
\hatcurPPmshortxxxxxF
\else
??????\fi
\fi
\fi
\fi
\fi
\fi
}
\newcommand{\hatcurPPperi}[1]{\ifnum#1=25 %
\hatcurPPperixxxxxA
\else
\ifnum#1=26 %
\hatcurPPperixxxxxB
\else
\ifnum#1=27 %
\hatcurPPperixxxxxxC
\else
\ifnum#1=28 %
\hatcurPPperixxxxxD
\else
\ifnum#1=29 %
\hatcurPPperixxxxxE
\else
\ifnum#1=30 %
\hatcurPPperixxxxxF
\else
??????\fi
\fi
\fi
\fi
\fi
\fi
}
\newcommand{\hatcurPPphiconj}[1]{\ifnum#1=25 %
\hatcurPPphiconjxxxxxA
\else
\ifnum#1=26 %
\hatcurPPphiconjxxxxxB
\else
\ifnum#1=27 %
\hatcurPPphiconjxxxxxxC
\else
\ifnum#1=28 %
\hatcurPPphiconjxxxxxD
\else
\ifnum#1=29 %
\hatcurPPphiconjxxxxxE
\else
\ifnum#1=30 %
\hatcurPPphiconjxxxxxF
\else
??????\fi
\fi
\fi
\fi
\fi
\fi
}
\newcommand{\hatcurPPr}[1]{\ifnum#1=25 %
\hatcurPPrxxxxxA
\else
\ifnum#1=26 %
\hatcurPPrxxxxxB
\else
\ifnum#1=27 %
\hatcurPPrxxxxxxC
\else
\ifnum#1=28 %
\hatcurPPrxxxxxD
\else
\ifnum#1=29 %
\hatcurPPrxxxxxE
\else
\ifnum#1=30 %
\hatcurPPrxxxxxF
\else
??????\fi
\fi
\fi
\fi
\fi
\fi
}
\newcommand{\hatcurPPre}[1]{\ifnum#1=25 %
\hatcurPPrexxxxxA
\else
\ifnum#1=26 %
\hatcurPPrexxxxxB
\else
\ifnum#1=27 %
\hatcurPPrexxxxxxC
\else
\ifnum#1=28 %
\hatcurPPrexxxxxD
\else
\ifnum#1=29 %
\hatcurPPrexxxxxE
\else
\ifnum#1=30 %
\hatcurPPrexxxxxF
\else
??????\fi
\fi
\fi
\fi
\fi
\fi
}
\newcommand{\hatcurPPrelong}[1]{\ifnum#1=25 %
\hatcurPPrelongxxxxxA
\else
\ifnum#1=26 %
\hatcurPPrelongxxxxxB
\else
\ifnum#1=27 %
\hatcurPPrelongxxxxxxC
\else
\ifnum#1=28 %
\hatcurPPrelongxxxxxD
\else
\ifnum#1=29 %
\hatcurPPrelongxxxxxE
\else
\ifnum#1=30 %
\hatcurPPrelongxxxxxF
\else
??????\fi
\fi
\fi
\fi
\fi
\fi
}
\newcommand{\hatcurPPreshort}[1]{\ifnum#1=25 %
\hatcurPPreshortxxxxxA
\else
\ifnum#1=26 %
\hatcurPPreshortxxxxxB
\else
\ifnum#1=27 %
\hatcurPPreshortxxxxxxC
\else
\ifnum#1=28 %
\hatcurPPreshortxxxxxD
\else
\ifnum#1=29 %
\hatcurPPreshortxxxxxE
\else
\ifnum#1=30 %
\hatcurPPreshortxxxxxF
\else
??????\fi
\fi
\fi
\fi
\fi
\fi
}
\newcommand{\hatcurPPrho}[1]{\ifnum#1=25 %
\hatcurPPrhoxxxxxA
\else
\ifnum#1=26 %
\hatcurPPrhoxxxxxB
\else
\ifnum#1=27 %
\hatcurPPrhoxxxxxxC
\else
\ifnum#1=28 %
\hatcurPPrhoxxxxxD
\else
\ifnum#1=29 %
\hatcurPPrhoxxxxxE
\else
\ifnum#1=30 %
\hatcurPPrhoxxxxxF
\else
??????\fi
\fi
\fi
\fi
\fi
\fi
}
\newcommand{\hatcurPPrlong}[1]{\ifnum#1=25 %
\hatcurPPrlongxxxxxA
\else
\ifnum#1=26 %
\hatcurPPrlongxxxxxB
\else
\ifnum#1=27 %
\hatcurPPrlongxxxxxxC
\else
\ifnum#1=28 %
\hatcurPPrlongxxxxxD
\else
\ifnum#1=29 %
\hatcurPPrlongxxxxxE
\else
\ifnum#1=30 %
\hatcurPPrlongxxxxxF
\else
??????\fi
\fi
\fi
\fi
\fi
\fi
}
\newcommand{\hatcurPPrshort}[1]{\ifnum#1=25 %
\hatcurPPrshortxxxxxA
\else
\ifnum#1=26 %
\hatcurPPrshortxxxxxB
\else
\ifnum#1=27 %
\hatcurPPrshortxxxxxxC
\else
\ifnum#1=28 %
\hatcurPPrshortxxxxxD
\else
\ifnum#1=29 %
\hatcurPPrshortxxxxxE
\else
\ifnum#1=30 %
\hatcurPPrshortxxxxxF
\else
??????\fi
\fi
\fi
\fi
\fi
\fi
}
\newcommand{\hatcurPPtcirc}[1]{\ifnum#1=25 %
\hatcurPPtcircxxxxxA
\else
\ifnum#1=26 %
\hatcurPPtcircxxxxxB
\else
\ifnum#1=27 %
\hatcurPPtcircxxxxxxC
\else
\ifnum#1=28 %
\hatcurPPtcircxxxxxD
\else
\ifnum#1=29 %
\hatcurPPtcircxxxxxE
\else
\ifnum#1=30 %
\hatcurPPtcircxxxxxF
\else
??????\fi
\fi
\fi
\fi
\fi
\fi
}
\newcommand{\hatcurPPteff}[1]{\ifnum#1=25 %
\hatcurPPteffxxxxxA
\else
\ifnum#1=26 %
\hatcurPPteffxxxxxB
\else
\ifnum#1=27 %
\hatcurPPteffxxxxxxC
\else
\ifnum#1=28 %
\hatcurPPteffxxxxxD
\else
\ifnum#1=29 %
\hatcurPPteffxxxxxE
\else
\ifnum#1=30 %
\hatcurPPteffxxxxxF
\else
??????\fi
\fi
\fi
\fi
\fi
\fi
}
\newcommand{\hatcurPPtheta}[1]{\ifnum#1=25 %
\hatcurPPthetaxxxxxA
\else
\ifnum#1=26 %
\hatcurPPthetaxxxxxB
\else
\ifnum#1=27 %
\hatcurPPthetaxxxxxxC
\else
\ifnum#1=28 %
\hatcurPPthetaxxxxxD
\else
\ifnum#1=29 %
\hatcurPPthetaxxxxxE
\else
\ifnum#1=30 %
\hatcurPPthetaxxxxxF
\else
??????\fi
\fi
\fi
\fi
\fi
\fi
}
\newcommand{\hatcurPPtinfall}[1]{\ifnum#1=25 %
\hatcurPPtinfallxxxxxA
\else
\ifnum#1=26 %
\hatcurPPtinfallxxxxxB
\else
\ifnum#1=27 %
\hatcurPPtinfallxxxxxxC
\else
\ifnum#1=28 %
\hatcurPPtinfallxxxxxD
\else
\ifnum#1=29 %
\hatcurPPtinfallxxxxxE
\else
\ifnum#1=30 %
\hatcurPPtinfallxxxxxF
\else
??????\fi
\fi
\fi
\fi
\fi
\fi
}
\newcommand{\hatcurRVeccen}[1]{\ifnum#1=25 %
\hatcurRVeccenxxxxxA
\else
\ifnum#1=26 %
\hatcurRVeccenxxxxxB
\else
\ifnum#1=27 %
\hatcurRVeccenxxxxxxC
\else
\ifnum#1=28 %
\hatcurRVeccenxxxxxD
\else
\ifnum#1=29 %
\hatcurRVeccenxxxxxE
\else
\ifnum#1=30 %
\hatcurRVeccenxxxxxF
\else
??????\fi
\fi
\fi
\fi
\fi
\fi
}
\newcommand{\hatcurRVeccentwosiglim}[1]{\ifnum#1=25 %
\hatcurRVeccentwosiglimxxxxxA
\else
\ifnum#1=26 %
\hatcurRVeccentwosiglimxxxxxB
\else
\ifnum#1=27 %
\hatcurRVeccentwosiglimxxxxxxC
\else
\ifnum#1=28 %
\hatcurRVeccentwosiglimxxxxxD
\else
\ifnum#1=29 %
\hatcurRVeccentwosiglimxxxxxE
\else
\ifnum#1=30 %
\hatcurRVeccentwosiglimxxxxxF
\else
??????\fi
\fi
\fi
\fi
\fi
\fi
}
\newcommand{\hatcurRVfitrms}[1]{\ifnum#1=25 %
\hatcurRVfitrmsxxxxxA
\else
\ifnum#1=28 %
\hatcurRVfitrmsxxxxxD
\else
??????\fi
\fi
}
\newcommand{\hatcurRVfitrmsA}[1]{\ifnum#1=26 %
\hatcurRVfitrmsAxxxxxB
\else
\ifnum#1=27 %
\hatcurRVfitrmsAxxxxxxC
\else
\ifnum#1=29 %
\hatcurRVfitrmsAxxxxxE
\else
\ifnum#1=30 %
\hatcurRVfitrmsAxxxxxF
\else
??????\fi
\fi
\fi
\fi
}
\newcommand{\hatcurRVfitrmsB}[1]{\ifnum#1=26 %
\hatcurRVfitrmsBxxxxxB
\else
\ifnum#1=27 %
\hatcurRVfitrmsBxxxxxxC
\else
\ifnum#1=29 %
\hatcurRVfitrmsBxxxxxE
\else
\ifnum#1=30 %
\hatcurRVfitrmsBxxxxxF
\else
??????\fi
\fi
\fi
\fi
}
\newcommand{\hatcurRVfitrmsC}[1]{\ifnum#1=26 %
\hatcurRVfitrmsCxxxxxB
\else
\ifnum#1=27 %
\hatcurRVfitrmsCxxxxxxC
\else
\ifnum#1=29 %
\hatcurRVfitrmsCxxxxxE
\else
??????\fi
\fi
\fi
}
\newcommand{\hatcurRVgamma}[1]{\ifnum#1=25 %
\hatcurRVgammaxxxxxA
\else
\ifnum#1=28 %
\hatcurRVgammaxxxxxD
\else
??????\fi
\fi
}
\newcommand{\hatcurRVgammaA}[1]{\ifnum#1=26 %
\hatcurRVgammaAxxxxxB
\else
\ifnum#1=27 %
\hatcurRVgammaAxxxxxxC
\else
\ifnum#1=29 %
\hatcurRVgammaAxxxxxE
\else
\ifnum#1=30 %
\hatcurRVgammaAxxxxxF
\else
??????\fi
\fi
\fi
\fi
}
\newcommand{\hatcurRVgammaB}[1]{\ifnum#1=26 %
\hatcurRVgammaBxxxxxB
\else
\ifnum#1=27 %
\hatcurRVgammaBxxxxxxC
\else
\ifnum#1=29 %
\hatcurRVgammaBxxxxxE
\else
\ifnum#1=30 %
\hatcurRVgammaBxxxxxF
\else
??????\fi
\fi
\fi
\fi
}
\newcommand{\hatcurRVgammaC}[1]{\ifnum#1=26 %
\hatcurRVgammaCxxxxxB
\else
\ifnum#1=27 %
\hatcurRVgammaCxxxxxxC
\else
\ifnum#1=29 %
\hatcurRVgammaCxxxxxE
\else
??????\fi
\fi
\fi
}
\newcommand{\hatcurRVh}[1]{\ifnum#1=25 %
\hatcurRVhxxxxxA
\else
\ifnum#1=26 %
\hatcurRVhxxxxxB
\else
\ifnum#1=27 %
\hatcurRVhxxxxxxC
\else
\ifnum#1=28 %
\hatcurRVhxxxxxD
\else
\ifnum#1=29 %
\hatcurRVhxxxxxE
\else
\ifnum#1=30 %
\hatcurRVhxxxxxF
\else
??????\fi
\fi
\fi
\fi
\fi
\fi
}
\newcommand{\hatcurRVjitter}[1]{\ifnum#1=25 %
\hatcurRVjitterxxxxxA
\else
\ifnum#1=28 %
\hatcurRVjitterxxxxxD
\else
??????\fi
\fi
}
\newcommand{\hatcurRVjitterA}[1]{\ifnum#1=26 %
\hatcurRVjitterAxxxxxB
\else
\ifnum#1=27 %
\hatcurRVjitterAxxxxxxC
\else
\ifnum#1=29 %
\hatcurRVjitterAxxxxxE
\else
\ifnum#1=30 %
\hatcurRVjitterAxxxxxF
\else
??????\fi
\fi
\fi
\fi
}
\newcommand{\hatcurRVjitterB}[1]{\ifnum#1=26 %
\hatcurRVjitterBxxxxxB
\else
\ifnum#1=27 %
\hatcurRVjitterBxxxxxxC
\else
\ifnum#1=29 %
\hatcurRVjitterBxxxxxE
\else
\ifnum#1=30 %
\hatcurRVjitterBxxxxxF
\else
??????\fi
\fi
\fi
\fi
}
\newcommand{\hatcurRVjitterC}[1]{\ifnum#1=26 %
\hatcurRVjitterCxxxxxB
\else
\ifnum#1=27 %
\hatcurRVjitterCxxxxxxC
\else
\ifnum#1=29 %
\hatcurRVjitterCxxxxxE
\else
??????\fi
\fi
\fi
}
\newcommand{\hatcurRVjittertwosiglim}[1]{\ifnum#1=25 %
\hatcurRVjittertwosiglimxxxxxA
\else
\ifnum#1=28 %
\hatcurRVjittertwosiglimxxxxxD
\else
??????\fi
\fi
}
\newcommand{\hatcurRVjittertwosiglimA}[1]{\ifnum#1=26 %
\hatcurRVjittertwosiglimAxxxxxB
\else
\ifnum#1=27 %
\hatcurRVjittertwosiglimAxxxxxxC
\else
\ifnum#1=29 %
\hatcurRVjittertwosiglimAxxxxxE
\else
\ifnum#1=30 %
\hatcurRVjittertwosiglimAxxxxxF
\else
??????\fi
\fi
\fi
\fi
}
\newcommand{\hatcurRVjittertwosiglimB}[1]{\ifnum#1=26 %
\hatcurRVjittertwosiglimBxxxxxB
\else
\ifnum#1=27 %
\hatcurRVjittertwosiglimBxxxxxxC
\else
\ifnum#1=29 %
\hatcurRVjittertwosiglimBxxxxxE
\else
\ifnum#1=30 %
\hatcurRVjittertwosiglimBxxxxxF
\else
??????\fi
\fi
\fi
\fi
}
\newcommand{\hatcurRVjittertwosiglimC}[1]{\ifnum#1=26 %
\hatcurRVjittertwosiglimCxxxxxB
\else
\ifnum#1=27 %
\hatcurRVjittertwosiglimCxxxxxxC
\else
\ifnum#1=29 %
\hatcurRVjittertwosiglimCxxxxxE
\else
??????\fi
\fi
\fi
}
\newcommand{\hatcurRVk}[1]{\ifnum#1=25 %
\hatcurRVkxxxxxA
\else
\ifnum#1=26 %
\hatcurRVkxxxxxB
\else
\ifnum#1=27 %
\hatcurRVkxxxxxxC
\else
\ifnum#1=28 %
\hatcurRVkxxxxxD
\else
\ifnum#1=29 %
\hatcurRVkxxxxxE
\else
\ifnum#1=30 %
\hatcurRVkxxxxxF
\else
??????\fi
\fi
\fi
\fi
\fi
\fi
}
\newcommand{\hatcurRVK}[1]{\ifnum#1=25 %
\hatcurRVKxxxxxA
\else
\ifnum#1=26 %
\hatcurRVKxxxxxB
\else
\ifnum#1=27 %
\hatcurRVKxxxxxxC
\else
\ifnum#1=28 %
\hatcurRVKxxxxxD
\else
\ifnum#1=29 %
\hatcurRVKxxxxxE
\else
\ifnum#1=30 %
\hatcurRVKxxxxxF
\else
??????\fi
\fi
\fi
\fi
\fi
\fi
}
\newcommand{\hatcurRVomega}[1]{\ifnum#1=25 %
\hatcurRVomegaxxxxxA
\else
\ifnum#1=26 %
\hatcurRVomegaxxxxxB
\else
\ifnum#1=27 %
\hatcurRVomegaxxxxxxC
\else
\ifnum#1=28 %
\hatcurRVomegaxxxxxD
\else
\ifnum#1=29 %
\hatcurRVomegaxxxxxE
\else
\ifnum#1=30 %
\hatcurRVomegaxxxxxF
\else
??????\fi
\fi
\fi
\fi
\fi
\fi
}
\newcommand{\hatcurRVrh}[1]{\ifnum#1=25 %
\hatcurRVrhxxxxxA
\else
\ifnum#1=26 %
\hatcurRVrhxxxxxB
\else
\ifnum#1=27 %
\hatcurRVrhxxxxxxC
\else
\ifnum#1=28 %
\hatcurRVrhxxxxxD
\else
\ifnum#1=29 %
\hatcurRVrhxxxxxE
\else
\ifnum#1=30 %
\hatcurRVrhxxxxxF
\else
??????\fi
\fi
\fi
\fi
\fi
\fi
}
\newcommand{\hatcurRVrk}[1]{\ifnum#1=25 %
\hatcurRVrkxxxxxA
\else
\ifnum#1=26 %
\hatcurRVrkxxxxxB
\else
\ifnum#1=27 %
\hatcurRVrkxxxxxxC
\else
\ifnum#1=28 %
\hatcurRVrkxxxxxD
\else
\ifnum#1=29 %
\hatcurRVrkxxxxxE
\else
\ifnum#1=30 %
\hatcurRVrkxxxxxF
\else
??????\fi
\fi
\fi
\fi
\fi
\fi
}
\newcommand{\hatcurRVtrone}[1]{\ifnum#1=25 %
\hatcurRVtronexxxxxA
\else
\ifnum#1=26 %
\hatcurRVtronexxxxxB
\else
\ifnum#1=27 %
\hatcurRVtronexxxxxxC
\else
\ifnum#1=28 %
\hatcurRVtronexxxxxD
\else
\ifnum#1=29 %
\hatcurRVtronexxxxxE
\else
\ifnum#1=30 %
\hatcurRVtronexxxxxF
\else
??????\fi
\fi
\fi
\fi
\fi
\fi
}
\newcommand{\hatcurRVtrtwo}[1]{\ifnum#1=25 %
\hatcurRVtrtwoxxxxxA
\else
\ifnum#1=26 %
\hatcurRVtrtwoxxxxxB
\else
\ifnum#1=27 %
\hatcurRVtrtwoxxxxxxC
\else
\ifnum#1=28 %
\hatcurRVtrtwoxxxxxD
\else
\ifnum#1=29 %
\hatcurRVtrtwoxxxxxE
\else
\ifnum#1=30 %
\hatcurRVtrtwoxxxxxF
\else
??????\fi
\fi
\fi
\fi
\fi
\fi
}
\newcommand{\hatcurSMEiilogg}[1]{\ifnum#1=25 %
\hatcurSMEiiloggxxxxxA
\else
\ifnum#1=26 %
\hatcurSMEiiloggxxxxxB
\else
\ifnum#1=28 %
\hatcurSMEiiloggxxxxxD
\else
\ifnum#1=29 %
\hatcurSMEiiloggxxxxxE
\else
\ifnum#1=30 %
\hatcurSMEiiloggxxxxxF
\else
??????\fi
\fi
\fi
\fi
\fi
}
\newcommand{\hatcurSMEiiteff}[1]{\ifnum#1=25 %
\hatcurSMEiiteffxxxxxA
\else
\ifnum#1=26 %
\hatcurSMEiiteffxxxxxB
\else
\ifnum#1=28 %
\hatcurSMEiiteffxxxxxD
\else
\ifnum#1=29 %
\hatcurSMEiiteffxxxxxE
\else
\ifnum#1=30 %
\hatcurSMEiiteffxxxxxF
\else
??????\fi
\fi
\fi
\fi
\fi
}
\newcommand{\hatcurSMEiivsin}[1]{\ifnum#1=25 %
\hatcurSMEiivsinxxxxxA
\else
\ifnum#1=26 %
\hatcurSMEiivsinxxxxxB
\else
\ifnum#1=28 %
\hatcurSMEiivsinxxxxxD
\else
\ifnum#1=29 %
\hatcurSMEiivsinxxxxxE
\else
\ifnum#1=30 %
\hatcurSMEiivsinxxxxxF
\else
??????\fi
\fi
\fi
\fi
\fi
}
\newcommand{\hatcurSMEiizfeh}[1]{\ifnum#1=25 %
\hatcurSMEiizfehxxxxxA
\else
\ifnum#1=26 %
\hatcurSMEiizfehxxxxxB
\else
\ifnum#1=28 %
\hatcurSMEiizfehxxxxxD
\else
\ifnum#1=29 %
\hatcurSMEiizfehxxxxxE
\else
\ifnum#1=30 %
\hatcurSMEiizfehxxxxxF
\else
??????\fi
\fi
\fi
\fi
\fi
}
\newcommand{\hatcurSMEiizfehshort}[1]{\ifnum#1=25 %
\hatcurSMEiizfehshortxxxxxA
\else
\ifnum#1=26 %
\hatcurSMEiizfehshortxxxxxB
\else
\ifnum#1=28 %
\hatcurSMEiizfehshortxxxxxD
\else
\ifnum#1=29 %
\hatcurSMEiizfehshortxxxxxE
\else
\ifnum#1=30 %
\hatcurSMEiizfehshortxxxxxF
\else
??????\fi
\fi
\fi
\fi
\fi
}
\newcommand{\hatcurSMEilogg}[1]{\ifnum#1=25 %
\hatcurSMEiloggxxxxxA
\else
\ifnum#1=26 %
\hatcurSMEiloggxxxxxB
\else
\ifnum#1=27 %
\hatcurSMEiloggxxxxxxC
\else
\ifnum#1=28 %
\hatcurSMEiloggxxxxxD
\else
\ifnum#1=29 %
\hatcurSMEiloggxxxxxE
\else
\ifnum#1=30 %
\hatcurSMEiloggxxxxxF
\else
??????\fi
\fi
\fi
\fi
\fi
\fi
}
\newcommand{\hatcurSMEiteff}[1]{\ifnum#1=25 %
\hatcurSMEiteffxxxxxA
\else
\ifnum#1=26 %
\hatcurSMEiteffxxxxxB
\else
\ifnum#1=27 %
\hatcurSMEiteffxxxxxxC
\else
\ifnum#1=28 %
\hatcurSMEiteffxxxxxD
\else
\ifnum#1=29 %
\hatcurSMEiteffxxxxxE
\else
\ifnum#1=30 %
\hatcurSMEiteffxxxxxF
\else
??????\fi
\fi
\fi
\fi
\fi
\fi
}
\newcommand{\hatcurSMEivmac}[1]{\ifnum#1=25 %
\hatcurSMEivmacxxxxxA
\else
\ifnum#1=26 %
\hatcurSMEivmacxxxxxB
\else
\ifnum#1=27 %
\hatcurSMEivmacxxxxxxC
\else
\ifnum#1=28 %
\hatcurSMEivmacxxxxxD
\else
\ifnum#1=29 %
\hatcurSMEivmacxxxxxE
\else
\ifnum#1=30 %
\hatcurSMEivmacxxxxxF
\else
??????\fi
\fi
\fi
\fi
\fi
\fi
}
\newcommand{\hatcurSMEivmic}[1]{\ifnum#1=25 %
\hatcurSMEivmicxxxxxA
\else
\ifnum#1=26 %
\hatcurSMEivmicxxxxxB
\else
\ifnum#1=27 %
\hatcurSMEivmicxxxxxxC
\else
\ifnum#1=28 %
\hatcurSMEivmicxxxxxD
\else
\ifnum#1=29 %
\hatcurSMEivmicxxxxxE
\else
\ifnum#1=30 %
\hatcurSMEivmicxxxxxF
\else
??????\fi
\fi
\fi
\fi
\fi
\fi
}
\newcommand{\hatcurSMEivsin}[1]{\ifnum#1=25 %
\hatcurSMEivsinxxxxxA
\else
\ifnum#1=26 %
\hatcurSMEivsinxxxxxB
\else
\ifnum#1=27 %
\hatcurSMEivsinxxxxxxC
\else
\ifnum#1=28 %
\hatcurSMEivsinxxxxxD
\else
\ifnum#1=29 %
\hatcurSMEivsinxxxxxE
\else
\ifnum#1=30 %
\hatcurSMEivsinxxxxxF
\else
??????\fi
\fi
\fi
\fi
\fi
\fi
}
\newcommand{\hatcurSMEizfeh}[1]{\ifnum#1=25 %
\hatcurSMEizfehxxxxxA
\else
\ifnum#1=26 %
\hatcurSMEizfehxxxxxB
\else
\ifnum#1=27 %
\hatcurSMEizfehxxxxxxC
\else
\ifnum#1=28 %
\hatcurSMEizfehxxxxxD
\else
\ifnum#1=29 %
\hatcurSMEizfehxxxxxE
\else
\ifnum#1=30 %
\hatcurSMEizfehxxxxxF
\else
??????\fi
\fi
\fi
\fi
\fi
\fi
}
\newcommand{\hatcurSMEizfehshort}[1]{\ifnum#1=25 %
\hatcurSMEizfehshortxxxxxA
\else
\ifnum#1=26 %
\hatcurSMEizfehshortxxxxxB
\else
\ifnum#1=27 %
\hatcurSMEizfehshortxxxxxxC
\else
\ifnum#1=28 %
\hatcurSMEizfehshortxxxxxD
\else
\ifnum#1=29 %
\hatcurSMEizfehshortxxxxxE
\else
\ifnum#1=30 %
\hatcurSMEizfehshortxxxxxF
\else
??????\fi
\fi
\fi
\fi
\fi
\fi
}
\newcommand{\hatcurXAv}[1]{\ifnum#1=25 %
\hatcurXAvxxxxxA
\else
\ifnum#1=26 %
\hatcurXAvxxxxxB
\else
\ifnum#1=27 %
\hatcurXAvxxxxxxC
\else
\ifnum#1=28 %
\hatcurXAvxxxxxD
\else
\ifnum#1=29 %
\hatcurXAvxxxxxE
\else
\ifnum#1=30 %
\hatcurXAvxxxxxF
\else
??????\fi
\fi
\fi
\fi
\fi
\fi
}
\newcommand{\hatcurXdist}[1]{\ifnum#1=25 %
\hatcurXdistxxxxxA
\else
\ifnum#1=26 %
\hatcurXdistxxxxxB
\else
\ifnum#1=27 %
\hatcurXdistxxxxxxC
\else
\ifnum#1=28 %
\hatcurXdistxxxxxD
\else
\ifnum#1=29 %
\hatcurXdistxxxxxE
\else
\ifnum#1=30 %
\hatcurXdistxxxxxF
\else
??????\fi
\fi
\fi
\fi
\fi
\fi
}
\newcommand{\hatcurXdistred}[1]{\ifnum#1=25 %
\hatcurXdistredxxxxxA
\else
\ifnum#1=26 %
\hatcurXdistredxxxxxB
\else
\ifnum#1=27 %
\hatcurXdistredxxxxxxC
\else
\ifnum#1=28 %
\hatcurXdistredxxxxxD
\else
\ifnum#1=29 %
\hatcurXdistredxxxxxE
\else
\ifnum#1=30 %
\hatcurXdistredxxxxxF
\else
??????\fi
\fi
\fi
\fi
\fi
\fi
}
\newcommand{\hatcurXEBV}[1]{\ifnum#1=25 %
\hatcurXEBVxxxxxA
\else
\ifnum#1=26 %
\hatcurXEBVxxxxxB
\else
\ifnum#1=27 %
\hatcurXEBVxxxxxxC
\else
\ifnum#1=28 %
\hatcurXEBVxxxxxD
\else
\ifnum#1=29 %
\hatcurXEBVxxxxxE
\else
\ifnum#1=30 %
\hatcurXEBVxxxxxF
\else
??????\fi
\fi
\fi
\fi
\fi
\fi
}
\newcommand{\hatcurXjhisored}[1]{\ifnum#1=25 %
\hatcurXjhisoredxxxxxA
\else
\ifnum#1=26 %
\hatcurXjhisoredxxxxxB
\else
\ifnum#1=27 %
\hatcurXjhisoredxxxxxxC
\else
\ifnum#1=28 %
\hatcurXjhisoredxxxxxD
\else
\ifnum#1=29 %
\hatcurXjhisoredxxxxxE
\else
\ifnum#1=30 %
\hatcurXjhisoredxxxxxF
\else
??????\fi
\fi
\fi
\fi
\fi
\fi
}
\newcommand{\hatcurXjkisored}[1]{\ifnum#1=25 %
\hatcurXjkisoredxxxxxA
\else
\ifnum#1=26 %
\hatcurXjkisoredxxxxxB
\else
\ifnum#1=27 %
\hatcurXjkisoredxxxxxxC
\else
\ifnum#1=28 %
\hatcurXjkisoredxxxxxD
\else
\ifnum#1=29 %
\hatcurXjkisoredxxxxxE
\else
\ifnum#1=30 %
\hatcurXjkisoredxxxxxF
\else
??????\fi
\fi
\fi
\fi
\fi
\fi
}
\newcommand{\hatcurXmhisored}[1]{\ifnum#1=25 %
\hatcurXmhisoredxxxxxA
\else
\ifnum#1=26 %
\hatcurXmhisoredxxxxxB
\else
\ifnum#1=27 %
\hatcurXmhisoredxxxxxxC
\else
\ifnum#1=28 %
\hatcurXmhisoredxxxxxD
\else
\ifnum#1=29 %
\hatcurXmhisoredxxxxxE
\else
\ifnum#1=30 %
\hatcurXmhisoredxxxxxF
\else
??????\fi
\fi
\fi
\fi
\fi
\fi
}
\newcommand{\hatcurXmiisored}[1]{\ifnum#1=25 %
\hatcurXmiisoredxxxxxA
\else
\ifnum#1=26 %
\hatcurXmiisoredxxxxxB
\else
\ifnum#1=27 %
\hatcurXmiisoredxxxxxxC
\else
\ifnum#1=28 %
\hatcurXmiisoredxxxxxD
\else
\ifnum#1=29 %
\hatcurXmiisoredxxxxxE
\else
\ifnum#1=30 %
\hatcurXmiisoredxxxxxF
\else
??????\fi
\fi
\fi
\fi
\fi
\fi
}
\newcommand{\hatcurXmjisored}[1]{\ifnum#1=25 %
\hatcurXmjisoredxxxxxA
\else
\ifnum#1=26 %
\hatcurXmjisoredxxxxxB
\else
\ifnum#1=27 %
\hatcurXmjisoredxxxxxxC
\else
\ifnum#1=28 %
\hatcurXmjisoredxxxxxD
\else
\ifnum#1=29 %
\hatcurXmjisoredxxxxxE
\else
\ifnum#1=30 %
\hatcurXmjisoredxxxxxF
\else
??????\fi
\fi
\fi
\fi
\fi
\fi
}
\newcommand{\hatcurXmkisored}[1]{\ifnum#1=25 %
\hatcurXmkisoredxxxxxA
\else
\ifnum#1=26 %
\hatcurXmkisoredxxxxxB
\else
\ifnum#1=27 %
\hatcurXmkisoredxxxxxxC
\else
\ifnum#1=28 %
\hatcurXmkisoredxxxxxD
\else
\ifnum#1=29 %
\hatcurXmkisoredxxxxxE
\else
\ifnum#1=30 %
\hatcurXmkisoredxxxxxF
\else
??????\fi
\fi
\fi
\fi
\fi
\fi
}
\newcommand{\hatcurXmvisored}[1]{\ifnum#1=25 %
\hatcurXmvisoredxxxxxA
\else
\ifnum#1=26 %
\hatcurXmvisoredxxxxxB
\else
\ifnum#1=27 %
\hatcurXmvisoredxxxxxxC
\else
\ifnum#1=28 %
\hatcurXmvisoredxxxxxD
\else
\ifnum#1=29 %
\hatcurXmvisoredxxxxxE
\else
\ifnum#1=30 %
\hatcurXmvisoredxxxxxF
\else
??????\fi
\fi
\fi
\fi
\fi
\fi
}
\newcommand{\hatcurXsecdur}[1]{\ifnum#1=25 %
\hatcurXsecdurxxxxxA
\else
\ifnum#1=26 %
\hatcurXsecdurxxxxxB
\else
\ifnum#1=27 %
\hatcurXsecdurxxxxxxC
\else
\ifnum#1=28 %
\hatcurXsecdurxxxxxD
\else
\ifnum#1=29 %
\hatcurXsecdurxxxxxE
\else
\ifnum#1=30 %
\hatcurXsecdurxxxxxF
\else
??????\fi
\fi
\fi
\fi
\fi
\fi
}
\newcommand{\hatcurXsecingdur}[1]{\ifnum#1=25 %
\hatcurXsecingdurxxxxxA
\else
\ifnum#1=26 %
\hatcurXsecingdurxxxxxB
\else
\ifnum#1=27 %
\hatcurXsecingdurxxxxxxC
\else
\ifnum#1=28 %
\hatcurXsecingdurxxxxxD
\else
\ifnum#1=29 %
\hatcurXsecingdurxxxxxE
\else
\ifnum#1=30 %
\hatcurXsecingdurxxxxxF
\else
??????\fi
\fi
\fi
\fi
\fi
\fi
}
\newcommand{\hatcurXsecondary}[1]{\ifnum#1=25 %
\hatcurXsecondaryxxxxxA
\else
\ifnum#1=26 %
\hatcurXsecondaryxxxxxB
\else
\ifnum#1=27 %
\hatcurXsecondaryxxxxxxC
\else
\ifnum#1=28 %
\hatcurXsecondaryxxxxxD
\else
\ifnum#1=29 %
\hatcurXsecondaryxxxxxE
\else
\ifnum#1=30 %
\hatcurXsecondaryxxxxxF
\else
??????\fi
\fi
\fi
\fi
\fi
\fi
}
\newcommand{\hatcurXsecphase}[1]{\ifnum#1=25 %
\hatcurXsecphasexxxxxA
\else
\ifnum#1=26 %
\hatcurXsecphasexxxxxB
\else
\ifnum#1=27 %
\hatcurXsecphasexxxxxxC
\else
\ifnum#1=28 %
\hatcurXsecphasexxxxxD
\else
\ifnum#1=29 %
\hatcurXsecphasexxxxxE
\else
\ifnum#1=30 %
\hatcurXsecphasexxxxxF
\else
??????\fi
\fi
\fi
\fi
\fi
\fi
}
\newcommand{\hatcurXviisored}[1]{\ifnum#1=25 %
\hatcurXviisoredxxxxxA
\else
\ifnum#1=26 %
\hatcurXviisoredxxxxxB
\else
\ifnum#1=27 %
\hatcurXviisoredxxxxxxC
\else
\ifnum#1=28 %
\hatcurXviisoredxxxxxD
\else
\ifnum#1=29 %
\hatcurXviisoredxxxxxE
\else
\ifnum#1=30 %
\hatcurXviisoredxxxxxF
\else
??????\fi
\fi
\fi
\fi
\fi
\fi
}
\newcommand{\hatcurXvkisored}[1]{\ifnum#1=25 %
\hatcurXvkisoredxxxxxA
\else
\ifnum#1=26 %
\hatcurXvkisoredxxxxxB
\else
\ifnum#1=27 %
\hatcurXvkisoredxxxxxxC
\else
\ifnum#1=28 %
\hatcurXvkisoredxxxxxD
\else
\ifnum#1=29 %
\hatcurXvkisoredxxxxxE
\else
\ifnum#1=30 %
\hatcurXvkisoredxxxxxF
\else
??????\fi
\fi
\fi
\fi
\fi
\fi
}

\newcommand{\hatcurhtreccenxxxxxA}{HATS568-001}                              
\newcommand{\hatcurfieldeccenxxxxxA}{\ensuremath{string}}                    
\newcommand{\hatcurCCraeccenxxxxxA}{\ensuremath{13^{\mathrm h}51^{\mathrm m}37.80{\mathrm s}}}                             
\newcommand{\hatcurCCdececcenxxxxxA}{\ensuremath{-23{\arcdeg}46{\arcmin}52.2{\arcsec}}}                            
\newcommand{\hatcurCCmageccenxxxxxA}{13.097}                                 
\newcommand{\hatcurCCtwomasseccenxxxxxA}{2MASS~13513786-2346522}             
\newcommand{\hatcurCCgsceccenxxxxxA}{GSC~6716-01190}                         
\newcommand{\hatcurCCtassmveccenxxxxxA}{\ensuremath{13.097\pm0.030}}         
\newcommand{\hatcurCCtassmvshorteccenxxxxxA}{\ensuremath{13.1}}              
\newcommand{\hatcurCCtassmBeccenxxxxxA}{\ensuremath{13.812\pm0.030}}         
\newcommand{\hatcurCCtassmBshorteccenxxxxxA}{\ensuremath{13.8}}              
\newcommand{\hatcurCCtassmIeccenxxxxxA}{\ensuremath{nff\pmnff}}              
\newcommand{\hatcurCCtassmIshorteccenxxxxxA}{\ensuremath{0.0}}               
\newcommand{\hatcurCCtassmgeccenxxxxxA}{\ensuremath{13.380\pm0.020}}         
\newcommand{\hatcurCCtassmgshorteccenxxxxxA}{\ensuremath{13.4}}              
\newcommand{\hatcurCCtassmreccenxxxxxA}{\ensuremath{12.909\pm0.040}}         
\newcommand{\hatcurCCtassmrshorteccenxxxxxA}{\ensuremath{12.9}}              
\newcommand{\hatcurCCtassmieccenxxxxxA}{\ensuremath{12.687\pm0.050}}         
\newcommand{\hatcurCCtassmishorteccenxxxxxA}{\ensuremath{12.7}}              
\newcommand{\hatcurCCtwomassJmageccenxxxxxA}{\ensuremath{11.788\pm0.022}}    
\newcommand{\hatcurCCtwomassHmageccenxxxxxA}{\ensuremath{11.487\pm0.024}}    
\newcommand{\hatcurCCtwomassKmageccenxxxxxA}{\ensuremath{11.416\pm0.021}}    
\newcommand{\hatcurCCcitJmageccenxxxxxA}{\ensuremath{11.805\pm0.022}}        
\newcommand{\hatcurCCcitHmageccenxxxxxA}{\ensuremath{11.482\pm0.024}}        
\newcommand{\hatcurCCcitKmageccenxxxxxA}{\ensuremath{11.440\pm0.021}}        
\newcommand{\hatcurCCbbJmageccenxxxxxA}{\ensuremath{11.854\pm0.024}}         
\newcommand{\hatcurCCbbHmageccenxxxxxA}{\ensuremath{11.503\pm0.025}}         
\newcommand{\hatcurCCbbKmageccenxxxxxA}{\ensuremath{11.460\pm0.021}}         
\newcommand{\hatcurCCesoJmageccenxxxxxA}{\ensuremath{11.857\pm0.025}}        
\newcommand{\hatcurCCesoHmageccenxxxxxA}{\ensuremath{11.498\pm0.028}}        
\newcommand{\hatcurCCesoKmageccenxxxxxA}{\ensuremath{11.459\pm0.022}}        
\newcommand{\hatcurCCesoJHmageccenxxxxxA}{\ensuremath{0.358\pm0.036}}        
\newcommand{\hatcurCCesoJKmageccenxxxxxA}{\ensuremath{0.398\pm0.033}}        
\newcommand{\hatcurCCesoHKmageccenxxxxxA}{\ensuremath{0.040\pm0.036}}        
\newcommand{\hatcurLCdipeccenxxxxxA}{\ensuremath{16.8}}                      
\newcommand{\hatcurLCrprstareccenxxxxxA}{\ensuremath{0.1174\pm0.0026}}       
\newcommand{\hatcurLCbsqeccenxxxxxA}{\ensuremath{0.315_{-0.094}^{+0.065}}}   
\newcommand{\hatcurLCimpeccenxxxxxA}{\ensuremath{0.561_{-0.091}^{+0.055}}}   
\newcommand{\hatcurLCzetaeccenxxxxxA}{\ensuremath{17.43\pm0.19}}             
\newcommand{\hatcurLCdureccenxxxxxA}{\ensuremath{0.1339\pm0.0024}}           
\newcommand{\hatcurLCdurshorteccenxxxxxA}{\ensuremath{0.1339}}               
\newcommand{\hatcurLCdurhreccenxxxxxA}{\ensuremath{3.214\pm0.057}}           
\newcommand{\hatcurLCdurhrshorteccenxxxxxA}{\ensuremath{3.214}}              
\newcommand{\hatcurLCqeccenxxxxxA}{\ensuremath{0.03120\pm0.00055}}           
\newcommand{\hatcurLCqshorteccenxxxxxA}{\ensuremath{0.031}}                  
\newcommand{\hatcurLCingdureccenxxxxxA}{\ensuremath{0.0198\pm0.0025}}        
\newcommand{\hatcurLCPeccenxxxxxA}{\ensuremath{4.2986444\pm0.0000044}}       
\newcommand{\hatcurLCPprececcenxxxxxA}{\ensuremath{4.2986444}}               
\newcommand{\hatcurLCPshorteccenxxxxxA}{\ensuremath{4.2986}}                 
\newcommand{\hatcurLCTeccenxxxxxA}{\ensuremath{2456866.07015\pm0.00055}}     
\newcommand{\hatcurLCTAeccenxxxxxA}{\ensuremath{2455645.2552\pm0.0014}}      
\newcommand{\hatcurLCTBeccenxxxxxA}{\ensuremath{2457098.19697\pm0.00060}}    
\newcommand{\hatcurLChatnetmeccenxxxxxA}{\ensuremath{12.864660\pm0.000098}}  
\newcommand{\hatcurLCiblendeccenxxxxxA}{\ensuremath{0.957\pm0.049}}          
\newcommand{\hatcurLCrhoeccenxxxxxA}{\ensuremath{1.07^{+0.34}_{-0.25}}}                    
\newcommand{\hatcurSMEiteffeccenxxxxxA}{\ensuremath{5860\pm180}}             
\newcommand{\hatcurSMEizfeheccenxxxxxA}{\ensuremath{0.100\pm0.080}}          
\newcommand{\hatcurSMEizfehshorteccenxxxxxA}{\ensuremath{0.10}}              
\newcommand{\hatcurSMEiloggeccenxxxxxA}{\ensuremath{4.55\pm0.17}}            
\newcommand{\hatcurSMEivsineccenxxxxxA}{\ensuremath{3.14\pm0.34}}            
\newcommand{\hatcurSMEivmaceccenxxxxxA}{\ensuremath{0.0}}                    
\newcommand{\hatcurSMEivmiceccenxxxxxA}{\ensuremath{0.0}}                    
\newcommand{\hatcurSMEiiteffeccenxxxxxA}{\ensuremath{5715\pm73}}             
\newcommand{\hatcurSMEiizfeheccenxxxxxA}{\ensuremath{0.020\pm0.050}}         
\newcommand{\hatcurSMEiizfehshorteccenxxxxxA}{\ensuremath{0.02}}             
\newcommand{\hatcurSMEiiloggeccenxxxxxA}{\ensuremath{4\pm0}}                 
\newcommand{\hatcurSMEiivsineccenxxxxxA}{\ensuremath{3.88\pm0.50}}           
\newcommand{\hatcurLBizeccenxxxxxA}{\ensuremath{0.2160}}                     
\newcommand{\hatcurLBiizeccenxxxxxA}{\ensuremath{0.3244}}                    
\newcommand{\hatcurLBiieccenxxxxxA}{\ensuremath{0.2774}}                     
\newcommand{\hatcurLBiiieccenxxxxxA}{\ensuremath{0.3246}}                    
\newcommand{\hatcurLBiIeccenxxxxxA}{\ensuremath{0.2567}}                     
\newcommand{\hatcurLBiiIeccenxxxxxA}{\ensuremath{0.3254}}                    
\newcommand{\hatcurLBigeccenxxxxxA}{\ensuremath{0.5644}}                     
\newcommand{\hatcurLBiigeccenxxxxxA}{\ensuremath{0.2277}}                    
\newcommand{\hatcurLBireccenxxxxxA}{\ensuremath{0.3674}}                     
\newcommand{\hatcurLBiireccenxxxxxA}{\ensuremath{0.3192}}                    
\newcommand{\hatcurLBiReccenxxxxxA}{\ensuremath{0.3425}}                     
\newcommand{\hatcurLBiiReccenxxxxxA}{\ensuremath{0.3217}}                    
\newcommand{\hatcurLBikepeccenxxxxxA}{\ensuremath{0.1000}}                   
\newcommand{\hatcurLBiikepeccenxxxxxA}{\ensuremath{0.1000}}                  
\newcommand{\hatcurISOmeccenxxxxxA}{\ensuremath{0.995\pm0.035}}              
\newcommand{\hatcurISOmshorteccenxxxxxA}{\ensuremath{1.00}}                  
\newcommand{\hatcurISOmlongeccenxxxxxA}{\ensuremath{0.995\pm0.035}}          
\newcommand{\hatcurISOreccenxxxxxA}{\ensuremath{1.100\pm0.087}}              
\newcommand{\hatcurISOrshorteccenxxxxxA}{\ensuremath{1.10}}                  
\newcommand{\hatcurISOrlongeccenxxxxxA}{\ensuremath{1.100\pm0.087}}          
\newcommand{\hatcurISOrhoeccenxxxxxA}{\ensuremath{1.05\pm0.26}}              
\newcommand{\hatcurISOrholongeccenxxxxxA}{\ensuremath{1.05\pm0.26}}          
\newcommand{\hatcurISOloggeccenxxxxxA}{\ensuremath{4.353\pm0.068}}           
\newcommand{\hatcurISOlumeccenxxxxxA}{\ensuremath{1.16\pm0.20}}              
\newcommand{\hatcurISOlumshorteccenxxxxxA}{\ensuremath{1.16}}                
\newcommand{\hatcurISOmveccenxxxxxA}{\ensuremath{4.68\pm0.19}}               
\newcommand{\hatcurISOvieccenxxxxxA}{\ensuremath{0.716\pm0.023}}             
\newcommand{\hatcurISOageeccenxxxxxA}{\ensuremath{7.3\pm2.3}}                
\newcommand{\hatcurISOsigmaeccenxxxxxA}{\ensuremath{0.00070\pm0.00019}}      
\newcommand{\hatcurISOMJeccenxxxxxA}{\ensuremath{3.52\pm0.18}}               
\newcommand{\hatcurISOMHeccenxxxxxA}{\ensuremath{3.16\pm0.17}}               
\newcommand{\hatcurISOMKeccenxxxxxA}{\ensuremath{3.11\pm0.17}}               
\newcommand{\hatcurISOJKeccenxxxxxA}{\ensuremath{0.410\pm0.020}}             
\newcommand{\hatcurISOspececcenxxxxxA}{G}                                    
\newcommand{\hatcurRVKeccenxxxxxA}{\ensuremath{76.3\pm7.1}}                  
\newcommand{\hatcurRVrkeccenxxxxxA}{\ensuremath{-0.03\pm0.18}}               
\newcommand{\hatcurRVrheccenxxxxxA}{\ensuremath{-0.06\pm0.18}}               
\newcommand{\hatcurRVkeccenxxxxxA}{\ensuremath{-0.003\pm0.061}}              
\newcommand{\hatcurRVheccenxxxxxA}{\ensuremath{-0.008_{-0.065}^{+0.041}}}    
\newcommand{\hatcurRVtroneeccenxxxxxA}{\ensuremath{0\pm0}}                   
\newcommand{\hatcurRVtrtwoeccenxxxxxA}{\ensuremath{0\pm0}}                   
\newcommand{\hatcurRVgammaeccenxxxxxA}{\ensuremath{31662.8\pm6.1}}           
\newcommand{\hatcurRVjittereccenxxxxxA}{\ensuremath{0.00\pm0.55}}            
\newcommand{\hatcurRVjittertwosiglimeccenxxxxxA}{\ensuremath{<1.4}}               
\newcommand{\hatcurRVfitrmseccenxxxxxA}{\ensuremath{.1fym}}                  %
\newcommand{\hatcurRVecceneccenxxxxxA}{\ensuremath{0.062\pm0.055}}           
\newcommand{\hatcurRVeccentwosiglimeccenxxxxxA}{\ensuremath{<0.176}}         
\newcommand{\hatcurRVomegaeccenxxxxxA}{\ensuremath{211\pm93}}                
\newcommand{\hatcurPPieccenxxxxxA}{\ensuremath{86.90\pm0.70}}                
\newcommand{\hatcurPPgeccenxxxxxA}{\ensuremath{9.6\pm2.3}}                   
\newcommand{\hatcurPPloggeccenxxxxxA}{\ensuremath{2.980\pm0.099}}            
\newcommand{\hatcurPPareccenxxxxxA}{\ensuremath{10.10\pm0.79}}               
\newcommand{\hatcurPPareleccenxxxxxA}{\ensuremath{0.05166\pm0.00060}}        
\newcommand{\hatcurPPrhoeccenxxxxxA}{\ensuremath{0.38\pm0.13}}               
\newcommand{\hatcurPPmeccenxxxxxA}{\ensuremath{0.609\pm0.057}}               
\newcommand{\hatcurPPmshorteccenxxxxxA}{\ensuremath{0.61}}                   
\newcommand{\hatcurPPmlongeccenxxxxxA}{\ensuremath{0.609\pm0.057}}           
\newcommand{\hatcurPPmeeccenxxxxxA}{\ensuremath{193\pm18}}                   
\newcommand{\hatcurPPmeshorteccenxxxxxA}{\ensuremath{193.4}}                 
\newcommand{\hatcurPPmelongeccenxxxxxA}{\ensuremath{193\pm18}}               
\newcommand{\hatcurPPreccenxxxxxA}{\ensuremath{1.26\pm0.12}}                 
\newcommand{\hatcurPPrshorteccenxxxxxA}{\ensuremath{1.26}}                   
\newcommand{\hatcurPPrlongeccenxxxxxA}{\ensuremath{1.26\pm0.12}}             
\newcommand{\hatcurPPreeccenxxxxxA}{\ensuremath{14.1\pm1.3}}                 
\newcommand{\hatcurPPreshorteccenxxxxxA}{\ensuremath{14.1}}                  
\newcommand{\hatcurPPrelongeccenxxxxxA}{\ensuremath{14.1\pm1.3}}             
\newcommand{\hatcurPPmrcorreccenxxxxxA}{\ensuremath{-0.27}}                  
\newcommand{\hatcurPPteffeccenxxxxxA}{\ensuremath{1274\pm52}}                
\newcommand{\hatcurPPthetaeccenxxxxxA}{\ensuremath{0.0501\pm0.0076}}         
\newcommand{\hatcurPPfluxperieccenxxxxxA}{\ensuremath{6.74_{-0.95}^{+1.40}}} 
\newcommand{\hatcurPPfluxperidimeccenxxxxxA}{\ensuremath{8}}                 
\newcommand{\hatcurPPfluxapeccenxxxxxA}{\ensuremath{5.3\pm1.0}}              
\newcommand{\hatcurPPfluxapdimeccenxxxxxA}{\ensuremath{8}}                   
\newcommand{\hatcurPPfluxavgeccenxxxxxA}{\ensuremath{5.94\pm0.98}}           
\newcommand{\hatcurPPfluxavgdimeccenxxxxxA}{\ensuremath{8}}                  
\newcommand{\hatcurPPfluxavglogeccenxxxxxA}{\ensuremath{8.774\pm0.071}}      
\newcommand{\hatcurXsecphaseeccenxxxxxA}{\ensuremath{0.498\pm0.039}}         
\newcommand{\hatcurXsecondaryeccenxxxxxA}{\ensuremath{2456868.21\pm0.17}}    
\newcommand{\hatcurXsecdureccenxxxxxA}{\ensuremath{0.132\pm0.011}}           
\newcommand{\hatcurXsecingdureccenxxxxxA}{\ensuremath{0.0184\pm0.0044}}      
\newcommand{\hatcurPPphiconjeccenxxxxxA}{\ensuremath{-0.06_{-0.28}^{+0.43}}} 
\newcommand{\hatcurPPperieccenxxxxxA}{\ensuremath{2456866.3\pm1.3}}          
\newcommand{\hatcurPPaequiveccenxxxxxA}{\ensuremath{0.0480\pm0.0040}}        
\newcommand{\hatcurPPtcirceccenxxxxxA}{\ensuremath{320_{-130}^{+190}}}       
\newcommand{\hatcurPPtinfalleccenxxxxxA}{\ensuremath{7000_{-2000}^{+3300}}}  
\newcommand{\hatcurXdisteccenxxxxxA}{\ensuremath{468\pm38}}                  
\newcommand{\hatcurXAveccenxxxxxA}{\ensuremath{0.082\pm0.061}}               
\newcommand{\hatcurXdistredeccenxxxxxA}{\ensuremath{463\pm37}}               
\newcommand{\hatcurXEBVeccenxxxxxA}{\ensuremath{0.026\pm0.020}}              
\newcommand{\hatcurXmvisoredeccenxxxxxA}{\ensuremath{13.096\pm0.029}}        
\newcommand{\hatcurXmiisoredeccenxxxxxA}{\ensuremath{12.336\pm0.016}}        
\newcommand{\hatcurXmjisoredeccenxxxxxA}{\ensuremath{11.868\pm0.014}}        
\newcommand{\hatcurXmhisoredeccenxxxxxA}{\ensuremath{11.508\pm0.015}}        
\newcommand{\hatcurXmkisoredeccenxxxxxA}{\ensuremath{11.445\pm0.016}}        
\newcommand{\hatcurXviisoredeccenxxxxxA}{\ensuremath{0.759\pm0.020}}         
\newcommand{\hatcurXvkisoredeccenxxxxxA}{\ensuremath{1.651\pm0.035}}         
\newcommand{\hatcurXjhisoredeccenxxxxxA}{\ensuremath{0.3600\pm0.0097}}       
\newcommand{\hatcurXjkisoredeccenxxxxxA}{\ensuremath{0.422\pm0.010}}         
\newcommand{\hatcurCCpmraeccenxxxxxA}{\ensuremath{-20.5\pm1.0}}              
\newcommand{\hatcurCCpmdececcenxxxxxA}{\ensuremath{-11.9\pm1.1}}             
\newcommand{\hatcurCCpmeccenxxxxxA}{\ensuremath{23.7\pm1.5}}                 

\newcommand{\hatcurhtreccenxxxxxB}{HATS606-007}                              
\newcommand{\hatcurfieldeccenxxxxxB}{\ensuremath{string}}                    
\newcommand{\hatcurCCraeccenxxxxxB}{\ensuremath{09^{\mathrm h}39^{\mathrm m}42.44{\mathrm s}}}                             
\newcommand{\hatcurCCdececcenxxxxxB}{\ensuremath{-28{\arcdeg}35{\arcmin}08.1{\arcsec}}}                            
\newcommand{\hatcurCCmageccenxxxxxB}{12.955}                                 
\newcommand{\hatcurCCtwomasseccenxxxxxB}{2MASS~09394244-2835081}             
\newcommand{\hatcurCCgsceccenxxxxxB}{GSC~6614-01083}                         
\newcommand{\hatcurCCtassmveccenxxxxxB}{\ensuremath{12.955\pm0.030}}         
\newcommand{\hatcurCCtassmvshorteccenxxxxxB}{\ensuremath{13.0}}              
\newcommand{\hatcurCCtassmBeccenxxxxxB}{\ensuremath{13.553\pm0.030}}         
\newcommand{\hatcurCCtassmBshorteccenxxxxxB}{\ensuremath{13.6}}              
\newcommand{\hatcurCCtassmIeccenxxxxxB}{\ensuremath{100\pm1000}}             
\newcommand{\hatcurCCtassmIshorteccenxxxxxB}{\ensuremath{100.0}}             
\newcommand{\hatcurCCtassmgeccenxxxxxB}{\ensuremath{13.229\pm0.010}}         
\newcommand{\hatcurCCtassmgshorteccenxxxxxB}{\ensuremath{13.2}}              
\newcommand{\hatcurCCtassmreccenxxxxxB}{\ensuremath{12.822\pm0.010}}         
\newcommand{\hatcurCCtassmrshorteccenxxxxxB}{\ensuremath{12.8}}              
\newcommand{\hatcurCCtassmieccenxxxxxB}{\ensuremath{12.695\pm0.030}}         
\newcommand{\hatcurCCtassmishorteccenxxxxxB}{\ensuremath{12.7}}              
\newcommand{\hatcurCCtwomassJmageccenxxxxxB}{\ensuremath{11.839\pm0.024}}    
\newcommand{\hatcurCCtwomassHmageccenxxxxxB}{\ensuremath{11.510\pm0.024}}    
\newcommand{\hatcurCCtwomassKmageccenxxxxxB}{\ensuremath{11.435\pm0.021}}    
\newcommand{\hatcurCCcitJmageccenxxxxxB}{\ensuremath{11.854\pm0.024}}        
\newcommand{\hatcurCCcitHmageccenxxxxxB}{\ensuremath{11.505\pm0.025}}        
\newcommand{\hatcurCCcitKmageccenxxxxxB}{\ensuremath{11.459\pm0.021}}        
\newcommand{\hatcurCCbbJmageccenxxxxxB}{\ensuremath{11.906\pm0.026}}         
\newcommand{\hatcurCCbbHmageccenxxxxxB}{\ensuremath{11.526\pm0.025}}         
\newcommand{\hatcurCCbbKmageccenxxxxxB}{\ensuremath{11.479\pm0.021}}         
\newcommand{\hatcurCCesoJmageccenxxxxxB}{\ensuremath{11.909\pm0.027}}        
\newcommand{\hatcurCCesoHmageccenxxxxxB}{\ensuremath{11.521\pm0.028}}        
\newcommand{\hatcurCCesoKmageccenxxxxxB}{\ensuremath{11.478\pm0.022}}        
\newcommand{\hatcurCCesoJHmageccenxxxxxB}{\ensuremath{0.388\pm0.011}}        
\newcommand{\hatcurCCesoJKmageccenxxxxxB}{\ensuremath{0.431\pm0.034}}        
\newcommand{\hatcurCCesoHKmageccenxxxxxB}{\ensuremath{0.043\pm0.035}}        
\newcommand{\hatcurLCdipeccenxxxxxB}{\ensuremath{8.3}}                       
\newcommand{\hatcurLCrprstareccenxxxxxB}{\ensuremath{0.0897\pm0.0051}}       
\newcommand{\hatcurLCbsqeccenxxxxxB}{\ensuremath{0.108_{-0.082}^{+0.108}}}   
\newcommand{\hatcurLCimpeccenxxxxxB}{\ensuremath{0.33_{-0.17}^{+0.14}}}      
\newcommand{\hatcurLCzetaeccenxxxxxB}{\ensuremath{10.17\pm0.15}}             
\newcommand{\hatcurLCdureccenxxxxxB}{\ensuremath{0.2162\pm0.0041}}           
\newcommand{\hatcurLCdurshorteccenxxxxxB}{\ensuremath{0.2162}}               
\newcommand{\hatcurLCdurhreccenxxxxxB}{\ensuremath{5.189\pm0.099}}           
\newcommand{\hatcurLCdurhrshorteccenxxxxxB}{\ensuremath{5.189}}              
\newcommand{\hatcurLCqeccenxxxxxB}{\ensuremath{0.0655\pm0.0012}}             
\newcommand{\hatcurLCqshorteccenxxxxxB}{\ensuremath{0.066}}                  
\newcommand{\hatcurLCingdureccenxxxxxB}{\ensuremath{0.0196\pm0.0033}}        
\newcommand{\hatcurLCPeccenxxxxxB}{\ensuremath{3.3023890\pm0.0000084}}       
\newcommand{\hatcurLCPprececcenxxxxxB}{\ensuremath{3.3023890}}               
\newcommand{\hatcurLCPshorteccenxxxxxB}{\ensuremath{3.3024}}                 
\newcommand{\hatcurLCTeccenxxxxxB}{\ensuremath{2456969.7974\pm0.0011}}       
\newcommand{\hatcurLCTAeccenxxxxxB}{\ensuremath{2455972.4758\pm0.0027}}      
\newcommand{\hatcurLCTBeccenxxxxxB}{\ensuremath{2457177.8479\pm0.0012}}      
\newcommand{\hatcurLChatnetmAeccenxxxxxB}{\ensuremath{-12486\pm17}}          
\newcommand{\hatcurLCiblendAeccenxxxxxB}{\ensuremath{0.01\pm0.29}}           
\newcommand{\hatcurLChatnetmBeccenxxxxxB}{\ensuremath{-12558.7\pm7.7}}       
\newcommand{\hatcurLCiblendBeccenxxxxxB}{\ensuremath{0.1\pm5.3}}             
\newcommand{\hatcurLCrhoeccenxxxxxB}{\ensuremath{0.200^{+0.044}_{-0.072}}}                    
\newcommand{\hatcurSMEiteffeccenxxxxxB}{\ensuremath{6062\pm98}}              
\newcommand{\hatcurSMEizfeheccenxxxxxB}{\ensuremath{-0.020\pm0.050}}         
\newcommand{\hatcurSMEizfehshorteccenxxxxxB}{\ensuremath{-0.02}}             
\newcommand{\hatcurSMEiloggeccenxxxxxB}{\ensuremath{3.97\pm0.19}}            
\newcommand{\hatcurSMEivsineccenxxxxxB}{\ensuremath{7.43\pm0.50}}            
\newcommand{\hatcurSMEivmaceccenxxxxxB}{\ensuremath{0.0}}                    
\newcommand{\hatcurSMEivmiceccenxxxxxB}{\ensuremath{0.0}}                    
\newcommand{\hatcurSMEiiteffeccenxxxxxB}{\ensuremath{6071\pm81}}             
\newcommand{\hatcurSMEiizfeheccenxxxxxB}{\ensuremath{-0.020\pm0.050}}        
\newcommand{\hatcurSMEiizfehshorteccenxxxxxB}{\ensuremath{-0.02}}            
\newcommand{\hatcurSMEiiloggeccenxxxxxB}{\ensuremath{3.951\pm0.042}}         
\newcommand{\hatcurSMEiivsineccenxxxxxB}{\ensuremath{7.48\pm0.50}}           
\newcommand{\hatcurLBizeccenxxxxxB}{\ensuremath{0.1616}}                     
\newcommand{\hatcurLBiizeccenxxxxxB}{\ensuremath{0.3511}}                    
\newcommand{\hatcurLBiieccenxxxxxB}{\ensuremath{0.2145}}                     
\newcommand{\hatcurLBiiieccenxxxxxB}{\ensuremath{0.3580}}                    
\newcommand{\hatcurLBiIeccenxxxxxB}{\ensuremath{0.1954}}                     
\newcommand{\hatcurLBiiIeccenxxxxxB}{\ensuremath{0.3572}}                    
\newcommand{\hatcurLBigeccenxxxxxB}{\ensuremath{0.4752}}                     
\newcommand{\hatcurLBiigeccenxxxxxB}{\ensuremath{0.2907}}                    
\newcommand{\hatcurLBireccenxxxxxB}{\ensuremath{0.2947}}                     
\newcommand{\hatcurLBiireccenxxxxxB}{\ensuremath{0.3611}}                    
\newcommand{\hatcurLBiReccenxxxxxB}{\ensuremath{0.2721}}                     
\newcommand{\hatcurLBiiReccenxxxxxB}{\ensuremath{0.3616}}                    
\newcommand{\hatcurLBikepeccenxxxxxB}{\ensuremath{0.1000}}                   
\newcommand{\hatcurLBiikepeccenxxxxxB}{\ensuremath{0.1000}}                  
\newcommand{\hatcurISOmeccenxxxxxB}{\ensuremath{1.342\pm0.096}}              
\newcommand{\hatcurISOmshorteccenxxxxxB}{\ensuremath{1.34}}                  
\newcommand{\hatcurISOmlongeccenxxxxxB}{\ensuremath{1.342\pm0.096}}          
\newcommand{\hatcurISOreccenxxxxxB}{\ensuremath{2.12_{-0.17}^{+0.39}}}       
\newcommand{\hatcurISOrshorteccenxxxxxB}{\ensuremath{2.12}}                  
\newcommand{\hatcurISOrlongeccenxxxxxB}{\ensuremath{2.12_{-0.17}^{+0.39}}}   
\newcommand{\hatcurISOrhoeccenxxxxxB}{\ensuremath{0.197_{-0.069}^{+0.047}}}  
\newcommand{\hatcurISOrholongeccenxxxxxB}{\ensuremath{0.197_{-0.069}^{+0.047}}} 
\newcommand{\hatcurISOloggeccenxxxxxB}{\ensuremath{3.909\pm0.087}}           
\newcommand{\hatcurISOlumeccenxxxxxB}{\ensuremath{5.51_{-0.98}^{+2.13}}}     
\newcommand{\hatcurISOlumshorteccenxxxxxB}{\ensuremath{5.51}}                
\newcommand{\hatcurISOmveccenxxxxxB}{\ensuremath{2.94\pm0.29}}               
\newcommand{\hatcurISOvieccenxxxxxB}{\ensuremath{0.608\pm0.023}}             
\newcommand{\hatcurISOageeccenxxxxxB}{\ensuremath{3.74\pm0.73}}              
\newcommand{\hatcurISOsigmaeccenxxxxxB}{\ensuremath{0.000200\pm0.000046}}    
\newcommand{\hatcurISOMJeccenxxxxxB}{\ensuremath{1.94\pm0.28}}               
\newcommand{\hatcurISOMHeccenxxxxxB}{\ensuremath{1.65\pm0.28}}               
\newcommand{\hatcurISOMKeccenxxxxxB}{\ensuremath{1.60\pm0.28}}               
\newcommand{\hatcurISOJKeccenxxxxxB}{\ensuremath{0.340\pm0.010}}             
\newcommand{\hatcurISOspececcenxxxxxB}{F}                                    
\newcommand{\hatcurRVKeccenxxxxxB}{\ensuremath{76.0\pm6.0}}                  
\newcommand{\hatcurRVrkeccenxxxxxB}{\ensuremath{0.10\pm0.14}}                
\newcommand{\hatcurRVrheccenxxxxxB}{\ensuremath{0.13\pm0.22}}                
\newcommand{\hatcurRVkeccenxxxxxB}{\ensuremath{0.025_{-0.033}^{+0.055}}}     
\newcommand{\hatcurRVheccenxxxxxB}{\ensuremath{0.027_{-0.045}^{+0.117}}}     
\newcommand{\hatcurRVtroneeccenxxxxxB}{\ensuremath{0\pm0}}                   
\newcommand{\hatcurRVtrtwoeccenxxxxxB}{\ensuremath{0\pm0}}                   
\newcommand{\hatcurRVgammaeccenxxxxxB}{\ensuremath{-12512.3\pm7.2}}          
\newcommand{\hatcurRVjittereccenxxxxxB}{\ensuremath{2.1\pm8.6}}              
\newcommand{\hatcurRVjittertwosiglimeccenxxxxxB}{\ensuremath{<23}}               
\newcommand{\hatcurRVfitrmseccenxxxxxB}{\ensuremath{.1fym}}                  %
\newcommand{\hatcurRVecceneccenxxxxxB}{\ensuremath{0.068\pm0.085}}           
\newcommand{\hatcurRVeccentwosiglimeccenxxxxxB}{\ensuremath{<0.245}}         
\newcommand{\hatcurRVomegaeccenxxxxxB}{\ensuremath{90\pm120}}                
\newcommand{\hatcurPPieccenxxxxxB}{\ensuremath{86.2_{-3.1}^{+1.9}}}          
\newcommand{\hatcurPPgeccenxxxxxB}{\ensuremath{4.9_{-1.3}^{+1.0}}}           
\newcommand{\hatcurPPloggeccenxxxxxB}{\ensuremath{2.687_{-0.139}^{+0.083}}}  
\newcommand{\hatcurPPareccenxxxxxB}{\ensuremath{4.84_{-0.64}^{+0.36}}}       
\newcommand{\hatcurPPareleccenxxxxxB}{\ensuremath{0.0479\pm0.0011}}          
\newcommand{\hatcurPPrhoeccenxxxxxB}{\ensuremath{0.133\pm0.050}}             
\newcommand{\hatcurPPmeccenxxxxxB}{\ensuremath{0.675\pm0.065}}               
\newcommand{\hatcurPPmshorteccenxxxxxB}{\ensuremath{0.68}}                   
\newcommand{\hatcurPPmlongeccenxxxxxB}{\ensuremath{0.675\pm0.065}}           
\newcommand{\hatcurPPmeeccenxxxxxB}{\ensuremath{215\pm21}}                   
\newcommand{\hatcurPPmeshorteccenxxxxxB}{\ensuremath{214.7}}                 
\newcommand{\hatcurPPmelongeccenxxxxxB}{\ensuremath{215\pm21}}               
\newcommand{\hatcurPPreccenxxxxxB}{\ensuremath{1.84_{-0.17}^{+0.39}}}        
\newcommand{\hatcurPPrshorteccenxxxxxB}{\ensuremath{1.84}}                   
\newcommand{\hatcurPPrlongeccenxxxxxB}{\ensuremath{1.84_{-0.17}^{+0.39}}}    
\newcommand{\hatcurPPreeccenxxxxxB}{\ensuremath{20.6_{-2.0}^{+4.4}}}         
\newcommand{\hatcurPPreshorteccenxxxxxB}{\ensuremath{20.6}}                  
\newcommand{\hatcurPPrelongeccenxxxxxB}{\ensuremath{20.6_{-2.0}^{+4.4}}}     
\newcommand{\hatcurPPmrcorreccenxxxxxB}{\ensuremath{0.56}}                   
\newcommand{\hatcurPPteffeccenxxxxxB}{\ensuremath{1953_{-75}^{+153}}}        
\newcommand{\hatcurPPthetaeccenxxxxxB}{\ensuremath{0.0255\pm0.0038}}         
\newcommand{\hatcurPPfluxperieccenxxxxxB}{\ensuremath{3.71_{-0.63}^{+2.69}}} 
\newcommand{\hatcurPPfluxperidimeccenxxxxxB}{\ensuremath{9}}                 
\newcommand{\hatcurPPfluxapeccenxxxxxB}{\ensuremath{2.87_{-0.37}^{+0.49}}}   
\newcommand{\hatcurPPfluxapdimeccenxxxxxB}{\ensuremath{9}}                   
\newcommand{\hatcurPPfluxavgeccenxxxxxB}{\ensuremath{3.29_{-0.48}^{+1.15}}}  
\newcommand{\hatcurPPfluxavgdimeccenxxxxxB}{\ensuremath{9}}                  
\newcommand{\hatcurPPfluxavglogeccenxxxxxB}{\ensuremath{9.517_{-0.068}^{+0.131}}} 
\newcommand{\hatcurXsecphaseeccenxxxxxB}{\ensuremath{0.516\pm0.031}}         
\newcommand{\hatcurXsecondaryeccenxxxxxB}{\ensuremath{2456971.50\pm0.10}}    
\newcommand{\hatcurXsecdureccenxxxxxB}{\ensuremath{0.228\pm0.041}}           
\newcommand{\hatcurXsecingdureccenxxxxxB}{\ensuremath{0.021\pm0.013}}        
\newcommand{\hatcurPPphiconjeccenxxxxxB}{\ensuremath{0.07_{-0.12}^{+0.24}}}  
\newcommand{\hatcurPPperieccenxxxxxB}{\ensuremath{2456969.58\pm0.64}}        
\newcommand{\hatcurPPaequiveccenxxxxxB}{\ensuremath{0.0204_{-0.0027}^{+0.0017}}} 
\newcommand{\hatcurPPtcirceccenxxxxxB}{\ensuremath{20\pm13}}                 
\newcommand{\hatcurPPtinfalleccenxxxxxB}{\ensuremath{163\pm82}}              
\newcommand{\hatcurXdisteccenxxxxxB}{\ensuremath{947_{-80}^{+169}}}          
\newcommand{\hatcurXAveccenxxxxxB}{\ensuremath{0.142\pm0.070}}               
\newcommand{\hatcurXdistredeccenxxxxxB}{\ensuremath{944_{-78}^{+169}}}       
\newcommand{\hatcurXEBVeccenxxxxxB}{\ensuremath{0.046\pm0.023}}              
\newcommand{\hatcurXmvisoredeccenxxxxxB}{\ensuremath{12.965\pm0.029}}        
\newcommand{\hatcurXmiisoredeccenxxxxxB}{\ensuremath{12.281\pm0.016}}        
\newcommand{\hatcurXmjisoredeccenxxxxxB}{\ensuremath{11.855\pm0.014}}        
\newcommand{\hatcurXmhisoredeccenxxxxxB}{\ensuremath{11.552\pm0.015}}        
\newcommand{\hatcurXmkisoredeccenxxxxxB}{\ensuremath{11.489\pm0.016}}        
\newcommand{\hatcurXviisoredeccenxxxxxB}{\ensuremath{0.683\pm0.022}}         
\newcommand{\hatcurXvkisoredeccenxxxxxB}{\ensuremath{1.475\pm0.035}}         
\newcommand{\hatcurXjhisoredeccenxxxxxB}{\ensuremath{0.3040\pm0.0087}}       
\newcommand{\hatcurXjkisoredeccenxxxxxB}{\ensuremath{0.3660\pm0.0083}}       
\newcommand{\hatcurCCpmraeccenxxxxxB}{\ensuremath{-1.3\pm1.4}}               
\newcommand{\hatcurCCpmdececcenxxxxxB}{\ensuremath{-6.1\pm1.3}}              
\newcommand{\hatcurCCpmeccenxxxxxB}{\ensuremath{6.2\pm1.9}}                  

\newcommand{\hatcurhtreccenxxxxxC}{HATS700-004}                        
\newcommand{\hatcurfieldeccenxxxxxC}{\ensuremath{string}}              
\newcommand{\hatcurCCraeccenxxxxxC}{\ensuremath{12^{\mathrm h}54^{\mathrm m}12.60{\mathrm s}}}                       
\newcommand{\hatcurCCdececcenxxxxxC}{\ensuremath{-46{\arcdeg}35{\arcmin}15.8{\arcsec}}}                      
\newcommand{\hatcurCCmageccenxxxxxC}{12.766}                           
\newcommand{\hatcurCCtwomasseccenxxxxxC}{2MASS~12541261-4635157}       
\newcommand{\hatcurCCgsceccenxxxxxC}{GSC~8245-02236}                   
\newcommand{\hatcurCCtassmveccenxxxxxC}{\ensuremath{12.766\pm0.040}}   
\newcommand{\hatcurCCtassmvshorteccenxxxxxC}{\ensuremath{12.8}}        
\newcommand{\hatcurCCtassmBeccenxxxxxC}{\ensuremath{13.239\pm0.050}}   
\newcommand{\hatcurCCtassmBshorteccenxxxxxC}{\ensuremath{13.2}}        
\newcommand{\hatcurCCtassmIeccenxxxxxC}{\ensuremath{100\pm100}}        
\newcommand{\hatcurCCtassmIshorteccenxxxxxC}{\ensuremath{100.0}}       
\newcommand{\hatcurCCtassmgeccenxxxxxC}{\ensuremath{12.927\pm0.040}}   
\newcommand{\hatcurCCtassmgshorteccenxxxxxC}{\ensuremath{12.9}}        
\newcommand{\hatcurCCtassmreccenxxxxxC}{\ensuremath{12.665\pm0.040}}   
\newcommand{\hatcurCCtassmrshorteccenxxxxxC}{\ensuremath{12.7}}        
\newcommand{\hatcurCCtassmieccenxxxxxC}{\ensuremath{12.515\pm0.080}}   
\newcommand{\hatcurCCtassmishorteccenxxxxxC}{\ensuremath{12.5}}        
\newcommand{\hatcurCCtwomassJmageccenxxxxxC}{\ensuremath{11.831\pm0.022}} 
\newcommand{\hatcurCCtwomassHmageccenxxxxxC}{\ensuremath{11.651\pm0.023}} 
\newcommand{\hatcurCCtwomassKmageccenxxxxxC}{\ensuremath{11.550\pm0.023}} 
\newcommand{\hatcurCCcitJmageccenxxxxxC}{\ensuremath{11.853\pm0.022}}  
\newcommand{\hatcurCCcitHmageccenxxxxxC}{\ensuremath{11.645\pm0.024}}  
\newcommand{\hatcurCCcitKmageccenxxxxxC}{\ensuremath{11.574\pm0.023}}  
\newcommand{\hatcurCCbbJmageccenxxxxxC}{\ensuremath{11.895\pm0.024}}   
\newcommand{\hatcurCCbbHmageccenxxxxxC}{\ensuremath{11.667\pm0.024}}   
\newcommand{\hatcurCCbbKmageccenxxxxxC}{\ensuremath{11.594\pm0.023}}   
\newcommand{\hatcurCCesoJmageccenxxxxxC}{\ensuremath{11.895\pm0.025}}  
\newcommand{\hatcurCCesoHmageccenxxxxxC}{\ensuremath{11.665\pm0.026}}  
\newcommand{\hatcurCCesoKmageccenxxxxxC}{\ensuremath{11.594\pm0.024}}  
\newcommand{\hatcurCCesoJHmageccenxxxxxC}{\ensuremath{0.231\pm0.035}}  
\newcommand{\hatcurCCesoJKmageccenxxxxxC}{\ensuremath{0.302\pm0.034}}  
\newcommand{\hatcurCCesoHKmageccenxxxxxC}{\ensuremath{0.070\pm0.012}}  
\newcommand{\hatcurLCdipeccenxxxxxC}{\ensuremath{8.2}}                 
\newcommand{\hatcurLCrprstareccenxxxxxC}{\ensuremath{0.0892\pm0.0042}} 
\newcommand{\hatcurLCbsqeccenxxxxxC}{\ensuremath{0.126_{-0.080}^{+0.075}}} 
\newcommand{\hatcurLCimpeccenxxxxxC}{\ensuremath{0.355_{-0.141}^{+0.093}}} 
\newcommand{\hatcurLCzetaeccenxxxxxC}{\ensuremath{11.04\pm0.14}}       
\newcommand{\hatcurLCdureccenxxxxxC}{\ensuremath{0.1996\pm0.0033}}     
\newcommand{\hatcurLCdurshorteccenxxxxxC}{\ensuremath{0.1996}}         
\newcommand{\hatcurLCdurhreccenxxxxxC}{\ensuremath{4.791\pm0.078}}     
\newcommand{\hatcurLCdurhrshorteccenxxxxxC}{\ensuremath{4.791}}        
\newcommand{\hatcurLCqeccenxxxxxC}{\ensuremath{0.04300\pm0.00071}}     
\newcommand{\hatcurLCqshorteccenxxxxxC}{\ensuremath{0.043}}            
\newcommand{\hatcurLCingdureccenxxxxxC}{\ensuremath{0.0182\pm0.0023}}  
\newcommand{\hatcurLCPeccenxxxxxC}{\ensuremath{4.637034\pm0.000012}}   
\newcommand{\hatcurLCPprececcenxxxxxC}{\ensuremath{4.6370340}}         
\newcommand{\hatcurLCPshorteccenxxxxxC}{\ensuremath{4.6370}}           
\newcommand{\hatcurLCTeccenxxxxxC}{\ensuremath{2456945.8704\pm0.0013}} 
\newcommand{\hatcurLCTAeccenxxxxxC}{\ensuremath{2455679.9599\pm0.0036}} 
\newcommand{\hatcurLCTBeccenxxxxxC}{\ensuremath{2457122.0778\pm0.0013}} 
\newcommand{\hatcurLChatnetmeccenxxxxxC}{\ensuremath{12.698730\pm0.000071}} 
\newcommand{\hatcurLCiblendeccenxxxxxC}{\ensuremath{0.794\pm0.084}}    
\newcommand{\hatcurLCrhoeccenxxxxxC}{\ensuremath{0.26\pm0.24}}         
\newcommand{\hatcurSMEiteffeccenxxxxxC}{\ensuremath{6438\pm64}}        
\newcommand{\hatcurSMEizfeheccenxxxxxC}{\ensuremath{0.090\pm0.040}}    
\newcommand{\hatcurSMEizfehshorteccenxxxxxC}{\ensuremath{0.09}}        
\newcommand{\hatcurSMEiloggeccenxxxxxC}{\ensuremath{4.07\pm0.10}}      
\newcommand{\hatcurSMEivsineccenxxxxxC}{\ensuremath{9.32\pm0.50}}      
\newcommand{\hatcurSMEivmaceccenxxxxxC}{\ensuremath{0.0}}              
\newcommand{\hatcurSMEivmiceccenxxxxxC}{\ensuremath{0.0}}              
\newcommand{\hatcurLBizeccenxxxxxC}{\ensuremath{0.1252}}               
\newcommand{\hatcurLBiizeccenxxxxxC}{\ensuremath{0.3688}}              
\newcommand{\hatcurLBiieccenxxxxxC}{\ensuremath{0.1754}}               
\newcommand{\hatcurLBiiieccenxxxxxC}{\ensuremath{0.3788}}              
\newcommand{\hatcurLBiIeccenxxxxxC}{\ensuremath{0.1568}}               
\newcommand{\hatcurLBiiIeccenxxxxxC}{\ensuremath{0.3769}}              
\newcommand{\hatcurLBigeccenxxxxxC}{\ensuremath{0.4175}}               
\newcommand{\hatcurLBiigeccenxxxxxC}{\ensuremath{0.3314}}              
\newcommand{\hatcurLBireccenxxxxxC}{\ensuremath{0.2511}}               
\newcommand{\hatcurLBiireccenxxxxxC}{\ensuremath{0.3857}}              
\newcommand{\hatcurLBiReccenxxxxxC}{\ensuremath{0.2295}}               
\newcommand{\hatcurLBiiReccenxxxxxC}{\ensuremath{0.3855}}              
\newcommand{\hatcurLBikepeccenxxxxxC}{\ensuremath{0.1000}}             
\newcommand{\hatcurLBiikepeccenxxxxxC}{\ensuremath{0.1000}}            
\newcommand{\hatcurISOmeccenxxxxxC}{\ensuremath{1.50_{-0.10}^{+0.18}}} 
\newcommand{\hatcurISOmshorteccenxxxxxC}{\ensuremath{1.50}}            
\newcommand{\hatcurISOmlongeccenxxxxxC}{\ensuremath{1.50_{-0.10}^{+0.18}}} 
\newcommand{\hatcurISOreccenxxxxxC}{\ensuremath{2.01_{-0.30}^{+0.59}}} 
\newcommand{\hatcurISOrshorteccenxxxxxC}{\ensuremath{2.01}}            
\newcommand{\hatcurISOrlongeccenxxxxxC}{\ensuremath{2.01_{-0.30}^{+0.59}}} 
\newcommand{\hatcurISOrhoeccenxxxxxC}{\ensuremath{0.26\pm0.14}}        
\newcommand{\hatcurISOrholongeccenxxxxxC}{\ensuremath{0.26\pm0.14}}    
\newcommand{\hatcurISOloggeccenxxxxxC}{\ensuremath{4.01\pm0.14}}       
\newcommand{\hatcurISOlumeccenxxxxxC}{\ensuremath{6.2_{-1.7}^{+4.2}}}  
\newcommand{\hatcurISOlumshorteccenxxxxxC}{\ensuremath{6.19}}          
\newcommand{\hatcurISOmveccenxxxxxC}{\ensuremath{2.75\pm0.44}}         
\newcommand{\hatcurISOvieccenxxxxxC}{\ensuremath{0.511\pm0.016}}       
\newcommand{\hatcurISOageeccenxxxxxC}{\ensuremath{2.11\pm0.33}}        
\newcommand{\hatcurISOsigmaeccenxxxxxC}{\ensuremath{0.00020\pm0.00010}} 
\newcommand{\hatcurISOMJeccenxxxxxC}{\ensuremath{1.93\pm0.43}}         
\newcommand{\hatcurISOMHeccenxxxxxC}{\ensuremath{1.71\pm0.43}}         
\newcommand{\hatcurISOMKeccenxxxxxC}{\ensuremath{1.66\pm0.43}}         
\newcommand{\hatcurISOJKeccenxxxxxC}{\ensuremath{0.21\pm0.11}}         
\newcommand{\hatcurISOspececcenxxxxxC}{F}                              
\newcommand{\hatcurRVKeccenxxxxxC}{\ensuremath{63\pm46}}               
\newcommand{\hatcurRVrkeccenxxxxxC}{\ensuremath{-0.41_{-0.18}^{+0.35}}} 
\newcommand{\hatcurRVrheccenxxxxxC}{\ensuremath{0.21_{-0.28}^{+0.20}}} 
\newcommand{\hatcurRVkeccenxxxxxC}{\ensuremath{-0.19\pm0.18}}          
\newcommand{\hatcurRVheccenxxxxxC}{\ensuremath{0.09_{-0.10}^{+0.16}}}  
\newcommand{\hatcurRVtroneeccenxxxxxC}{\ensuremath{0\pm0}}             
\newcommand{\hatcurRVtrtwoeccenxxxxxC}{\ensuremath{0\pm0}}             
\newcommand{\hatcurRVgammaAeccenxxxxxC}{\ensuremath{-3544\pm23}}       
\newcommand{\hatcurRVjitterAeccenxxxxxC}{\ensuremath{65\pm17}}         
\newcommand{\hatcurRVjittertwosiglimAeccenxxxxxC}{\ensuremath{<99.7}}  
\newcommand{\hatcurRVfitrmsAeccenxxxxxC}{\ensuremath{0.0}}             
\newcommand{\hatcurRVgammaBeccenxxxxxC}{\ensuremath{-3531\pm49}}       
\newcommand{\hatcurRVjitterBeccenxxxxxC}{\ensuremath{1\pm76}}          
\newcommand{\hatcurRVjittertwosiglimBeccenxxxxxC}{\ensuremath{<135.8}} 
\newcommand{\hatcurRVfitrmsBeccenxxxxxC}{\ensuremath{0.0}}             
\newcommand{\hatcurRVgammaCeccenxxxxxC}{\ensuremath{-3591\pm16}}       
\newcommand{\hatcurRVjitterCeccenxxxxxC}{\ensuremath{20\pm16}}         
\newcommand{\hatcurRVjittertwosiglimCeccenxxxxxC}{\ensuremath{<47.3}}  
\newcommand{\hatcurRVfitrmsCeccenxxxxxC}{\ensuremath{0.0}}             
\newcommand{\hatcurRVecceneccenxxxxxC}{\ensuremath{0.24\pm0.17}}       
\newcommand{\hatcurRVeccentwosiglimeccenxxxxxC}{\ensuremath{<0.581}}   
\newcommand{\hatcurRVomegaeccenxxxxxC}{\ensuremath{154\pm52}}          
\newcommand{\hatcurPPieccenxxxxxC}{\ensuremath{86.5_{-2.5}^{+1.5}}}    
\newcommand{\hatcurPPgeccenxxxxxC}{\ensuremath{5.9\pm2.0}}             
\newcommand{\hatcurPPloggeccenxxxxxC}{\ensuremath{2.77_{-0.17}^{+0.12}}} 
\newcommand{\hatcurPPareccenxxxxxC}{\ensuremath{6.7\pm1.1}}            
\newcommand{\hatcurPPareleccenxxxxxC}{\ensuremath{0.0623_{-0.0015}^{+0.0024}}} 
\newcommand{\hatcurPPrhoeccenxxxxxC}{\ensuremath{0.163\pm0.082}}       
\newcommand{\hatcurPPmeccenxxxxxC}{\ensuremath{0.65_{-0.17}^{+0.52}}}  
\newcommand{\hatcurPPmshorteccenxxxxxC}{\ensuremath{0.65}}             
\newcommand{\hatcurPPmlongeccenxxxxxC}{\ensuremath{0.65_{-0.17}^{+0.52}}} 
\newcommand{\hatcurPPmeeccenxxxxxC}{\ensuremath{208_{-55}^{+165}}}     
\newcommand{\hatcurPPmeshorteccenxxxxxC}{\ensuremath{208.0}}           
\newcommand{\hatcurPPmelongeccenxxxxxC}{\ensuremath{208_{-55}^{+165}}} 
\newcommand{\hatcurPPreccenxxxxxC}{\ensuremath{1.74_{-0.28}^{+0.52}}}  
\newcommand{\hatcurPPrshorteccenxxxxxC}{\ensuremath{1.74}}             
\newcommand{\hatcurPPrlongeccenxxxxxC}{\ensuremath{1.74_{-0.28}^{+0.52}}} 
\newcommand{\hatcurPPreeccenxxxxxC}{\ensuremath{19.5_{-3.2}^{+5.8}}}   
\newcommand{\hatcurPPreshorteccenxxxxxC}{\ensuremath{19.5}}            
\newcommand{\hatcurPPrelongeccenxxxxxC}{\ensuremath{19.5_{-3.2}^{+5.8}}} 
\newcommand{\hatcurPPmrcorreccenxxxxxC}{\ensuremath{0.67}}             
\newcommand{\hatcurPPteffeccenxxxxxC}{\ensuremath{1770_{-130}^{+260}}} 
\newcommand{\hatcurPPthetaeccenxxxxxC}{\ensuremath{0.034\pm0.012}}     
\newcommand{\hatcurPPfluxperieccenxxxxxC}{\ensuremath{3.7_{-1.6}^{+8.4}}} 
\newcommand{\hatcurPPfluxperidimeccenxxxxxC}{\ensuremath{9}}           
\newcommand{\hatcurPPfluxapeccenxxxxxC}{\ensuremath{1.49\pm0.32}}      
\newcommand{\hatcurPPfluxapdimeccenxxxxxC}{\ensuremath{9}}             
\newcommand{\hatcurPPfluxavgeccenxxxxxC}{\ensuremath{2.24_{-0.58}^{+1.60}}} 
\newcommand{\hatcurPPfluxavgdimeccenxxxxxC}{\ensuremath{9}}            
\newcommand{\hatcurPPfluxavglogeccenxxxxxC}{\ensuremath{9.35_{-0.13}^{+0.23}}} 
\newcommand{\hatcurXsecphaseeccenxxxxxC}{\ensuremath{0.38\pm0.11}}     
\newcommand{\hatcurXsecondaryeccenxxxxxC}{\ensuremath{2456947.63\pm0.52}} 
\newcommand{\hatcurXsecdureccenxxxxxC}{\ensuremath{0.232\pm0.054}}     
\newcommand{\hatcurXsecingdureccenxxxxxC}{\ensuremath{0.023\pm0.017}}  
\newcommand{\hatcurPPphiconjeccenxxxxxC}{\ensuremath{-0.080_{-0.116}^{+0.056}}} 
\newcommand{\hatcurPPperieccenxxxxxC}{\ensuremath{2456946.24\pm0.69}}  
\newcommand{\hatcurPPaequiveccenxxxxxC}{\ensuremath{0.0250\pm0.0042}}  
\newcommand{\hatcurPPtcirceccenxxxxxC}{\ensuremath{87_{-75}^{+112}}}   
\newcommand{\hatcurPPtinfalleccenxxxxxC}{\ensuremath{1400_{-1100}^{+1900}}} 
\newcommand{\hatcurXdisteccenxxxxxC}{\ensuremath{970_{-150}^{+280}}}   
\newcommand{\hatcurXAveccenxxxxxC}{\ensuremath{0.088\pm0.066}}         
\newcommand{\hatcurXdistredeccenxxxxxC}{\ensuremath{970_{-150}^{+280}}} 
\newcommand{\hatcurXEBVeccenxxxxxC}{\ensuremath{0.028\pm0.021}}        
\newcommand{\hatcurXmvisoredeccenxxxxxC}{\ensuremath{12.773\pm0.036}}  
\newcommand{\hatcurXmiisoredeccenxxxxxC}{\ensuremath{12.215\pm0.020}}  
\newcommand{\hatcurXmjisoredeccenxxxxxC}{\ensuremath{11.888\pm0.014}}  
\newcommand{\hatcurXmhisoredeccenxxxxxC}{\ensuremath{11.656\pm0.015}}  
\newcommand{\hatcurXmkisoredeccenxxxxxC}{\ensuremath{11.606\pm0.016}}  
\newcommand{\hatcurXviisoredeccenxxxxxC}{\ensuremath{0.557\pm0.026}}   
\newcommand{\hatcurXvkisoredeccenxxxxxC}{\ensuremath{1.167\pm0.042}}   
\newcommand{\hatcurXjhisoredeccenxxxxxC}{\ensuremath{0.2320\pm0.0094}} 
\newcommand{\hatcurXjkisoredeccenxxxxxC}{\ensuremath{0.2820\pm0.0099}} 
\newcommand{\hatcurCCpmraeccenxxxxxC}{\ensuremath{-10.2\pm1.1}}        
\newcommand{\hatcurCCpmdececcenxxxxxC}{\ensuremath{4.7\pm1.1}}         
\newcommand{\hatcurCCpmeccenxxxxxC}{\ensuremath{11.2\pm1.6}}           

\newcommand{\hatcurhtreccenxxxxxD}{HATS746-005}                      
\newcommand{\hatcurfieldeccenxxxxxD}{\ensuremath{string}}            
\newcommand{\hatcurCCraeccenxxxxxD}{\ensuremath{18^{\mathrm h}57^{\mathrm m}36.00{\mathrm s}}}                     
\newcommand{\hatcurCCdececcenxxxxxD}{\ensuremath{-49{\arcdeg}08{\arcmin}18.5{\arcsec}}}                    
\newcommand{\hatcurCCmageccenxxxxxD}{13.934}                         
\newcommand{\hatcurCCtwomasseccenxxxxxD}{2MASS~18573592-4908184}     
\newcommand{\hatcurCCgsceccenxxxxxD}{GSC~8382-00661}                 
\newcommand{\hatcurCCtassmveccenxxxxxD}{\ensuremath{13.934\pm0.080}} 
\newcommand{\hatcurCCtassmvshorteccenxxxxxD}{\ensuremath{13.9}}      
\newcommand{\hatcurCCtassmBeccenxxxxxD}{\ensuremath{14.697\pm0.020}} 
\newcommand{\hatcurCCtassmBshorteccenxxxxxD}{\ensuremath{14.7}}      
\newcommand{\hatcurCCtassmIeccenxxxxxD}{\ensuremath{nff\pmnff}}      
\newcommand{\hatcurCCtassmIshorteccenxxxxxD}{\ensuremath{0.0}}       
\newcommand{\hatcurCCtassmgeccenxxxxxD}{\ensuremath{14.274\pm0.030}} 
\newcommand{\hatcurCCtassmgshorteccenxxxxxD}{\ensuremath{14.3}}      
\newcommand{\hatcurCCtassmreccenxxxxxD}{\ensuremath{13.717\pm0.010}} 
\newcommand{\hatcurCCtassmrshorteccenxxxxxD}{\ensuremath{13.7}}      
\newcommand{\hatcurCCtassmieccenxxxxxD}{\ensuremath{13.615\pm0.010}} 
\newcommand{\hatcurCCtassmishorteccenxxxxxD}{\ensuremath{13.6}}      
\newcommand{\hatcurCCtwomassJmageccenxxxxxD}{\ensuremath{12.522\pm0.026}} 
\newcommand{\hatcurCCtwomassHmageccenxxxxxD}{\ensuremath{12.188\pm0.025}} 
\newcommand{\hatcurCCtwomassKmageccenxxxxxD}{\ensuremath{12.086\pm0.029}} 
\newcommand{\hatcurCCcitJmageccenxxxxxD}{\ensuremath{12.536\pm0.026}} 
\newcommand{\hatcurCCcitHmageccenxxxxxD}{\ensuremath{12.182\pm0.025}} 
\newcommand{\hatcurCCcitKmageccenxxxxxD}{\ensuremath{12.110\pm0.029}} 
\newcommand{\hatcurCCbbJmageccenxxxxxD}{\ensuremath{12.590\pm0.028}} 
\newcommand{\hatcurCCbbHmageccenxxxxxD}{\ensuremath{12.204\pm0.026}} 
\newcommand{\hatcurCCbbKmageccenxxxxxD}{\ensuremath{12.130\pm0.029}} 
\newcommand{\hatcurCCesoJmageccenxxxxxD}{\ensuremath{12.593\pm0.030}} 
\newcommand{\hatcurCCesoHmageccenxxxxxD}{\ensuremath{12.200\pm0.031}} 
\newcommand{\hatcurCCesoKmageccenxxxxxD}{\ensuremath{12.129\pm0.030}} 
\newcommand{\hatcurCCesoJHmageccenxxxxxD}{\ensuremath{0.392\pm0.040}} 
\newcommand{\hatcurCCesoJKmageccenxxxxxD}{\ensuremath{0.465\pm0.042}} 
\newcommand{\hatcurCCesoHKmageccenxxxxxD}{\ensuremath{0.073\pm0.043}} 
\newcommand{\hatcurLCdipeccenxxxxxD}{\ensuremath{17.6}}              
\newcommand{\hatcurLCrprstareccenxxxxxD}{\ensuremath{0.1331\pm0.0028}} 
\newcommand{\hatcurLCbsqeccenxxxxxD}{\ensuremath{0.418_{-0.060}^{+0.054}}} 
\newcommand{\hatcurLCimpeccenxxxxxD}{\ensuremath{0.646_{-0.048}^{+0.040}}} 
\newcommand{\hatcurLCzetaeccenxxxxxD}{\ensuremath{24.84\pm0.28}}     
\newcommand{\hatcurLCdureccenxxxxxD}{\ensuremath{0.0982\pm0.0018}}   
\newcommand{\hatcurLCdurshorteccenxxxxxD}{\ensuremath{0.0982}}       
\newcommand{\hatcurLCdurhreccenxxxxxD}{\ensuremath{2.357\pm0.043}}   
\newcommand{\hatcurLCdurhrshorteccenxxxxxD}{\ensuremath{2.357}}      
\newcommand{\hatcurLCqeccenxxxxxD}{\ensuremath{0.03090\pm0.00057}}   
\newcommand{\hatcurLCqshorteccenxxxxxD}{\ensuremath{0.031}}          
\newcommand{\hatcurLCingdureccenxxxxxD}{\ensuremath{0.0186\pm0.0020}} 
\newcommand{\hatcurLCPeccenxxxxxD}{\ensuremath{3.1810785\pm0.0000040}} 
\newcommand{\hatcurLCPprececcenxxxxxD}{\ensuremath{3.1810785}}       
\newcommand{\hatcurLCPshorteccenxxxxxD}{\ensuremath{3.1811}}         
\newcommand{\hatcurLCTeccenxxxxxD}{\ensuremath{2457040.64523\pm0.00046}} 
\newcommand{\hatcurLCTAeccenxxxxxD}{\ensuremath{2456366.25656\pm0.00093}} 
\newcommand{\hatcurLCTBeccenxxxxxD}{\ensuremath{2457269.68289\pm0.00056}} 
\newcommand{\hatcurLChatnetmeccenxxxxxD}{\ensuremath{13.681480\pm0.000077}} 
\newcommand{\hatcurLCiblendeccenxxxxxD}{\ensuremath{0.896\pm0.042}}  
\newcommand{\hatcurLCrhoeccenxxxxxD}{\ensuremath{1.74^{+0.83}_{-0.48}}}                    
\newcommand{\hatcurSMEiteffeccenxxxxxD}{\ensuremath{5600\pm110}}     
\newcommand{\hatcurSMEizfeheccenxxxxxD}{\ensuremath{0.080\pm0.080}}  
\newcommand{\hatcurSMEizfehshorteccenxxxxxD}{\ensuremath{0.08}}      
\newcommand{\hatcurSMEiloggeccenxxxxxD}{\ensuremath{4.60\pm0.14}}    
\newcommand{\hatcurSMEivsineccenxxxxxD}{\ensuremath{1.8\pm1.2}}      
\newcommand{\hatcurSMEivmaceccenxxxxxD}{\ensuremath{0.0}}            
\newcommand{\hatcurSMEivmiceccenxxxxxD}{\ensuremath{0.0}}            
\newcommand{\hatcurSMEiiteffeccenxxxxxD}{\ensuremath{5498\pm84}}     
\newcommand{\hatcurSMEiizfeheccenxxxxxD}{\ensuremath{0.010\pm0.060}} 
\newcommand{\hatcurSMEiizfehshorteccenxxxxxD}{\ensuremath{0.01}}     
\newcommand{\hatcurSMEiiloggeccenxxxxxD}{\ensuremath{4.477\pm0.035}} 
\newcommand{\hatcurSMEiivsineccenxxxxxD}{\ensuremath{2.6\pm1.0}}     
\newcommand{\hatcurLBizeccenxxxxxD}{\ensuremath{0.2476}}             
\newcommand{\hatcurLBiizeccenxxxxxD}{\ensuremath{0.3079}}            
\newcommand{\hatcurLBiieccenxxxxxD}{\ensuremath{0.3148}}             
\newcommand{\hatcurLBiiieccenxxxxxD}{\ensuremath{0.3030}}            
\newcommand{\hatcurLBiIeccenxxxxxD}{\ensuremath{0.2926}}             
\newcommand{\hatcurLBiiIeccenxxxxxD}{\ensuremath{0.3051}}            
\newcommand{\hatcurLBigeccenxxxxxD}{\ensuremath{0.6241}}             
\newcommand{\hatcurLBiigeccenxxxxxD}{\ensuremath{0.1819}}            
\newcommand{\hatcurLBireccenxxxxxD}{\ensuremath{0.4137}}             
\newcommand{\hatcurLBiireccenxxxxxD}{\ensuremath{0.2900}}            
\newcommand{\hatcurLBiReccenxxxxxD}{\ensuremath{0.3864}}             
\newcommand{\hatcurLBiiReccenxxxxxD}{\ensuremath{0.2944}}            
\newcommand{\hatcurLBikepeccenxxxxxD}{\ensuremath{0.1000}}           
\newcommand{\hatcurLBiikepeccenxxxxxD}{\ensuremath{0.1000}}          
\newcommand{\hatcurISOmeccenxxxxxD}{\ensuremath{0.929\pm0.034}}      
\newcommand{\hatcurISOmshorteccenxxxxxD}{\ensuremath{0.93}}          
\newcommand{\hatcurISOmlongeccenxxxxxD}{\ensuremath{0.929\pm0.034}}  
\newcommand{\hatcurISOreccenxxxxxD}{\ensuremath{0.933_{-0.054}^{+0.080}}} 
\newcommand{\hatcurISOrshorteccenxxxxxD}{\ensuremath{0.93}}          
\newcommand{\hatcurISOrlongeccenxxxxxD}{\ensuremath{0.933_{-0.054}^{+0.080}}} 
\newcommand{\hatcurISOrhoeccenxxxxxD}{\ensuremath{1.60\pm0.35}}      
\newcommand{\hatcurISOrholongeccenxxxxxD}{\ensuremath{1.60\pm0.35}}  
\newcommand{\hatcurISOloggeccenxxxxxD}{\ensuremath{4.464\pm0.073}}   
\newcommand{\hatcurISOlumeccenxxxxxD}{\ensuremath{0.72_{-0.10}^{+0.15}}} 
\newcommand{\hatcurISOlumshorteccenxxxxxD}{\ensuremath{0.72}}        
\newcommand{\hatcurISOmveccenxxxxxD}{\ensuremath{5.25\pm0.23}}       
\newcommand{\hatcurISOvieccenxxxxxD}{\ensuremath{0.781\pm0.023}}     
\newcommand{\hatcurISOageeccenxxxxxD}{\ensuremath{7.0\pm3.3}}        
\newcommand{\hatcurISOsigmaeccenxxxxxD}{\ensuremath{0.00110\pm0.00039}} 
\newcommand{\hatcurISOMJeccenxxxxxD}{\ensuremath{3.96\pm0.20}}       
\newcommand{\hatcurISOMHeccenxxxxxD}{\ensuremath{3.57\pm0.19}}       
\newcommand{\hatcurISOMKeccenxxxxxD}{\ensuremath{3.50\pm0.19}}       
\newcommand{\hatcurISOJKeccenxxxxxD}{\ensuremath{0.460\pm0.020}}     
\newcommand{\hatcurISOspececcenxxxxxD}{G}                            
\newcommand{\hatcurRVKeccenxxxxxD}{\ensuremath{96\pm13}}             
\newcommand{\hatcurRVrkeccenxxxxxD}{\ensuremath{0.16_{-0.18}^{+0.12}}} 
\newcommand{\hatcurRVrheccenxxxxxD}{\ensuremath{0.04\pm0.21}}        
\newcommand{\hatcurRVkeccenxxxxxD}{\ensuremath{0.039\pm0.053}}       
\newcommand{\hatcurRVheccenxxxxxD}{\ensuremath{0.006_{-0.054}^{+0.083}}} 
\newcommand{\hatcurRVtroneeccenxxxxxD}{\ensuremath{0\pm0}}           
\newcommand{\hatcurRVtrtwoeccenxxxxxD}{\ensuremath{0\pm0}}           
\newcommand{\hatcurRVgammaeccenxxxxxD}{\ensuremath{-8650.1\pm9.3}}   
\newcommand{\hatcurRVjittereccenxxxxxD}{\ensuremath{32.9\pm8.5}}     
\newcommand{\hatcurRVjittertwosiglimeccenxxxxxD}{\ensuremath{<49}}               
\newcommand{\hatcurRVfitrmseccenxxxxxD}{\ensuremath{.1fym}}          %
\newcommand{\hatcurRVecceneccenxxxxxD}{\ensuremath{0.075\pm0.066}}   
\newcommand{\hatcurRVeccentwosiglimeccenxxxxxD}{\ensuremath{<0.202}} 
\newcommand{\hatcurRVomegaeccenxxxxxD}{\ensuremath{120\pm140}}       
\newcommand{\hatcurPPieccenxxxxxD}{\ensuremath{86.07_{-0.76}^{+0.51}}} 
\newcommand{\hatcurPPgeccenxxxxxD}{\ensuremath{11.0\pm2.4}}          
\newcommand{\hatcurPPloggeccenxxxxxD}{\ensuremath{3.040\pm0.098}}    
\newcommand{\hatcurPPareccenxxxxxD}{\ensuremath{9.50\pm0.75}}        
\newcommand{\hatcurPPareleccenxxxxxD}{\ensuremath{0.04130\pm0.00051}} 
\newcommand{\hatcurPPrhoeccenxxxxxD}{\ensuremath{0.45\pm0.13}}       
\newcommand{\hatcurPPmeccenxxxxxD}{\ensuremath{0.659\pm0.088}}       
\newcommand{\hatcurPPmshorteccenxxxxxD}{\ensuremath{0.66}}           
\newcommand{\hatcurPPmlongeccenxxxxxD}{\ensuremath{0.659\pm0.088}}   
\newcommand{\hatcurPPmeeccenxxxxxD}{\ensuremath{209\pm28}}           
\newcommand{\hatcurPPmeshorteccenxxxxxD}{\ensuremath{209.3}}         
\newcommand{\hatcurPPmelongeccenxxxxxD}{\ensuremath{209\pm28}}       
\newcommand{\hatcurPPreccenxxxxxD}{\ensuremath{1.210_{-0.078}^{+0.113}}} 
\newcommand{\hatcurPPrshorteccenxxxxxD}{\ensuremath{1.21}}           
\newcommand{\hatcurPPrlongeccenxxxxxD}{\ensuremath{1.210_{-0.078}^{+0.113}}} 
\newcommand{\hatcurPPreeccenxxxxxD}{\ensuremath{13.56_{-0.88}^{+1.26}}} 
\newcommand{\hatcurPPreshorteccenxxxxxD}{\ensuremath{13.6}}          
\newcommand{\hatcurPPrelongeccenxxxxxD}{\ensuremath{13.56_{-0.88}^{+1.26}}} 
\newcommand{\hatcurPPmrcorreccenxxxxxD}{\ensuremath{-0.03}}          
\newcommand{\hatcurPPteffeccenxxxxxD}{\ensuremath{1263_{-46}^{+61}}} 
\newcommand{\hatcurPPthetaeccenxxxxxD}{\ensuremath{0.0478\pm0.0077}} 
\newcommand{\hatcurPPfluxperieccenxxxxxD}{\ensuremath{6.59_{-0.99}^{+2.14}}} 
\newcommand{\hatcurPPfluxperidimeccenxxxxxD}{\ensuremath{8}}         
\newcommand{\hatcurPPfluxapeccenxxxxxD}{\ensuremath{5.06\pm0.82}}    
\newcommand{\hatcurPPfluxapdimeccenxxxxxD}{\ensuremath{8}}           
\newcommand{\hatcurPPfluxavgeccenxxxxxD}{\ensuremath{5.75_{-0.80}^{+1.20}}} 
\newcommand{\hatcurPPfluxavgdimeccenxxxxxD}{\ensuremath{8}}          
\newcommand{\hatcurPPfluxavglogeccenxxxxxD}{\ensuremath{8.760\pm0.085}} 
\newcommand{\hatcurXsecphaseeccenxxxxxD}{\ensuremath{0.525\pm0.034}} 
\newcommand{\hatcurXsecondaryeccenxxxxxD}{\ensuremath{2457042.31\pm0.11}} 
\newcommand{\hatcurXsecdureccenxxxxxD}{\ensuremath{0.0991\pm0.0099}} 
\newcommand{\hatcurXsecingdureccenxxxxxD}{\ensuremath{0.019\pm0.012}} 
\newcommand{\hatcurPPphiconjeccenxxxxxD}{\ensuremath{0.15\pm0.21}}   
\newcommand{\hatcurPPperieccenxxxxxD}{\ensuremath{2457040.18\pm0.67}} 
\newcommand{\hatcurPPaequiveccenxxxxxD}{\ensuremath{0.0488\pm0.0044}} 
\newcommand{\hatcurPPtcirceccenxxxxxD}{\ensuremath{104\pm46}}        
\newcommand{\hatcurPPtinfalleccenxxxxxD}{\ensuremath{3300\pm1300}}   
\newcommand{\hatcurXdisteccenxxxxxD}{\ensuremath{532_{-33}^{+47}}}   
\newcommand{\hatcurXAveccenxxxxxD}{\ensuremath{0.058_{-0.058}^{+0.125}}} 
\newcommand{\hatcurXdistredeccenxxxxxD}{\ensuremath{528_{-33}^{+46}}} 
\newcommand{\hatcurXEBVeccenxxxxxD}{\ensuremath{0.019_{-0.019}^{+0.040}}} 
\newcommand{\hatcurXmvisoredeccenxxxxxD}{\ensuremath{13.942\pm0.061}} 
\newcommand{\hatcurXmiisoredeccenxxxxxD}{\ensuremath{13.121\pm0.032}} 
\newcommand{\hatcurXmjisoredeccenxxxxxD}{\ensuremath{12.601\pm0.019}} 
\newcommand{\hatcurXmhisoredeccenxxxxxD}{\ensuremath{12.195\pm0.018}} 
\newcommand{\hatcurXmkisoredeccenxxxxxD}{\ensuremath{12.124\pm0.020}} 
\newcommand{\hatcurXviisoredeccenxxxxxD}{\ensuremath{0.816_{-0.028}^{+0.047}}} 
\newcommand{\hatcurXvkisoredeccenxxxxxD}{\ensuremath{1.819\pm0.068}} 
\newcommand{\hatcurXjhisoredeccenxxxxxD}{\ensuremath{0.406\pm0.015}} 
\newcommand{\hatcurXjkisoredeccenxxxxxD}{\ensuremath{0.478\pm0.017}} 
\newcommand{\hatcurCCpmraeccenxxxxxD}{\ensuremath{10.3\pm1.6}}       
\newcommand{\hatcurCCpmdececcenxxxxxD}{\ensuremath{-2.4\pm1.4}}      
\newcommand{\hatcurCCpmeccenxxxxxD}{\ensuremath{10.6\pm2.1}}         

\newcommand{\hatcurhtreccenxxxxxE}{HATS747-002}                      
\newcommand{\hatcurfieldeccenxxxxxE}{\ensuremath{string}}            
\newcommand{\hatcurCCraeccenxxxxxE}{\ensuremath{19^{\mathrm h}00^{\mathrm m}23.04{\mathrm s}}}                     
\newcommand{\hatcurCCdececcenxxxxxE}{\ensuremath{-54{\arcdeg}53{\arcmin}35.5{\arcsec}}}                    
\newcommand{\hatcurCCmageccenxxxxxE}{12.612}                         
\newcommand{\hatcurCCtwomasseccenxxxxxE}{2MASS~19002314-5453354}     
\newcommand{\hatcurCCgsceccenxxxxxE}{GSC~8763-00475}                 
\newcommand{\hatcurCCtassmveccenxxxxxE}{\ensuremath{12.612\pm0.010}} 
\newcommand{\hatcurCCtassmvshorteccenxxxxxE}{\ensuremath{12.6}}      
\newcommand{\hatcurCCtassmBeccenxxxxxE}{\ensuremath{13.361\pm0.010}} 
\newcommand{\hatcurCCtassmBshorteccenxxxxxE}{\ensuremath{13.4}}      
\newcommand{\hatcurCCtassmIeccenxxxxxE}{\ensuremath{100\pm1000}}     
\newcommand{\hatcurCCtassmIshorteccenxxxxxE}{\ensuremath{100.0}}     
\newcommand{\hatcurCCtassmgeccenxxxxxE}{\ensuremath{12.950\pm0.010}} 
\newcommand{\hatcurCCtassmgshorteccenxxxxxE}{\ensuremath{12.9}}      
\newcommand{\hatcurCCtassmreccenxxxxxE}{\ensuremath{12.430\pm0.010}} 
\newcommand{\hatcurCCtassmrshorteccenxxxxxE}{\ensuremath{12.4}}      
\newcommand{\hatcurCCtassmieccenxxxxxE}{\ensuremath{12.154\pm0.010}} 
\newcommand{\hatcurCCtassmishorteccenxxxxxE}{\ensuremath{12.2}}      
\newcommand{\hatcurCCtwomassJmageccenxxxxxE}{\ensuremath{11.286\pm0.026}} 
\newcommand{\hatcurCCtwomassHmageccenxxxxxE}{\ensuremath{10.933\pm0.021}} 
\newcommand{\hatcurCCtwomassKmageccenxxxxxE}{\ensuremath{10.877\pm0.019}} 
\newcommand{\hatcurCCcitJmageccenxxxxxE}{\ensuremath{11.301\pm0.026}} 
\newcommand{\hatcurCCcitHmageccenxxxxxE}{\ensuremath{10.928\pm0.022}} 
\newcommand{\hatcurCCcitKmageccenxxxxxE}{\ensuremath{10.901\pm0.019}} 
\newcommand{\hatcurCCbbJmageccenxxxxxE}{\ensuremath{11.353\pm0.028}} 
\newcommand{\hatcurCCbbHmageccenxxxxxE}{\ensuremath{10.950\pm0.022}} 
\newcommand{\hatcurCCbbKmageccenxxxxxE}{\ensuremath{10.921\pm0.019}} 
\newcommand{\hatcurCCesoJmageccenxxxxxE}{\ensuremath{11.356\pm0.030}} 
\newcommand{\hatcurCCesoHmageccenxxxxxE}{\ensuremath{10.943\pm0.025}} 
\newcommand{\hatcurCCesoKmageccenxxxxxE}{\ensuremath{10.920\pm0.020}} 
\newcommand{\hatcurCCesoJHmageccenxxxxxE}{\ensuremath{0.412\pm0.036}} 
\newcommand{\hatcurCCesoJKmageccenxxxxxE}{\ensuremath{0.437\pm0.035}} 
\newcommand{\hatcurCCesoHKmageccenxxxxxE}{\ensuremath{0.024\pm0.032}} 
\newcommand{\hatcurLCdipeccenxxxxxE}{\ensuremath{15.6}}              
\newcommand{\hatcurLCrprstareccenxxxxxE}{\ensuremath{0.1206\pm0.0032}} 
\newcommand{\hatcurLCbsqeccenxxxxxE}{\ensuremath{0.270_{-0.062}^{+0.052}}} 
\newcommand{\hatcurLCimpeccenxxxxxE}{\ensuremath{0.519_{-0.064}^{+0.048}}} 
\newcommand{\hatcurLCzetaeccenxxxxxE}{\ensuremath{17.32\pm0.11}}     
\newcommand{\hatcurLCdureccenxxxxxE}{\ensuremath{0.1340\pm0.0018}}   
\newcommand{\hatcurLCdurshorteccenxxxxxE}{\ensuremath{0.1340}}       
\newcommand{\hatcurLCdurhreccenxxxxxE}{\ensuremath{3.216\pm0.044}}   
\newcommand{\hatcurLCdurhrshorteccenxxxxxE}{\ensuremath{3.216}}      
\newcommand{\hatcurLCqeccenxxxxxE}{\ensuremath{0.02910\pm0.00040}}   
\newcommand{\hatcurLCqshorteccenxxxxxE}{\ensuremath{0.029}}          
\newcommand{\hatcurLCingdureccenxxxxxE}{\ensuremath{0.0190\pm0.0019}} 
\newcommand{\hatcurLCPeccenxxxxxE}{\ensuremath{4.6058762\pm0.0000066}} 
\newcommand{\hatcurLCPprececcenxxxxxE}{\ensuremath{4.6058762}}       
\newcommand{\hatcurLCPshorteccenxxxxxE}{\ensuremath{4.6059}}         
\newcommand{\hatcurLCTeccenxxxxxE}{\ensuremath{2457041.16805\pm0.00037}} 
\newcommand{\hatcurLCTAeccenxxxxxE}{\ensuremath{2456368.7102\pm0.0010}} 
\newcommand{\hatcurLCTBeccenxxxxxE}{\ensuremath{2457174.73845\pm0.00043}} 
\newcommand{\hatcurLChatnetmeccenxxxxxE}{\ensuremath{12.463780\pm0.000045}} 
\newcommand{\hatcurLCiblendeccenxxxxxE}{\ensuremath{0.861\pm0.044}}  
\newcommand{\hatcurLCrhoeccenxxxxxE}{\ensuremath{1.53^{+0.55}_{-0.36}}}                    
\newcommand{\hatcurSMEiteffeccenxxxxxE}{\ensuremath{5680\pm120}}     
\newcommand{\hatcurSMEizfeheccenxxxxxE}{\ensuremath{0.190\pm0.080}}  
\newcommand{\hatcurSMEizfehshorteccenxxxxxE}{\ensuremath{0.19}}      
\newcommand{\hatcurSMEiloggeccenxxxxxE}{\ensuremath{4.39\pm0.20}}    
\newcommand{\hatcurSMEivsineccenxxxxxE}{\ensuremath{2.50\pm0.94}}    
\newcommand{\hatcurSMEivmaceccenxxxxxE}{\ensuremath{0.0}}            
\newcommand{\hatcurSMEivmiceccenxxxxxE}{\ensuremath{0.0}}            
\newcommand{\hatcurSMEiiteffeccenxxxxxE}{\ensuremath{5670\pm110}}    
\newcommand{\hatcurSMEiizfeheccenxxxxxE}{\ensuremath{0.160\pm0.080}} 
\newcommand{\hatcurSMEiizfehshorteccenxxxxxE}{\ensuremath{0.16}}     
\newcommand{\hatcurSMEiiloggeccenxxxxxE}{\ensuremath{4.382\pm0.046}} 
\newcommand{\hatcurSMEiivsineccenxxxxxE}{\ensuremath{2.35\pm0.80}}   
\newcommand{\hatcurLBizeccenxxxxxE}{\ensuremath{0.2248}}             
\newcommand{\hatcurLBiizeccenxxxxxE}{\ensuremath{0.3233}}            
\newcommand{\hatcurLBiieccenxxxxxE}{\ensuremath{0.2914}}             
\newcommand{\hatcurLBiiieccenxxxxxE}{\ensuremath{0.3213}}            
\newcommand{\hatcurLBiIeccenxxxxxE}{\ensuremath{0.2690}}             
\newcommand{\hatcurLBiiIeccenxxxxxE}{\ensuremath{0.3226}}            
\newcommand{\hatcurLBigeccenxxxxxE}{\ensuremath{0.5937}}             
\newcommand{\hatcurLBiigeccenxxxxxE}{\ensuremath{0.2066}}            
\newcommand{\hatcurLBireccenxxxxxE}{\ensuremath{0.3875}}             
\newcommand{\hatcurLBiireccenxxxxxE}{\ensuremath{0.3096}}            
\newcommand{\hatcurLBiReccenxxxxxE}{\ensuremath{0.3607}}             
\newcommand{\hatcurLBiiReccenxxxxxE}{\ensuremath{0.3138}}            
\newcommand{\hatcurLBikepeccenxxxxxE}{\ensuremath{0.1000}}           
\newcommand{\hatcurLBiikepeccenxxxxxE}{\ensuremath{0.1000}}          
\newcommand{\hatcurISOmeccenxxxxxE}{\ensuremath{1.029\pm0.047}}      
\newcommand{\hatcurISOmshorteccenxxxxxE}{\ensuremath{1.03}}          
\newcommand{\hatcurISOmlongeccenxxxxxE}{\ensuremath{1.029\pm0.047}}  
\newcommand{\hatcurISOreccenxxxxxE}{\ensuremath{1.024\pm0.067}}      
\newcommand{\hatcurISOrshorteccenxxxxxE}{\ensuremath{1.02}}          
\newcommand{\hatcurISOrlongeccenxxxxxE}{\ensuremath{1.024\pm0.067}}  
\newcommand{\hatcurISOrhoeccenxxxxxE}{\ensuremath{1.34\pm0.25}}      
\newcommand{\hatcurISOrholongeccenxxxxxE}{\ensuremath{1.34\pm0.25}}  
\newcommand{\hatcurISOloggeccenxxxxxE}{\ensuremath{4.429\pm0.053}}   
\newcommand{\hatcurISOlumeccenxxxxxE}{\ensuremath{0.97\pm0.17}}      
\newcommand{\hatcurISOlumshorteccenxxxxxE}{\ensuremath{0.97}}        
\newcommand{\hatcurISOmveccenxxxxxE}{\ensuremath{4.88\pm0.21}}       
\newcommand{\hatcurISOvieccenxxxxxE}{\ensuremath{0.738\pm0.034}}     
\newcommand{\hatcurISOageeccenxxxxxE}{\ensuremath{4.3\pm2.5}}        
\newcommand{\hatcurISOsigmaeccenxxxxxE}{\ensuremath{0.00090\pm0.00019}} 
\newcommand{\hatcurISOMJeccenxxxxxE}{\ensuremath{3.68\pm0.17}}       
\newcommand{\hatcurISOMHeccenxxxxxE}{\ensuremath{3.32\pm0.15}}       
\newcommand{\hatcurISOMKeccenxxxxxE}{\ensuremath{3.27\pm0.15}}       
\newcommand{\hatcurISOJKeccenxxxxxE}{\ensuremath{0.420\pm0.030}}     
\newcommand{\hatcurISOspececcenxxxxxE}{G}                            
\newcommand{\hatcurRVKeccenxxxxxE}{\ensuremath{79.1\pm6.4}}          
\newcommand{\hatcurRVrkeccenxxxxxE}{\ensuremath{0.09\pm0.11}}        
\newcommand{\hatcurRVrheccenxxxxxE}{\ensuremath{-0.20_{-0.13}^{+0.23}}} 
\newcommand{\hatcurRVkeccenxxxxxE}{\ensuremath{0.024_{-0.026}^{+0.036}}} 
\newcommand{\hatcurRVheccenxxxxxE}{\ensuremath{-0.046_{-0.065}^{+0.049}}} 
\newcommand{\hatcurRVtroneeccenxxxxxE}{\ensuremath{0\pm0}}           
\newcommand{\hatcurRVtrtwoeccenxxxxxE}{\ensuremath{0\pm0}}           
\newcommand{\hatcurRVgammaAeccenxxxxxE}{\ensuremath{-19696.0\pm7.1}} 
\newcommand{\hatcurRVjitterAeccenxxxxxE}{\ensuremath{0.0\pm5.3}}     
\newcommand{\hatcurRVjittertwosiglimAeccenxxxxxE}{\ensuremath{<12}}               
\newcommand{\hatcurRVfitrmsAeccenxxxxxE}{\ensuremath{0.0}}           
\newcommand{\hatcurRVgammaBeccenxxxxxE}{\ensuremath{-19721.0\pm7.7}} 
\newcommand{\hatcurRVjitterBeccenxxxxxE}{\ensuremath{0.01\pm0.71}}   
\newcommand{\hatcurRVjittertwosiglimBeccenxxxxxE}{\ensuremath{<1.9}}               
\newcommand{\hatcurRVfitrmsBeccenxxxxxE}{\ensuremath{0.0}}           
\newcommand{\hatcurRVgammaCeccenxxxxxE}{\ensuremath{-19725\pm15}}    
\newcommand{\hatcurRVjitterCeccenxxxxxE}{\ensuremath{35\pm14}}       
\newcommand{\hatcurRVjittertwosiglimCeccenxxxxxE}{\ensuremath{<67}}               
\newcommand{\hatcurRVfitrmsCeccenxxxxxE}{\ensuremath{0.0}}           
\newcommand{\hatcurRVecceneccenxxxxxE}{\ensuremath{0.069\pm0.046}}   
\newcommand{\hatcurRVeccentwosiglimeccenxxxxxE}{\ensuremath{<0.158}} 
\newcommand{\hatcurRVomegaeccenxxxxxE}{\ensuremath{283\pm99}}        
\newcommand{\hatcurPPieccenxxxxxE}{\ensuremath{87.56\pm0.42}}        
\newcommand{\hatcurPPgeccenxxxxxE}{\ensuremath{11.2\pm2.0}}          
\newcommand{\hatcurPPloggeccenxxxxxE}{\ensuremath{3.048\pm0.075}}    
\newcommand{\hatcurPPareccenxxxxxE}{\ensuremath{11.46\pm0.70}}       
\newcommand{\hatcurPPareleccenxxxxxE}{\ensuremath{0.05470\pm0.00084}} 
\newcommand{\hatcurPPrhoeccenxxxxxE}{\ensuremath{0.47\pm0.12}}       
\newcommand{\hatcurPPmeccenxxxxxE}{\ensuremath{0.655\pm0.057}}       
\newcommand{\hatcurPPmshorteccenxxxxxE}{\ensuremath{0.66}}           
\newcommand{\hatcurPPmlongeccenxxxxxE}{\ensuremath{0.655\pm0.057}}   
\newcommand{\hatcurPPmeeccenxxxxxE}{\ensuremath{208\pm18}}           
\newcommand{\hatcurPPmeshorteccenxxxxxE}{\ensuremath{208.3}}         
\newcommand{\hatcurPPmelongeccenxxxxxE}{\ensuremath{208\pm18}}       
\newcommand{\hatcurPPreccenxxxxxE}{\ensuremath{1.198_{-0.083}^{+0.111}}} 
\newcommand{\hatcurPPrshorteccenxxxxxE}{\ensuremath{1.20}}           
\newcommand{\hatcurPPrlongeccenxxxxxE}{\ensuremath{1.198_{-0.083}^{+0.111}}} 
\newcommand{\hatcurPPreeccenxxxxxE}{\ensuremath{13.43_{-0.93}^{+1.25}}} 
\newcommand{\hatcurPPreshorteccenxxxxxE}{\ensuremath{13.4}}          
\newcommand{\hatcurPPrelongeccenxxxxxE}{\ensuremath{13.43_{-0.93}^{+1.25}}} 
\newcommand{\hatcurPPmrcorreccenxxxxxE}{\ensuremath{0.06}}           
\newcommand{\hatcurPPteffeccenxxxxxE}{\ensuremath{1184\pm46}}        
\newcommand{\hatcurPPthetaeccenxxxxxE}{\ensuremath{0.0578\pm0.0067}} 
\newcommand{\hatcurPPfluxperieccenxxxxxE}{\ensuremath{5.15\pm0.70}}  
\newcommand{\hatcurPPfluxperidimeccenxxxxxE}{\ensuremath{8}}         
\newcommand{\hatcurPPfluxapeccenxxxxxE}{\ensuremath{3.90\pm0.82}}    
\newcommand{\hatcurPPfluxapdimeccenxxxxxE}{\ensuremath{8}}           
\newcommand{\hatcurPPfluxavgeccenxxxxxE}{\ensuremath{4.43\pm0.70}}   
\newcommand{\hatcurPPfluxavgdimeccenxxxxxE}{\ensuremath{8}}          
\newcommand{\hatcurPPfluxavglogeccenxxxxxE}{\ensuremath{8.647\pm0.067}} 
\newcommand{\hatcurXsecphaseeccenxxxxxE}{\ensuremath{0.515\pm0.021}} 
\newcommand{\hatcurXsecondaryeccenxxxxxE}{\ensuremath{2457043.541\pm0.095}} 
\newcommand{\hatcurXsecdureccenxxxxxE}{\ensuremath{0.125\pm0.011}}   
\newcommand{\hatcurXsecingdureccenxxxxxE}{\ensuremath{0.0165\pm0.0028}} 
\newcommand{\hatcurPPphiconjeccenxxxxxE}{\ensuremath{0.361_{-0.575}^{+0.097}}} 
\newcommand{\hatcurPPperieccenxxxxxE}{\ensuremath{2457039.5\pm1.6}}  
\newcommand{\hatcurPPaequiveccenxxxxxE}{\ensuremath{0.0556\pm0.0044}} 
\newcommand{\hatcurPPtcirceccenxxxxxE}{\ensuremath{580_{-190}^{+280}}} 
\newcommand{\hatcurPPtinfalleccenxxxxxE}{\ensuremath{13600\pm4400}}  
\newcommand{\hatcurXdisteccenxxxxxE}{\ensuremath{339\pm24}}          
\newcommand{\hatcurXAveccenxxxxxE}{\ensuremath{0.097\pm0.079}}       
\newcommand{\hatcurXdistredeccenxxxxxE}{\ensuremath{335\pm24}}       
\newcommand{\hatcurXEBVeccenxxxxxE}{\ensuremath{0.031\pm0.026}}      
\newcommand{\hatcurXmvisoredeccenxxxxxE}{\ensuremath{12.614\pm0.012}} 
\newcommand{\hatcurXmiisoredeccenxxxxxE}{\ensuremath{11.824\pm0.014}} 
\newcommand{\hatcurXmjisoredeccenxxxxxE}{\ensuremath{11.339\pm0.015}} 
\newcommand{\hatcurXmhisoredeccenxxxxxE}{\ensuremath{10.971\pm0.022}} 
\newcommand{\hatcurXmkisoredeccenxxxxxE}{\ensuremath{10.905\pm0.023}} 
\newcommand{\hatcurXviisoredeccenxxxxxE}{\ensuremath{0.790\pm0.014}} 
\newcommand{\hatcurXvkisoredeccenxxxxxE}{\ensuremath{1.708\pm0.031}} 
\newcommand{\hatcurXjhisoredeccenxxxxxE}{\ensuremath{0.369\pm0.015}} 
\newcommand{\hatcurXjkisoredeccenxxxxxE}{\ensuremath{0.434_{-0.010}^{+0.014}}} 
\newcommand{\hatcurCCpmraeccenxxxxxE}{\ensuremath{2.8\pm1.3}}        
\newcommand{\hatcurCCpmdececcenxxxxxE}{\ensuremath{-37.1\pm3.7}}     
\newcommand{\hatcurCCpmeccenxxxxxE}{\ensuremath{37.2\pm3.9}}         

\newcommand{\hatcurhtreccenxxxxxF}{HATS754-005}                              
\newcommand{\hatcurfieldeccenxxxxxF}{\ensuremath{string}}                    
\newcommand{\hatcurCCraeccenxxxxxF}{\ensuremath{00^{\mathrm h}22^{\mathrm m}28.49{\mathrm s}}}                             
\newcommand{\hatcurCCdececcenxxxxxF}{\ensuremath{-59{\arcdeg}56{\arcmin}33.2{\arcsec}}}                            
\newcommand{\hatcurCCmageccenxxxxxF}{12.192}                                 
\newcommand{\hatcurCCtwomasseccenxxxxxF}{2MASS~00222848-5956331}             
\newcommand{\hatcurCCgsceccenxxxxxF}{GSC~8471-00231}                         
\newcommand{\hatcurCCtassmveccenxxxxxF}{\ensuremath{12.192\pm0.010}}         
\newcommand{\hatcurCCtassmvshorteccenxxxxxF}{\ensuremath{12.2}}              
\newcommand{\hatcurCCtassmBeccenxxxxxF}{\ensuremath{12.790\pm0.010}}         
\newcommand{\hatcurCCtassmBshorteccenxxxxxF}{\ensuremath{12.8}}              
\newcommand{\hatcurCCtassmIeccenxxxxxF}{\ensuremath{100\pm1000}}             
\newcommand{\hatcurCCtassmIshorteccenxxxxxF}{\ensuremath{100.0}}             
\newcommand{\hatcurCCtassmgeccenxxxxxF}{\ensuremath{12.439\pm0.010}}         
\newcommand{\hatcurCCtassmgshorteccenxxxxxF}{\ensuremath{12.4}}              
\newcommand{\hatcurCCtassmreccenxxxxxF}{\ensuremath{12.046\pm0.010}}         
\newcommand{\hatcurCCtassmrshorteccenxxxxxF}{\ensuremath{12.0}}              
\newcommand{\hatcurCCtassmieccenxxxxxF}{\ensuremath{11.935\pm0.010}}         
\newcommand{\hatcurCCtassmishorteccenxxxxxF}{\ensuremath{11.9}}              
\newcommand{\hatcurCCtwomassJmageccenxxxxxF}{\ensuremath{11.129\pm0.024}}    
\newcommand{\hatcurCCtwomassHmageccenxxxxxF}{\ensuremath{10.826\pm0.024}}    
\newcommand{\hatcurCCtwomassKmageccenxxxxxF}{\ensuremath{10.793\pm0.019}}    
\newcommand{\hatcurCCcitJmageccenxxxxxF}{\ensuremath{11.147\pm0.024}}        
\newcommand{\hatcurCCcitHmageccenxxxxxF}{\ensuremath{10.822\pm0.024}}        
\newcommand{\hatcurCCcitKmageccenxxxxxF}{\ensuremath{10.817\pm0.019}}        
\newcommand{\hatcurCCbbJmageccenxxxxxF}{\ensuremath{11.193\pm0.026}}         
\newcommand{\hatcurCCbbHmageccenxxxxxF}{\ensuremath{10.842\pm0.025}}         
\newcommand{\hatcurCCbbKmageccenxxxxxF}{\ensuremath{10.837\pm0.019}}         
\newcommand{\hatcurCCesoJmageccenxxxxxF}{\ensuremath{11.196\pm0.027}}        
\newcommand{\hatcurCCesoHmageccenxxxxxF}{\ensuremath{10.835\pm0.027}}        
\newcommand{\hatcurCCesoKmageccenxxxxxF}{\ensuremath{10.836\pm0.020}}        
\newcommand{\hatcurCCesoJHmageccenxxxxxF}{\ensuremath{0.3610\pm0.0080}}      
\newcommand{\hatcurCCesoJKmageccenxxxxxF}{\ensuremath{0.360\pm0.033}}        
\newcommand{\hatcurCCesoHKmageccenxxxxxF}{\ensuremath{-0.001\pm0.033}}       
\newcommand{\hatcurLCdipeccenxxxxxF}{\ensuremath{14.0}}                      
\newcommand{\hatcurLCrprstareccenxxxxxF}{\ensuremath{0.1139\pm0.0016}}       
\newcommand{\hatcurLCbsqeccenxxxxxF}{\ensuremath{0.259_{-0.061}^{+0.040}}}   
\newcommand{\hatcurLCimpeccenxxxxxF}{\ensuremath{0.508_{-0.064}^{+0.038}}}   
\newcommand{\hatcurLCzetaeccenxxxxxF}{\ensuremath{19.99\pm0.14}}             
\newcommand{\hatcurLCdureccenxxxxxF}{\ensuremath{0.1152\pm0.0014}}           
\newcommand{\hatcurLCdurshorteccenxxxxxF}{\ensuremath{0.1152}}               
\newcommand{\hatcurLCdurhreccenxxxxxF}{\ensuremath{2.764\pm0.034}}           
\newcommand{\hatcurLCdurhrshorteccenxxxxxF}{\ensuremath{2.764}}              
\newcommand{\hatcurLCqeccenxxxxxF}{\ensuremath{0.03630\pm0.00044}}           
\newcommand{\hatcurLCqshorteccenxxxxxF}{\ensuremath{0.036}}                  
\newcommand{\hatcurLCingdureccenxxxxxF}{\ensuremath{0.0154\pm0.0012}}        
\newcommand{\hatcurLCPeccenxxxxxF}{\ensuremath{3.1743508\pm0.0000029}}       
\newcommand{\hatcurLCPprececcenxxxxxF}{\ensuremath{3.1743508}}               
\newcommand{\hatcurLCPshorteccenxxxxxF}{\ensuremath{3.1744}}                 
\newcommand{\hatcurLCTeccenxxxxxF}{\ensuremath{2456629.76160\pm0.00035}}     
\newcommand{\hatcurLCTAeccenxxxxxF}{\ensuremath{2455759.98952\pm0.00079}}    
\newcommand{\hatcurLCTBeccenxxxxxF}{\ensuremath{2456953.54539\pm0.00051}}    
\newcommand{\hatcurLChatnetmAeccenxxxxxF}{\ensuremath{12.184500\pm0.000061}} 
\newcommand{\hatcurLCiblendAeccenxxxxxF}{\ensuremath{0.963\pm0.027}}         
\newcommand{\hatcurLChatnetmBeccenxxxxxF}{\ensuremath{12.184750\pm0.000065}} 
\newcommand{\hatcurLCiblendBeccenxxxxxF}{\ensuremath{0.822\pm0.032}}         
\newcommand{\hatcurLChatnetmCeccenxxxxxF}{\ensuremath{12.184520\pm0.000057}} 
\newcommand{\hatcurLCiblendCeccenxxxxxF}{\ensuremath{0.967\pm0.027}}         
\newcommand{\hatcurLCrhoeccenxxxxxF}{\ensuremath{1.57\pm0.33}}                    
\newcommand{\hatcurSMEiteffeccenxxxxxF}{\ensuremath{5947\pm70}}              
\newcommand{\hatcurSMEizfeheccenxxxxxF}{\ensuremath{0.040\pm0.040}}          
\newcommand{\hatcurSMEizfehshorteccenxxxxxF}{\ensuremath{0.04}}              
\newcommand{\hatcurSMEiloggeccenxxxxxF}{\ensuremath{4.330\pm0.070}}          
\newcommand{\hatcurSMEivsineccenxxxxxF}{\ensuremath{4.24\pm0.50}}            
\newcommand{\hatcurSMEivmaceccenxxxxxF}{\ensuremath{0.0}}                    
\newcommand{\hatcurSMEivmiceccenxxxxxF}{\ensuremath{0.0}}                    
\newcommand{\hatcurSMEiiteffeccenxxxxxF}{\ensuremath{5943\pm70}}             
\newcommand{\hatcurSMEiizfeheccenxxxxxF}{\ensuremath{0.060\pm0.050}}         
\newcommand{\hatcurSMEiizfehshorteccenxxxxxF}{\ensuremath{0.06}}             
\newcommand{\hatcurSMEiiloggeccenxxxxxF}{\ensuremath{4.426\pm0.033}}         
\newcommand{\hatcurSMEiivsineccenxxxxxF}{\ensuremath{4.11\pm0.50}}           
\newcommand{\hatcurLBizeccenxxxxxF}{\ensuremath{0.1876}}                     
\newcommand{\hatcurLBiizeccenxxxxxF}{\ensuremath{0.3386}}                    
\newcommand{\hatcurLBiieccenxxxxxF}{\ensuremath{0.2450}}                     
\newcommand{\hatcurLBiiieccenxxxxxF}{\ensuremath{0.3430}}                    
\newcommand{\hatcurLBiIeccenxxxxxF}{\ensuremath{0.2254}}                     
\newcommand{\hatcurLBiiIeccenxxxxxF}{\ensuremath{0.3425}}                    
\newcommand{\hatcurLBigeccenxxxxxF}{\ensuremath{0.5117}}                     
\newcommand{\hatcurLBiigeccenxxxxxF}{\ensuremath{0.2670}}                    
\newcommand{\hatcurLBireccenxxxxxF}{\ensuremath{0.3275}}                     
\newcommand{\hatcurLBiireccenxxxxxF}{\ensuremath{0.3438}}                    
\newcommand{\hatcurLBiReccenxxxxxF}{\ensuremath{0.3045}}                     
\newcommand{\hatcurLBiiReccenxxxxxF}{\ensuremath{0.3447}}                    
\newcommand{\hatcurLBikepeccenxxxxxF}{\ensuremath{0.1000}}                   
\newcommand{\hatcurLBiikepeccenxxxxxF}{\ensuremath{0.1000}}                  
\newcommand{\hatcurISOmeccenxxxxxF}{\ensuremath{1.092\pm0.031}}              
\newcommand{\hatcurISOmshorteccenxxxxxF}{\ensuremath{1.09}}                  
\newcommand{\hatcurISOmlongeccenxxxxxF}{\ensuremath{1.092\pm0.031}}          
\newcommand{\hatcurISOreccenxxxxxF}{\ensuremath{1.041\pm0.045}}              
\newcommand{\hatcurISOrshorteccenxxxxxF}{\ensuremath{1.04}}                  
\newcommand{\hatcurISOrlongeccenxxxxxF}{\ensuremath{1.041\pm0.045}}          
\newcommand{\hatcurISOrhoeccenxxxxxF}{\ensuremath{1.37\pm0.16}}              
\newcommand{\hatcurISOrholongeccenxxxxxF}{\ensuremath{1.37\pm0.16}}          
\newcommand{\hatcurISOloggeccenxxxxxF}{\ensuremath{4.442\pm0.034}}           
\newcommand{\hatcurISOlumeccenxxxxxF}{\ensuremath{1.20\pm0.14}}              
\newcommand{\hatcurISOlumshorteccenxxxxxF}{\ensuremath{1.20}}                
\newcommand{\hatcurISOmveccenxxxxxF}{\ensuremath{4.61\pm0.13}}               
\newcommand{\hatcurISOvieccenxxxxxF}{\ensuremath{0.654\pm0.020}}             
\newcommand{\hatcurISOageeccenxxxxxF}{\ensuremath{1.9\pm1.3}}                
\newcommand{\hatcurISOsigmaeccenxxxxxF}{\ensuremath{0.00100\pm0.00017}}      
\newcommand{\hatcurISOMJeccenxxxxxF}{\ensuremath{3.53\pm0.11}}               
\newcommand{\hatcurISOMHeccenxxxxxF}{\ensuremath{3.22\pm0.10}}               
\newcommand{\hatcurISOMKeccenxxxxxF}{\ensuremath{3.17\pm0.10}}               
\newcommand{\hatcurISOJKeccenxxxxxF}{\ensuremath{0.370\pm0.010}}             
\newcommand{\hatcurISOspececcenxxxxxF}{G}                                    
\newcommand{\hatcurRVKeccenxxxxxF}{\ensuremath{91.8\pm6.2}}                  
\newcommand{\hatcurRVrkeccenxxxxxF}{\ensuremath{-0.111_{-0.063}^{+0.088}}}   
\newcommand{\hatcurRVrheccenxxxxxF}{\ensuremath{-0.13_{-0.11}^{+0.18}}}      
\newcommand{\hatcurRVkeccenxxxxxF}{\ensuremath{-0.022\pm0.018}}              
\newcommand{\hatcurRVheccenxxxxxF}{\ensuremath{-0.025\pm0.034}}              
\newcommand{\hatcurRVtroneeccenxxxxxF}{\ensuremath{0\pm0}}                   
\newcommand{\hatcurRVtrtwoeccenxxxxxF}{\ensuremath{0\pm0}}                   
\newcommand{\hatcurRVgammaAeccenxxxxxF}{\ensuremath{-78.6\pm3.7}}            
\newcommand{\hatcurRVjitterAeccenxxxxxF}{\ensuremath{0.0\pm2.1}}             
\newcommand{\hatcurRVjittertwosiglimAeccenxxxxxF}{\ensuremath{<6.9}}               
\newcommand{\hatcurRVfitrmsAeccenxxxxxF}{\ensuremath{0.0}}                   
\newcommand{\hatcurRVgammaBeccenxxxxxF}{\ensuremath{-112.5\pm7.3}}           
\newcommand{\hatcurRVjitterBeccenxxxxxF}{\ensuremath{1\pm13}}                
\newcommand{\hatcurRVjittertwosiglimBeccenxxxxxF}{\ensuremath{<27}}               
\newcommand{\hatcurRVfitrmsBeccenxxxxxF}{\ensuremath{0.0}}                   
\newcommand{\hatcurRVecceneccenxxxxxF}{\ensuremath{0.040\pm0.026}}           
\newcommand{\hatcurRVeccentwosiglimeccenxxxxxF}{\ensuremath{<0.096}}         
\newcommand{\hatcurRVomegaeccenxxxxxF}{\ensuremath{230\pm58}}                
\newcommand{\hatcurPPieccenxxxxxF}{\ensuremath{86.91\pm0.42}}                
\newcommand{\hatcurPPgeccenxxxxxF}{\ensuremath{13.0_{-1.3}^{+1.8}}}          
\newcommand{\hatcurPPloggeccenxxxxxF}{\ensuremath{3.115\pm0.053}}            
\newcommand{\hatcurPPareccenxxxxxF}{\ensuremath{9.00\pm0.35}}                
\newcommand{\hatcurPPareleccenxxxxxF}{\ensuremath{0.04353\pm0.00041}}        
\newcommand{\hatcurPPrhoeccenxxxxxF}{\ensuremath{0.564_{-0.077}^{+0.108}}}   
\newcommand{\hatcurPPmeccenxxxxxF}{\ensuremath{0.706\pm0.049}}               
\newcommand{\hatcurPPmshorteccenxxxxxF}{\ensuremath{0.71}}                   
\newcommand{\hatcurPPmlongeccenxxxxxF}{\ensuremath{0.706\pm0.049}}           
\newcommand{\hatcurPPmeeccenxxxxxF}{\ensuremath{224\pm15}}                   
\newcommand{\hatcurPPmeshorteccenxxxxxF}{\ensuremath{224.3}}                 
\newcommand{\hatcurPPmelongeccenxxxxxF}{\ensuremath{224\pm15}}               
\newcommand{\hatcurPPreccenxxxxxF}{\ensuremath{1.155\pm0.058}}               
\newcommand{\hatcurPPrshorteccenxxxxxF}{\ensuremath{1.15}}                   
\newcommand{\hatcurPPrlongeccenxxxxxF}{\ensuremath{1.155\pm0.058}}           
\newcommand{\hatcurPPreeccenxxxxxF}{\ensuremath{12.95\pm0.65}}               
\newcommand{\hatcurPPreshorteccenxxxxxF}{\ensuremath{12.9}}                  
\newcommand{\hatcurPPrelongeccenxxxxxF}{\ensuremath{12.95\pm0.65}}           
\newcommand{\hatcurPPmrcorreccenxxxxxF}{\ensuremath{-0.06}}                  
\newcommand{\hatcurPPteffeccenxxxxxF}{\ensuremath{1400\pm36}}                
\newcommand{\hatcurPPthetaeccenxxxxxF}{\ensuremath{0.0484\pm0.0044}}         
\newcommand{\hatcurPPfluxperieccenxxxxxF}{\ensuremath{9.5\pm1.0}}            
\newcommand{\hatcurPPfluxperidimeccenxxxxxF}{\ensuremath{8}}                 
\newcommand{\hatcurPPfluxapeccenxxxxxF}{\ensuremath{7.98\pm0.98}}            
\newcommand{\hatcurPPfluxapdimeccenxxxxxF}{\ensuremath{8}}                   
\newcommand{\hatcurPPfluxavgeccenxxxxxF}{\ensuremath{8.67\pm0.90}}           
\newcommand{\hatcurPPfluxavgdimeccenxxxxxF}{\ensuremath{8}}                  
\newcommand{\hatcurPPfluxavglogeccenxxxxxF}{\ensuremath{8.938\pm0.044}}      
\newcommand{\hatcurXsecphaseeccenxxxxxF}{\ensuremath{0.486\pm0.012}}         
\newcommand{\hatcurXsecondaryeccenxxxxxF}{\ensuremath{2456631.305\pm0.037}}  
\newcommand{\hatcurXsecdureccenxxxxxF}{\ensuremath{0.1111\pm0.0055}}         
\newcommand{\hatcurXsecingdureccenxxxxxF}{\ensuremath{0.0142\pm0.0015}}      
\newcommand{\hatcurPPphiconjeccenxxxxxF}{\ensuremath{-0.363_{-0.087}^{+0.250}}} 
\newcommand{\hatcurPPperieccenxxxxxF}{\ensuremath{2456630.92\pm0.80}}        
\newcommand{\hatcurPPaequiveccenxxxxxF}{\ensuremath{0.0397\pm0.0020}}        
\newcommand{\hatcurPPtcirceccenxxxxxF}{\ensuremath{163\pm41}}                
\newcommand{\hatcurPPtinfalleccenxxxxxF}{\ensuremath{2730\pm510}}            
\newcommand{\hatcurXdisteccenxxxxxF}{\ensuremath{342\pm17}}                  
\newcommand{\hatcurXAveccenxxxxxF}{\ensuremath{0.0000\pm0.0055}}             
\newcommand{\hatcurXdistredeccenxxxxxF}{\ensuremath{332\pm18}}               
\newcommand{\hatcurXEBVeccenxxxxxF}{\ensuremath{0.0000\pm0.0018}}            
\newcommand{\hatcurXmvisoredeccenxxxxxF}{\ensuremath{12.218\pm0.015}}        
\newcommand{\hatcurXmiisoredeccenxxxxxF}{\ensuremath{11.564\pm0.011}}        
\newcommand{\hatcurXmjisoredeccenxxxxxF}{\ensuremath{11.140\pm0.018}}        
\newcommand{\hatcurXmhisoredeccenxxxxxF}{\ensuremath{10.821\pm0.027}}        
\newcommand{\hatcurXmkisoredeccenxxxxxF}{\ensuremath{10.769\pm0.028}}        
\newcommand{\hatcurXviisoredeccenxxxxxF}{\ensuremath{0.654\pm0.019}}         
\newcommand{\hatcurXvkisoredeccenxxxxxF}{\ensuremath{1.449\pm0.040}}         
\newcommand{\hatcurXjhisoredeccenxxxxxF}{\ensuremath{0.319\pm0.010}}         
\newcommand{\hatcurXjkisoredeccenxxxxxF}{\ensuremath{0.371\pm0.011}}         
\newcommand{\hatcurCCpmraeccenxxxxxF}{\ensuremath{-25.3\pm1.0}}              
\newcommand{\hatcurCCpmdececcenxxxxxF}{\ensuremath{-8.2\pm1.0}}              
\newcommand{\hatcurCCpmeccenxxxxxF}{\ensuremath{26.6\pm1.4}}                 

\newcommand{\hatcurCCbbHmageccen}[1]{\ifnum#1=25 %
        \hatcurCCbbHmageccenxxxxxA
                \else
                \ifnum#1=26 %
                \hatcurCCbbHmageccenxxxxxB
                \else
                \ifnum#1=27 %
                \hatcurCCbbHmageccenxxxxxC
                \else
                \ifnum#1=28 %
                \hatcurCCbbHmageccenxxxxxD
                \else
                \ifnum#1=29 %
                \hatcurCCbbHmageccenxxxxxE
                \else
                \ifnum#1=30 %
                \hatcurCCbbHmageccenxxxxxF
                \else
                ??????\fi
                \fi
                \fi
                \fi
                \fi
                \fi
}
\newcommand{\hatcurCCbbJmageccen}[1]{\ifnum#1=25 %
        \hatcurCCbbJmageccenxxxxxA
                \else
                \ifnum#1=26 %
                \hatcurCCbbJmageccenxxxxxB
                \else
                \ifnum#1=27 %
                \hatcurCCbbJmageccenxxxxxC
                \else
                \ifnum#1=28 %
                \hatcurCCbbJmageccenxxxxxD
                \else
                \ifnum#1=29 %
                \hatcurCCbbJmageccenxxxxxE
                \else
                \ifnum#1=30 %
                \hatcurCCbbJmageccenxxxxxF
                \else
                ??????\fi
                \fi
                \fi
                \fi
                \fi
                \fi
}
\newcommand{\hatcurCCbbKmageccen}[1]{\ifnum#1=25 %
        \hatcurCCbbKmageccenxxxxxA
                \else
                \ifnum#1=26 %
                \hatcurCCbbKmageccenxxxxxB
                \else
                \ifnum#1=27 %
                \hatcurCCbbKmageccenxxxxxC
                \else
                \ifnum#1=28 %
                \hatcurCCbbKmageccenxxxxxD
                \else
                \ifnum#1=29 %
                \hatcurCCbbKmageccenxxxxxE
                \else
                \ifnum#1=30 %
                \hatcurCCbbKmageccenxxxxxF
                \else
                ??????\fi
                \fi
                \fi
                \fi
                \fi
                \fi
}
\newcommand{\hatcurCCcitHmageccen}[1]{\ifnum#1=25 %
        \hatcurCCcitHmageccenxxxxxA
                \else
                \ifnum#1=26 %
                \hatcurCCcitHmageccenxxxxxB
                \else
                \ifnum#1=27 %
                \hatcurCCcitHmageccenxxxxxC
                \else
                \ifnum#1=28 %
                \hatcurCCcitHmageccenxxxxxD
                \else
                \ifnum#1=29 %
                \hatcurCCcitHmageccenxxxxxE
                \else
                \ifnum#1=30 %
                \hatcurCCcitHmageccenxxxxxF
                \else
                ??????\fi
                \fi
                \fi
                \fi
                \fi
                \fi
}
\newcommand{\hatcurCCcitJmageccen}[1]{\ifnum#1=25 %
        \hatcurCCcitJmageccenxxxxxA
                \else
                \ifnum#1=26 %
                \hatcurCCcitJmageccenxxxxxB
                \else
                \ifnum#1=27 %
                \hatcurCCcitJmageccenxxxxxC
                \else
                \ifnum#1=28 %
                \hatcurCCcitJmageccenxxxxxD
                \else
                \ifnum#1=29 %
                \hatcurCCcitJmageccenxxxxxE
                \else
                \ifnum#1=30 %
                \hatcurCCcitJmageccenxxxxxF
                \else
                ??????\fi
                \fi
                \fi
                \fi
                \fi
                \fi
}
\newcommand{\hatcurCCcitKmageccen}[1]{\ifnum#1=25 %
        \hatcurCCcitKmageccenxxxxxA
                \else
                \ifnum#1=26 %
                \hatcurCCcitKmageccenxxxxxB
                \else
                \ifnum#1=27 %
                \hatcurCCcitKmageccenxxxxxC
                \else
                \ifnum#1=28 %
                \hatcurCCcitKmageccenxxxxxD
                \else
                \ifnum#1=29 %
                \hatcurCCcitKmageccenxxxxxE
                \else
                \ifnum#1=30 %
                \hatcurCCcitKmageccenxxxxxF
                \else
                ??????\fi
                \fi
                \fi
                \fi
                \fi
                \fi
}
\newcommand{\hatcurCCdececcen}[1]{\ifnum#1=25 %
        \hatcurCCdececcenxxxxxA
                \else
                \ifnum#1=26 %
                \hatcurCCdececcenxxxxxB
                \else
                \ifnum#1=27 %
                \hatcurCCdececcenxxxxxC
                \else
                \ifnum#1=28 %
                \hatcurCCdececcenxxxxxD
                \else
                \ifnum#1=29 %
                \hatcurCCdececcenxxxxxE
                \else
                \ifnum#1=30 %
                \hatcurCCdececcenxxxxxF
                \else
                ??????\fi
                \fi
                \fi
                \fi
                \fi
                \fi
}
\newcommand{\hatcurCCesoHKmageccen}[1]{\ifnum#1=25 %
        \hatcurCCesoHKmageccenxxxxxA
                \else
                \ifnum#1=26 %
                \hatcurCCesoHKmageccenxxxxxB
                \else
                \ifnum#1=27 %
                \hatcurCCesoHKmageccenxxxxxC
                \else
                \ifnum#1=28 %
                \hatcurCCesoHKmageccenxxxxxD
                \else
                \ifnum#1=29 %
                \hatcurCCesoHKmageccenxxxxxE
                \else
                \ifnum#1=30 %
                \hatcurCCesoHKmageccenxxxxxF
                \else
                ??????\fi
                \fi
                \fi
                \fi
                \fi
                \fi
}
\newcommand{\hatcurCCesoHmageccen}[1]{\ifnum#1=25 %
        \hatcurCCesoHmageccenxxxxxA
                \else
                \ifnum#1=26 %
                \hatcurCCesoHmageccenxxxxxB
                \else
                \ifnum#1=27 %
                \hatcurCCesoHmageccenxxxxxC
                \else
                \ifnum#1=28 %
                \hatcurCCesoHmageccenxxxxxD
                \else
                \ifnum#1=29 %
                \hatcurCCesoHmageccenxxxxxE
                \else
                \ifnum#1=30 %
                \hatcurCCesoHmageccenxxxxxF
                \else
                ??????\fi
                \fi
                \fi
                \fi
                \fi
                \fi
}
\newcommand{\hatcurCCesoJHmageccen}[1]{\ifnum#1=25 %
        \hatcurCCesoJHmageccenxxxxxA
                \else
                \ifnum#1=26 %
                \hatcurCCesoJHmageccenxxxxxB
                \else
                \ifnum#1=27 %
                \hatcurCCesoJHmageccenxxxxxC
                \else
                \ifnum#1=28 %
                \hatcurCCesoJHmageccenxxxxxD
                \else
                \ifnum#1=29 %
                \hatcurCCesoJHmageccenxxxxxE
                \else
                \ifnum#1=30 %
                \hatcurCCesoJHmageccenxxxxxF
                \else
                ??????\fi
                \fi
                \fi
                \fi
                \fi
                \fi
}
\newcommand{\hatcurCCesoJKmageccen}[1]{\ifnum#1=25 %
        \hatcurCCesoJKmageccenxxxxxA
                \else
                \ifnum#1=26 %
                \hatcurCCesoJKmageccenxxxxxB
                \else
                \ifnum#1=27 %
                \hatcurCCesoJKmageccenxxxxxC
                \else
                \ifnum#1=28 %
                \hatcurCCesoJKmageccenxxxxxD
                \else
                \ifnum#1=29 %
                \hatcurCCesoJKmageccenxxxxxE
                \else
                \ifnum#1=30 %
                \hatcurCCesoJKmageccenxxxxxF
                \else
                ??????\fi
                \fi
                \fi
                \fi
                \fi
                \fi
}
\newcommand{\hatcurCCesoJmageccen}[1]{\ifnum#1=25 %
        \hatcurCCesoJmageccenxxxxxA
                \else
                \ifnum#1=26 %
                \hatcurCCesoJmageccenxxxxxB
                \else
                \ifnum#1=27 %
                \hatcurCCesoJmageccenxxxxxC
                \else
                \ifnum#1=28 %
                \hatcurCCesoJmageccenxxxxxD
                \else
                \ifnum#1=29 %
                \hatcurCCesoJmageccenxxxxxE
                \else
                \ifnum#1=30 %
                \hatcurCCesoJmageccenxxxxxF
                \else
                ??????\fi
                \fi
                \fi
                \fi
                \fi
                \fi
}
\newcommand{\hatcurCCesoKmageccen}[1]{\ifnum#1=25 %
        \hatcurCCesoKmageccenxxxxxA
                \else
                \ifnum#1=26 %
                \hatcurCCesoKmageccenxxxxxB
                \else
                \ifnum#1=27 %
                \hatcurCCesoKmageccenxxxxxC
                \else
                \ifnum#1=28 %
                \hatcurCCesoKmageccenxxxxxD
                \else
                \ifnum#1=29 %
                \hatcurCCesoKmageccenxxxxxE
                \else
                \ifnum#1=30 %
                \hatcurCCesoKmageccenxxxxxF
                \else
                ??????\fi
                \fi
                \fi
                \fi
                \fi
                \fi
}
\newcommand{\hatcurCCgsceccen}[1]{\ifnum#1=25 %
        \hatcurCCgsceccenxxxxxA
                \else
                \ifnum#1=26 %
                \hatcurCCgsceccenxxxxxB
                \else
                \ifnum#1=27 %
                \hatcurCCgsceccenxxxxxC
                \else
                \ifnum#1=28 %
                \hatcurCCgsceccenxxxxxD
                \else
                \ifnum#1=29 %
                \hatcurCCgsceccenxxxxxE
                \else
                \ifnum#1=30 %
                \hatcurCCgsceccenxxxxxF
                \else
                ??????\fi
                \fi
                \fi
                \fi
                \fi
                \fi
}
\newcommand{\hatcurCCmageccen}[1]{\ifnum#1=25 %
        \hatcurCCmageccenxxxxxA
                \else
                \ifnum#1=26 %
                \hatcurCCmageccenxxxxxB
                \else
                \ifnum#1=27 %
                \hatcurCCmageccenxxxxxC
                \else
                \ifnum#1=28 %
                \hatcurCCmageccenxxxxxD
                \else
                \ifnum#1=29 %
                \hatcurCCmageccenxxxxxE
                \else
                \ifnum#1=30 %
                \hatcurCCmageccenxxxxxF
                \else
                ??????\fi
                \fi
                \fi
                \fi
                \fi
                \fi
}
\newcommand{\hatcurCCpmdececcen}[1]{\ifnum#1=25 %
        \hatcurCCpmdececcenxxxxxA
                \else
                \ifnum#1=26 %
                \hatcurCCpmdececcenxxxxxB
                \else
                \ifnum#1=27 %
                \hatcurCCpmdececcenxxxxxC
                \else
                \ifnum#1=28 %
                \hatcurCCpmdececcenxxxxxD
                \else
                \ifnum#1=29 %
                \hatcurCCpmdececcenxxxxxE
                \else
                \ifnum#1=30 %
                \hatcurCCpmdececcenxxxxxF
                \else
                ??????\fi
                \fi
                \fi
                \fi
                \fi
                \fi
}
\newcommand{\hatcurCCpmeccen}[1]{\ifnum#1=25 %
        \hatcurCCpmeccenxxxxxA
                \else
                \ifnum#1=26 %
                \hatcurCCpmeccenxxxxxB
                \else
                \ifnum#1=27 %
                \hatcurCCpmeccenxxxxxC
                \else
                \ifnum#1=28 %
                \hatcurCCpmeccenxxxxxD
                \else
                \ifnum#1=29 %
                \hatcurCCpmeccenxxxxxE
                \else
                \ifnum#1=30 %
                \hatcurCCpmeccenxxxxxF
                \else
                ??????\fi
                \fi
                \fi
                \fi
                \fi
                \fi
}
\newcommand{\hatcurCCpmraeccen}[1]{\ifnum#1=25 %
        \hatcurCCpmraeccenxxxxxA
                \else
                \ifnum#1=26 %
                \hatcurCCpmraeccenxxxxxB
                \else
                \ifnum#1=27 %
                \hatcurCCpmraeccenxxxxxC
                \else
                \ifnum#1=28 %
                \hatcurCCpmraeccenxxxxxD
                \else
                \ifnum#1=29 %
                \hatcurCCpmraeccenxxxxxE
                \else
                \ifnum#1=30 %
                \hatcurCCpmraeccenxxxxxF
                \else
                ??????\fi
                \fi
                \fi
                \fi
                \fi
                \fi
}
\newcommand{\hatcurCCraeccen}[1]{\ifnum#1=25 %
        \hatcurCCraeccenxxxxxA
                \else
                \ifnum#1=26 %
                \hatcurCCraeccenxxxxxB
                \else
                \ifnum#1=27 %
                \hatcurCCraeccenxxxxxC
                \else
                \ifnum#1=28 %
                \hatcurCCraeccenxxxxxD
                \else
                \ifnum#1=29 %
                \hatcurCCraeccenxxxxxE
                \else
                \ifnum#1=30 %
                \hatcurCCraeccenxxxxxF
                \else
                ??????\fi
                \fi
                \fi
                \fi
                \fi
                \fi
}
\newcommand{\hatcurCCtassmBeccen}[1]{\ifnum#1=25 %
        \hatcurCCtassmBeccenxxxxxA
                \else
                \ifnum#1=26 %
                \hatcurCCtassmBeccenxxxxxB
                \else
                \ifnum#1=27 %
                \hatcurCCtassmBeccenxxxxxC
                \else
                \ifnum#1=28 %
                \hatcurCCtassmBeccenxxxxxD
                \else
                \ifnum#1=29 %
                \hatcurCCtassmBeccenxxxxxE
                \else
                \ifnum#1=30 %
                \hatcurCCtassmBeccenxxxxxF
                \else
                ??????\fi
                \fi
                \fi
                \fi
                \fi
                \fi
}
\newcommand{\hatcurCCtassmBshorteccen}[1]{\ifnum#1=25 %
        \hatcurCCtassmBshorteccenxxxxxA
                \else
                \ifnum#1=26 %
                \hatcurCCtassmBshorteccenxxxxxB
                \else
                \ifnum#1=27 %
                \hatcurCCtassmBshorteccenxxxxxC
                \else
                \ifnum#1=28 %
                \hatcurCCtassmBshorteccenxxxxxD
                \else
                \ifnum#1=29 %
                \hatcurCCtassmBshorteccenxxxxxE
                \else
                \ifnum#1=30 %
                \hatcurCCtassmBshorteccenxxxxxF
                \else
                ??????\fi
                \fi
                \fi
                \fi
                \fi
                \fi
}
\newcommand{\hatcurCCtassmgeccen}[1]{\ifnum#1=25 %
        \hatcurCCtassmgeccenxxxxxA
                \else
                \ifnum#1=26 %
                \hatcurCCtassmgeccenxxxxxB
                \else
                \ifnum#1=27 %
                \hatcurCCtassmgeccenxxxxxC
                \else
                \ifnum#1=28 %
                \hatcurCCtassmgeccenxxxxxD
                \else
                \ifnum#1=29 %
                \hatcurCCtassmgeccenxxxxxE
                \else
                \ifnum#1=30 %
                \hatcurCCtassmgeccenxxxxxF
                \else
                ??????\fi
                \fi
                \fi
                \fi
                \fi
                \fi
}
\newcommand{\hatcurCCtassmgshorteccen}[1]{\ifnum#1=25 %
        \hatcurCCtassmgshorteccenxxxxxA
                \else
                \ifnum#1=26 %
                \hatcurCCtassmgshorteccenxxxxxB
                \else
                \ifnum#1=27 %
                \hatcurCCtassmgshorteccenxxxxxC
                \else
                \ifnum#1=28 %
                \hatcurCCtassmgshorteccenxxxxxD
                \else
                \ifnum#1=29 %
                \hatcurCCtassmgshorteccenxxxxxE
                \else
                \ifnum#1=30 %
                \hatcurCCtassmgshorteccenxxxxxF
                \else
                ??????\fi
                \fi
                \fi
                \fi
                \fi
                \fi
}
\newcommand{\hatcurCCtassmieccen}[1]{\ifnum#1=25 %
        \hatcurCCtassmieccenxxxxxA
                \else
                \ifnum#1=26 %
                \hatcurCCtassmieccenxxxxxB
                \else
                \ifnum#1=27 %
                \hatcurCCtassmieccenxxxxxC
                \else
                \ifnum#1=28 %
                \hatcurCCtassmieccenxxxxxD
                \else
                \ifnum#1=29 %
                \hatcurCCtassmieccenxxxxxE
                \else
                \ifnum#1=30 %
                \hatcurCCtassmieccenxxxxxF
                \else
                ??????\fi
                \fi
                \fi
                \fi
                \fi
                \fi
}
\newcommand{\hatcurCCtassmIeccen}[1]{\ifnum#1=25 %
        \hatcurCCtassmIeccenxxxxxA
                \else
                \ifnum#1=26 %
                \hatcurCCtassmIeccenxxxxxB
                \else
                \ifnum#1=27 %
                \hatcurCCtassmIeccenxxxxxC
                \else
                \ifnum#1=28 %
                \hatcurCCtassmIeccenxxxxxD
                \else
                \ifnum#1=29 %
                \hatcurCCtassmIeccenxxxxxE
                \else
                \ifnum#1=30 %
                \hatcurCCtassmIeccenxxxxxF
                \else
                ??????\fi
                \fi
                \fi
                \fi
                \fi
                \fi
}
\newcommand{\hatcurCCtassmishorteccen}[1]{\ifnum#1=25 %
        \hatcurCCtassmishorteccenxxxxxA
                \else
                \ifnum#1=26 %
                \hatcurCCtassmishorteccenxxxxxB
                \else
                \ifnum#1=27 %
                \hatcurCCtassmishorteccenxxxxxC
                \else
                \ifnum#1=28 %
                \hatcurCCtassmishorteccenxxxxxD
                \else
                \ifnum#1=29 %
                \hatcurCCtassmishorteccenxxxxxE
                \else
                \ifnum#1=30 %
                \hatcurCCtassmishorteccenxxxxxF
                \else
                ??????\fi
                \fi
                \fi
                \fi
                \fi
                \fi
}
\newcommand{\hatcurCCtassmIshorteccen}[1]{\ifnum#1=25 %
        \hatcurCCtassmIshorteccenxxxxxA
                \else
                \ifnum#1=26 %
                \hatcurCCtassmIshorteccenxxxxxB
                \else
                \ifnum#1=27 %
                \hatcurCCtassmIshorteccenxxxxxC
                \else
                \ifnum#1=28 %
                \hatcurCCtassmIshorteccenxxxxxD
                \else
                \ifnum#1=29 %
                \hatcurCCtassmIshorteccenxxxxxE
                \else
                \ifnum#1=30 %
                \hatcurCCtassmIshorteccenxxxxxF
                \else
                ??????\fi
                \fi
                \fi
                \fi
                \fi
                \fi
}
\newcommand{\hatcurCCtassmreccen}[1]{\ifnum#1=25 %
        \hatcurCCtassmreccenxxxxxA
                \else
                \ifnum#1=26 %
                \hatcurCCtassmreccenxxxxxB
                \else
                \ifnum#1=27 %
                \hatcurCCtassmreccenxxxxxC
                \else
                \ifnum#1=28 %
                \hatcurCCtassmreccenxxxxxD
                \else
                \ifnum#1=29 %
                \hatcurCCtassmreccenxxxxxE
                \else
                \ifnum#1=30 %
                \hatcurCCtassmreccenxxxxxF
                \else
                ??????\fi
                \fi
                \fi
                \fi
                \fi
                \fi
}
\newcommand{\hatcurCCtassmrshorteccen}[1]{\ifnum#1=25 %
        \hatcurCCtassmrshorteccenxxxxxA
                \else
                \ifnum#1=26 %
                \hatcurCCtassmrshorteccenxxxxxB
                \else
                \ifnum#1=27 %
                \hatcurCCtassmrshorteccenxxxxxC
                \else
                \ifnum#1=28 %
                \hatcurCCtassmrshorteccenxxxxxD
                \else
                \ifnum#1=29 %
                \hatcurCCtassmrshorteccenxxxxxE
                \else
                \ifnum#1=30 %
                \hatcurCCtassmrshorteccenxxxxxF
                \else
                ??????\fi
                \fi
                \fi
                \fi
                \fi
                \fi
}
\newcommand{\hatcurCCtassmveccen}[1]{\ifnum#1=25 %
        \hatcurCCtassmveccenxxxxxA
                \else
                \ifnum#1=26 %
                \hatcurCCtassmveccenxxxxxB
                \else
                \ifnum#1=27 %
                \hatcurCCtassmveccenxxxxxC
                \else
                \ifnum#1=28 %
                \hatcurCCtassmveccenxxxxxD
                \else
                \ifnum#1=29 %
                \hatcurCCtassmveccenxxxxxE
                \else
                \ifnum#1=30 %
                \hatcurCCtassmveccenxxxxxF
                \else
                ??????\fi
                \fi
                \fi
                \fi
                \fi
                \fi
}
\newcommand{\hatcurCCtassmvshorteccen}[1]{\ifnum#1=25 %
        \hatcurCCtassmvshorteccenxxxxxA
                \else
                \ifnum#1=26 %
                \hatcurCCtassmvshorteccenxxxxxB
                \else
                \ifnum#1=27 %
                \hatcurCCtassmvshorteccenxxxxxC
                \else
                \ifnum#1=28 %
                \hatcurCCtassmvshorteccenxxxxxD
                \else
                \ifnum#1=29 %
                \hatcurCCtassmvshorteccenxxxxxE
                \else
                \ifnum#1=30 %
                \hatcurCCtassmvshorteccenxxxxxF
                \else
                ??????\fi
                \fi
                \fi
                \fi
                \fi
                \fi
}
\newcommand{\hatcurCCtwomasseccen}[1]{\ifnum#1=25 %
        \hatcurCCtwomasseccenxxxxxA
                \else
                \ifnum#1=26 %
                \hatcurCCtwomasseccenxxxxxB
                \else
                \ifnum#1=27 %
                \hatcurCCtwomasseccenxxxxxC
                \else
                \ifnum#1=28 %
                \hatcurCCtwomasseccenxxxxxD
                \else
                \ifnum#1=29 %
                \hatcurCCtwomasseccenxxxxxE
                \else
                \ifnum#1=30 %
                \hatcurCCtwomasseccenxxxxxF
                \else
                ??????\fi
                \fi
                \fi
                \fi
                \fi
                \fi
}
\newcommand{\hatcurCCtwomassHmageccen}[1]{\ifnum#1=25 %
        \hatcurCCtwomassHmageccenxxxxxA
                \else
                \ifnum#1=26 %
                \hatcurCCtwomassHmageccenxxxxxB
                \else
                \ifnum#1=27 %
                \hatcurCCtwomassHmageccenxxxxxC
                \else
                \ifnum#1=28 %
                \hatcurCCtwomassHmageccenxxxxxD
                \else
                \ifnum#1=29 %
                \hatcurCCtwomassHmageccenxxxxxE
                \else
                \ifnum#1=30 %
                \hatcurCCtwomassHmageccenxxxxxF
                \else
                ??????\fi
                \fi
                \fi
                \fi
                \fi
                \fi
}
\newcommand{\hatcurCCtwomassJmageccen}[1]{\ifnum#1=25 %
        \hatcurCCtwomassJmageccenxxxxxA
                \else
                \ifnum#1=26 %
                \hatcurCCtwomassJmageccenxxxxxB
                \else
                \ifnum#1=27 %
                \hatcurCCtwomassJmageccenxxxxxC
                \else
                \ifnum#1=28 %
                \hatcurCCtwomassJmageccenxxxxxD
                \else
                \ifnum#1=29 %
                \hatcurCCtwomassJmageccenxxxxxE
                \else
                \ifnum#1=30 %
                \hatcurCCtwomassJmageccenxxxxxF
                \else
                ??????\fi
                \fi
                \fi
                \fi
                \fi
                \fi
}
\newcommand{\hatcurCCtwomassKmageccen}[1]{\ifnum#1=25 %
        \hatcurCCtwomassKmageccenxxxxxA
                \else
                \ifnum#1=26 %
                \hatcurCCtwomassKmageccenxxxxxB
                \else
                \ifnum#1=27 %
                \hatcurCCtwomassKmageccenxxxxxC
                \else
                \ifnum#1=28 %
                \hatcurCCtwomassKmageccenxxxxxD
                \else
                \ifnum#1=29 %
                \hatcurCCtwomassKmageccenxxxxxE
                \else
                \ifnum#1=30 %
                \hatcurCCtwomassKmageccenxxxxxF
                \else
                ??????\fi
                \fi
                \fi
                \fi
                \fi
                \fi
}
\newcommand{\hatcurfieldeccen}[1]{\ifnum#1=25 %
        \hatcurfieldeccenxxxxxA
                \else
                \ifnum#1=26 %
                \hatcurfieldeccenxxxxxB
                \else
                \ifnum#1=27 %
                \hatcurfieldeccenxxxxxC
                \else
                \ifnum#1=28 %
                \hatcurfieldeccenxxxxxD
                \else
                \ifnum#1=29 %
                \hatcurfieldeccenxxxxxE
                \else
                \ifnum#1=30 %
                \hatcurfieldeccenxxxxxF
                \else
                ??????\fi
                \fi
                \fi
                \fi
                \fi
                \fi
}
\newcommand{\hatcurhtreccen}[1]{\ifnum#1=25 %
        \hatcurhtreccenxxxxxA
                \else
                \ifnum#1=26 %
                \hatcurhtreccenxxxxxB
                \else
                \ifnum#1=27 %
                \hatcurhtreccenxxxxxC
                \else
                \ifnum#1=28 %
                \hatcurhtreccenxxxxxD
                \else
                \ifnum#1=29 %
                \hatcurhtreccenxxxxxE
                \else
                \ifnum#1=30 %
                \hatcurhtreccenxxxxxF
                \else
                ??????\fi
                \fi
                \fi
                \fi
                \fi
                \fi
}
\newcommand{\hatcurISOageeccen}[1]{\ifnum#1=25 %
        \hatcurISOageeccenxxxxxA
                \else
                \ifnum#1=26 %
                \hatcurISOageeccenxxxxxB
                \else
                \ifnum#1=27 %
                \hatcurISOageeccenxxxxxC
                \else
                \ifnum#1=28 %
                \hatcurISOageeccenxxxxxD
                \else
                \ifnum#1=29 %
                \hatcurISOageeccenxxxxxE
                \else
                \ifnum#1=30 %
                \hatcurISOageeccenxxxxxF
                \else
                ??????\fi
                \fi
                \fi
                \fi
                \fi
                \fi
}
\newcommand{\hatcurISOJKeccen}[1]{\ifnum#1=25 %
        \hatcurISOJKeccenxxxxxA
                \else
                \ifnum#1=26 %
                \hatcurISOJKeccenxxxxxB
                \else
                \ifnum#1=27 %
                \hatcurISOJKeccenxxxxxC
                \else
                \ifnum#1=28 %
                \hatcurISOJKeccenxxxxxD
                \else
                \ifnum#1=29 %
                \hatcurISOJKeccenxxxxxE
                \else
                \ifnum#1=30 %
                \hatcurISOJKeccenxxxxxF
                \else
                ??????\fi
                \fi
                \fi
                \fi
                \fi
                \fi
}
\newcommand{\hatcurISOloggeccen}[1]{\ifnum#1=25 %
        \hatcurISOloggeccenxxxxxA
                \else
                \ifnum#1=26 %
                \hatcurISOloggeccenxxxxxB
                \else
                \ifnum#1=27 %
                \hatcurISOloggeccenxxxxxC
                \else
                \ifnum#1=28 %
                \hatcurISOloggeccenxxxxxD
                \else
                \ifnum#1=29 %
                \hatcurISOloggeccenxxxxxE
                \else
                \ifnum#1=30 %
                \hatcurISOloggeccenxxxxxF
                \else
                ??????\fi
                \fi
                \fi
                \fi
                \fi
                \fi
}
\newcommand{\hatcurISOlumeccen}[1]{\ifnum#1=25 %
        \hatcurISOlumeccenxxxxxA
                \else
                \ifnum#1=26 %
                \hatcurISOlumeccenxxxxxB
                \else
                \ifnum#1=27 %
                \hatcurISOlumeccenxxxxxC
                \else
                \ifnum#1=28 %
                \hatcurISOlumeccenxxxxxD
                \else
                \ifnum#1=29 %
                \hatcurISOlumeccenxxxxxE
                \else
                \ifnum#1=30 %
                \hatcurISOlumeccenxxxxxF
                \else
                ??????\fi
                \fi
                \fi
                \fi
                \fi
                \fi
}
\newcommand{\hatcurISOlumshorteccen}[1]{\ifnum#1=25 %
        \hatcurISOlumshorteccenxxxxxA
                \else
                \ifnum#1=26 %
                \hatcurISOlumshorteccenxxxxxB
                \else
                \ifnum#1=27 %
                \hatcurISOlumshorteccenxxxxxC
                \else
                \ifnum#1=28 %
                \hatcurISOlumshorteccenxxxxxD
                \else
                \ifnum#1=29 %
                \hatcurISOlumshorteccenxxxxxE
                \else
                \ifnum#1=30 %
                \hatcurISOlumshorteccenxxxxxF
                \else
                ??????\fi
                \fi
                \fi
                \fi
                \fi
                \fi
}
\newcommand{\hatcurISOmeccen}[1]{\ifnum#1=25 %
        \hatcurISOmeccenxxxxxA
                \else
                \ifnum#1=26 %
                \hatcurISOmeccenxxxxxB
                \else
                \ifnum#1=27 %
                \hatcurISOmeccenxxxxxC
                \else
                \ifnum#1=28 %
                \hatcurISOmeccenxxxxxD
                \else
                \ifnum#1=29 %
                \hatcurISOmeccenxxxxxE
                \else
                \ifnum#1=30 %
                \hatcurISOmeccenxxxxxF
                \else
                ??????\fi
                \fi
                \fi
                \fi
                \fi
                \fi
}
\newcommand{\hatcurISOMHeccen}[1]{\ifnum#1=25 %
        \hatcurISOMHeccenxxxxxA
                \else
                \ifnum#1=26 %
                \hatcurISOMHeccenxxxxxB
                \else
                \ifnum#1=27 %
                \hatcurISOMHeccenxxxxxC
                \else
                \ifnum#1=28 %
                \hatcurISOMHeccenxxxxxD
                \else
                \ifnum#1=29 %
                \hatcurISOMHeccenxxxxxE
                \else
                \ifnum#1=30 %
                \hatcurISOMHeccenxxxxxF
                \else
                ??????\fi
                \fi
                \fi
                \fi
                \fi
                \fi
}
\newcommand{\hatcurISOMJeccen}[1]{\ifnum#1=25 %
        \hatcurISOMJeccenxxxxxA
                \else
                \ifnum#1=26 %
                \hatcurISOMJeccenxxxxxB
                \else
                \ifnum#1=27 %
                \hatcurISOMJeccenxxxxxC
                \else
                \ifnum#1=28 %
                \hatcurISOMJeccenxxxxxD
                \else
                \ifnum#1=29 %
                \hatcurISOMJeccenxxxxxE
                \else
                \ifnum#1=30 %
                \hatcurISOMJeccenxxxxxF
                \else
                ??????\fi
                \fi
                \fi
                \fi
                \fi
                \fi
}
\newcommand{\hatcurISOMKeccen}[1]{\ifnum#1=25 %
        \hatcurISOMKeccenxxxxxA
                \else
                \ifnum#1=26 %
                \hatcurISOMKeccenxxxxxB
                \else
                \ifnum#1=27 %
                \hatcurISOMKeccenxxxxxC
                \else
                \ifnum#1=28 %
                \hatcurISOMKeccenxxxxxD
                \else
                \ifnum#1=29 %
                \hatcurISOMKeccenxxxxxE
                \else
                \ifnum#1=30 %
                \hatcurISOMKeccenxxxxxF
                \else
                ??????\fi
                \fi
                \fi
                \fi
                \fi
                \fi
}
\newcommand{\hatcurISOmlongeccen}[1]{\ifnum#1=25 %
        \hatcurISOmlongeccenxxxxxA
                \else
                \ifnum#1=26 %
                \hatcurISOmlongeccenxxxxxB
                \else
                \ifnum#1=27 %
                \hatcurISOmlongeccenxxxxxC
                \else
                \ifnum#1=28 %
                \hatcurISOmlongeccenxxxxxD
                \else
                \ifnum#1=29 %
                \hatcurISOmlongeccenxxxxxE
                \else
                \ifnum#1=30 %
                \hatcurISOmlongeccenxxxxxF
                \else
                ??????\fi
                \fi
                \fi
                \fi
                \fi
                \fi
}
\newcommand{\hatcurISOmshorteccen}[1]{\ifnum#1=25 %
        \hatcurISOmshorteccenxxxxxA
                \else
                \ifnum#1=26 %
                \hatcurISOmshorteccenxxxxxB
                \else
                \ifnum#1=27 %
                \hatcurISOmshorteccenxxxxxC
                \else
                \ifnum#1=28 %
                \hatcurISOmshorteccenxxxxxD
                \else
                \ifnum#1=29 %
                \hatcurISOmshorteccenxxxxxE
                \else
                \ifnum#1=30 %
                \hatcurISOmshorteccenxxxxxF
                \else
                ??????\fi
                \fi
                \fi
                \fi
                \fi
                \fi
}
\newcommand{\hatcurISOmveccen}[1]{\ifnum#1=25 %
        \hatcurISOmveccenxxxxxA
                \else
                \ifnum#1=26 %
                \hatcurISOmveccenxxxxxB
                \else
                \ifnum#1=27 %
                \hatcurISOmveccenxxxxxC
                \else
                \ifnum#1=28 %
                \hatcurISOmveccenxxxxxD
                \else
                \ifnum#1=29 %
                \hatcurISOmveccenxxxxxE
                \else
                \ifnum#1=30 %
                \hatcurISOmveccenxxxxxF
                \else
                ??????\fi
                \fi
                \fi
                \fi
                \fi
                \fi
}
\newcommand{\hatcurISOreccen}[1]{\ifnum#1=25 %
        \hatcurISOreccenxxxxxA
                \else
                \ifnum#1=26 %
                \hatcurISOreccenxxxxxB
                \else
                \ifnum#1=27 %
                \hatcurISOreccenxxxxxC
                \else
                \ifnum#1=28 %
                \hatcurISOreccenxxxxxD
                \else
                \ifnum#1=29 %
                \hatcurISOreccenxxxxxE
                \else
                \ifnum#1=30 %
                \hatcurISOreccenxxxxxF
                \else
                ??????\fi
                \fi
                \fi
                \fi
                \fi
                \fi
}
\newcommand{\hatcurISOrhoeccen}[1]{\ifnum#1=25 %
        \hatcurISOrhoeccenxxxxxA
                \else
                \ifnum#1=26 %
                \hatcurISOrhoeccenxxxxxB
                \else
                \ifnum#1=27 %
                \hatcurISOrhoeccenxxxxxC
                \else
                \ifnum#1=28 %
                \hatcurISOrhoeccenxxxxxD
                \else
                \ifnum#1=29 %
                \hatcurISOrhoeccenxxxxxE
                \else
                \ifnum#1=30 %
                \hatcurISOrhoeccenxxxxxF
                \else
                ??????\fi
                \fi
                \fi
                \fi
                \fi
                \fi
}
\newcommand{\hatcurISOrholongeccen}[1]{\ifnum#1=25 %
        \hatcurISOrholongeccenxxxxxA
                \else
                \ifnum#1=26 %
                \hatcurISOrholongeccenxxxxxB
                \else
                \ifnum#1=27 %
                \hatcurISOrholongeccenxxxxxC
                \else
                \ifnum#1=28 %
                \hatcurISOrholongeccenxxxxxD
                \else
                \ifnum#1=29 %
                \hatcurISOrholongeccenxxxxxE
                \else
                \ifnum#1=30 %
                \hatcurISOrholongeccenxxxxxF
                \else
                ??????\fi
                \fi
                \fi
                \fi
                \fi
                \fi
}
\newcommand{\hatcurISOrlongeccen}[1]{\ifnum#1=25 %
        \hatcurISOrlongeccenxxxxxA
                \else
                \ifnum#1=26 %
                \hatcurISOrlongeccenxxxxxB
                \else
                \ifnum#1=27 %
                \hatcurISOrlongeccenxxxxxC
                \else
                \ifnum#1=28 %
                \hatcurISOrlongeccenxxxxxD
                \else
                \ifnum#1=29 %
                \hatcurISOrlongeccenxxxxxE
                \else
                \ifnum#1=30 %
                \hatcurISOrlongeccenxxxxxF
                \else
                ??????\fi
                \fi
                \fi
                \fi
                \fi
                \fi
}
\newcommand{\hatcurISOrshorteccen}[1]{\ifnum#1=25 %
        \hatcurISOrshorteccenxxxxxA
                \else
                \ifnum#1=26 %
                \hatcurISOrshorteccenxxxxxB
                \else
                \ifnum#1=27 %
                \hatcurISOrshorteccenxxxxxC
                \else
                \ifnum#1=28 %
                \hatcurISOrshorteccenxxxxxD
                \else
                \ifnum#1=29 %
                \hatcurISOrshorteccenxxxxxE
                \else
                \ifnum#1=30 %
                \hatcurISOrshorteccenxxxxxF
                \else
                ??????\fi
                \fi
                \fi
                \fi
                \fi
                \fi
}
\newcommand{\hatcurISOsigmaeccen}[1]{\ifnum#1=25 %
        \hatcurISOsigmaeccenxxxxxA
                \else
                \ifnum#1=26 %
                \hatcurISOsigmaeccenxxxxxB
                \else
                \ifnum#1=27 %
                \hatcurISOsigmaeccenxxxxxC
                \else
                \ifnum#1=28 %
                \hatcurISOsigmaeccenxxxxxD
                \else
                \ifnum#1=29 %
                \hatcurISOsigmaeccenxxxxxE
                \else
                \ifnum#1=30 %
                \hatcurISOsigmaeccenxxxxxF
                \else
                ??????\fi
                \fi
                \fi
                \fi
                \fi
                \fi
}
\newcommand{\hatcurISOspececcen}[1]{\ifnum#1=25 %
        \hatcurISOspececcenxxxxxA
                \else
                \ifnum#1=26 %
                \hatcurISOspececcenxxxxxB
                \else
                \ifnum#1=27 %
                \hatcurISOspececcenxxxxxC
                \else
                \ifnum#1=28 %
                \hatcurISOspececcenxxxxxD
                \else
                \ifnum#1=29 %
                \hatcurISOspececcenxxxxxE
                \else
                \ifnum#1=30 %
                \hatcurISOspececcenxxxxxF
                \else
                ??????\fi
                \fi
                \fi
                \fi
                \fi
                \fi
}
\newcommand{\hatcurISOvieccen}[1]{\ifnum#1=25 %
        \hatcurISOvieccenxxxxxA
                \else
                \ifnum#1=26 %
                \hatcurISOvieccenxxxxxB
                \else
                \ifnum#1=27 %
                \hatcurISOvieccenxxxxxC
                \else
                \ifnum#1=28 %
                \hatcurISOvieccenxxxxxD
                \else
                \ifnum#1=29 %
                \hatcurISOvieccenxxxxxE
                \else
                \ifnum#1=30 %
                \hatcurISOvieccenxxxxxF
                \else
                ??????\fi
                \fi
                \fi
                \fi
                \fi
                \fi
}
\newcommand{\hatcurLBigeccen}[1]{\ifnum#1=25 %
        \hatcurLBigeccenxxxxxA
                \else
                \ifnum#1=26 %
                \hatcurLBigeccenxxxxxB
                \else
                \ifnum#1=27 %
                \hatcurLBigeccenxxxxxC
                \else
                \ifnum#1=28 %
                \hatcurLBigeccenxxxxxD
                \else
                \ifnum#1=29 %
                \hatcurLBigeccenxxxxxE
                \else
                \ifnum#1=30 %
                \hatcurLBigeccenxxxxxF
                \else
                ??????\fi
                \fi
                \fi
                \fi
                \fi
                \fi
}
\newcommand{\hatcurLBiieccen}[1]{\ifnum#1=25 %
        \hatcurLBiieccenxxxxxA
                \else
                \ifnum#1=26 %
                \hatcurLBiieccenxxxxxB
                \else
                \ifnum#1=27 %
                \hatcurLBiieccenxxxxxC
                \else
                \ifnum#1=28 %
                \hatcurLBiieccenxxxxxD
                \else
                \ifnum#1=29 %
                \hatcurLBiieccenxxxxxE
                \else
                \ifnum#1=30 %
                \hatcurLBiieccenxxxxxF
                \else
                ??????\fi
                \fi
                \fi
                \fi
                \fi
                \fi
}
\newcommand{\hatcurLBiIeccen}[1]{\ifnum#1=25 %
        \hatcurLBiIeccenxxxxxA
                \else
                \ifnum#1=26 %
                \hatcurLBiIeccenxxxxxB
                \else
                \ifnum#1=27 %
                \hatcurLBiIeccenxxxxxC
                \else
                \ifnum#1=28 %
                \hatcurLBiIeccenxxxxxD
                \else
                \ifnum#1=29 %
                \hatcurLBiIeccenxxxxxE
                \else
                \ifnum#1=30 %
                \hatcurLBiIeccenxxxxxF
                \else
                ??????\fi
                \fi
                \fi
                \fi
                \fi
                \fi
}
\newcommand{\hatcurLBiigeccen}[1]{\ifnum#1=25 %
        \hatcurLBiigeccenxxxxxA
                \else
                \ifnum#1=26 %
                \hatcurLBiigeccenxxxxxB
                \else
                \ifnum#1=27 %
                \hatcurLBiigeccenxxxxxC
                \else
                \ifnum#1=28 %
                \hatcurLBiigeccenxxxxxD
                \else
                \ifnum#1=29 %
                \hatcurLBiigeccenxxxxxE
                \else
                \ifnum#1=30 %
                \hatcurLBiigeccenxxxxxF
                \else
                ??????\fi
                \fi
                \fi
                \fi
                \fi
                \fi
}
\newcommand{\hatcurLBiiieccen}[1]{\ifnum#1=25 %
        \hatcurLBiiieccenxxxxxA
                \else
                \ifnum#1=26 %
                \hatcurLBiiieccenxxxxxB
                \else
                \ifnum#1=27 %
                \hatcurLBiiieccenxxxxxC
                \else
                \ifnum#1=28 %
                \hatcurLBiiieccenxxxxxD
                \else
                \ifnum#1=29 %
                \hatcurLBiiieccenxxxxxE
                \else
                \ifnum#1=30 %
                \hatcurLBiiieccenxxxxxF
                \else
                ??????\fi
                \fi
                \fi
                \fi
                \fi
                \fi
}
\newcommand{\hatcurLBiiIeccen}[1]{\ifnum#1=25 %
        \hatcurLBiiIeccenxxxxxA
                \else
                \ifnum#1=26 %
                \hatcurLBiiIeccenxxxxxB
                \else
                \ifnum#1=27 %
                \hatcurLBiiIeccenxxxxxC
                \else
                \ifnum#1=28 %
                \hatcurLBiiIeccenxxxxxD
                \else
                \ifnum#1=29 %
                \hatcurLBiiIeccenxxxxxE
                \else
                \ifnum#1=30 %
                \hatcurLBiiIeccenxxxxxF
                \else
                ??????\fi
                \fi
                \fi
                \fi
                \fi
                \fi
}
\newcommand{\hatcurLBiikepeccen}[1]{\ifnum#1=25 %
        \hatcurLBiikepeccenxxxxxA
                \else
                \ifnum#1=26 %
                \hatcurLBiikepeccenxxxxxB
                \else
                \ifnum#1=27 %
                \hatcurLBiikepeccenxxxxxC
                \else
                \ifnum#1=28 %
                \hatcurLBiikepeccenxxxxxD
                \else
                \ifnum#1=29 %
                \hatcurLBiikepeccenxxxxxE
                \else
                \ifnum#1=30 %
                \hatcurLBiikepeccenxxxxxF
                \else
                ??????\fi
                \fi
                \fi
                \fi
                \fi
                \fi
}
\newcommand{\hatcurLBiireccen}[1]{\ifnum#1=25 %
        \hatcurLBiireccenxxxxxA
                \else
                \ifnum#1=26 %
                \hatcurLBiireccenxxxxxB
                \else
                \ifnum#1=27 %
                \hatcurLBiireccenxxxxxC
                \else
                \ifnum#1=28 %
                \hatcurLBiireccenxxxxxD
                \else
                \ifnum#1=29 %
                \hatcurLBiireccenxxxxxE
                \else
                \ifnum#1=30 %
                \hatcurLBiireccenxxxxxF
                \else
                ??????\fi
                \fi
                \fi
                \fi
                \fi
                \fi
}
\newcommand{\hatcurLBiiReccen}[1]{\ifnum#1=25 %
        \hatcurLBiiReccenxxxxxA
                \else
                \ifnum#1=26 %
                \hatcurLBiiReccenxxxxxB
                \else
                \ifnum#1=27 %
                \hatcurLBiiReccenxxxxxC
                \else
                \ifnum#1=28 %
                \hatcurLBiiReccenxxxxxD
                \else
                \ifnum#1=29 %
                \hatcurLBiiReccenxxxxxE
                \else
                \ifnum#1=30 %
                \hatcurLBiiReccenxxxxxF
                \else
                ??????\fi
                \fi
                \fi
                \fi
                \fi
                \fi
}
\newcommand{\hatcurLBiizeccen}[1]{\ifnum#1=25 %
        \hatcurLBiizeccenxxxxxA
                \else
                \ifnum#1=26 %
                \hatcurLBiizeccenxxxxxB
                \else
                \ifnum#1=27 %
                \hatcurLBiizeccenxxxxxC
                \else
                \ifnum#1=28 %
                \hatcurLBiizeccenxxxxxD
                \else
                \ifnum#1=29 %
                \hatcurLBiizeccenxxxxxE
                \else
                \ifnum#1=30 %
                \hatcurLBiizeccenxxxxxF
                \else
                ??????\fi
                \fi
                \fi
                \fi
                \fi
                \fi
}
\newcommand{\hatcurLBikepeccen}[1]{\ifnum#1=25 %
        \hatcurLBikepeccenxxxxxA
                \else
                \ifnum#1=26 %
                \hatcurLBikepeccenxxxxxB
                \else
                \ifnum#1=27 %
                \hatcurLBikepeccenxxxxxC
                \else
                \ifnum#1=28 %
                \hatcurLBikepeccenxxxxxD
                \else
                \ifnum#1=29 %
                \hatcurLBikepeccenxxxxxE
                \else
                \ifnum#1=30 %
                \hatcurLBikepeccenxxxxxF
                \else
                ??????\fi
                \fi
                \fi
                \fi
                \fi
                \fi
}
\newcommand{\hatcurLBireccen}[1]{\ifnum#1=25 %
        \hatcurLBireccenxxxxxA
                \else
                \ifnum#1=26 %
                \hatcurLBireccenxxxxxB
                \else
                \ifnum#1=27 %
                \hatcurLBireccenxxxxxC
                \else
                \ifnum#1=28 %
                \hatcurLBireccenxxxxxD
                \else
                \ifnum#1=29 %
                \hatcurLBireccenxxxxxE
                \else
                \ifnum#1=30 %
                \hatcurLBireccenxxxxxF
                \else
                ??????\fi
                \fi
                \fi
                \fi
                \fi
                \fi
}
\newcommand{\hatcurLBiReccen}[1]{\ifnum#1=25 %
        \hatcurLBiReccenxxxxxA
                \else
                \ifnum#1=26 %
                \hatcurLBiReccenxxxxxB
                \else
                \ifnum#1=27 %
                \hatcurLBiReccenxxxxxC
                \else
                \ifnum#1=28 %
                \hatcurLBiReccenxxxxxD
                \else
                \ifnum#1=29 %
                \hatcurLBiReccenxxxxxE
                \else
                \ifnum#1=30 %
                \hatcurLBiReccenxxxxxF
                \else
                ??????\fi
                \fi
                \fi
                \fi
                \fi
                \fi
}
\newcommand{\hatcurLBizeccen}[1]{\ifnum#1=25 %
        \hatcurLBizeccenxxxxxA
                \else
                \ifnum#1=26 %
                \hatcurLBizeccenxxxxxB
                \else
                \ifnum#1=27 %
                \hatcurLBizeccenxxxxxC
                \else
                \ifnum#1=28 %
                \hatcurLBizeccenxxxxxD
                \else
                \ifnum#1=29 %
                \hatcurLBizeccenxxxxxE
                \else
                \ifnum#1=30 %
                \hatcurLBizeccenxxxxxF
                \else
                ??????\fi
                \fi
                \fi
                \fi
                \fi
                \fi
}
\newcommand{\hatcurLCbsqeccen}[1]{\ifnum#1=25 %
        \hatcurLCbsqeccenxxxxxA
                \else
                \ifnum#1=26 %
                \hatcurLCbsqeccenxxxxxB
                \else
                \ifnum#1=27 %
                \hatcurLCbsqeccenxxxxxC
                \else
                \ifnum#1=28 %
                \hatcurLCbsqeccenxxxxxD
                \else
                \ifnum#1=29 %
                \hatcurLCbsqeccenxxxxxE
                \else
                \ifnum#1=30 %
                \hatcurLCbsqeccenxxxxxF
                \else
                ??????\fi
                \fi
                \fi
                \fi
                \fi
                \fi
}
\newcommand{\hatcurLCdipeccen}[1]{\ifnum#1=25 %
        \hatcurLCdipeccenxxxxxA
                \else
                \ifnum#1=26 %
                \hatcurLCdipeccenxxxxxB
                \else
                \ifnum#1=27 %
                \hatcurLCdipeccenxxxxxC
                \else
                \ifnum#1=28 %
                \hatcurLCdipeccenxxxxxD
                \else
                \ifnum#1=29 %
                \hatcurLCdipeccenxxxxxE
                \else
                \ifnum#1=30 %
                \hatcurLCdipeccenxxxxxF
                \else
                ??????\fi
                \fi
                \fi
                \fi
                \fi
                \fi
}
\newcommand{\hatcurLCdureccen}[1]{\ifnum#1=25 %
        \hatcurLCdureccenxxxxxA
                \else
                \ifnum#1=26 %
                \hatcurLCdureccenxxxxxB
                \else
                \ifnum#1=27 %
                \hatcurLCdureccenxxxxxC
                \else
                \ifnum#1=28 %
                \hatcurLCdureccenxxxxxD
                \else
                \ifnum#1=29 %
                \hatcurLCdureccenxxxxxE
                \else
                \ifnum#1=30 %
                \hatcurLCdureccenxxxxxF
                \else
                ??????\fi
                \fi
                \fi
                \fi
                \fi
                \fi
}
\newcommand{\hatcurLCdurhreccen}[1]{\ifnum#1=25 %
        \hatcurLCdurhreccenxxxxxA
                \else
                \ifnum#1=26 %
                \hatcurLCdurhreccenxxxxxB
                \else
                \ifnum#1=27 %
                \hatcurLCdurhreccenxxxxxC
                \else
                \ifnum#1=28 %
                \hatcurLCdurhreccenxxxxxD
                \else
                \ifnum#1=29 %
                \hatcurLCdurhreccenxxxxxE
                \else
                \ifnum#1=30 %
                \hatcurLCdurhreccenxxxxxF
                \else
                ??????\fi
                \fi
                \fi
                \fi
                \fi
                \fi
}
\newcommand{\hatcurLCdurhrshorteccen}[1]{\ifnum#1=25 %
        \hatcurLCdurhrshorteccenxxxxxA
                \else
                \ifnum#1=26 %
                \hatcurLCdurhrshorteccenxxxxxB
                \else
                \ifnum#1=27 %
                \hatcurLCdurhrshorteccenxxxxxC
                \else
                \ifnum#1=28 %
                \hatcurLCdurhrshorteccenxxxxxD
                \else
                \ifnum#1=29 %
                \hatcurLCdurhrshorteccenxxxxxE
                \else
                \ifnum#1=30 %
                \hatcurLCdurhrshorteccenxxxxxF
                \else
                ??????\fi
                \fi
                \fi
                \fi
                \fi
                \fi
}
\newcommand{\hatcurLCdurshorteccen}[1]{\ifnum#1=25 %
        \hatcurLCdurshorteccenxxxxxA
                \else
                \ifnum#1=26 %
                \hatcurLCdurshorteccenxxxxxB
                \else
                \ifnum#1=27 %
                \hatcurLCdurshorteccenxxxxxC
                \else
                \ifnum#1=28 %
                \hatcurLCdurshorteccenxxxxxD
                \else
                \ifnum#1=29 %
                \hatcurLCdurshorteccenxxxxxE
                \else
                \ifnum#1=30 %
                \hatcurLCdurshorteccenxxxxxF
                \else
                ??????\fi
                \fi
                \fi
                \fi
                \fi
                \fi
}
\newcommand{\hatcurLChatnetmAeccen}[1]{\ifnum#1=26 %
        \hatcurLChatnetmAeccenxxxxxB
                \else
                \ifnum#1=30 %
                \hatcurLChatnetmAeccenxxxxxF
                \else
                ??????\fi
                \fi
}
\newcommand{\hatcurLChatnetmBeccen}[1]{\ifnum#1=26 %
        \hatcurLChatnetmBeccenxxxxxB
                \else
                \ifnum#1=30 %
                \hatcurLChatnetmBeccenxxxxxF
                \else
                ??????\fi
                \fi
}
\newcommand{\hatcurLChatnetmCeccen}[1]{\ifnum#1=30 %
        \hatcurLChatnetmCeccenxxxxxF
                \else
                ??????\fi
}
\newcommand{\hatcurLChatnetmeccen}[1]{\ifnum#1=25 %
        \hatcurLChatnetmeccenxxxxxA
                \else
                \ifnum#1=27 %
                \hatcurLChatnetmeccenxxxxxC
                \else
                \ifnum#1=28 %
                \hatcurLChatnetmeccenxxxxxD
                \else
                \ifnum#1=29 %
                \hatcurLChatnetmeccenxxxxxE
                \else
                ??????\fi
                \fi
                \fi
                \fi
}
\newcommand{\hatcurLCiblendAeccen}[1]{\ifnum#1=26 %
        \hatcurLCiblendAeccenxxxxxB
                \else
                \ifnum#1=30 %
                \hatcurLCiblendAeccenxxxxxF
                \else
                ??????\fi
                \fi
}
\newcommand{\hatcurLCiblendBeccen}[1]{\ifnum#1=26 %
        \hatcurLCiblendBeccenxxxxxB
                \else
                \ifnum#1=30 %
                \hatcurLCiblendBeccenxxxxxF
                \else
                ??????\fi
                \fi
}
\newcommand{\hatcurLCiblendCeccen}[1]{\ifnum#1=30 %
        \hatcurLCiblendCeccenxxxxxF
                \else
                ??????\fi
}
\newcommand{\hatcurLCiblendeccen}[1]{\ifnum#1=25 %
        \hatcurLCiblendeccenxxxxxA
                \else
                \ifnum#1=27 %
                \hatcurLCiblendeccenxxxxxC
                \else
                \ifnum#1=28 %
                \hatcurLCiblendeccenxxxxxD
                \else
                \ifnum#1=29 %
                \hatcurLCiblendeccenxxxxxE
                \else
                ??????\fi
                \fi
                \fi
                \fi
}
\newcommand{\hatcurLCimpeccen}[1]{\ifnum#1=25 %
        \hatcurLCimpeccenxxxxxA
                \else
                \ifnum#1=26 %
                \hatcurLCimpeccenxxxxxB
                \else
                \ifnum#1=27 %
                \hatcurLCimpeccenxxxxxC
                \else
                \ifnum#1=28 %
                \hatcurLCimpeccenxxxxxD
                \else
                \ifnum#1=29 %
                \hatcurLCimpeccenxxxxxE
                \else
                \ifnum#1=30 %
                \hatcurLCimpeccenxxxxxF
                \else
                ??????\fi
                \fi
                \fi
                \fi
                \fi
                \fi
}
\newcommand{\hatcurLCingdureccen}[1]{\ifnum#1=25 %
        \hatcurLCingdureccenxxxxxA
                \else
                \ifnum#1=26 %
                \hatcurLCingdureccenxxxxxB
                \else
                \ifnum#1=27 %
                \hatcurLCingdureccenxxxxxC
                \else
                \ifnum#1=28 %
                \hatcurLCingdureccenxxxxxD
                \else
                \ifnum#1=29 %
                \hatcurLCingdureccenxxxxxE
                \else
                \ifnum#1=30 %
                \hatcurLCingdureccenxxxxxF
                \else
                ??????\fi
                \fi
                \fi
                \fi
                \fi
                \fi
}
\newcommand{\hatcurLCPeccen}[1]{\ifnum#1=25 %
        \hatcurLCPeccenxxxxxA
                \else
                \ifnum#1=26 %
                \hatcurLCPeccenxxxxxB
                \else
                \ifnum#1=27 %
                \hatcurLCPeccenxxxxxC
                \else
                \ifnum#1=28 %
                \hatcurLCPeccenxxxxxD
                \else
                \ifnum#1=29 %
                \hatcurLCPeccenxxxxxE
                \else
                \ifnum#1=30 %
                \hatcurLCPeccenxxxxxF
                \else
                ??????\fi
                \fi
                \fi
                \fi
                \fi
                \fi
}
\newcommand{\hatcurLCPprececcen}[1]{\ifnum#1=25 %
        \hatcurLCPprececcenxxxxxA
                \else
                \ifnum#1=26 %
                \hatcurLCPprececcenxxxxxB
                \else
                \ifnum#1=27 %
                \hatcurLCPprececcenxxxxxC
                \else
                \ifnum#1=28 %
                \hatcurLCPprececcenxxxxxD
                \else
                \ifnum#1=29 %
                \hatcurLCPprececcenxxxxxE
                \else
                \ifnum#1=30 %
                \hatcurLCPprececcenxxxxxF
                \else
                ??????\fi
                \fi
                \fi
                \fi
                \fi
                \fi
}
\newcommand{\hatcurLCPshorteccen}[1]{\ifnum#1=25 %
        \hatcurLCPshorteccenxxxxxA
                \else
                \ifnum#1=26 %
                \hatcurLCPshorteccenxxxxxB
                \else
                \ifnum#1=27 %
                \hatcurLCPshorteccenxxxxxC
                \else
                \ifnum#1=28 %
                \hatcurLCPshorteccenxxxxxD
                \else
                \ifnum#1=29 %
                \hatcurLCPshorteccenxxxxxE
                \else
                \ifnum#1=30 %
                \hatcurLCPshorteccenxxxxxF
                \else
                ??????\fi
                \fi
                \fi
                \fi
                \fi
                \fi
}
\newcommand{\hatcurLCqeccen}[1]{\ifnum#1=25 %
        \hatcurLCqeccenxxxxxA
                \else
                \ifnum#1=26 %
                \hatcurLCqeccenxxxxxB
                \else
                \ifnum#1=27 %
                \hatcurLCqeccenxxxxxC
                \else
                \ifnum#1=28 %
                \hatcurLCqeccenxxxxxD
                \else
                \ifnum#1=29 %
                \hatcurLCqeccenxxxxxE
                \else
                \ifnum#1=30 %
                \hatcurLCqeccenxxxxxF
                \else
                ??????\fi
                \fi
                \fi
                \fi
                \fi
                \fi
}
\newcommand{\hatcurLCqshorteccen}[1]{\ifnum#1=25 %
        \hatcurLCqshorteccenxxxxxA
                \else
                \ifnum#1=26 %
                \hatcurLCqshorteccenxxxxxB
                \else
                \ifnum#1=27 %
                \hatcurLCqshorteccenxxxxxC
                \else
                \ifnum#1=28 %
                \hatcurLCqshorteccenxxxxxD
                \else
                \ifnum#1=29 %
                \hatcurLCqshorteccenxxxxxE
                \else
                \ifnum#1=30 %
                \hatcurLCqshorteccenxxxxxF
                \else
                ??????\fi
                \fi
                \fi
                \fi
                \fi
                \fi
}
\newcommand{\hatcurLCrhoeccen}[1]{\ifnum#1=25 %
        \hatcurLCrhoeccenxxxxxA
                \else
                \ifnum#1=26 %
                \hatcurLCrhoeccenxxxxxB
                \else
                \ifnum#1=27 %
                \hatcurLCrhoeccenxxxxxC
                \else
                \ifnum#1=28 %
                \hatcurLCrhoeccenxxxxxD
                \else
                \ifnum#1=29 %
                \hatcurLCrhoeccenxxxxxE
                \else
                \ifnum#1=30 %
                \hatcurLCrhoeccenxxxxxF
                \else
                ??????\fi
                \fi
                \fi
                \fi
                \fi
                \fi
}
\newcommand{\hatcurLCrprstareccen}[1]{\ifnum#1=25 %
        \hatcurLCrprstareccenxxxxxA
                \else
                \ifnum#1=26 %
                \hatcurLCrprstareccenxxxxxB
                \else
                \ifnum#1=27 %
                \hatcurLCrprstareccenxxxxxC
                \else
                \ifnum#1=28 %
                \hatcurLCrprstareccenxxxxxD
                \else
                \ifnum#1=29 %
                \hatcurLCrprstareccenxxxxxE
                \else
                \ifnum#1=30 %
                \hatcurLCrprstareccenxxxxxF
                \else
                ??????\fi
                \fi
                \fi
                \fi
                \fi
                \fi
}
\newcommand{\hatcurLCTAeccen}[1]{\ifnum#1=25 %
        \hatcurLCTAeccenxxxxxA
                \else
                \ifnum#1=26 %
                \hatcurLCTAeccenxxxxxB
                \else
                \ifnum#1=27 %
                \hatcurLCTAeccenxxxxxC
                \else
                \ifnum#1=28 %
                \hatcurLCTAeccenxxxxxD
                \else
                \ifnum#1=29 %
                \hatcurLCTAeccenxxxxxE
                \else
                \ifnum#1=30 %
                \hatcurLCTAeccenxxxxxF
                \else
                ??????\fi
                \fi
                \fi
                \fi
                \fi
                \fi
}
\newcommand{\hatcurLCTBeccen}[1]{\ifnum#1=25 %
        \hatcurLCTBeccenxxxxxA
                \else
                \ifnum#1=26 %
                \hatcurLCTBeccenxxxxxB
                \else
                \ifnum#1=27 %
                \hatcurLCTBeccenxxxxxC
                \else
                \ifnum#1=28 %
                \hatcurLCTBeccenxxxxxD
                \else
                \ifnum#1=29 %
                \hatcurLCTBeccenxxxxxE
                \else
                \ifnum#1=30 %
                \hatcurLCTBeccenxxxxxF
                \else
                ??????\fi
                \fi
                \fi
                \fi
                \fi
                \fi
}
\newcommand{\hatcurLCTeccen}[1]{\ifnum#1=25 %
        \hatcurLCTeccenxxxxxA
                \else
                \ifnum#1=26 %
                \hatcurLCTeccenxxxxxB
                \else
                \ifnum#1=27 %
                \hatcurLCTeccenxxxxxC
                \else
                \ifnum#1=28 %
                \hatcurLCTeccenxxxxxD
                \else
                \ifnum#1=29 %
                \hatcurLCTeccenxxxxxE
                \else
                \ifnum#1=30 %
                \hatcurLCTeccenxxxxxF
                \else
                ??????\fi
                \fi
                \fi
                \fi
                \fi
                \fi
}
\newcommand{\hatcurLCzetaeccen}[1]{\ifnum#1=25 %
        \hatcurLCzetaeccenxxxxxA
                \else
                \ifnum#1=26 %
                \hatcurLCzetaeccenxxxxxB
                \else
                \ifnum#1=27 %
                \hatcurLCzetaeccenxxxxxC
                \else
                \ifnum#1=28 %
                \hatcurLCzetaeccenxxxxxD
                \else
                \ifnum#1=29 %
                \hatcurLCzetaeccenxxxxxE
                \else
                \ifnum#1=30 %
                \hatcurLCzetaeccenxxxxxF
                \else
                ??????\fi
                \fi
                \fi
                \fi
                \fi
                \fi
}
\newcommand{\hatcurPPaequiveccen}[1]{\ifnum#1=25 %
        \hatcurPPaequiveccenxxxxxA
                \else
                \ifnum#1=26 %
                \hatcurPPaequiveccenxxxxxB
                \else
                \ifnum#1=27 %
                \hatcurPPaequiveccenxxxxxC
                \else
                \ifnum#1=28 %
                \hatcurPPaequiveccenxxxxxD
                \else
                \ifnum#1=29 %
                \hatcurPPaequiveccenxxxxxE
                \else
                \ifnum#1=30 %
                \hatcurPPaequiveccenxxxxxF
                \else
                ??????\fi
                \fi
                \fi
                \fi
                \fi
                \fi
}
\newcommand{\hatcurPPareccen}[1]{\ifnum#1=25 %
        \hatcurPPareccenxxxxxA
                \else
                \ifnum#1=26 %
                \hatcurPPareccenxxxxxB
                \else
                \ifnum#1=27 %
                \hatcurPPareccenxxxxxC
                \else
                \ifnum#1=28 %
                \hatcurPPareccenxxxxxD
                \else
                \ifnum#1=29 %
                \hatcurPPareccenxxxxxE
                \else
                \ifnum#1=30 %
                \hatcurPPareccenxxxxxF
                \else
                ??????\fi
                \fi
                \fi
                \fi
                \fi
                \fi
}
\newcommand{\hatcurPPareleccen}[1]{\ifnum#1=25 %
        \hatcurPPareleccenxxxxxA
                \else
                \ifnum#1=26 %
                \hatcurPPareleccenxxxxxB
                \else
                \ifnum#1=27 %
                \hatcurPPareleccenxxxxxC
                \else
                \ifnum#1=28 %
                \hatcurPPareleccenxxxxxD
                \else
                \ifnum#1=29 %
                \hatcurPPareleccenxxxxxE
                \else
                \ifnum#1=30 %
                \hatcurPPareleccenxxxxxF
                \else
                ??????\fi
                \fi
                \fi
                \fi
                \fi
                \fi
}
\newcommand{\hatcurPPfluxapdimeccen}[1]{\ifnum#1=25 %
        \hatcurPPfluxapdimeccenxxxxxA
                \else
                \ifnum#1=26 %
                \hatcurPPfluxapdimeccenxxxxxB
                \else
                \ifnum#1=27 %
                \hatcurPPfluxapdimeccenxxxxxC
                \else
                \ifnum#1=28 %
                \hatcurPPfluxapdimeccenxxxxxD
                \else
                \ifnum#1=29 %
                \hatcurPPfluxapdimeccenxxxxxE
                \else
                \ifnum#1=30 %
                \hatcurPPfluxapdimeccenxxxxxF
                \else
                ??????\fi
                \fi
                \fi
                \fi
                \fi
                \fi
}
\newcommand{\hatcurPPfluxapeccen}[1]{\ifnum#1=25 %
        \hatcurPPfluxapeccenxxxxxA
                \else
                \ifnum#1=26 %
                \hatcurPPfluxapeccenxxxxxB
                \else
                \ifnum#1=27 %
                \hatcurPPfluxapeccenxxxxxC
                \else
                \ifnum#1=28 %
                \hatcurPPfluxapeccenxxxxxD
                \else
                \ifnum#1=29 %
                \hatcurPPfluxapeccenxxxxxE
                \else
                \ifnum#1=30 %
                \hatcurPPfluxapeccenxxxxxF
                \else
                ??????\fi
                \fi
                \fi
                \fi
                \fi
                \fi
}
\newcommand{\hatcurPPfluxavgdimeccen}[1]{\ifnum#1=25 %
        \hatcurPPfluxavgdimeccenxxxxxA
                \else
                \ifnum#1=26 %
                \hatcurPPfluxavgdimeccenxxxxxB
                \else
                \ifnum#1=27 %
                \hatcurPPfluxavgdimeccenxxxxxC
                \else
                \ifnum#1=28 %
                \hatcurPPfluxavgdimeccenxxxxxD
                \else
                \ifnum#1=29 %
                \hatcurPPfluxavgdimeccenxxxxxE
                \else
                \ifnum#1=30 %
                \hatcurPPfluxavgdimeccenxxxxxF
                \else
                ??????\fi
                \fi
                \fi
                \fi
                \fi
                \fi
}
\newcommand{\hatcurPPfluxavgeccen}[1]{\ifnum#1=25 %
        \hatcurPPfluxavgeccenxxxxxA
                \else
                \ifnum#1=26 %
                \hatcurPPfluxavgeccenxxxxxB
                \else
                \ifnum#1=27 %
                \hatcurPPfluxavgeccenxxxxxC
                \else
                \ifnum#1=28 %
                \hatcurPPfluxavgeccenxxxxxD
                \else
                \ifnum#1=29 %
                \hatcurPPfluxavgeccenxxxxxE
                \else
                \ifnum#1=30 %
                \hatcurPPfluxavgeccenxxxxxF
                \else
                ??????\fi
                \fi
                \fi
                \fi
                \fi
                \fi
}
\newcommand{\hatcurPPfluxavglogeccen}[1]{\ifnum#1=25 %
        \hatcurPPfluxavglogeccenxxxxxA
                \else
                \ifnum#1=26 %
                \hatcurPPfluxavglogeccenxxxxxB
                \else
                \ifnum#1=27 %
                \hatcurPPfluxavglogeccenxxxxxC
                \else
                \ifnum#1=28 %
                \hatcurPPfluxavglogeccenxxxxxD
                \else
                \ifnum#1=29 %
                \hatcurPPfluxavglogeccenxxxxxE
                \else
                \ifnum#1=30 %
                \hatcurPPfluxavglogeccenxxxxxF
                \else
                ??????\fi
                \fi
                \fi
                \fi
                \fi
                \fi
}
\newcommand{\hatcurPPfluxperidimeccen}[1]{\ifnum#1=25 %
        \hatcurPPfluxperidimeccenxxxxxA
                \else
                \ifnum#1=26 %
                \hatcurPPfluxperidimeccenxxxxxB
                \else
                \ifnum#1=27 %
                \hatcurPPfluxperidimeccenxxxxxC
                \else
                \ifnum#1=28 %
                \hatcurPPfluxperidimeccenxxxxxD
                \else
                \ifnum#1=29 %
                \hatcurPPfluxperidimeccenxxxxxE
                \else
                \ifnum#1=30 %
                \hatcurPPfluxperidimeccenxxxxxF
                \else
                ??????\fi
                \fi
                \fi
                \fi
                \fi
                \fi
}
\newcommand{\hatcurPPfluxperieccen}[1]{\ifnum#1=25 %
        \hatcurPPfluxperieccenxxxxxA
                \else
                \ifnum#1=26 %
                \hatcurPPfluxperieccenxxxxxB
                \else
                \ifnum#1=27 %
                \hatcurPPfluxperieccenxxxxxC
                \else
                \ifnum#1=28 %
                \hatcurPPfluxperieccenxxxxxD
                \else
                \ifnum#1=29 %
                \hatcurPPfluxperieccenxxxxxE
                \else
                \ifnum#1=30 %
                \hatcurPPfluxperieccenxxxxxF
                \else
                ??????\fi
                \fi
                \fi
                \fi
                \fi
                \fi
}
\newcommand{\hatcurPPgeccen}[1]{\ifnum#1=25 %
        \hatcurPPgeccenxxxxxA
                \else
                \ifnum#1=26 %
                \hatcurPPgeccenxxxxxB
                \else
                \ifnum#1=27 %
                \hatcurPPgeccenxxxxxC
                \else
                \ifnum#1=28 %
                \hatcurPPgeccenxxxxxD
                \else
                \ifnum#1=29 %
                \hatcurPPgeccenxxxxxE
                \else
                \ifnum#1=30 %
                \hatcurPPgeccenxxxxxF
                \else
                ??????\fi
                \fi
                \fi
                \fi
                \fi
                \fi
}
\newcommand{\hatcurPPieccen}[1]{\ifnum#1=25 %
        \hatcurPPieccenxxxxxA
                \else
                \ifnum#1=26 %
                \hatcurPPieccenxxxxxB
                \else
                \ifnum#1=27 %
                \hatcurPPieccenxxxxxC
                \else
                \ifnum#1=28 %
                \hatcurPPieccenxxxxxD
                \else
                \ifnum#1=29 %
                \hatcurPPieccenxxxxxE
                \else
                \ifnum#1=30 %
                \hatcurPPieccenxxxxxF
                \else
                ??????\fi
                \fi
                \fi
                \fi
                \fi
                \fi
}
\newcommand{\hatcurPPloggeccen}[1]{\ifnum#1=25 %
        \hatcurPPloggeccenxxxxxA
                \else
                \ifnum#1=26 %
                \hatcurPPloggeccenxxxxxB
                \else
                \ifnum#1=27 %
                \hatcurPPloggeccenxxxxxC
                \else
                \ifnum#1=28 %
                \hatcurPPloggeccenxxxxxD
                \else
                \ifnum#1=29 %
                \hatcurPPloggeccenxxxxxE
                \else
                \ifnum#1=30 %
                \hatcurPPloggeccenxxxxxF
                \else
                ??????\fi
                \fi
                \fi
                \fi
                \fi
                \fi
}
\newcommand{\hatcurPPmeccen}[1]{\ifnum#1=25 %
        \hatcurPPmeccenxxxxxA
                \else
                \ifnum#1=26 %
                \hatcurPPmeccenxxxxxB
                \else
                \ifnum#1=27 %
                \hatcurPPmeccenxxxxxC
                \else
                \ifnum#1=28 %
                \hatcurPPmeccenxxxxxD
                \else
                \ifnum#1=29 %
                \hatcurPPmeccenxxxxxE
                \else
                \ifnum#1=30 %
                \hatcurPPmeccenxxxxxF
                \else
                ??????\fi
                \fi
                \fi
                \fi
                \fi
                \fi
}
\newcommand{\hatcurPPmeeccen}[1]{\ifnum#1=25 %
        \hatcurPPmeeccenxxxxxA
                \else
                \ifnum#1=26 %
                \hatcurPPmeeccenxxxxxB
                \else
                \ifnum#1=27 %
                \hatcurPPmeeccenxxxxxC
                \else
                \ifnum#1=28 %
                \hatcurPPmeeccenxxxxxD
                \else
                \ifnum#1=29 %
                \hatcurPPmeeccenxxxxxE
                \else
                \ifnum#1=30 %
                \hatcurPPmeeccenxxxxxF
                \else
                ??????\fi
                \fi
                \fi
                \fi
                \fi
                \fi
}
\newcommand{\hatcurPPmelongeccen}[1]{\ifnum#1=25 %
        \hatcurPPmelongeccenxxxxxA
                \else
                \ifnum#1=26 %
                \hatcurPPmelongeccenxxxxxB
                \else
                \ifnum#1=27 %
                \hatcurPPmelongeccenxxxxxC
                \else
                \ifnum#1=28 %
                \hatcurPPmelongeccenxxxxxD
                \else
                \ifnum#1=29 %
                \hatcurPPmelongeccenxxxxxE
                \else
                \ifnum#1=30 %
                \hatcurPPmelongeccenxxxxxF
                \else
                ??????\fi
                \fi
                \fi
                \fi
                \fi
                \fi
}
\newcommand{\hatcurPPmeshorteccen}[1]{\ifnum#1=25 %
        \hatcurPPmeshorteccenxxxxxA
                \else
                \ifnum#1=26 %
                \hatcurPPmeshorteccenxxxxxB
                \else
                \ifnum#1=27 %
                \hatcurPPmeshorteccenxxxxxC
                \else
                \ifnum#1=28 %
                \hatcurPPmeshorteccenxxxxxD
                \else
                \ifnum#1=29 %
                \hatcurPPmeshorteccenxxxxxE
                \else
                \ifnum#1=30 %
                \hatcurPPmeshorteccenxxxxxF
                \else
                ??????\fi
                \fi
                \fi
                \fi
                \fi
                \fi
}
\newcommand{\hatcurPPmlongeccen}[1]{\ifnum#1=25 %
        \hatcurPPmlongeccenxxxxxA
                \else
                \ifnum#1=26 %
                \hatcurPPmlongeccenxxxxxB
                \else
                \ifnum#1=27 %
                \hatcurPPmlongeccenxxxxxC
                \else
                \ifnum#1=28 %
                \hatcurPPmlongeccenxxxxxD
                \else
                \ifnum#1=29 %
                \hatcurPPmlongeccenxxxxxE
                \else
                \ifnum#1=30 %
                \hatcurPPmlongeccenxxxxxF
                \else
                ??????\fi
                \fi
                \fi
                \fi
                \fi
                \fi
}
\newcommand{\hatcurPPmrcorreccen}[1]{\ifnum#1=25 %
        \hatcurPPmrcorreccenxxxxxA
                \else
                \ifnum#1=26 %
                \hatcurPPmrcorreccenxxxxxB
                \else
                \ifnum#1=27 %
                \hatcurPPmrcorreccenxxxxxC
                \else
                \ifnum#1=28 %
                \hatcurPPmrcorreccenxxxxxD
                \else
                \ifnum#1=29 %
                \hatcurPPmrcorreccenxxxxxE
                \else
                \ifnum#1=30 %
                \hatcurPPmrcorreccenxxxxxF
                \else
                ??????\fi
                \fi
                \fi
                \fi
                \fi
                \fi
}
\newcommand{\hatcurPPmshorteccen}[1]{\ifnum#1=25 %
        \hatcurPPmshorteccenxxxxxA
                \else
                \ifnum#1=26 %
                \hatcurPPmshorteccenxxxxxB
                \else
                \ifnum#1=27 %
                \hatcurPPmshorteccenxxxxxC
                \else
                \ifnum#1=28 %
                \hatcurPPmshorteccenxxxxxD
                \else
                \ifnum#1=29 %
                \hatcurPPmshorteccenxxxxxE
                \else
                \ifnum#1=30 %
                \hatcurPPmshorteccenxxxxxF
                \else
                ??????\fi
                \fi
                \fi
                \fi
                \fi
                \fi
}
\newcommand{\hatcurPPperieccen}[1]{\ifnum#1=25 %
        \hatcurPPperieccenxxxxxA
                \else
                \ifnum#1=26 %
                \hatcurPPperieccenxxxxxB
                \else
                \ifnum#1=27 %
                \hatcurPPperieccenxxxxxC
                \else
                \ifnum#1=28 %
                \hatcurPPperieccenxxxxxD
                \else
                \ifnum#1=29 %
                \hatcurPPperieccenxxxxxE
                \else
                \ifnum#1=30 %
                \hatcurPPperieccenxxxxxF
                \else
                ??????\fi
                \fi
                \fi
                \fi
                \fi
                \fi
}
\newcommand{\hatcurPPphiconjeccen}[1]{\ifnum#1=25 %
        \hatcurPPphiconjeccenxxxxxA
                \else
                \ifnum#1=26 %
                \hatcurPPphiconjeccenxxxxxB
                \else
                \ifnum#1=27 %
                \hatcurPPphiconjeccenxxxxxC
                \else
                \ifnum#1=28 %
                \hatcurPPphiconjeccenxxxxxD
                \else
                \ifnum#1=29 %
                \hatcurPPphiconjeccenxxxxxE
                \else
                \ifnum#1=30 %
                \hatcurPPphiconjeccenxxxxxF
                \else
                ??????\fi
                \fi
                \fi
                \fi
                \fi
                \fi
}
\newcommand{\hatcurPPreccen}[1]{\ifnum#1=25 %
        \hatcurPPreccenxxxxxA
                \else
                \ifnum#1=26 %
                \hatcurPPreccenxxxxxB
                \else
                \ifnum#1=27 %
                \hatcurPPreccenxxxxxC
                \else
                \ifnum#1=28 %
                \hatcurPPreccenxxxxxD
                \else
                \ifnum#1=29 %
                \hatcurPPreccenxxxxxE
                \else
                \ifnum#1=30 %
                \hatcurPPreccenxxxxxF
                \else
                ??????\fi
                \fi
                \fi
                \fi
                \fi
                \fi
}
\newcommand{\hatcurPPreeccen}[1]{\ifnum#1=25 %
        \hatcurPPreeccenxxxxxA
                \else
                \ifnum#1=26 %
                \hatcurPPreeccenxxxxxB
                \else
                \ifnum#1=27 %
                \hatcurPPreeccenxxxxxC
                \else
                \ifnum#1=28 %
                \hatcurPPreeccenxxxxxD
                \else
                \ifnum#1=29 %
                \hatcurPPreeccenxxxxxE
                \else
                \ifnum#1=30 %
                \hatcurPPreeccenxxxxxF
                \else
                ??????\fi
                \fi
                \fi
                \fi
                \fi
                \fi
}
\newcommand{\hatcurPPrelongeccen}[1]{\ifnum#1=25 %
        \hatcurPPrelongeccenxxxxxA
                \else
                \ifnum#1=26 %
                \hatcurPPrelongeccenxxxxxB
                \else
                \ifnum#1=27 %
                \hatcurPPrelongeccenxxxxxC
                \else
                \ifnum#1=28 %
                \hatcurPPrelongeccenxxxxxD
                \else
                \ifnum#1=29 %
                \hatcurPPrelongeccenxxxxxE
                \else
                \ifnum#1=30 %
                \hatcurPPrelongeccenxxxxxF
                \else
                ??????\fi
                \fi
                \fi
                \fi
                \fi
                \fi
}
\newcommand{\hatcurPPreshorteccen}[1]{\ifnum#1=25 %
        \hatcurPPreshorteccenxxxxxA
                \else
                \ifnum#1=26 %
                \hatcurPPreshorteccenxxxxxB
                \else
                \ifnum#1=27 %
                \hatcurPPreshorteccenxxxxxC
                \else
                \ifnum#1=28 %
                \hatcurPPreshorteccenxxxxxD
                \else
                \ifnum#1=29 %
                \hatcurPPreshorteccenxxxxxE
                \else
                \ifnum#1=30 %
                \hatcurPPreshorteccenxxxxxF
                \else
                ??????\fi
                \fi
                \fi
                \fi
                \fi
                \fi
}
\newcommand{\hatcurPPrhoeccen}[1]{\ifnum#1=25 %
        \hatcurPPrhoeccenxxxxxA
                \else
                \ifnum#1=26 %
                \hatcurPPrhoeccenxxxxxB
                \else
                \ifnum#1=27 %
                \hatcurPPrhoeccenxxxxxC
                \else
                \ifnum#1=28 %
                \hatcurPPrhoeccenxxxxxD
                \else
                \ifnum#1=29 %
                \hatcurPPrhoeccenxxxxxE
                \else
                \ifnum#1=30 %
                \hatcurPPrhoeccenxxxxxF
                \else
                ??????\fi
                \fi
                \fi
                \fi
                \fi
                \fi
}
\newcommand{\hatcurPPrlongeccen}[1]{\ifnum#1=25 %
        \hatcurPPrlongeccenxxxxxA
                \else
                \ifnum#1=26 %
                \hatcurPPrlongeccenxxxxxB
                \else
                \ifnum#1=27 %
                \hatcurPPrlongeccenxxxxxC
                \else
                \ifnum#1=28 %
                \hatcurPPrlongeccenxxxxxD
                \else
                \ifnum#1=29 %
                \hatcurPPrlongeccenxxxxxE
                \else
                \ifnum#1=30 %
                \hatcurPPrlongeccenxxxxxF
                \else
                ??????\fi
                \fi
                \fi
                \fi
                \fi
                \fi
}
\newcommand{\hatcurPPrshorteccen}[1]{\ifnum#1=25 %
        \hatcurPPrshorteccenxxxxxA
                \else
                \ifnum#1=26 %
                \hatcurPPrshorteccenxxxxxB
                \else
                \ifnum#1=27 %
                \hatcurPPrshorteccenxxxxxC
                \else
                \ifnum#1=28 %
                \hatcurPPrshorteccenxxxxxD
                \else
                \ifnum#1=29 %
                \hatcurPPrshorteccenxxxxxE
                \else
                \ifnum#1=30 %
                \hatcurPPrshorteccenxxxxxF
                \else
                ??????\fi
                \fi
                \fi
                \fi
                \fi
                \fi
}
\newcommand{\hatcurPPtcirceccen}[1]{\ifnum#1=25 %
        \hatcurPPtcirceccenxxxxxA
                \else
                \ifnum#1=26 %
                \hatcurPPtcirceccenxxxxxB
                \else
                \ifnum#1=27 %
                \hatcurPPtcirceccenxxxxxC
                \else
                \ifnum#1=28 %
                \hatcurPPtcirceccenxxxxxD
                \else
                \ifnum#1=29 %
                \hatcurPPtcirceccenxxxxxE
                \else
                \ifnum#1=30 %
                \hatcurPPtcirceccenxxxxxF
                \else
                ??????\fi
                \fi
                \fi
                \fi
                \fi
                \fi
}
\newcommand{\hatcurPPteffeccen}[1]{\ifnum#1=25 %
        \hatcurPPteffeccenxxxxxA
                \else
                \ifnum#1=26 %
                \hatcurPPteffeccenxxxxxB
                \else
                \ifnum#1=27 %
                \hatcurPPteffeccenxxxxxC
                \else
                \ifnum#1=28 %
                \hatcurPPteffeccenxxxxxD
                \else
                \ifnum#1=29 %
                \hatcurPPteffeccenxxxxxE
                \else
                \ifnum#1=30 %
                \hatcurPPteffeccenxxxxxF
                \else
                ??????\fi
                \fi
                \fi
                \fi
                \fi
                \fi
}
\newcommand{\hatcurPPthetaeccen}[1]{\ifnum#1=25 %
        \hatcurPPthetaeccenxxxxxA
                \else
                \ifnum#1=26 %
                \hatcurPPthetaeccenxxxxxB
                \else
                \ifnum#1=27 %
                \hatcurPPthetaeccenxxxxxC
                \else
                \ifnum#1=28 %
                \hatcurPPthetaeccenxxxxxD
                \else
                \ifnum#1=29 %
                \hatcurPPthetaeccenxxxxxE
                \else
                \ifnum#1=30 %
                \hatcurPPthetaeccenxxxxxF
                \else
                ??????\fi
                \fi
                \fi
                \fi
                \fi
                \fi
}
\newcommand{\hatcurPPtinfalleccen}[1]{\ifnum#1=25 %
        \hatcurPPtinfalleccenxxxxxA
                \else
                \ifnum#1=26 %
                \hatcurPPtinfalleccenxxxxxB
                \else
                \ifnum#1=27 %
                \hatcurPPtinfalleccenxxxxxC
                \else
                \ifnum#1=28 %
                \hatcurPPtinfalleccenxxxxxD
                \else
                \ifnum#1=29 %
                \hatcurPPtinfalleccenxxxxxE
                \else
                \ifnum#1=30 %
                \hatcurPPtinfalleccenxxxxxF
                \else
                ??????\fi
                \fi
                \fi
                \fi
                \fi
                \fi
}
\newcommand{\hatcurRVecceneccen}[1]{\ifnum#1=25 %
        \hatcurRVecceneccenxxxxxA
                \else
                \ifnum#1=26 %
                \hatcurRVecceneccenxxxxxB
                \else
                \ifnum#1=27 %
                \hatcurRVecceneccenxxxxxC
                \else
                \ifnum#1=28 %
                \hatcurRVecceneccenxxxxxD
                \else
                \ifnum#1=29 %
                \hatcurRVecceneccenxxxxxE
                \else
                \ifnum#1=30 %
                \hatcurRVecceneccenxxxxxF
                \else
                ??????\fi
                \fi
                \fi
                \fi
                \fi
                \fi
}
\newcommand{\hatcurRVeccentwosiglimeccen}[1]{\ifnum#1=25 %
        \hatcurRVeccentwosiglimeccenxxxxxA
                \else
                \ifnum#1=26 %
                \hatcurRVeccentwosiglimeccenxxxxxB
                \else
                \ifnum#1=27 %
                \hatcurRVeccentwosiglimeccenxxxxxC
                \else
                \ifnum#1=28 %
                \hatcurRVeccentwosiglimeccenxxxxxD
                \else
                \ifnum#1=29 %
                \hatcurRVeccentwosiglimeccenxxxxxE
                \else
                \ifnum#1=30 %
                \hatcurRVeccentwosiglimeccenxxxxxF
                \else
                ??????\fi
                \fi
                \fi
                \fi
                \fi
                \fi
}
\newcommand{\hatcurRVfitrmsAeccen}[1]{\ifnum#1=27 %
        \hatcurRVfitrmsAeccenxxxxxC
                \else
                \ifnum#1=29 %
                \hatcurRVfitrmsAeccenxxxxxE
                \else
                \ifnum#1=30 %
                \hatcurRVfitrmsAeccenxxxxxF
                \else
                ??????\fi
                \fi
                \fi
}
\newcommand{\hatcurRVfitrmsBeccen}[1]{\ifnum#1=27 %
        \hatcurRVfitrmsBeccenxxxxxC
                \else
                \ifnum#1=29 %
                \hatcurRVfitrmsBeccenxxxxxE
                \else
                \ifnum#1=30 %
                \hatcurRVfitrmsBeccenxxxxxF
                \else
                ??????\fi
                \fi
                \fi
}
\newcommand{\hatcurRVfitrmsCeccen}[1]{\ifnum#1=27 %
        \hatcurRVfitrmsCeccenxxxxxC
                \else
                \ifnum#1=29 %
                \hatcurRVfitrmsCeccenxxxxxE
                \else
                ??????\fi
                \fi
}
\newcommand{\hatcurRVfitrmseccen}[1]{\ifnum#1=25 %
        \hatcurRVfitrmseccenxxxxxA
                \else
                \ifnum#1=26 %
                \hatcurRVfitrmseccenxxxxxB
                \else
                \ifnum#1=28 %
                \hatcurRVfitrmseccenxxxxxD
                \else
                ??????\fi
                \fi
                \fi
}
\newcommand{\hatcurRVgammaAeccen}[1]{\ifnum#1=27 %
        \hatcurRVgammaAeccenxxxxxC
                \else
                \ifnum#1=29 %
                \hatcurRVgammaAeccenxxxxxE
                \else
                \ifnum#1=30 %
                \hatcurRVgammaAeccenxxxxxF
                \else
                ??????\fi
                \fi
                \fi
}
\newcommand{\hatcurRVgammaBeccen}[1]{\ifnum#1=27 %
        \hatcurRVgammaBeccenxxxxxC
                \else
                \ifnum#1=29 %
                \hatcurRVgammaBeccenxxxxxE
                \else
                \ifnum#1=30 %
                \hatcurRVgammaBeccenxxxxxF
                \else
                ??????\fi
                \fi
                \fi
}
\newcommand{\hatcurRVgammaCeccen}[1]{\ifnum#1=27 %
        \hatcurRVgammaCeccenxxxxxC
                \else
                \ifnum#1=29 %
                \hatcurRVgammaCeccenxxxxxE
                \else
                ??????\fi
                \fi
}
\newcommand{\hatcurRVgammaeccen}[1]{\ifnum#1=25 %
        \hatcurRVgammaeccenxxxxxA
                \else
                \ifnum#1=26 %
                \hatcurRVgammaeccenxxxxxB
                \else
                \ifnum#1=28 %
                \hatcurRVgammaeccenxxxxxD
                \else
                ??????\fi
                \fi
                \fi
}
\newcommand{\hatcurRVheccen}[1]{\ifnum#1=25 %
        \hatcurRVheccenxxxxxA
                \else
                \ifnum#1=26 %
                \hatcurRVheccenxxxxxB
                \else
                \ifnum#1=27 %
                \hatcurRVheccenxxxxxC
                \else
                \ifnum#1=28 %
                \hatcurRVheccenxxxxxD
                \else
                \ifnum#1=29 %
                \hatcurRVheccenxxxxxE
                \else
                \ifnum#1=30 %
                \hatcurRVheccenxxxxxF
                \else
                ??????\fi
                \fi
                \fi
                \fi
                \fi
                \fi
}
\newcommand{\hatcurRVjitterAeccen}[1]{\ifnum#1=27 %
        \hatcurRVjitterAeccenxxxxxC
                \else
                \ifnum#1=29 %
                \hatcurRVjitterAeccenxxxxxE
                \else
                \ifnum#1=30 %
                \hatcurRVjitterAeccenxxxxxF
                \else
                ??????\fi
                \fi
                \fi
}
\newcommand{\hatcurRVjitterBeccen}[1]{\ifnum#1=27 %
        \hatcurRVjitterBeccenxxxxxC
                \else
                \ifnum#1=29 %
                \hatcurRVjitterBeccenxxxxxE
                \else
                \ifnum#1=30 %
                \hatcurRVjitterBeccenxxxxxF
                \else
                ??????\fi
                \fi
                \fi
}
\newcommand{\hatcurRVjitterCeccen}[1]{\ifnum#1=27 %
        \hatcurRVjitterCeccenxxxxxC
                \else
                \ifnum#1=29 %
                \hatcurRVjitterCeccenxxxxxE
                \else
                ??????\fi
                \fi
}
\newcommand{\hatcurRVjittereccen}[1]{\ifnum#1=25 %
        \hatcurRVjittereccenxxxxxA
                \else
                \ifnum#1=26 %
                \hatcurRVjittereccenxxxxxB
                \else
                \ifnum#1=28 %
                \hatcurRVjittereccenxxxxxD
                \else
                ??????\fi
                \fi
                \fi
}
\newcommand{\hatcurRVjittertwosiglimAeccen}[1]{\ifnum#1=27 %
        \hatcurRVjittertwosiglimAeccenxxxxxC
                \else
                \ifnum#1=29 %
                \hatcurRVjittertwosiglimAeccenxxxxxE
                \else
                \ifnum#1=30 %
                \hatcurRVjittertwosiglimAeccenxxxxxF
                \else
                ??????\fi
                \fi
                \fi
}
\newcommand{\hatcurRVjittertwosiglimBeccen}[1]{\ifnum#1=27 %
        \hatcurRVjittertwosiglimBeccenxxxxxC
                \else
                \ifnum#1=29 %
                \hatcurRVjittertwosiglimBeccenxxxxxE
                \else
                \ifnum#1=30 %
                \hatcurRVjittertwosiglimBeccenxxxxxF
                \else
                ??????\fi
                \fi
                \fi
}
\newcommand{\hatcurRVjittertwosiglimCeccen}[1]{\ifnum#1=27 %
        \hatcurRVjittertwosiglimCeccenxxxxxC
                \else
                \ifnum#1=29 %
                \hatcurRVjittertwosiglimCeccenxxxxxE
                \else
                ??????\fi
                \fi
}
\newcommand{\hatcurRVjittertwosiglimeccen}[1]{\ifnum#1=25 %
        \hatcurRVjittertwosiglimeccenxxxxxA
                \else
                \ifnum#1=26 %
                \hatcurRVjittertwosiglimeccenxxxxxB
                \else
                \ifnum#1=28 %
                \hatcurRVjittertwosiglimeccenxxxxxD
                \else
                ??????\fi
                \fi
                \fi
}
\newcommand{\hatcurRVkeccen}[1]{\ifnum#1=25 %
        \hatcurRVkeccenxxxxxA
                \else
                \ifnum#1=26 %
                \hatcurRVkeccenxxxxxB
                \else
                \ifnum#1=27 %
                \hatcurRVkeccenxxxxxC
                \else
                \ifnum#1=28 %
                \hatcurRVkeccenxxxxxD
                \else
                \ifnum#1=29 %
                \hatcurRVkeccenxxxxxE
                \else
                \ifnum#1=30 %
                \hatcurRVkeccenxxxxxF
                \else
                ??????\fi
                \fi
                \fi
                \fi
                \fi
                \fi
}
\newcommand{\hatcurRVKeccen}[1]{\ifnum#1=25 %
        \hatcurRVKeccenxxxxxA
                \else
                \ifnum#1=26 %
                \hatcurRVKeccenxxxxxB
                \else
                \ifnum#1=27 %
                \hatcurRVKeccenxxxxxC
                \else
                \ifnum#1=28 %
                \hatcurRVKeccenxxxxxD
                \else
                \ifnum#1=29 %
                \hatcurRVKeccenxxxxxE
                \else
                \ifnum#1=30 %
                \hatcurRVKeccenxxxxxF
                \else
                ??????\fi
                \fi
                \fi
                \fi
                \fi
                \fi
}
\newcommand{\hatcurRVomegaeccen}[1]{\ifnum#1=25 %
        \hatcurRVomegaeccenxxxxxA
                \else
                \ifnum#1=26 %
                \hatcurRVomegaeccenxxxxxB
                \else
                \ifnum#1=27 %
                \hatcurRVomegaeccenxxxxxC
                \else
                \ifnum#1=28 %
                \hatcurRVomegaeccenxxxxxD
                \else
                \ifnum#1=29 %
                \hatcurRVomegaeccenxxxxxE
                \else
                \ifnum#1=30 %
                \hatcurRVomegaeccenxxxxxF
                \else
                ??????\fi
                \fi
                \fi
                \fi
                \fi
                \fi
}
\newcommand{\hatcurRVrheccen}[1]{\ifnum#1=25 %
        \hatcurRVrheccenxxxxxA
                \else
                \ifnum#1=26 %
                \hatcurRVrheccenxxxxxB
                \else
                \ifnum#1=27 %
                \hatcurRVrheccenxxxxxC
                \else
                \ifnum#1=28 %
                \hatcurRVrheccenxxxxxD
                \else
                \ifnum#1=29 %
                \hatcurRVrheccenxxxxxE
                \else
                \ifnum#1=30 %
                \hatcurRVrheccenxxxxxF
                \else
                ??????\fi
                \fi
                \fi
                \fi
                \fi
                \fi
}
\newcommand{\hatcurRVrkeccen}[1]{\ifnum#1=25 %
        \hatcurRVrkeccenxxxxxA
                \else
                \ifnum#1=26 %
                \hatcurRVrkeccenxxxxxB
                \else
                \ifnum#1=27 %
                \hatcurRVrkeccenxxxxxC
                \else
                \ifnum#1=28 %
                \hatcurRVrkeccenxxxxxD
                \else
                \ifnum#1=29 %
                \hatcurRVrkeccenxxxxxE
                \else
                \ifnum#1=30 %
                \hatcurRVrkeccenxxxxxF
                \else
                ??????\fi
                \fi
                \fi
                \fi
                \fi
                \fi
}
\newcommand{\hatcurRVtroneeccen}[1]{\ifnum#1=25 %
        \hatcurRVtroneeccenxxxxxA
                \else
                \ifnum#1=26 %
                \hatcurRVtroneeccenxxxxxB
                \else
                \ifnum#1=27 %
                \hatcurRVtroneeccenxxxxxC
                \else
                \ifnum#1=28 %
                \hatcurRVtroneeccenxxxxxD
                \else
                \ifnum#1=29 %
                \hatcurRVtroneeccenxxxxxE
                \else
                \ifnum#1=30 %
                \hatcurRVtroneeccenxxxxxF
                \else
                ??????\fi
                \fi
                \fi
                \fi
                \fi
                \fi
}
\newcommand{\hatcurRVtrtwoeccen}[1]{\ifnum#1=25 %
        \hatcurRVtrtwoeccenxxxxxA
                \else
                \ifnum#1=26 %
                \hatcurRVtrtwoeccenxxxxxB
                \else
                \ifnum#1=27 %
                \hatcurRVtrtwoeccenxxxxxC
                \else
                \ifnum#1=28 %
                \hatcurRVtrtwoeccenxxxxxD
                \else
                \ifnum#1=29 %
                \hatcurRVtrtwoeccenxxxxxE
                \else
                \ifnum#1=30 %
                \hatcurRVtrtwoeccenxxxxxF
                \else
                ??????\fi
                \fi
                \fi
                \fi
                \fi
                \fi
}
\newcommand{\hatcurSMEiiloggeccen}[1]{\ifnum#1=25 %
        \hatcurSMEiiloggeccenxxxxxA
                \else
                \ifnum#1=26 %
                \hatcurSMEiiloggeccenxxxxxB
                \else
                \ifnum#1=28 %
                \hatcurSMEiiloggeccenxxxxxD
                \else
                \ifnum#1=29 %
                \hatcurSMEiiloggeccenxxxxxE
                \else
                \ifnum#1=30 %
                \hatcurSMEiiloggeccenxxxxxF
                \else
                ??????\fi
                \fi
                \fi
                \fi
                \fi
}
\newcommand{\hatcurSMEiiteffeccen}[1]{\ifnum#1=25 %
        \hatcurSMEiiteffeccenxxxxxA
                \else
                \ifnum#1=26 %
                \hatcurSMEiiteffeccenxxxxxB
                \else
                \ifnum#1=28 %
                \hatcurSMEiiteffeccenxxxxxD
                \else
                \ifnum#1=29 %
                \hatcurSMEiiteffeccenxxxxxE
                \else
                \ifnum#1=30 %
                \hatcurSMEiiteffeccenxxxxxF
                \else
                ??????\fi
                \fi
                \fi
                \fi
                \fi
}
\newcommand{\hatcurSMEiivsineccen}[1]{\ifnum#1=25 %
        \hatcurSMEiivsineccenxxxxxA
                \else
                \ifnum#1=26 %
                \hatcurSMEiivsineccenxxxxxB
                \else
                \ifnum#1=28 %
                \hatcurSMEiivsineccenxxxxxD
                \else
                \ifnum#1=29 %
                \hatcurSMEiivsineccenxxxxxE
                \else
                \ifnum#1=30 %
                \hatcurSMEiivsineccenxxxxxF
                \else
                ??????\fi
                \fi
                \fi
                \fi
                \fi
}
\newcommand{\hatcurSMEiizfeheccen}[1]{\ifnum#1=25 %
        \hatcurSMEiizfeheccenxxxxxA
                \else
                \ifnum#1=26 %
                \hatcurSMEiizfeheccenxxxxxB
                \else
                \ifnum#1=28 %
                \hatcurSMEiizfeheccenxxxxxD
                \else
                \ifnum#1=29 %
                \hatcurSMEiizfeheccenxxxxxE
                \else
                \ifnum#1=30 %
                \hatcurSMEiizfeheccenxxxxxF
                \else
                ??????\fi
                \fi
                \fi
                \fi
                \fi
}
\newcommand{\hatcurSMEiizfehshorteccen}[1]{\ifnum#1=25 %
        \hatcurSMEiizfehshorteccenxxxxxA
                \else
                \ifnum#1=26 %
                \hatcurSMEiizfehshorteccenxxxxxB
                \else
                \ifnum#1=28 %
                \hatcurSMEiizfehshorteccenxxxxxD
                \else
                \ifnum#1=29 %
                \hatcurSMEiizfehshorteccenxxxxxE
                \else
                \ifnum#1=30 %
                \hatcurSMEiizfehshorteccenxxxxxF
                \else
                ??????\fi
                \fi
                \fi
                \fi
                \fi
}
\newcommand{\hatcurSMEiloggeccen}[1]{\ifnum#1=25 %
        \hatcurSMEiloggeccenxxxxxA
                \else
                \ifnum#1=26 %
                \hatcurSMEiloggeccenxxxxxB
                \else
                \ifnum#1=27 %
                \hatcurSMEiloggeccenxxxxxC
                \else
                \ifnum#1=28 %
                \hatcurSMEiloggeccenxxxxxD
                \else
                \ifnum#1=29 %
                \hatcurSMEiloggeccenxxxxxE
                \else
                \ifnum#1=30 %
                \hatcurSMEiloggeccenxxxxxF
                \else
                ??????\fi
                \fi
                \fi
                \fi
                \fi
                \fi
}
\newcommand{\hatcurSMEiteffeccen}[1]{\ifnum#1=25 %
        \hatcurSMEiteffeccenxxxxxA
                \else
                \ifnum#1=26 %
                \hatcurSMEiteffeccenxxxxxB
                \else
                \ifnum#1=27 %
                \hatcurSMEiteffeccenxxxxxC
                \else
                \ifnum#1=28 %
                \hatcurSMEiteffeccenxxxxxD
                \else
                \ifnum#1=29 %
                \hatcurSMEiteffeccenxxxxxE
                \else
                \ifnum#1=30 %
                \hatcurSMEiteffeccenxxxxxF
                \else
                ??????\fi
                \fi
                \fi
                \fi
                \fi
                \fi
}
\newcommand{\hatcurSMEivmaceccen}[1]{\ifnum#1=25 %
        \hatcurSMEivmaceccenxxxxxA
                \else
                \ifnum#1=26 %
                \hatcurSMEivmaceccenxxxxxB
                \else
                \ifnum#1=27 %
                \hatcurSMEivmaceccenxxxxxC
                \else
                \ifnum#1=28 %
                \hatcurSMEivmaceccenxxxxxD
                \else
                \ifnum#1=29 %
                \hatcurSMEivmaceccenxxxxxE
                \else
                \ifnum#1=30 %
                \hatcurSMEivmaceccenxxxxxF
                \else
                ??????\fi
                \fi
                \fi
                \fi
                \fi
                \fi
}
\newcommand{\hatcurSMEivmiceccen}[1]{\ifnum#1=25 %
        \hatcurSMEivmiceccenxxxxxA
                \else
                \ifnum#1=26 %
                \hatcurSMEivmiceccenxxxxxB
                \else
                \ifnum#1=27 %
                \hatcurSMEivmiceccenxxxxxC
                \else
                \ifnum#1=28 %
                \hatcurSMEivmiceccenxxxxxD
                \else
                \ifnum#1=29 %
                \hatcurSMEivmiceccenxxxxxE
                \else
                \ifnum#1=30 %
                \hatcurSMEivmiceccenxxxxxF
                \else
                ??????\fi
                \fi
                \fi
                \fi
                \fi
                \fi
}
\newcommand{\hatcurSMEivsineccen}[1]{\ifnum#1=25 %
        \hatcurSMEivsineccenxxxxxA
                \else
                \ifnum#1=26 %
                \hatcurSMEivsineccenxxxxxB
                \else
                \ifnum#1=27 %
                \hatcurSMEivsineccenxxxxxC
                \else
                \ifnum#1=28 %
                \hatcurSMEivsineccenxxxxxD
                \else
                \ifnum#1=29 %
                \hatcurSMEivsineccenxxxxxE
                \else
                \ifnum#1=30 %
                \hatcurSMEivsineccenxxxxxF
                \else
                ??????\fi
                \fi
                \fi
                \fi
                \fi
                \fi
}
\newcommand{\hatcurSMEizfeheccen}[1]{\ifnum#1=25 %
        \hatcurSMEizfeheccenxxxxxA
                \else
                \ifnum#1=26 %
                \hatcurSMEizfeheccenxxxxxB
                \else
                \ifnum#1=27 %
                \hatcurSMEizfeheccenxxxxxC
                \else
                \ifnum#1=28 %
                \hatcurSMEizfeheccenxxxxxD
                \else
                \ifnum#1=29 %
                \hatcurSMEizfeheccenxxxxxE
                \else
                \ifnum#1=30 %
                \hatcurSMEizfeheccenxxxxxF
                \else
                ??????\fi
                \fi
                \fi
                \fi
                \fi
                \fi
}
\newcommand{\hatcurSMEizfehshorteccen}[1]{\ifnum#1=25 %
        \hatcurSMEizfehshorteccenxxxxxA
                \else
                \ifnum#1=26 %
                \hatcurSMEizfehshorteccenxxxxxB
                \else
                \ifnum#1=27 %
                \hatcurSMEizfehshorteccenxxxxxC
                \else
                \ifnum#1=28 %
                \hatcurSMEizfehshorteccenxxxxxD
                \else
                \ifnum#1=29 %
                \hatcurSMEizfehshorteccenxxxxxE
                \else
                \ifnum#1=30 %
                \hatcurSMEizfehshorteccenxxxxxF
                \else
                ??????\fi
                \fi
                \fi
                \fi
                \fi
                \fi
}
\newcommand{\hatcurXAveccen}[1]{\ifnum#1=25 %
        \hatcurXAveccenxxxxxA
                \else
                \ifnum#1=26 %
                \hatcurXAveccenxxxxxB
                \else
                \ifnum#1=27 %
                \hatcurXAveccenxxxxxC
                \else
                \ifnum#1=28 %
                \hatcurXAveccenxxxxxD
                \else
                \ifnum#1=29 %
                \hatcurXAveccenxxxxxE
                \else
                \ifnum#1=30 %
                \hatcurXAveccenxxxxxF
                \else
                ??????\fi
                \fi
                \fi
                \fi
                \fi
                \fi
}
\newcommand{\hatcurXdisteccen}[1]{\ifnum#1=25 %
        \hatcurXdisteccenxxxxxA
                \else
                \ifnum#1=26 %
                \hatcurXdisteccenxxxxxB
                \else
                \ifnum#1=27 %
                \hatcurXdisteccenxxxxxC
                \else
                \ifnum#1=28 %
                \hatcurXdisteccenxxxxxD
                \else
                \ifnum#1=29 %
                \hatcurXdisteccenxxxxxE
                \else
                \ifnum#1=30 %
                \hatcurXdisteccenxxxxxF
                \else
                ??????\fi
                \fi
                \fi
                \fi
                \fi
                \fi
}
\newcommand{\hatcurXdistredeccen}[1]{\ifnum#1=25 %
        \hatcurXdistredeccenxxxxxA
                \else
                \ifnum#1=26 %
                \hatcurXdistredeccenxxxxxB
                \else
                \ifnum#1=27 %
                \hatcurXdistredeccenxxxxxC
                \else
                \ifnum#1=28 %
                \hatcurXdistredeccenxxxxxD
                \else
                \ifnum#1=29 %
                \hatcurXdistredeccenxxxxxE
                \else
                \ifnum#1=30 %
                \hatcurXdistredeccenxxxxxF
                \else
                ??????\fi
                \fi
                \fi
                \fi
                \fi
                \fi
}
\newcommand{\hatcurXEBVeccen}[1]{\ifnum#1=25 %
        \hatcurXEBVeccenxxxxxA
                \else
                \ifnum#1=26 %
                \hatcurXEBVeccenxxxxxB
                \else
                \ifnum#1=27 %
                \hatcurXEBVeccenxxxxxC
                \else
                \ifnum#1=28 %
                \hatcurXEBVeccenxxxxxD
                \else
                \ifnum#1=29 %
                \hatcurXEBVeccenxxxxxE
                \else
                \ifnum#1=30 %
                \hatcurXEBVeccenxxxxxF
                \else
                ??????\fi
                \fi
                \fi
                \fi
                \fi
                \fi
}
\newcommand{\hatcurXjhisoredeccen}[1]{\ifnum#1=25 %
        \hatcurXjhisoredeccenxxxxxA
                \else
                \ifnum#1=26 %
                \hatcurXjhisoredeccenxxxxxB
                \else
                \ifnum#1=27 %
                \hatcurXjhisoredeccenxxxxxC
                \else
                \ifnum#1=28 %
                \hatcurXjhisoredeccenxxxxxD
                \else
                \ifnum#1=29 %
                \hatcurXjhisoredeccenxxxxxE
                \else
                \ifnum#1=30 %
                \hatcurXjhisoredeccenxxxxxF
                \else
                ??????\fi
                \fi
                \fi
                \fi
                \fi
                \fi
}
\newcommand{\hatcurXjkisoredeccen}[1]{\ifnum#1=25 %
        \hatcurXjkisoredeccenxxxxxA
                \else
                \ifnum#1=26 %
                \hatcurXjkisoredeccenxxxxxB
                \else
                \ifnum#1=27 %
                \hatcurXjkisoredeccenxxxxxC
                \else
                \ifnum#1=28 %
                \hatcurXjkisoredeccenxxxxxD
                \else
                \ifnum#1=29 %
                \hatcurXjkisoredeccenxxxxxE
                \else
                \ifnum#1=30 %
                \hatcurXjkisoredeccenxxxxxF
                \else
                ??????\fi
                \fi
                \fi
                \fi
                \fi
                \fi
}
\newcommand{\hatcurXmhisoredeccen}[1]{\ifnum#1=25 %
        \hatcurXmhisoredeccenxxxxxA
                \else
                \ifnum#1=26 %
                \hatcurXmhisoredeccenxxxxxB
                \else
                \ifnum#1=27 %
                \hatcurXmhisoredeccenxxxxxC
                \else
                \ifnum#1=28 %
                \hatcurXmhisoredeccenxxxxxD
                \else
                \ifnum#1=29 %
                \hatcurXmhisoredeccenxxxxxE
                \else
                \ifnum#1=30 %
                \hatcurXmhisoredeccenxxxxxF
                \else
                ??????\fi
                \fi
                \fi
                \fi
                \fi
                \fi
}
\newcommand{\hatcurXmiisoredeccen}[1]{\ifnum#1=25 %
        \hatcurXmiisoredeccenxxxxxA
                \else
                \ifnum#1=26 %
                \hatcurXmiisoredeccenxxxxxB
                \else
                \ifnum#1=27 %
                \hatcurXmiisoredeccenxxxxxC
                \else
                \ifnum#1=28 %
                \hatcurXmiisoredeccenxxxxxD
                \else
                \ifnum#1=29 %
                \hatcurXmiisoredeccenxxxxxE
                \else
                \ifnum#1=30 %
                \hatcurXmiisoredeccenxxxxxF
                \else
                ??????\fi
                \fi
                \fi
                \fi
                \fi
                \fi
}
\newcommand{\hatcurXmjisoredeccen}[1]{\ifnum#1=25 %
        \hatcurXmjisoredeccenxxxxxA
                \else
                \ifnum#1=26 %
                \hatcurXmjisoredeccenxxxxxB
                \else
                \ifnum#1=27 %
                \hatcurXmjisoredeccenxxxxxC
                \else
                \ifnum#1=28 %
                \hatcurXmjisoredeccenxxxxxD
                \else
                \ifnum#1=29 %
                \hatcurXmjisoredeccenxxxxxE
                \else
                \ifnum#1=30 %
                \hatcurXmjisoredeccenxxxxxF
                \else
                ??????\fi
                \fi
                \fi
                \fi
                \fi
                \fi
}
\newcommand{\hatcurXmkisoredeccen}[1]{\ifnum#1=25 %
        \hatcurXmkisoredeccenxxxxxA
                \else
                \ifnum#1=26 %
                \hatcurXmkisoredeccenxxxxxB
                \else
                \ifnum#1=27 %
                \hatcurXmkisoredeccenxxxxxC
                \else
                \ifnum#1=28 %
                \hatcurXmkisoredeccenxxxxxD
                \else
                \ifnum#1=29 %
                \hatcurXmkisoredeccenxxxxxE
                \else
                \ifnum#1=30 %
                \hatcurXmkisoredeccenxxxxxF
                \else
                ??????\fi
                \fi
                \fi
                \fi
                \fi
                \fi
}
\newcommand{\hatcurXmvisoredeccen}[1]{\ifnum#1=25 %
        \hatcurXmvisoredeccenxxxxxA
                \else
                \ifnum#1=26 %
                \hatcurXmvisoredeccenxxxxxB
                \else
                \ifnum#1=27 %
                \hatcurXmvisoredeccenxxxxxC
                \else
                \ifnum#1=28 %
                \hatcurXmvisoredeccenxxxxxD
                \else
                \ifnum#1=29 %
                \hatcurXmvisoredeccenxxxxxE
                \else
                \ifnum#1=30 %
                \hatcurXmvisoredeccenxxxxxF
                \else
                ??????\fi
                \fi
                \fi
                \fi
                \fi
                \fi
}
\newcommand{\hatcurXsecdureccen}[1]{\ifnum#1=25 %
        \hatcurXsecdureccenxxxxxA
                \else
                \ifnum#1=26 %
                \hatcurXsecdureccenxxxxxB
                \else
                \ifnum#1=27 %
                \hatcurXsecdureccenxxxxxC
                \else
                \ifnum#1=28 %
                \hatcurXsecdureccenxxxxxD
                \else
                \ifnum#1=29 %
                \hatcurXsecdureccenxxxxxE
                \else
                \ifnum#1=30 %
                \hatcurXsecdureccenxxxxxF
                \else
                ??????\fi
                \fi
                \fi
                \fi
                \fi
                \fi
}
\newcommand{\hatcurXsecingdureccen}[1]{\ifnum#1=25 %
        \hatcurXsecingdureccenxxxxxA
                \else
                \ifnum#1=26 %
                \hatcurXsecingdureccenxxxxxB
                \else
                \ifnum#1=27 %
                \hatcurXsecingdureccenxxxxxC
                \else
                \ifnum#1=28 %
                \hatcurXsecingdureccenxxxxxD
                \else
                \ifnum#1=29 %
                \hatcurXsecingdureccenxxxxxE
                \else
                \ifnum#1=30 %
                \hatcurXsecingdureccenxxxxxF
                \else
                ??????\fi
                \fi
                \fi
                \fi
                \fi
                \fi
}
\newcommand{\hatcurXsecondaryeccen}[1]{\ifnum#1=25 %
        \hatcurXsecondaryeccenxxxxxA
                \else
                \ifnum#1=26 %
                \hatcurXsecondaryeccenxxxxxB
                \else
                \ifnum#1=27 %
                \hatcurXsecondaryeccenxxxxxC
                \else
                \ifnum#1=28 %
                \hatcurXsecondaryeccenxxxxxD
                \else
                \ifnum#1=29 %
                \hatcurXsecondaryeccenxxxxxE
                \else
                \ifnum#1=30 %
                \hatcurXsecondaryeccenxxxxxF
                \else
                ??????\fi
                \fi
                \fi
                \fi
                \fi
                \fi
}
\newcommand{\hatcurXsecphaseeccen}[1]{\ifnum#1=25 %
        \hatcurXsecphaseeccenxxxxxA
                \else
                \ifnum#1=26 %
                \hatcurXsecphaseeccenxxxxxB
                \else
                \ifnum#1=27 %
                \hatcurXsecphaseeccenxxxxxC
                \else
                \ifnum#1=28 %
                \hatcurXsecphaseeccenxxxxxD
                \else
                \ifnum#1=29 %
                \hatcurXsecphaseeccenxxxxxE
                \else
                \ifnum#1=30 %
                \hatcurXsecphaseeccenxxxxxF
                \else
                ??????\fi
                \fi
                \fi
                \fi
                \fi
                \fi
}
\newcommand{\hatcurXviisoredeccen}[1]{\ifnum#1=25 %
        \hatcurXviisoredeccenxxxxxA
                \else
                \ifnum#1=26 %
                \hatcurXviisoredeccenxxxxxB
                \else
                \ifnum#1=27 %
                \hatcurXviisoredeccenxxxxxC
                \else
                \ifnum#1=28 %
                \hatcurXviisoredeccenxxxxxD
                \else
                \ifnum#1=29 %
                \hatcurXviisoredeccenxxxxxE
                \else
                \ifnum#1=30 %
                \hatcurXviisoredeccenxxxxxF
                \else
                ??????\fi
                \fi
                \fi
                \fi
                \fi
                \fi
}
\newcommand{\hatcurXvkisoredeccen}[1]{\ifnum#1=25 %
        \hatcurXvkisoredeccenxxxxxA
                \else
                \ifnum#1=26 %
                \hatcurXvkisoredeccenxxxxxB
                \else
                \ifnum#1=27 %
                \hatcurXvkisoredeccenxxxxxC
                \else
                \ifnum#1=28 %
                \hatcurXvkisoredeccenxxxxxD
                \else
                \ifnum#1=29 %
                \hatcurXvkisoredeccenxxxxxE
                \else
                \ifnum#1=30 %
                \hatcurXvkisoredeccenxxxxxF
                \else
                ??????\fi
                \fi
                \fi
                \fi
                \fi
                \fi
}

\newcommand{\hatcurxxxxxA}{HATS-25}
\newcommand{\hatcurbxxxxxA}{HATS-25b}
\newcommand{\hatcurcxxxxxA}{HATS-25c}

\newcommand{\hatcurplanetnumxxxxxA}{25}

\newcommand{\hatcurRVgammaabsxxxxxA}{\hatcurRVgamma{\hatcurplanetnumxxxxxA}}                           

\newcommand{\hatcurRVgammaabsinstxxxxxA}{HARPS}                           

\newcommand{\hatcurRVgammarelxxxxxA}{\hatcurRVgamma{\hatcurplanetnumxxxxxA}}                           

\newcommand{\hatcurCCtassvixxxxxA}{NULL}                  

\newcommand{\hatcurSMEversionxxxxxA}{ii}                                       

\newcommand{\hatcurisoshortxxxxxA}{YY}
\newcommand{\hatcurisofullxxxxxA}{Yonsei-Yale (YY)}
\newcommand{\hatcurisocitexxxxxA}{yi:2001}

\newcommand{\hatcurlumindxxxxxA}{\arstar}

\newcommand{\hatcurjhkfilsetxxxxxA}{ESO}

\newcommand{\hatcurSMEteffxxxxxA}{\ifthenelse{\equal{\hatcurSMEversionxxxxxA}{i}}{\hatcurSMEiteff{\hatcurplanetnumxxxxxA}}{\hatcurSMEiiteff{\hatcurplanetnumxxxxxA}}}
\newcommand{\hatcurSMEzfehxxxxxA}{\ifthenelse{\equal{\hatcurSMEversionxxxxxA}{i}}{\hatcurSMEizfeh{\hatcurplanetnumxxxxxA}}{\hatcurSMEiizfeh{\hatcurplanetnumxxxxxA}}}
\newcommand{\hatcurSMEzfehshortxxxxxA}{\ifthenelse{\equal{\hatcurSMEversionxxxxxA}{i}}{\hatcurSMEizfehshort{\hatcurplanetnumxxxxxA}}{\hatcurSMEiizfehshort{\hatcurplanetnumxxxxxA}}}
\newcommand{\hatcurSMEloggxxxxxA}{\ifthenelse{\equal{\hatcurSMEversionxxxxxA}{i}}{\hatcurSMEilogg{\hatcurplanetnumxxxxxA}}{\hatcurSMEiilogg{\hatcurplanetnumxxxxxA}}}
\newcommand{\hatcurSMEvsinxxxxxA}{\ifthenelse{\equal{\hatcurSMEversionxxxxxA}{i}}{\hatcurSMEivsin{\hatcurplanetnumxxxxxA}}{\hatcurSMEiivsin{\hatcurplanetnumxxxxxA}}}
\newcommand{\hatcurSMEvmacxxxxxA}{\ifthenelse{\equal{\hatcurSMEversionxxxxxA}{i}}{\hatcurSMEivmac{\hatcurplanetnumxxxxxA}}{\hatcurSMEiivmac{\hatcurplanetnumxxxxxA}}}
\newcommand{\hatcurSMEvmicxxxxxA}{\ifthenelse{\equal{\hatcurSMEversionxxxxxA}{i}}{\hatcurSMEivmic{\hatcurplanetnumxxxxxA}}{\hatcurSMEiivmic{\hatcurplanetnumxxxxxA}}}

\newcommand{\hatcurxxxxxB}{HATS-26}
\newcommand{\hatcurbxxxxxB}{HATS-26b}
\newcommand{\hatcurcxxxxxB}{HATS-26c}

\newcommand{\hatcurplanetnumxxxxxB}{26}

\newcommand{\hatcurRVgammaabsxxxxxB}{\hatcurRVgammaA{\hatcurplanetnumxxxxxB}}                           

\newcommand{\hatcurRVgammaabsinstxxxxxB}{FEROS}                           

\newcommand{\hatcurRVgammarelxxxxxB}{\hatcurRVgammaA{\hatcurplanetnumxxxxxB}}                           

\newcommand{\hatcurCCtassvixxxxxB}{NULL}                  

\newcommand{\hatcurSMEversionxxxxxB}{ii}                                       

\newcommand{\hatcurisoshortxxxxxB}{YY}
\newcommand{\hatcurisofullxxxxxB}{Yonsei-Yale (YY)}
\newcommand{\hatcurisocitexxxxxB}{yi:2001}

\newcommand{\hatcurlumindxxxxxB}{\arstar}

\newcommand{\hatcurjhkfilsetxxxxxB}{ESO}

\newcommand{\hatcurSMEteffxxxxxB}{\ifthenelse{\equal{\hatcurSMEversionxxxxxB}{i}}{\hatcurSMEiteff{\hatcurplanetnumxxxxxB}}{\hatcurSMEiiteff{\hatcurplanetnumxxxxxB}}}
\newcommand{\hatcurSMEzfehxxxxxB}{\ifthenelse{\equal{\hatcurSMEversionxxxxxB}{i}}{\hatcurSMEizfeh{\hatcurplanetnumxxxxxB}}{\hatcurSMEiizfeh{\hatcurplanetnumxxxxxB}}}
\newcommand{\hatcurSMEzfehshortxxxxxB}{\ifthenelse{\equal{\hatcurSMEversionxxxxxB}{i}}{\hatcurSMEizfehshort{\hatcurplanetnumxxxxxB}}{\hatcurSMEiizfehshort{\hatcurplanetnumxxxxxB}}}
\newcommand{\hatcurSMEloggxxxxxB}{\ifthenelse{\equal{\hatcurSMEversionxxxxxB}{i}}{\hatcurSMEilogg{\hatcurplanetnumxxxxxB}}{\hatcurSMEiilogg{\hatcurplanetnumxxxxxB}}}
\newcommand{\hatcurSMEvsinxxxxxB}{\ifthenelse{\equal{\hatcurSMEversionxxxxxB}{i}}{\hatcurSMEivsin{\hatcurplanetnumxxxxxB}}{\hatcurSMEiivsin{\hatcurplanetnumxxxxxB}}}
\newcommand{\hatcurSMEvmacxxxxxB}{\ifthenelse{\equal{\hatcurSMEversionxxxxxB}{i}}{\hatcurSMEivmac{\hatcurplanetnumxxxxxB}}{\hatcurSMEiivmac{\hatcurplanetnumxxxxxB}}}
\newcommand{\hatcurSMEvmicxxxxxB}{\ifthenelse{\equal{\hatcurSMEversionxxxxxB}{i}}{\hatcurSMEivmic{\hatcurplanetnumxxxxxB}}{\hatcurSMEiivmic{\hatcurplanetnumxxxxxB}}}

\newcommand{\hatcurxxxxxC}{HATS-27}
\newcommand{\hatcurbxxxxxC}{HATS-27b}
\newcommand{\hatcurcxxxxxC}{HATS-27c}

\newcommand{\hatcurplanetnumxxxxxC}{27}

\newcommand{\hatcurRVgammaabsxxxxxC}{\hatcurRVgammaC{\hatcurplanetnumxxxxxC}}                           

\newcommand{\hatcurRVgammaabsinstxxxxxC}{HARPS}                           

\newcommand{\hatcurRVgammarelxxxxxC}{\hatcurRVgammaC{\hatcurplanetnumxxxxxC}}                           

\newcommand{\hatcurCCtassvixxxxxC}{NULL}                  

\newcommand{\hatcurSMEversionxxxxxC}{i}                                       

\newcommand{\hatcurisoshortxxxxxC}{YY}
\newcommand{\hatcurisofullxxxxxC}{Yonsei-Yale (YY)}
\newcommand{\hatcurisocitexxxxxC}{yi:2001}

\newcommand{\hatcurlumindxxxxxC}{\arstar}

\newcommand{\hatcurjhkfilsetxxxxxC}{ESO}

\newcommand{\hatcurSMEteffxxxxxC}{\ifthenelse{\equal{\hatcurSMEversionxxxxxC}{i}}{\hatcurSMEiteff{\hatcurplanetnumxxxxxC}}{\hatcurSMEiiteff{\hatcurplanetnumxxxxxC}}}
\newcommand{\hatcurSMEzfehxxxxxC}{\ifthenelse{\equal{\hatcurSMEversionxxxxxC}{i}}{\hatcurSMEizfeh{\hatcurplanetnumxxxxxC}}{\hatcurSMEiizfeh{\hatcurplanetnumxxxxxC}}}
\newcommand{\hatcurSMEzfehshortxxxxxC}{\ifthenelse{\equal{\hatcurSMEversionxxxxxC}{i}}{\hatcurSMEizfehshort{\hatcurplanetnumxxxxxC}}{\hatcurSMEiizfehshort{\hatcurplanetnumxxxxxC}}}
\newcommand{\hatcurSMEloggxxxxxC}{\ifthenelse{\equal{\hatcurSMEversionxxxxxC}{i}}{\hatcurSMEilogg{\hatcurplanetnumxxxxxC}}{\hatcurSMEiilogg{\hatcurplanetnumxxxxxC}}}
\newcommand{\hatcurSMEvsinxxxxxC}{\ifthenelse{\equal{\hatcurSMEversionxxxxxC}{i}}{\hatcurSMEivsin{\hatcurplanetnumxxxxxC}}{\hatcurSMEiivsin{\hatcurplanetnumxxxxxC}}}
\newcommand{\hatcurSMEvmacxxxxxC}{\ifthenelse{\equal{\hatcurSMEversionxxxxxC}{i}}{\hatcurSMEivmac{\hatcurplanetnumxxxxxC}}{\hatcurSMEiivmac{\hatcurplanetnumxxxxxC}}}
\newcommand{\hatcurSMEvmicxxxxxC}{\ifthenelse{\equal{\hatcurSMEversionxxxxxC}{i}}{\hatcurSMEivmic{\hatcurplanetnumxxxxxC}}{\hatcurSMEiivmic{\hatcurplanetnumxxxxxC}}}

\newcommand{\hatcurxxxxxD}{HATS-28}
\newcommand{\hatcurbxxxxxD}{HATS-28b}
\newcommand{\hatcurcxxxxxD}{HATS-28c}

\newcommand{\hatcurplanetnumxxxxxD}{28}

\newcommand{\hatcurRVgammaabsxxxxxD}{\hatcurRVgamma{\hatcurplanetnumxxxxxD}}                           

\newcommand{\hatcurRVgammaabsinstxxxxxD}{FEROS}                           

\newcommand{\hatcurRVgammarelxxxxxD}{\hatcurRVgamma{\hatcurplanetnumxxxxxD}}                           

\newcommand{\hatcurCCtassvixxxxxD}{NULL}                  

\newcommand{\hatcurSMEversionxxxxxD}{ii}                                       

\newcommand{\hatcurisoshortxxxxxD}{YY}
\newcommand{\hatcurisofullxxxxxD}{Yonsei-Yale (YY)}
\newcommand{\hatcurisocitexxxxxD}{yi:2001}

\newcommand{\hatcurlumindxxxxxD}{\arstar}

\newcommand{\hatcurjhkfilsetxxxxxD}{ESO}

\newcommand{\hatcurSMEteffxxxxxD}{\ifthenelse{\equal{\hatcurSMEversionxxxxxD}{i}}{\hatcurSMEiteff{\hatcurplanetnumxxxxxD}}{\hatcurSMEiiteff{\hatcurplanetnumxxxxxD}}}
\newcommand{\hatcurSMEzfehxxxxxD}{\ifthenelse{\equal{\hatcurSMEversionxxxxxD}{i}}{\hatcurSMEizfeh{\hatcurplanetnumxxxxxD}}{\hatcurSMEiizfeh{\hatcurplanetnumxxxxxD}}}
\newcommand{\hatcurSMEzfehshortxxxxxD}{\ifthenelse{\equal{\hatcurSMEversionxxxxxD}{i}}{\hatcurSMEizfehshort{\hatcurplanetnumxxxxxD}}{\hatcurSMEiizfehshort{\hatcurplanetnumxxxxxD}}}
\newcommand{\hatcurSMEloggxxxxxD}{\ifthenelse{\equal{\hatcurSMEversionxxxxxD}{i}}{\hatcurSMEilogg{\hatcurplanetnumxxxxxD}}{\hatcurSMEiilogg{\hatcurplanetnumxxxxxD}}}
\newcommand{\hatcurSMEvsinxxxxxD}{\ifthenelse{\equal{\hatcurSMEversionxxxxxD}{i}}{\hatcurSMEivsin{\hatcurplanetnumxxxxxD}}{\hatcurSMEiivsin{\hatcurplanetnumxxxxxD}}}
\newcommand{\hatcurSMEvmacxxxxxD}{\ifthenelse{\equal{\hatcurSMEversionxxxxxD}{i}}{\hatcurSMEivmac{\hatcurplanetnumxxxxxD}}{\hatcurSMEiivmac{\hatcurplanetnumxxxxxD}}}
\newcommand{\hatcurSMEvmicxxxxxD}{\ifthenelse{\equal{\hatcurSMEversionxxxxxD}{i}}{\hatcurSMEivmic{\hatcurplanetnumxxxxxD}}{\hatcurSMEiivmic{\hatcurplanetnumxxxxxD}}}

\newcommand{\hatcurxxxxxE}{HATS-29}
\newcommand{\hatcurbxxxxxE}{HATS-29b}
\newcommand{\hatcurcxxxxxE}{HATS-29c}

\newcommand{\hatcurplanetnumxxxxxE}{29}

\newcommand{\hatcurRVgammaabsxxxxxE}{\hatcurRVgammaB{\hatcurplanetnumxxxxxE}}                           

\newcommand{\hatcurRVgammaabsinstxxxxxE}{HARPS}                           

\newcommand{\hatcurRVgammarelxxxxxE}{\hatcurRVgammaB{\hatcurplanetnumxxxxxE}}                           

\newcommand{\hatcurCCtassvixxxxxE}{NULL}                  

\newcommand{\hatcurSMEversionxxxxxE}{ii}                                       

\newcommand{\hatcurisoshortxxxxxE}{YY}
\newcommand{\hatcurisofullxxxxxE}{Yonsei-Yale (YY)}
\newcommand{\hatcurisocitexxxxxE}{yi:2001}

\newcommand{\hatcurlumindxxxxxE}{\arstar}

\newcommand{\hatcurjhkfilsetxxxxxE}{ESO}

\newcommand{\hatcurSMEteffxxxxxE}{\ifthenelse{\equal{\hatcurSMEversionxxxxxE}{i}}{\hatcurSMEiteff{\hatcurplanetnumxxxxxE}}{\hatcurSMEiiteff{\hatcurplanetnumxxxxxE}}}
\newcommand{\hatcurSMEzfehxxxxxE}{\ifthenelse{\equal{\hatcurSMEversionxxxxxE}{i}}{\hatcurSMEizfeh{\hatcurplanetnumxxxxxE}}{\hatcurSMEiizfeh{\hatcurplanetnumxxxxxE}}}
\newcommand{\hatcurSMEzfehshortxxxxxE}{\ifthenelse{\equal{\hatcurSMEversionxxxxxE}{i}}{\hatcurSMEizfehshort{\hatcurplanetnumxxxxxE}}{\hatcurSMEiizfehshort{\hatcurplanetnumxxxxxE}}}
\newcommand{\hatcurSMEloggxxxxxE}{\ifthenelse{\equal{\hatcurSMEversionxxxxxE}{i}}{\hatcurSMEilogg{\hatcurplanetnumxxxxxE}}{\hatcurSMEiilogg{\hatcurplanetnumxxxxxE}}}
\newcommand{\hatcurSMEvsinxxxxxE}{\ifthenelse{\equal{\hatcurSMEversionxxxxxE}{i}}{\hatcurSMEivsin{\hatcurplanetnumxxxxxE}}{\hatcurSMEiivsin{\hatcurplanetnumxxxxxE}}}
\newcommand{\hatcurSMEvmacxxxxxE}{\ifthenelse{\equal{\hatcurSMEversionxxxxxE}{i}}{\hatcurSMEivmac{\hatcurplanetnumxxxxxE}}{\hatcurSMEiivmac{\hatcurplanetnumxxxxxE}}}
\newcommand{\hatcurSMEvmicxxxxxE}{\ifthenelse{\equal{\hatcurSMEversionxxxxxE}{i}}{\hatcurSMEivmic{\hatcurplanetnumxxxxxE}}{\hatcurSMEiivmic{\hatcurplanetnumxxxxxE}}}

\newcommand{\hatcurxxxxxF}{HATS-30}
\newcommand{\hatcurbxxxxxF}{HATS-30b}
\newcommand{\hatcurcxxxxxF}{HATS-30c}

\newcommand{\hatcurplanetnumxxxxxF}{30}

\newcommand{\hatcurRVgammaabsxxxxxF}{\hatcurRVgammaA{\hatcurplanetnumxxxxxF}}                           

\newcommand{\hatcurRVgammaabsinstxxxxxF}{FEROS}                           

\newcommand{\hatcurRVgammarelxxxxxF}{\hatcurRVgammaA{\hatcurplanetnumxxxxxF}}                           

\newcommand{\hatcurCCtassvixxxxxF}{NULL}                  

\newcommand{\hatcurSMEversionxxxxxF}{ii}                                       

\newcommand{\hatcurisoshortxxxxxF}{YY}
\newcommand{\hatcurisofullxxxxxF}{Yonsei-Yale (YY)}
\newcommand{\hatcurisocitexxxxxF}{yi:2001}

\newcommand{\hatcurlumindxxxxxF}{\arstar}

\newcommand{\hatcurjhkfilsetxxxxxF}{ESO}

\newcommand{\hatcurSMEteffxxxxxF}{\ifthenelse{\equal{\hatcurSMEversionxxxxxF}{i}}{\hatcurSMEiteff{\hatcurplanetnumxxxxxF}}{\hatcurSMEiiteff{\hatcurplanetnumxxxxxF}}}
\newcommand{\hatcurSMEzfehxxxxxF}{\ifthenelse{\equal{\hatcurSMEversionxxxxxF}{i}}{\hatcurSMEizfeh{\hatcurplanetnumxxxxxF}}{\hatcurSMEiizfeh{\hatcurplanetnumxxxxxF}}}
\newcommand{\hatcurSMEzfehshortxxxxxF}{\ifthenelse{\equal{\hatcurSMEversionxxxxxF}{i}}{\hatcurSMEizfehshort{\hatcurplanetnumxxxxxF}}{\hatcurSMEiizfehshort{\hatcurplanetnumxxxxxF}}}
\newcommand{\hatcurSMEloggxxxxxF}{\ifthenelse{\equal{\hatcurSMEversionxxxxxF}{i}}{\hatcurSMEilogg{\hatcurplanetnumxxxxxF}}{\hatcurSMEiilogg{\hatcurplanetnumxxxxxF}}}
\newcommand{\hatcurSMEvsinxxxxxF}{\ifthenelse{\equal{\hatcurSMEversionxxxxxF}{i}}{\hatcurSMEivsin{\hatcurplanetnumxxxxxF}}{\hatcurSMEiivsin{\hatcurplanetnumxxxxxF}}}
\newcommand{\hatcurSMEvmacxxxxxF}{\ifthenelse{\equal{\hatcurSMEversionxxxxxF}{i}}{\hatcurSMEivmac{\hatcurplanetnumxxxxxF}}{\hatcurSMEiivmac{\hatcurplanetnumxxxxxF}}}
\newcommand{\hatcurSMEvmicxxxxxF}{\ifthenelse{\equal{\hatcurSMEversionxxxxxF}{i}}{\hatcurSMEivmic{\hatcurplanetnumxxxxxF}}{\hatcurSMEiivmic{\hatcurplanetnumxxxxxF}}}

\newcommand{\hatcur}[1]{\ifnum#1=25 %
\hatcurxxxxxA
\else
\ifnum#1=26 %
\hatcurxxxxxB
\else
\ifnum#1=27 %
\hatcurxxxxxC
\else
\ifnum#1=28 %
\hatcurxxxxxD
\else
\ifnum#1=29 %
\hatcurxxxxxE
\else
\ifnum#1=30 %
\hatcurxxxxxF
\else
??????\fi
\fi
\fi
\fi
\fi
\fi
}
\newcommand{\hatcurb}[1]{\ifnum#1=25 %
\hatcurbxxxxxA
\else
\ifnum#1=26 %
\hatcurbxxxxxB
\else
\ifnum#1=27 %
\hatcurbxxxxxC
\else
\ifnum#1=28 %
\hatcurbxxxxxD
\else
\ifnum#1=29 %
\hatcurbxxxxxE
\else
\ifnum#1=30 %
\hatcurbxxxxxF
\else
??????\fi
\fi
\fi
\fi
\fi
\fi
}
\newcommand{\hatcurc}[1]{\ifnum#1=25 %
\hatcurcxxxxxA
\else
\ifnum#1=26 %
\hatcurcxxxxxB
\else
\ifnum#1=27 %
\hatcurcxxxxxC
\else
\ifnum#1=28 %
\hatcurcxxxxxD
\else
\ifnum#1=29 %
\hatcurcxxxxxE
\else
\ifnum#1=30 %
\hatcurcxxxxxF
\else
??????\fi
\fi
\fi
\fi
\fi
\fi
}
\newcommand{\hatcurCCtassvi}[1]{\ifnum#1=25 %
\hatcurCCtassvixxxxxA
\else
\ifnum#1=26 %
\hatcurCCtassvixxxxxB
\else
\ifnum#1=27 %
\hatcurCCtassvixxxxxC
\else
\ifnum#1=28 %
\hatcurCCtassvixxxxxD
\else
\ifnum#1=29 %
\hatcurCCtassvixxxxxE
\else
\ifnum#1=30 %
\hatcurCCtassvixxxxxF
\else
??????\fi
\fi
\fi
\fi
\fi
\fi
}
\newcommand{\hatcurisocite}[1]{\ifnum#1=25 %
\hatcurisocitexxxxxA
\else
\ifnum#1=26 %
\hatcurisocitexxxxxB
\else
\ifnum#1=27 %
\hatcurisocitexxxxxC
\else
\ifnum#1=28 %
\hatcurisocitexxxxxD
\else
\ifnum#1=29 %
\hatcurisocitexxxxxE
\else
\ifnum#1=30 %
\hatcurisocitexxxxxF
\else
??????\fi
\fi
\fi
\fi
\fi
\fi
}
\newcommand{\hatcurisofull}[1]{\ifnum#1=25 %
\hatcurisofullxxxxxA
\else
\ifnum#1=26 %
\hatcurisofullxxxxxB
\else
\ifnum#1=27 %
\hatcurisofullxxxxxC
\else
\ifnum#1=28 %
\hatcurisofullxxxxxD
\else
\ifnum#1=29 %
\hatcurisofullxxxxxE
\else
\ifnum#1=30 %
\hatcurisofullxxxxxF
\else
??????\fi
\fi
\fi
\fi
\fi
\fi
}
\newcommand{\hatcurisoshort}[1]{\ifnum#1=25 %
\hatcurisoshortxxxxxA
\else
\ifnum#1=26 %
\hatcurisoshortxxxxxB
\else
\ifnum#1=27 %
\hatcurisoshortxxxxxC
\else
\ifnum#1=28 %
\hatcurisoshortxxxxxD
\else
\ifnum#1=29 %
\hatcurisoshortxxxxxE
\else
\ifnum#1=30 %
\hatcurisoshortxxxxxF
\else
??????\fi
\fi
\fi
\fi
\fi
\fi
}
\newcommand{\hatcurjhkfilset}[1]{\ifnum#1=25 %
\hatcurjhkfilsetxxxxxA
\else
\ifnum#1=26 %
\hatcurjhkfilsetxxxxxB
\else
\ifnum#1=27 %
\hatcurjhkfilsetxxxxxC
\else
\ifnum#1=28 %
\hatcurjhkfilsetxxxxxD
\else
\ifnum#1=29 %
\hatcurjhkfilsetxxxxxE
\else
\ifnum#1=30 %
\hatcurjhkfilsetxxxxxF
\else
??????\fi
\fi
\fi
\fi
\fi
\fi
}
\newcommand{\hatcurlumind}[1]{\ifnum#1=25 %
\hatcurlumindxxxxxA
\else
\ifnum#1=26 %
\hatcurlumindxxxxxB
\else
\ifnum#1=27 %
\hatcurlumindxxxxxC
\else
\ifnum#1=28 %
\hatcurlumindxxxxxD
\else
\ifnum#1=29 %
\hatcurlumindxxxxxE
\else
\ifnum#1=30 %
\hatcurlumindxxxxxF
\else
??????\fi
\fi
\fi
\fi
\fi
\fi
}
\newcommand{\hatcurplanetnum}[1]{\ifnum#1=25 %
\hatcurplanetnumxxxxxA
\else
\ifnum#1=26 %
\hatcurplanetnumxxxxxB
\else
\ifnum#1=27 %
\hatcurplanetnumxxxxxC
\else
\ifnum#1=28 %
\hatcurplanetnumxxxxxD
\else
\ifnum#1=29 %
\hatcurplanetnumxxxxxE
\else
\ifnum#1=30 %
\hatcurplanetnumxxxxxF
\else
??????\fi
\fi
\fi
\fi
\fi
\fi
}
\newcommand{\hatcurRVgammaabs}[1]{\ifnum#1=25 %
\hatcurRVgammaabsxxxxxA
\else
\ifnum#1=26 %
\hatcurRVgammaabsxxxxxB
\else
\ifnum#1=27 %
\hatcurRVgammaabsxxxxxC
\else
\ifnum#1=28 %
\hatcurRVgammaabsxxxxxD
\else
\ifnum#1=29 %
\hatcurRVgammaabsxxxxxE
\else
\ifnum#1=30 %
\hatcurRVgammaabsxxxxxF
\else
??????\fi
\fi
\fi
\fi
\fi
\fi
}
\newcommand{\hatcurRVgammaabsinst}[1]{\ifnum#1=25 %
\hatcurRVgammaabsinstxxxxxA
\else
\ifnum#1=26 %
\hatcurRVgammaabsinstxxxxxB
\else
\ifnum#1=27 %
\hatcurRVgammaabsinstxxxxxC
\else
\ifnum#1=28 %
\hatcurRVgammaabsinstxxxxxD
\else
\ifnum#1=29 %
\hatcurRVgammaabsinstxxxxxE
\else
\ifnum#1=30 %
\hatcurRVgammaabsinstxxxxxF
\else
??????\fi
\fi
\fi
\fi
\fi
\fi
}
\newcommand{\hatcurRVgammarel}[1]{\ifnum#1=25 %
\hatcurRVgammarelxxxxxA
\else
\ifnum#1=26 %
\hatcurRVgammarelxxxxxB
\else
\ifnum#1=27 %
\hatcurRVgammarelxxxxxC
\else
\ifnum#1=28 %
\hatcurRVgammarelxxxxxD
\else
\ifnum#1=29 %
\hatcurRVgammarelxxxxxE
\else
\ifnum#1=30 %
\hatcurRVgammarelxxxxxF
\else
??????\fi
\fi
\fi
\fi
\fi
\fi
}
\newcommand{\hatcurSMElogg}[1]{\ifnum#1=25 %
\hatcurSMEloggxxxxxA
\else
\ifnum#1=26 %
\hatcurSMEloggxxxxxB
\else
\ifnum#1=27 %
\hatcurSMEloggxxxxxC
\else
\ifnum#1=28 %
\hatcurSMEloggxxxxxD
\else
\ifnum#1=29 %
\hatcurSMEloggxxxxxE
\else
\ifnum#1=30 %
\hatcurSMEloggxxxxxF
\else
??????\fi
\fi
\fi
\fi
\fi
\fi
}
\newcommand{\hatcurSMEteff}[1]{\ifnum#1=25 %
\hatcurSMEteffxxxxxA
\else
\ifnum#1=26 %
\hatcurSMEteffxxxxxB
\else
\ifnum#1=27 %
\hatcurSMEteffxxxxxC
\else
\ifnum#1=28 %
\hatcurSMEteffxxxxxD
\else
\ifnum#1=29 %
\hatcurSMEteffxxxxxE
\else
\ifnum#1=30 %
\hatcurSMEteffxxxxxF
\else
??????\fi
\fi
\fi
\fi
\fi
\fi
}
\newcommand{\hatcurSMEversion}[1]{\ifnum#1=25 %
\hatcurSMEversionxxxxxA
\else
\ifnum#1=26 %
\hatcurSMEversionxxxxxB
\else
\ifnum#1=27 %
\hatcurSMEversionxxxxxC
\else
\ifnum#1=28 %
\hatcurSMEversionxxxxxD
\else
\ifnum#1=29 %
\hatcurSMEversionxxxxxE
\else
\ifnum#1=30 %
\hatcurSMEversionxxxxxF
\else
??????\fi
\fi
\fi
\fi
\fi
\fi
}
\newcommand{\hatcurSMEvmac}[1]{\ifnum#1=25 %
\hatcurSMEvmacxxxxxA
\else
\ifnum#1=26 %
\hatcurSMEvmacxxxxxB
\else
\ifnum#1=27 %
\hatcurSMEvmacxxxxxC
\else
\ifnum#1=28 %
\hatcurSMEvmacxxxxxD
\else
\ifnum#1=29 %
\hatcurSMEvmacxxxxxE
\else
\ifnum#1=30 %
\hatcurSMEvmacxxxxxF
\else
??????\fi
\fi
\fi
\fi
\fi
\fi
}
\newcommand{\hatcurSMEvmic}[1]{\ifnum#1=25 %
\hatcurSMEvmicxxxxxA
\else
\ifnum#1=26 %
\hatcurSMEvmicxxxxxB
\else
\ifnum#1=27 %
\hatcurSMEvmicxxxxxC
\else
\ifnum#1=28 %
\hatcurSMEvmicxxxxxD
\else
\ifnum#1=29 %
\hatcurSMEvmicxxxxxE
\else
\ifnum#1=30 %
\hatcurSMEvmicxxxxxF
\else
??????\fi
\fi
\fi
\fi
\fi
\fi
}
\newcommand{\hatcurSMEvsin}[1]{\ifnum#1=25 %
\hatcurSMEvsinxxxxxA
\else
\ifnum#1=26 %
\hatcurSMEvsinxxxxxB
\else
\ifnum#1=27 %
\hatcurSMEvsinxxxxxC
\else
\ifnum#1=28 %
\hatcurSMEvsinxxxxxD
\else
\ifnum#1=29 %
\hatcurSMEvsinxxxxxE
\else
\ifnum#1=30 %
\hatcurSMEvsinxxxxxF
\else
??????\fi
\fi
\fi
\fi
\fi
\fi
}
\newcommand{\hatcurSMEzfeh}[1]{\ifnum#1=25 %
\hatcurSMEzfehxxxxxA
\else
\ifnum#1=26 %
\hatcurSMEzfehxxxxxB
\else
\ifnum#1=27 %
\hatcurSMEzfehxxxxxC
\else
\ifnum#1=28 %
\hatcurSMEzfehxxxxxD
\else
\ifnum#1=29 %
\hatcurSMEzfehxxxxxE
\else
\ifnum#1=30 %
\hatcurSMEzfehxxxxxF
\else
??????\fi
\fi
\fi
\fi
\fi
\fi
}
\newcommand{\hatcurSMEzfehshort}[1]{\ifnum#1=25 %
\hatcurSMEzfehshortxxxxxA
\else
\ifnum#1=26 %
\hatcurSMEzfehshortxxxxxB
\else
\ifnum#1=27 %
\hatcurSMEzfehshortxxxxxC
\else
\ifnum#1=28 %
\hatcurSMEzfehshortxxxxxD
\else
\ifnum#1=29 %
\hatcurSMEzfehshortxxxxxE
\else
\ifnum#1=30 %
\hatcurSMEzfehshortxxxxxF
\else
??????\fi
\fi
\fi
\fi
\fi
\fi
}

\newcounter{planetcounter}

\newboolean{emulateapj}
\setboolean{emulateapj}{true}

\newboolean{rvtablelong}
\setboolean{rvtablelong}{true}

\newboolean{astroph}
\setboolean{astroph}{true}

\newlength{\plotwidthtwo}

\shortauthors{Espinoza et al.}
\shorttitle{\hatcur{25}\lowercase{b}--\hatcur{30}\lowercase{b}}
\ifthenelse{\boolean{emulateapj}}{
    \newcommand{\titledag}{$\dagger$}
}{
    \newcommand{\titledag}{\dagger}
}

\begin{document}

\title{
\hatcur{25}\lowercase{b} through \hatcur{30}\lowercase{b}: a half-dozen new inflated transiting Hot Jupiters from the HATSouth survey\altaffilmark{\titledag}
}

\author{N. Espinoza\altaffilmark{1,2}, D. Bayliss\altaffilmark{3}, J. D. Hartman\altaffilmark{4}, G. \'A. Bakos\altaffilmark{4,*,**}, A. Jord\'an\altaffilmark{1,2},
G. Zhou\altaffilmark{5}, L. Mancini\altaffilmark{6}, R. Brahm\altaffilmark{1,2}, S. Ciceri\altaffilmark{6}, W. Bhatti\altaffilmark{4}, Z. Csubry\altaffilmark{4},
M. Rabus\altaffilmark{1,6}, K. Penev\altaffilmark{4}, J. Bento\altaffilmark{7}, M. de Val-Borro\altaffilmark{4}, T. Henning\altaffilmark{6}, B. Schmidt\altaffilmark{7}, 
V. Suc\altaffilmark{1}, D. J. Wright\altaffilmark{8,9}, C.G. Tinney\altaffilmark{8,9}, T.G. Tan\altaffilmark{10}, R. Noyes\altaffilmark{5}
}
\altaffiltext{1}{Instituto de Astrof\'isica, Facultad de F\'isica,
    Pontificia Universidad Cat\'olica de Chile, Av.\ Vicu\~na Mackenna
    4860, 7820436 Macul, Santiago, Chile; nespino@astro.puc.cl.}
\altaffiltext{2}{Millennium Institute of Astrophysics, Av.\ Vicu\~na Mackenna
    4860, 782-0436 Macul, Santiago, Chile.}
\altaffiltext{3}{Observatoire Astronomique de l'Universit\'e de Geneve, 51 ch. des Maillettes, 1290 Versoix, Switzerland.}
\altaffiltext{4}{Department of Astrophysical Sciences, Princeton University, NJ 08544, USA.}
\altaffiltext{5}{Harvard-Smithsonian Center for Astrophysics, 60 Garden St., Cambridge, MA 02138, USA.}
\altaffiltext{6}{Max Planck Institute for Astronomy, Heidelberg, Germany.}
\altaffiltext{7}{Research School of Astronomy and Astrophysics, Australian National University, Canberra, ACT 2611, Australia.}
\altaffiltext{8}{Exoplanetary Science at UNSW, School of Physics, UNSW Australia, 20152, Australia.}
\altaffiltext{9}{Australian Centre for Astrobiology, UNSW Australia, 20152, Australia.}
\altaffiltext{10}{Perth Exoplanet Survey Telescope, Perth, Australia.}
\altaffiltext{*}{Alfred P. Sloan Research Fellow.}
\altaffiltext{**}{Packard Fellow.}
\altaffiltext{$\dagger$}{
 The HATSouth network is operated by a collaboration consisting of
Princeton University (PU), the Max Planck Institute f\"ur Astronomie
(MPIA), the Australian National University (ANU), and the Pontificia
Universidad Cat\'olica de Chile (PUC).  The station at Las Campanas
Observatory (LCO) of the Carnegie Institute is operated by PU in
conjunction with PUC, the station at the High Energy Spectroscopic
Survey (H.E.S.S.) site is operated in conjunction with MPIA, and the
station at Siding Spring Observatory (SSO) is operated jointly with
ANU.
 Based in
 part on observations made with the MPG~2.2\,m Telescope at the ESO
 Observatory in La Silla.
}


\begin{abstract}

\setcounter{footnote}{1}
We report six new inflated hot Jupiters (HATS-25b through HATS-30b) discovered using the HATSouth global network of automated
telescopes. The planets orbit stars with $V$ magnitudes in the range $\sim 12-14$ and have masses in the largely populated $0.5M_J-0.7M_J$ 
region of parameter space but span a wide variety of radii, from $1.17R_J$ to $1.75 R_J$. HATS-25b, HATS-28b, HATS-29b and 
HATS-30b are typical inflated hot Jupiters ($R_p = 1.17-1.26R_J$) orbiting G-type stars in short period ($P=3.2-4.6$ days) orbits. However, 
HATS-26b ($R_p = \hatcurPPrshort{26}R_J$, $P = \hatcurLCPshort{26}$ days) and HATS-27b ($R_p=\hatcurPPrshort{27}R_J$, 
$P=\hatcurLCPshort{27}$ days) stand out as highly inflated planets orbiting slightly evolved F stars just after and in the turn-off points, respectively, which are among the least 
dense hot Jupiters, with densities of $0.153$ \gcmc and $0.180$ \gcmc, respectively. All the presented exoplanets 
but HATS-27b are good targets for future atmospheric characterization studies, while HATS-27b is a prime target for 
Rossiter-McLaughlin monitoring in order to determine its spin-orbit alignment given the brightness ($V = 12.8$) and stellar rotational 
velocity ($v \sin i \approx 9.3$ km/s) of the host star. These discoveries significantly increase the number of inflated hot Jupiters known, contributing
to our understanding of the mechanism(s) responsible for hot Jupiter inflation.
\setcounter{footnote}{0}
\end{abstract}

\keywords{
    planetary systems ---
    stars: individual (
\setcounter{planetcounter}{1}
\hatcur{25},
\hatcurCCgsc{25}\loopcommanoperiod
\setcounter{planetcounter}{2}
\hatcur{26},
\hatcurCCgsc{26}\loopcommanoperiod
\setcounter{planetcounter}{3}
\hatcur{27},
\hatcurCCgsc{27}\loopcommanoperiod
\setcounter{planetcounter}{4}
\hatcur{28},
\hatcurCCgsc{28}\loopcommanoperiod
\setcounter{planetcounter}{5}
\hatcur{29},
\hatcurCCgsc{29}\loopcommanoperiod
\setcounter{planetcounter}{6}
\hatcur{30},
\hatcurCCgsc{30}\loopcommanoperiod
\setcounter{planetcounter}{4}
) 
    techniques: spectroscopic, photometric
}


\section{Introduction}
\label{sec:introduction}
Since the observation of the transit of HD209458b, the first exoplanet to be observed to transit 
its host star by \cite{charbonneau:2000} and \cite{henry:2000}, the field of transiting extrasolar 
planets has evolved tremendously. Transiting planets not only allow us to study the distribution of 
exoplanetary sizes, but in combination with mass measurements allow us to unveil the wide 
range of densities for these distant worlds. This is critical data that delivers physical characterisation 
of these systems. In addition, these systems allow the study of atmospheric properties 
\citep[see, e.g., ][ and references therein]{crossfield:2015} and the relationship between the orbits of 
these systems and the spin of their host stars  \citep{queloz:2000,ohta:2005,winn:2007}. 

The so-called ``hot Jupiters" (i.e. planets with masses and radii similar to Jupiter, but with periods $P<10$ days) 
have been amongst the most studied exoplanets. Their observed sizes, orbits and compositions have presented 
mutliple theoretical challenges. One of the most substantial challenges has been to explain 
the observed ``inflated" nature of most of these systems \citep[i.e. the fact that their radii are typically larger than what is expected from 
models of irradiated planets see, e.g.,][]{baraffe:2003,fortney:2007}. This inflation suggests that additional processes must be at hand helping to 
avoid the gravitational contraction that self-gravitating bodies are subject to \citep[see, e.g.,][for a comprehensive 
review of the subject]{spiegel:2013}. 

Another long-lasting puzzle is the exact way in which these exoplanets acquire 
such close-in orbits. Core-accretion theory predicts these planets would form from a 
solid $\sim 10 M_\Earth$ embryo that then accumulates large amounts of gas from 
the protoplanetary disk at several AU from the host star \citep{lissauer:2007}. Once formed they migrate inwards, with the two main mechanisms 
proposed as driving this migration being the planet's interaction with the protoplanetary disk 
\citep{goldreich:1980} and/or interaction of the planet with other planetary or stellar objects in the system \citep[see, e.g., ][]{rasio:1996, 
wu:2011, fabrycky:2007,petrovich:2015}. 

The transiting nature of these systems allows observational characterisation to make powerful tests of a variety of models proposed for them. 
For example, one popular model explaining the inflated nature of hot Jupiters is 
Ohmic dissipation \citep{batygin:2010,perna:2010,batygin:2011,huang:2012,wu:2013}. 
However, many of the physical parameters that underlie these models -- such as wind speeds 
and planetary magnetic fields -- are largely unknown and are 
only just beginning to be constrained via detailed photometric \citep[see, e.g., ][ and references therein]{kataria:2016} and spectroscopic 
\citep{kislayakova:2014,louden:2015} characterization of transiting systems. Other models 
\citep[e.g., increased opacities in the atmospheres of hot Jupiters][]{burrows:2007}, can be 
tested by detailed spectral characterization of exoplanet atmospheres, which to date has mainly been provided through the technique of transmission spectroscopy. Interestingly, the composition of exoplanets inferred from 
studying their atmospheres is not only relevant for the problem of inflation or the study of atmospheric abundances in hot Jupiters \cite[see, e.g.,][]{sing:2016}, but can also constrain 
proposed migration mechanisms through the estimation of carbon-to-oxygen ratios \citep{madhusudhan:2014,benneke:2015}. Detection of more of these characterizable systems 
is thus critical to build the large samples required to test physical models.

In this work, we report the discovery of six new, well-characterized transiting hot Jupiters using the HATSouth global network of automated 
telescopes \citep{bakos:2013:hatsouth}, all of which are inflated and amenable for future atmospheric or Rossiter McLaughlin 
characterization: HATS-25b, HATS-26b, HATS-27b, HATS-28b, HATS-29b and HATS-30b. The structure of the paper is as follows. In 
\refsecl{obs} we summarize the detection of the photometric transit signal and the subsequent spectroscopic and photometric 
observations of each star to confirm and characterize the planets. In \refsecl{analysis} we analyze the data to rule out false positive scenarios, and 
to determine the stellar and planetary parameters. Our findings are discussed in \refsecl{discussion}.

\section{Observations}
\label{sec:obs}

\subsection{Photometric detection}
\label{sec:detection}
In Table \ref{tab:photobs} we summarise the HATSouth discovery data of the six exoplanets presented
in this work, all of which used data from the three HATSouth sites, namely, the site at Las Campanas 
Observatory in Chile (LCO, whose stations are designated HS-1 and HS-2), the site at of the HESS in 
Namibia (whose stations are designated HS-3 and HS-4) and the site at the Siding Spring 
Observatory (SSO, whose stations are designated HS-5 and HS-6). The large number of 
observations for HATS-28 and HATS-29 are due to them being observed as part of the HATSouth ``super-fields" program, where observations of the same field are taken with two telescopes 
from each HATSouth site. The large number of observations for HATS-30 are due to overlaps 
between its field and adjacent HATSouth fields. 

The observations, reductions and analysis of the data were carried out as detailed in \cite{bakos:2013:hatsouth}. In summary, the acquired 
images were obtained with a cadence of $\approx 300$ s using a $r$ SDSS filter on each of the sites. The 
images were then reduced and the resulting lightcurves detrended using the methods described in 
\cite{hartman:2015:hats6}. Finally, a Box Least Squares \citep[BLS, ][]{kovacs:2002:BLS} algorithm was ran on 
the lightcurves in order to search for periodic transit signatures. The discovery lightcurves of each of these stars, phased 
around the best-fit period of the transiting planet candidates, are depicted in Figure~\ref{fig:hatsouth}.

In addition to these detections, we also searched for additional signals in the lightcurves 
in order to search for variability, activity and/or additional transit signals in the candidate systems. To this end, 
we ran BLS and Generalised Lomb Scargle \citep[GLS,][]{zechmeister:2009:gls} algorithms on the residuals 
of each lightcurve, exploring each of the significant peaks (which we defined as peaks with false alarm probabilities 
lower than $0.1\%$) in each of the periodograms by fitting boxes and sinusoids, respectively, at those peaks and also inspecting 
visually the phased lightcurves. By analysing the periodograms along with the window functions, all the significant peaks are near 
prominent sampling frequencies in the window function, or their harmonics, and are likely to be instrumental in origin. We thus conclude 
that all of the lightcurves do not show any additional signs of variability, activity and/or additional transit signals at 
least at the mmag level.
%
%
\ifthenelse{\boolean{emulateapj}}{
    \begin{figure*}[!ht]
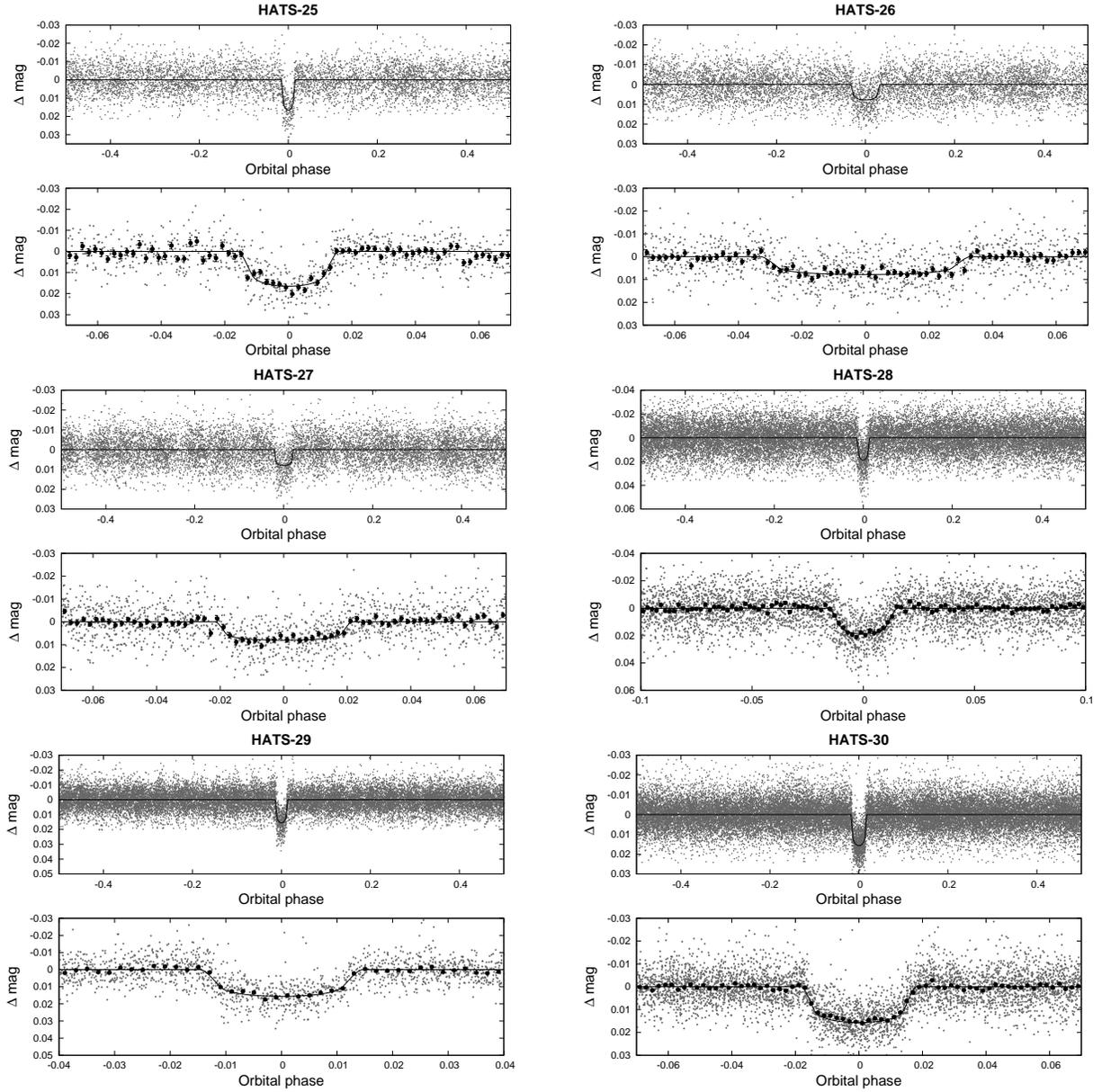

}{
    \begin{figure}[!ht]
}
\plottwo{\hatcurhtr{25}-hs.eps}{\hatcurhtr{26}-hs.eps}
\plottwo{\hatcurhtr{27}-hs.eps}{\hatcurhtr{28}-hs.eps}
\plottwo{\hatcurhtr{29}-hs.eps}{\hatcurhtr{30}-hs.eps}
\caption[]{
    Phase-folded unbinned HATSouth light curves for the six new transiting planet systems. In each case we show two panels. The
    top panel shows the full light curve, while the bottom panel shows
    the light curve zoomed-in on the transit. The solid lines show the
    model fits to the light curves. The dark filled circles in the
    bottom panels show the light curves binned in phase with a bin
    size of 0.002.
\label{fig:hatsouth}}
\ifthenelse{\boolean{emulateapj}}{
    \end{figure*}
}{
    \end{figure}
}

\ifthenelse{\boolean{emulateapj}}{
    \begin{deluxetable*}{llrrrr}
}{
    \begin{deluxetable}{llrrrr}
}
\tablewidth{0pc}
\tabletypesize{\scriptsize}
\tablecaption{
    Summary of photometric observations
    \label{tab:photobs}
}
\tablehead{
    \multicolumn{1}{c}{Instrument/Field\tablenotemark{a}} &
    \multicolumn{1}{c}{Date(s)} &
    \multicolumn{1}{c}{\# Images} &
    \multicolumn{1}{c}{Cadence\tablenotemark{b}} &
    \multicolumn{1}{c}{Filter} &
    \multicolumn{1}{c}{Precision\tablenotemark{c}} \\
    \multicolumn{1}{c}{} &
    \multicolumn{1}{c}{} &
    \multicolumn{1}{c}{} &
    \multicolumn{1}{c}{(s)} &
    \multicolumn{1}{c}{} &
    \multicolumn{1}{c}{(mmag)}
}
\startdata
\sidehead{\textbf{\hatcur{25}}}
~~~~HS-2.1/G568 & 2011 Mar--2011 Aug & 5055 & 290 & $r$ & 6.9 \\
~~~~HS-4.1/G568 & 2011 Jul--2011 Aug & 841 & 301 & $r$ & 7.8 \\
~~~~HS-6.1/G568 & 2011 May & 131 & 289 & $r$ & 9.3 \\
~~~~LCOGT~1\,m+CTIO/sinistro & 2015 Feb 23 & 70 & 226 & $i$ & 1.1 \\
~~~~LCOGT~1\,m+SSO/SBIG & 2015 Mar 16 & 104 & 196 & $i$ & 2.3 \\
\sidehead{\textbf{\hatcur{26}}}
~~~~HS-2.3/G606 & 2012 Feb--2012 Jun & 3134 & 291 & $r$ & 7.0 \\
~~~~HS-4.3/G606 & 2012 Feb--2012 Jun & 2761 & 300 & $r$ & 7.1 \\
~~~~HS-6.3/G606 & 2012 Feb--2012 Jun & 1170 & 299 & $r$ & 6.8 \\
~~~~LCOGT~1\,m+SAAO/SBIG & 2015 Mar 16 & 30 & 199 & $i$ & 1.8 \\
~~~~LCOGT~1\,m+SAAO/SBIG & 2015 Mar 26 & 46 & 137 & $i$ & 2.0 \\
~~~~LCOGT~1\,m+CTIO/sinistro & 2015 Apr 19 & 93 & 166 & $i$ & 1.0 \\
~~~~LCOGT~1\,m+CTIO/sinistro & 2015 May 21 & 40 & 165 & $i$ & 1.7 \\
~~~~LCOGT~1\,m+SSO/SBIG & 2015 Jun 04 & 110 & 73 & $i$ & 2.9 \\
\sidehead{\textbf{\hatcur{27}}}
~~~~HS-2.1/G700 & 2011 Apr--2012 Jul & 4603 & 292 & $r$ & 6.3 \\
~~~~HS-4.1/G700 & 2011 Jul--2012 Jul & 3851 & 301 & $r$ & 7.5 \\
~~~~HS-6.1/G700 & 2011 May--2012 Jul & 1512 & 300 & $r$ & 7.1 \\
~~~~PEST~0.3\,m & 2015 Mar 12 & 141 & 132 & $R_{C}$ & 4.1 \\
~~~~LCOGT~1\,m+SSO/SBIG & 2015 Apr 09 & 282 & 75 & $i$ & 2.3 \\
\sidehead{\textbf{\hatcur{28}}}
~~~~HS-1.2/G747 & 2013 Mar--2013 Oct & 4086 & 287 & $r$ & 12.8 \\
~~~~HS-2.2/G747 & 2013 Sep--2013 Oct & 650 & 287 & $r$ & 11.5 \\
~~~~HS-3.2/G747 & 2013 Apr--2013 Nov & 9051 & 297 & $r$ & 12.1 \\
~~~~HS-4.2/G747 & 2013 Sep--2013 Nov & 1464 & 297 & $r$ & 12.5 \\
~~~~HS-5.2/G747 & 2013 Mar--2013 Nov & 6018 & 297 & $r$ & 10.7 \\
~~~~HS-6.2/G747 & 2013 Sep--2013 Nov & 1576 & 290 & $r$ & 11.4 \\
~~~~LCOGT~1\,m+CTIO/sinistro & 2015 Aug 31 & 38 & 223 & $i$ & 1.4 \\
~~~~LCOGT~1\,m+CTIO/sinistro & 2015 Sep 03 & 55 & 223 & $i$ & 1.4 \\
\sidehead{\textbf{\hatcur{29}}}
~~~~HS-1.1/G747 & 2013 Apr--2013 May & 828 & 289 & $r$ & 7.2 \\
~~~~HS-2.1/G747 & 2013 Sep--2013 Oct & 1331 & 287 & $r$ & 7.5 \\
~~~~HS-3.1/G747 & 2013 Apr--2013 Nov & 9121 & 297 & $r$ & 6.1 \\
~~~~HS-4.1/G747 & 2013 Sep--2013 Nov & 1505 & 297 & $r$ & 8.2 \\
~~~~HS-5.1/G747 & 2013 Mar--2013 Nov & 6045 & 297 & $r$ & 6.4 \\
~~~~HS-6.1/G747 & 2013 Sep--2013 Nov & 1544 & 290 & $r$ & 7.2 \\
~~~~LCOGT~1\,m+CTIO/sinistro & 2015 Jun 01 & 90 & 166 & $i$ & 1.2 \\
~~~~LCOGT~1\,m+CTIO/sinistro & 2015 Jun 24 & 36 & 162 & $i$ & 1.0 \\
\sidehead{\textbf{\hatcur{30}}}
~~~~HS-2.3/G754 & 2012 Sep--2012 Dec & 3869 & 282 & $r$ & 6.1 \\
~~~~HS-6.3/G754 & 2012 Sep--2012 Dec & 3000 & 285 & $r$ & 6.2 \\
~~~~HS-2.4/G754 & 2012 Sep--2012 Dec & 3801 & 282 & $r$ & 6.0 \\
~~~~HS-4.4/G754 & 2012 Sep--2013 Jan & 2820 & 292 & $r$ & 6.6 \\
~~~~HS-6.4/G754 & 2012 Sep--2012 Dec & 2977 & 285 & $r$ & 5.7 \\
~~~~HS-1.1/G755 & 2011 Jul--2012 Oct & 5180 & 291 & $r$ & 9.2 \\
~~~~HS-3.1/G755 & 2011 Jul--2012 Oct & 4204 & 287 & $r$ & 7.4 \\
~~~~HS-5.1/G755 & 2011 Jul--2012 Oct & 4904 & 296 & $r$ & 6.5 \\
~~~~LCOGT~1\,m+SAAO/SBIG & 2014 Oct 19 & 50 & 196 & $i$ & 1.2 \\
~~~~LCOGT~1\,m+CTIO/sinistro & 2014 Oct 23 & 56 & 226 & $i$ & 1.0 \\
\enddata
\tablenotetext{a}{
    For HATSouth data we list the HATSouth unit, CCD and field name
    from which the observations are taken. HS-1 and -2 are located at
    Las Campanas Observatory in Chile, HS-3 and -4 are located at the
    H.E.S.S. site in Namibia, and HS-5 and -6 are located at Siding
    Spring Observatory in Australia. Each unit has 4 ccds. Each field
    corresponds to one of 838 fixed pointings used to cover the full
    4$\pi$ celestial sphere. All data from a given HATSouth field and
    CCD number are reduced together, while detrending through External
    Parameter Decorrelation (EPD) is done independently for each
    unique unit+CCD+field combination.
}
\tablenotetext{b}{
    The median time between consecutive images rounded to the nearest
    second. Due to factors such as weather, the day--night cycle,
    guiding and focus corrections the cadence is only approximately
    uniform over short timescales.
}
\tablenotetext{c}{
    The RMS of the residuals from the best-fit model.
}
\ifthenelse{\boolean{emulateapj}}{
    \end{deluxetable*}
}{
    \end{deluxetable}
}

\subsection{Spectroscopic Observations}
\label{sec:obsspec}

The spectroscopic observation of our planetary candidates is a two-step process. The 
first step is ``reconnaissance"  spectroscopy, which consists of observations used both to rule out 
false positive scenarios produced by certain configurations of stellar binaries that could mimic the 
detected transit features, and to estimate rough spectral parameters in order to estimate the physical 
and orbital parameters of the transiting planet candidates. The second step consists of spectroscopic 
observations that allow us to both confirm the planetary nature of the companion by radial velocity (RV) 
variations of the star due to the reflex motion produced by the planetary companion (which allows us to 
estimate its mass) and also to obtain precise stellar parameters from spectroscopic observables in order 
to derive absolute parameters of the planetary companion. The spectroscopic observations are 
summarized in \reftabl{specobs}, and are detailed below. 

\subsubsection{Reconnaissance spectroscopy}

The reconnaissance spectroscopy of our candidates was made using the Wide Field Spectrograph \citep[WiFeS, ][]{dopita:2007}, 
located on the ANU $2.3$m telescope. Details of the observing strategy, reduction methods and the processing of 
the spectra for this instrument can be found in \cite{bayliss:2013:hats3}. In summary, the observing strategy usually consists in taking 
data with two resolutions: $R = \lambda/\Delta \lambda = 7000$ (medium) and $R = 3000$ (low). The former are used to search for RV variations at the $\sim 2$ km/s level in order to rule out possible stellar companions, while the latter are used to estimate the spectroscopic parameters of the host stars. The results for each star were as follows:

\begin{itemize}
\item \textit{HATS-25:} four medium resolution spectra and one low resolution spectrum were obtained. From these, a 
temperature of $5830 \pm 300$ K, $\log(g)$ of $4.4\pm 0.3$, metallicity of $[\textnormal{Fe/H}] = 0.0 \pm 0.5$ 
was derived, implying that the star was a G-type star. No RV variations at the $\sim 2$ km/s level were found.
\item \textit{HATS-26:} two medium resolution spectra and one low resolution spectrum were obtained. No RV variation at 
the $\sim 2$ km/s level was found, and a temperature of $6333 \pm 300$ K, $\log(g)$ of $4.1\pm 0.3$ and a metallicity of 
$[\textnormal{Fe/H}] = 0.0 \pm 0.5$ was derived, which pointed to an F-type star. 
\item \textit{HATS-27:} three medium resolution and one low resolution spectra were obtained. We found no variation at the $\sim 2$ km/s level, and 
a temperature of $6683 \pm 300$ K, $\log(g)$ of $4.5\pm 0.3$  and a metallicity of $[\textnormal{Fe/H}] = 0.0\pm 0.5$ was derived for this star, implying 
it was consistent with being an F-type star. 
\item \textit{HATS-28:} only one low resolution spectrum was obtained. With it, we derived a temperature of $5800 \pm 300$ K, $\log(g)$ of $4.5\pm 0.3$ 
and a metallicity of $[\textnormal{Fe/H}] = 0.0 \pm 0.5$, hinting that this star was a G-type star. 
\item \textit{HATS-29:} four medium resolution spectra and one low resolution spectrum were obtained. No variations 
at the $\sim 2$ km/s level were found, and we derived a temperature of $5658 \pm 300$ K, $\log(g)$ of $4.5\pm 0.3$ and a metallicity of $[\textnormal{Fe/H}] = 0.0\pm 0.5$, 
for this star, finding it to be a G-type star. 
\item \textit{HATS-30:} three medium resolution spectra and one low resolution spectrum were obtained. No variations at the 
$\sim 2$ km/s level in the RVs were found. A temperature of $6155 \pm 300$ K, $\log(g)$ of $4.6\pm 0.3$ and a metallicity of 
$[\textnormal{Fe/H}] = 0.0\pm 0.5$ was derived, which suggested the star was either a hot G-type or a cool F-type star.
\end{itemize}

Given these results, our planet candidates were then promoted to our list requiring high-resolution spectroscopy and high precision 
photometric follow-up observations, which we now detail.

\subsubsection{High-precision spectroscopy}

High-precision spectroscopy was obtained for our targets with different instruments. Several $R=115000$ spectra were taken with 
the High Accuracy Radial Velocity Planet Searcher \citep[HARPS,][]{mayor:2003} on the ESO $3.6$m telescope at La Silla Observatory 
(LSO) between February 2015 and March 2016 in order to obtain high-precision RVs for HATS-25, HATS-26, HATS-27 and HATS-29. 
 Spectra with $R=48000$ were also taken with the FEROS spectrograph \citep{kaufer:1998} mounted on the MPG $2.2$m telescope at LSO between 
July 2014 and July 2015 in order to both extract precise spectroscopic parameters of the host stars (see \refsecl{analysis}) and obtain 
precise RVs for all of our targets. In addition, $R=60000$ spectra were also taken with the CORALIE \citep{queloz:2001} spectrograph 
mounted on the 1.2m Euler telescope at LSO between June and November of 2014 for HATS-26, HATS-27, HATS-29 and HATS-30. The reduction of the CORALIE, FEROS and HARPS spectra followed the procedures described in 
\cite{jordan:2014:hats4} for CORALIE, and adapted to FEROS and HARPS. Finally, eight $R=70000$ spectra were obtained for HATS-29 on May 2015 to measure RVs, using the 
CYCLOPS2 fibre feed with the UCLES spectrograph on the $3.9$m Anglo-Australian Telescope (AAT); the data was reduced following the methods detailed in \cite{addison:2013}.

The phased high-precision RV and bisector span (BS) measurements are shown for each system in \reffigl{rvbis}, while the data are 
listed in \reftabl{rvs}. It is important to note that the large observed scatter and errorbars on the RVs obtained from FEROS for HATS-27 
are both due to the hot temperature of the star and due to contamination by scattered moonlight. Despite of this, it is evident that all the candidates show 
RV variations that are in phase with the photometric ephemeris. In addition, computed correlation coefficients between the RV and 
the BS measurements are all consistent with zero.

\ifthenelse{\boolean{emulateapj}}{
    \begin{deluxetable*}{llrrrrr}
}{
    \begin{deluxetable}{llrrrrrrrr}
}
\tablewidth{0pc}
\tabletypesize{\scriptsize}
\tablecaption{
    Summary of spectroscopy observations 
     \label{tab:specobs}
}
\tablehead{
    \multicolumn{1}{c}{Instrument}          &
    \multicolumn{1}{c}{UT Date(s)}             &
    \multicolumn{1}{c}{\# Spec.}   &
    \multicolumn{1}{c}{Res.}          &
    \multicolumn{1}{c}{S/N Range\tablenotemark{a}}           &
    \multicolumn{1}{c}{$\gamma_{\rm RV}$\tablenotemark{b}} &
    \multicolumn{1}{c}{RV Precision\tablenotemark{c}} \\
    &
    &
    &
    \multicolumn{1}{c}{$\Delta \lambda$/$\lambda$/1000} &
    &
    \multicolumn{1}{c}{(\kms)}              &
    \multicolumn{1}{c}{(\ms)}
}
\startdata
%
\sidehead{\textbf{\hatcur{25}}}
ANU~2.3\,m/WiFeS & 2014 Jun--Aug & 4 & 7 & 26--152 & 30.0 & 4000 \\
ANU~2.3\,m/WiFeS & 2014 Aug 5 & 1 & 3 & 88 & $\cdots$ & $\cdots$ \\
ESO~3.6\,m/HARPS & 2015 Feb--Apr & 8 & 115 & 11--23 & 31.663 & 8.8 \\
MPG~2.2\,m/FEROS & 2015 Apr 9 & 1 & 48 & 64 & 31.649 & 20 \\
\sidehead{\textbf{\hatcur{26}}}
ANU~2.3\,m/WiFeS & 2014 Jun 3--5 & 2 & 7 & 95--107 & -14.4 & 4000 \\
ANU~2.3\,m/WiFeS & 2014 Jun 4 & 1 & 3 & 121 & $\cdots$ & $\cdots$ \\
Euler~1.2\,m/Coralie & 2014 Jun 19--21 & 2 & 60 & 17--19 & -12.489 & 5.2 \\
MPG~2.2\,m/FEROS & 2015 Jan--Feb & 8 & 48 & 56--74 & -12.516 & 21.0 \\
ESO~3.6\,m/HARPS & 2015 Feb 14--19 & 4 & 115 & 19--23 & -12.561 & 21.3 \\
\sidehead{\textbf{\hatcur{27}}}
ANU~2.3\,m/WiFeS & 2014 Jun 2 & 1 & 3 & 50 & $\cdots$ & $\cdots$ \\
ANU~2.3\,m/WiFeS & 2014 Jun 3--5 & 3 & 7 & 4.6--12 & -7.6 & 4000 \\
Euler~1.2\,m/Coralie & 2014 Jun 20--21 & 3 & 60 & 21--22 & -3.521 & 66 \\
MPG~2.2\,m/FEROS & 2014 Jul--2015 Apr & 15 & 48 & 18--92 & -3.525 & 78 \\
ESO~3.6\,m/HARPS & 2014 Aug--2016 Mar & 11 & 115 & 4--25 & -3.582 & 35 \\
\sidehead{\textbf{\hatcur{28}}}
ANU~2.3\,m/WiFeS & 2015 Jun 1 & 1 & 3 & 38 & $\cdots$ & $\cdots$ \\
MPG~2.2\,m/FEROS & 2015 Jun--Jul Apr & 18 & 48 & 17--52 & -8.651 & 38 \\
\sidehead{\textbf{\hatcur{29}}}
ANU~2.3\,m/WiFeS & 2014 Dec--2015 Mar & 4 & 7 & 3.1--31 & -17.5 & 4000 \\
ANU~2.3\,m/WiFeS & 2015 Mar 2 & 1 & 3 & 45 & $\cdots$ & $\cdots$ \\
ESO~3.6\,m/HARPS & 2015 Apr 6--8 & 3 & 115 & 12--23 & -19.719 & 18 \\
AAT~3.9\,m/CYCLOPS & 2015 May 6--9 & 9 & 70 & 16--30 & -19.722 & 40 \\
Euler~1.2\,m/Coralie & 2014 Jun 20--21 & 4 & 60 & 16--19 & -19.698 & 11 \\
MPG~2.2\,m/FEROS & 2015 Jun 13 & 3 & 48 & 48--50 & -19.670 & 20 \\
\sidehead{\textbf{\hatcur{30}}}
MPG~2.2\,m/FEROS & 2014 Oct--Dec & 7 & 48 & 60--96 & -0.079 & 8.3 \\
ANU~2.3\,m/WiFeS & 2014 Oct 4 & 1 & 3 & 233 & $\cdots$ & $\cdots$ \\
ANU~2.3\,m/WiFeS & 2014 Oct 4--10 & 3 & 7 & 87--118 & 1.4 & 4000 \\
Euler~1.2\,m/Coralie & 2014 Oct--Nov & 6 & 60 & 22--30 & -0.112 & 22 \\
\enddata 
\tablenotetext{a}{
    S/N per resolution element near 5180\,\AA for all instruments but CYCLOPS, for which the S/N per resolution 
    element near 5220\,\AA is presented.
}
\tablenotetext{b}{
    For high-precision RV observations included in the orbit
    determination this is the zero-point RV from the best-fit
    orbit. For other instruments it is the mean value. We do not
    provide this quantity for the lower resolution WiFeS observations
    which were only used to measure stellar atmospheric parameters.
}
\tablenotetext{c}{
    For high-precision RV observations included in the orbit
    determination this is the scatter in the RV residuals from the
    best-fit orbit (which may include astrophysical jitter), for other
    instruments this is either an estimate of the precision (not
    including jitter), or the measured standard deviation. We do not
    provide this quantity for low-resolution observations from the
    ANU~2.3\,m/WiFeS.
}
\ifthenelse{\boolean{emulateapj}}{
    \end{deluxetable*}
}{
    \end{deluxetable}
}

%
\setcounter{planetcounter}{1}
%
\ifthenelse{\boolean{emulateapj}}{
    \begin{figure*} [ht]
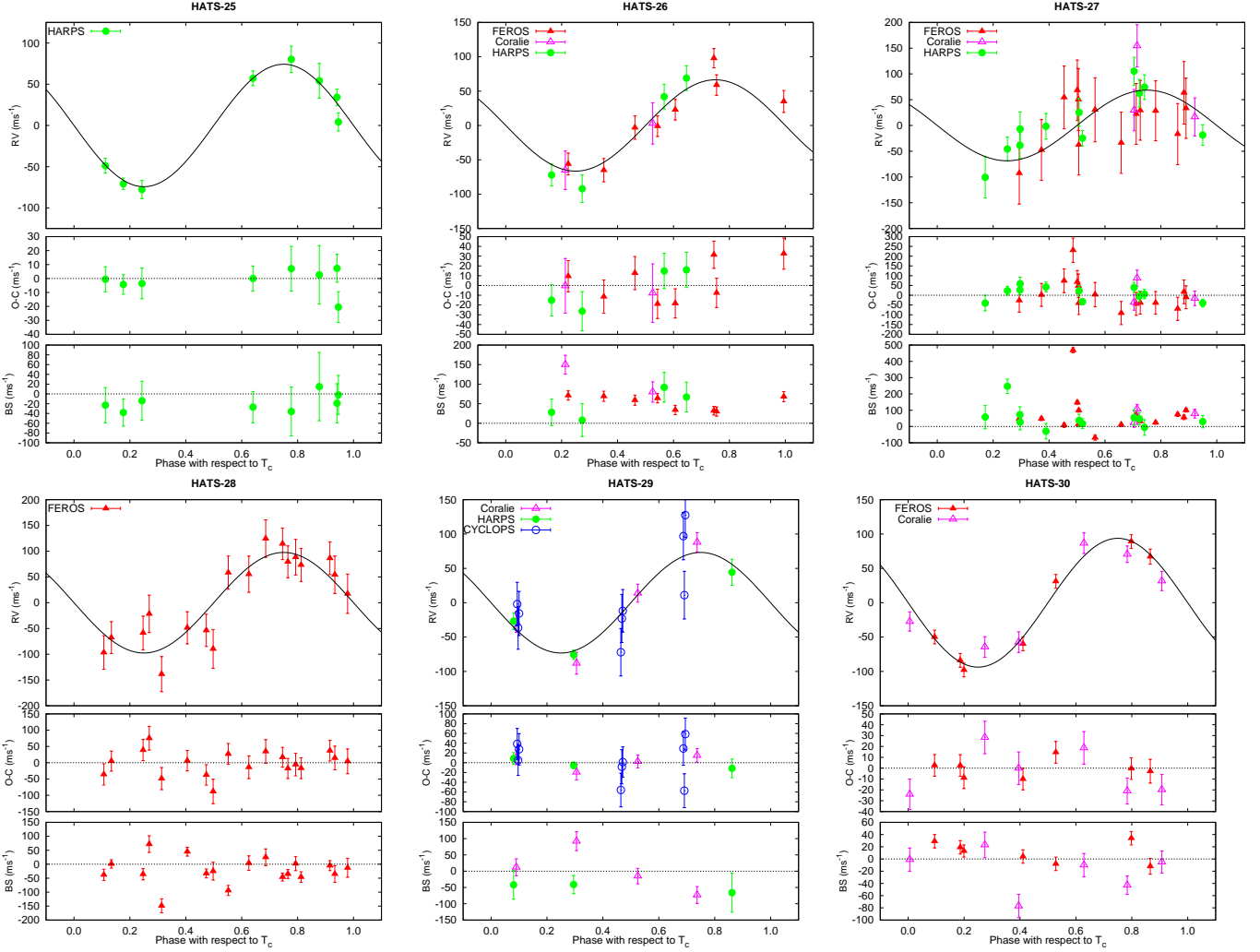

}{
    \begin{figure}[ht]
}
{
\centering
\setlength{\plotwidthtwo}{0.31\linewidth}
\includegraphics[width={\plotwidthtwo}]{\hatcurhtr{25}-rv.eps}
\hfil
\includegraphics[width={\plotwidthtwo}]{\hatcurhtr{26}-rv.eps}
\hfil
\includegraphics[width={\plotwidthtwo}]{\hatcurhtr{27}-rv.eps}
}
{
\centering
\setlength{\plotwidthtwo}{0.31\linewidth}
\includegraphics[width={\plotwidthtwo}]{\hatcurhtr{28}-rv.eps}
\hfil
\includegraphics[width={\plotwidthtwo}]{\hatcurhtr{29}-rv.eps}
\hfil
\includegraphics[width={\plotwidthtwo}]{\hatcurhtr{30}-rv.eps}
}
\caption{
    Phased high-precision RV measurements for the six new transiting planet systems. The instruments used are labelled in the plots. In each case we show three panels. The top panel shows the phased measurements together with our best-fit circular-orbit model (see \reftabl{planetparam}) for each system. Zero-phase corresponds to the time of mid-transit. The center-of-mass velocity has been subtracted. The second panel shows the velocity $O\!-\!C$ residuals from the best fit. The error bars include the jitter terms listed in Tables~\ref{tab:planetparam}~and~\ref{tab:planetparamtwo} added in quadrature to the formal errors for each instrument. The third panel shows the bisector spans (BS). Note the different vertical scales of the panels.
}
\label{fig:rvbis}
\ifthenelse{\boolean{emulateapj}}{
    \end{figure*}
}{
    \end{figure}
}


\subsection{Photometric follow-up observations}
\label{sec:phot}

\setcounter{planetcounter}{1}
%
\begin{figure*}[!ht]
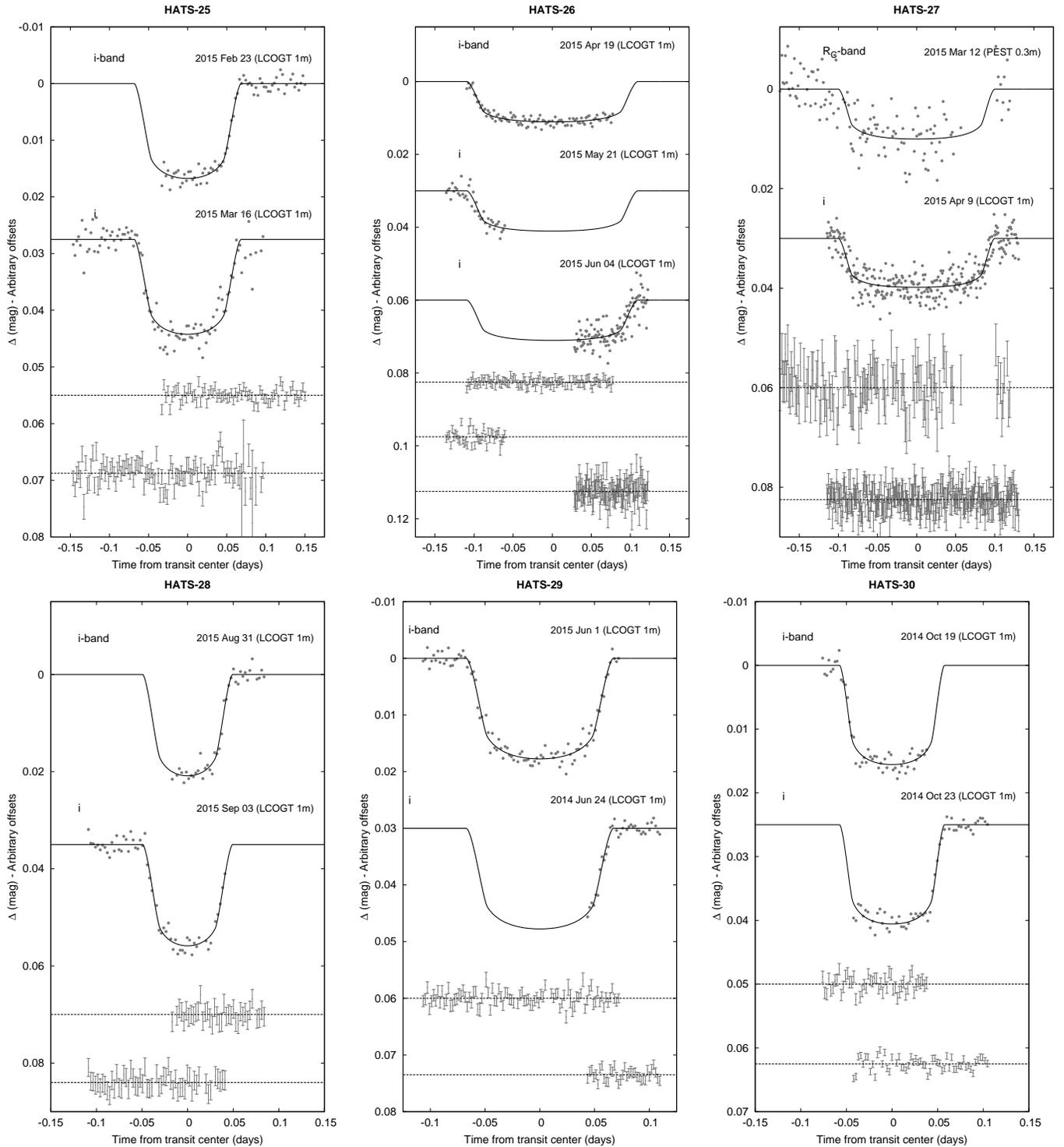

{
\centering
\setlength{\plotwidthtwo}{0.31\linewidth}
\includegraphics[width={\plotwidthtwo}]{\hatcurhtr{25}-lc.eps}
\hfil
\includegraphics[width={\plotwidthtwo}]{\hatcurhtr{26}-lc.eps}
\hfil
\includegraphics[width={\plotwidthtwo}]{\hatcurhtr{27}-lc.eps}
}
{
\centering
\setlength{\plotwidthtwo}{0.31\linewidth}
\includegraphics[width={\plotwidthtwo}]{\hatcurhtr{28}-lc.eps}
\hfil
\includegraphics[width={\plotwidthtwo}]{\hatcurhtr{29}-lc.eps}
\hfil
\includegraphics[width={\plotwidthtwo}]{\hatcurhtr{30}-lc.eps}
}
\caption{
    Unbinned transit \lcs{} for the six new transiting planet systems.  The light curves have been
    corrected for quadratic trends in time fitted simultaneously with
    the transit model, and for correlations with up to three parameters describing the shape of the PSF.
    The dates of the events, filters and instruments used are
    indicated.  Light curves following the first are displaced
    vertically for clarity.  Our best fit from the global modeling
    described in \refsecl{globmod} is shown by the solid lines. The
    residuals from the best-fit model are shown below in the same
    order as the original light curves.  The error bars represent the
    photon and background shot noise, plus the readout noise. Note the differing vertical and horizontal scales used for each system. For \hatcur{25} we do not show the LCOGT~1\,m light curves from UT 2015 Mar 16 and 26 which were taken entirely out of transit.
}
\label{fig:lc}
\end{figure*}
 
Photometric follow-up for the six systems was obtained in order to (1) rule out possible false 
positive scenarios not identified in our reconnaissance spectroscopy (e.g., blended eclipsing binaries, 
hierarchical triples) that would leave signatures in the transit events (e.g., significantly different depths 
between different bands), (2) refine the ephemerides and (3) refine the derived transit parameters obtained 
from the HATSouth discovery lightcurves. Our photometric follow-up observations are summarized in 
Table \ref{tab:photobs} and plotted in \reffigl{lc}.

Photometry for these six systems was obtained mainly from 1m-class telescopes at different sites of 
the Las Cumbres Observatory Global Telescope (LCOGT) network \citep{brown:2013:lcogt}, using the $i$ filter (each of the 
sites used are indicated in Table \ref{tab:photobs}). In particular, one partial transit and a full transit was observed for 
HATS-25b on February 2015 and March 2015 respectively, three partial transits were observed for 
HATS-26b on April, May and June of 2015, one full transit was observed for HATS-27b on April 2015, 
two partial transits were observed for HATS-28b on August and September of 2015, one full transit and 
a partial transit was observed for HATS-29b on June 2015 and 2014, respectively, and two partial transits 
were observed for HATS-30b on October 2014. In addition, one full transit of HATS-27b was observed 
using the $0.3m$ Perth Exoplanet Survey Telescope (PEST) on March of 2015. The instrument specifications, 
observing strategies and reduction of the data have been previously described in \cite{bayliss:2015:hats8} for the 
LCOGT data and in \cite{zhou:2014:hats5} for the PEST data.

\subsection{Lucky imaging observations}
\label{sec:luckyimaging}

\begin{figure*}
\setlength{\plotwidthtwo}{0.31\linewidth}
\plottwo{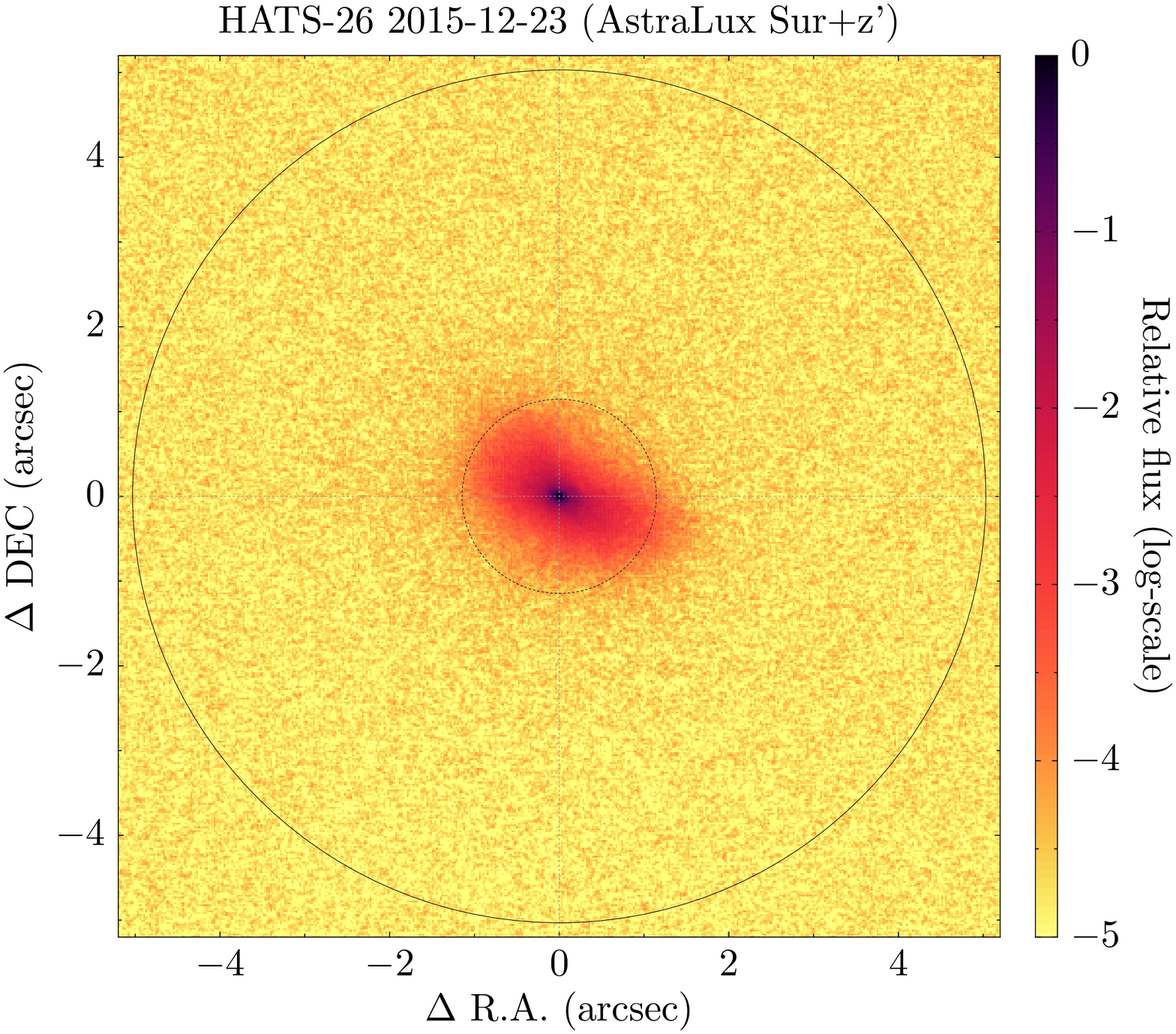}{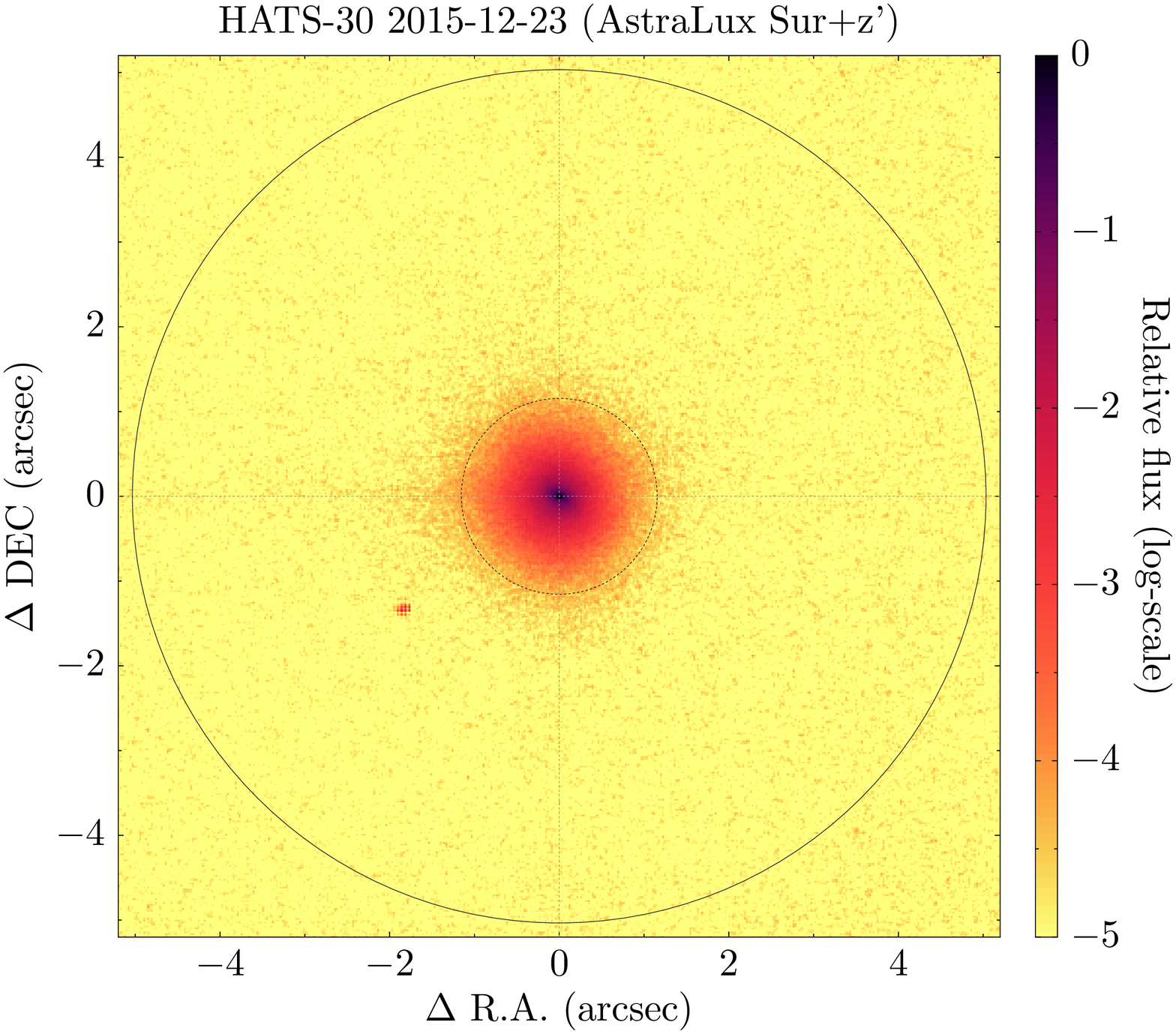}
\caption{(\textit{Left}) AstraLux Sur $z'$-band observations of HATS-26. Circles of $1''$ radius 
(approximately the mean FWHM measured for the image) and $5''$ radius are shown for reference 
on the images. The central lines indicate the fitted center of the star with our PSF modelling (see text). 
(\textit{Right}) Same image but for HATS-30. Note the difference in the shape of the PSF, which 
is a purely instrumental effect.
 \label{fig:lucky-images}}
\end{figure*}

\begin{figure}
\setlength{\plotwidthtwo}{0.31\linewidth}
\plotone{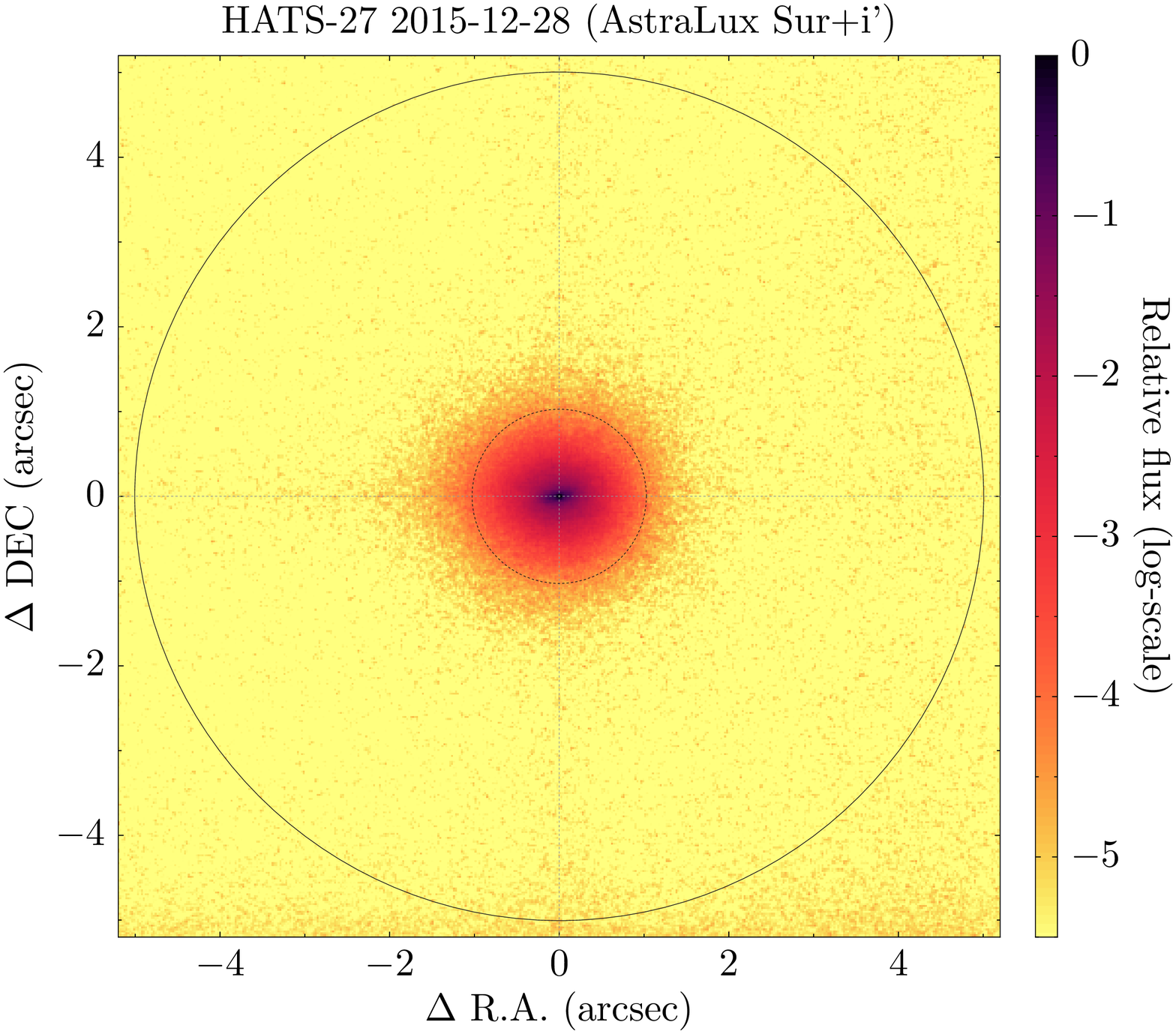}
\caption{AstraLux Sur $i'$-band observations of HATS-27. The circles and lines indicate the same distances and positions as 
the ones described in Figure~\ref{fig:lucky-images}.}
\label{fig:lucky-image}
\end{figure}

As part of a systematic program of obtaining high spatial resolution imaging for HATSouth candidates, ``lucky'' imaging 
observations were obtained for \hatcur{26}, \hatcur{27} and \hatcur{30} using the Astralux Sur camera \citep{hippler:2009} mounted on 
the New Technology Telescope (NTT) at La Silla Observatory, in Chile on December 23 and 28, 2015.

Both the \hatcur{26} and \hatcur{30} datasets, obtained on December 23, were obtained using the SDSS $z'$ filter, while the 
\hatcur{27} dataset, obtained on December 28, was obtained using the SDSS $i'$ filter. A drizzle algorithm \citep{drizzle2002} was used to combine 
the images, selecting the best of them from the set of $\sim10^4$ exposures taken for each target ($10^4$ images with an exposure time of 40 ms 
each for HATS-26, $2\times 10^4$ images with an exposure time of 15 ms each for HATS-27 and $2\times 10^4$ images with 
an exposure time of 15 ms each for HATS-30). Figure~\ref{fig:lucky-images} shows the resulting images for HATS-26 and HATS-30 
and Figure~\ref{fig:lucky-image} shows the resulting image for HATS-27, all of which are the combination of the best 
$10\%$ of the images acquired for each target. The resulting images show an asymmetric extended profile for HATS-26 (a purely 
instrumental effect as confirmed by taking images of other targets on different nights), whereas the profile is fairly symmetric for HATS-27 
and HATS-30 (we note that the latter shows an instrumental artefact close to $(-2,-2)$ arcsecs from the target star). As can be seen from our 
images, no obvious companions were detected out to a $5\arcsec$ radius. 

In order to extract quantitative information from these images, we generated $5\sigma$ contrast curves for each of our targets, which required us 
to model the Point Spread Functions (PSFs). We decided to model the PSFs of our targets as a weighted sum of a Moffat profile (which models 
the central part of the PSF) and an asymmetric Gaussian (to model asymmetries in the PSF wings). The full width at half maximum (FWHM) of the 
full model was measured numerically at 100 different angles by finding the points at which the model has half of the peak flux, and the median of 
these measurements (the ``effective" FWHM, $\rm{FWHM}_{\rm eff}$) is taken as the resolution limit of our observations. For HATS-26, we found 
$\rm{FWHM}_{\rm eff} = 3.27\pm 0.35$ pixels, which given the pixel scale of $23$ milli-arcseconds (mas) per pixel, gives a resolution limit of $75\pm 8$ 
mas. For HATS-27, we found $\rm{FWHM}_{\rm eff} = 3.17\pm0.28$ pixels, which implies a resolution limit of 
$72\pm 6$ mas. Finally, for HATS-30, $\rm{FWHM}_{\rm eff} = 3.55\pm0.29$ pixels, which implies a resolution limit of 
$81\pm 7$ mas. All the effective FWHMs are close to the diffraction limit of the instrument, which is $\sim 50$ mas \citep{hippler:2009}. 

\begin{figure}
\setlength{\plotwidthtwo}{0.31\linewidth}
\epsscale{1.0}
\plotone{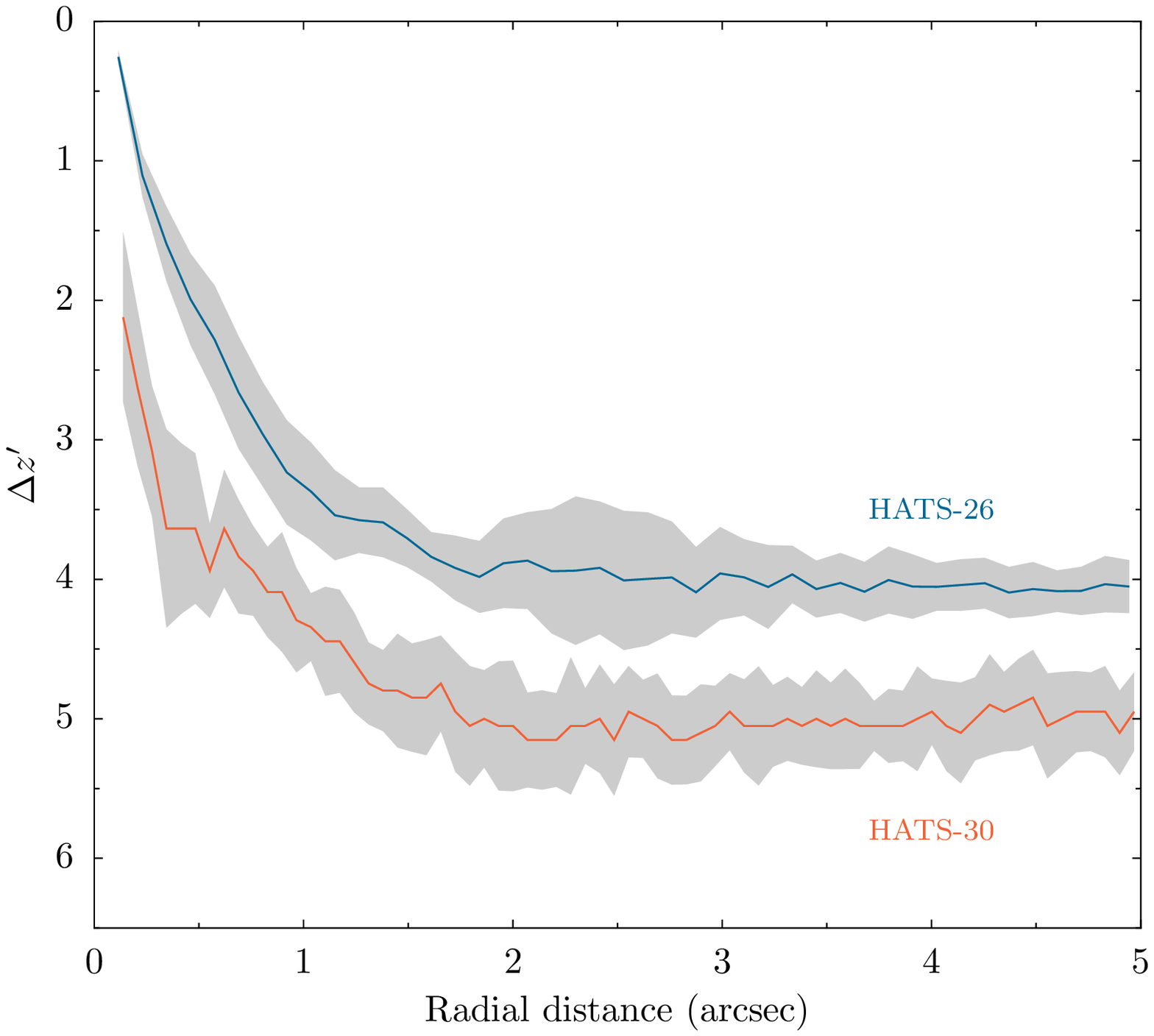}
\caption{Contrast curves generated for HATS-26 (blue) and HATS-30 (orange) using our AstraLux Sur $z'$-band observations. Gray bands 
show the uncertainty given by the scatter in the contrast in the azimuthal direction at a given radius (see text for details).}
\label{fig:lucky-contrast}
\end{figure}

\begin{figure}
\setlength{\plotwidthtwo}{0.31\linewidth}
\epsscale{1.0}
\plotone{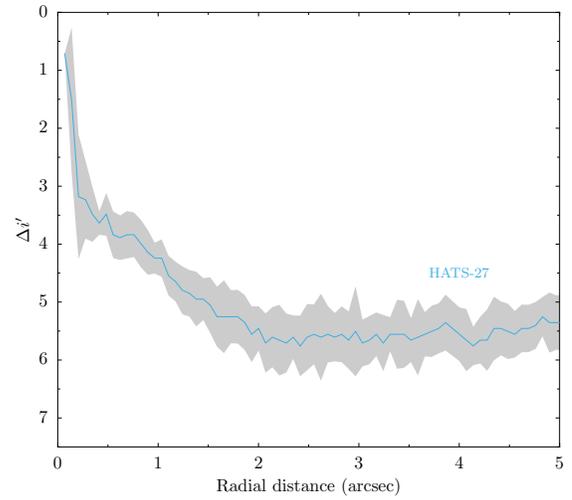}
\caption{Contrast curve generated for HATS-27 using our AstraLux Sur $i'$-band observations. Gray bands 
show the uncertainty given by the scatter in the contrast in the azimuthal direction at a given radius (see text for details).}
\label{fig:lucky-contrast2}
\end{figure}

Once modelled, we subtracted the PSF of the target stars from the images and generated the contrast curves by an ``injection and recovery" approach, in which 
we injected signals with the same fitted PSF parameters at different positions ($r$,$\theta$) in the image, where $r$ is the distance from the target star and 
$\theta$ is the azimuthal angle around it. We sampled $r$ in steps of $\rm{FWHM}_{\rm eff}$, while the angles are sampled at each radius covering $2\pi$ radians 
with independent regions of arc-length equal to $\rm{FWHM}_{\rm eff}$. The injected sources 
were scaled in order to simulate a wide range of contrasts, exploring from $\Delta z' = 0$ to $\Delta z' = 10$ in $0.01$ steps, where $\Delta z'$ is the magnitude contrast 
with respect to the target star. We considered an injected source to be detectable if five or more pixels were $5\sigma$ above the noise level, which was estimated as the 
standard deviation in a box of size $\rm{FWHM}_{\rm eff}\times \rm{FWHM}_{\rm eff}$ at each position in the residual image at which the signals were injected. Finally, the 
contrast at each radius was obtained by averaging the azimuthal contrasts and the standard deviation of these azimuthal contrasts was taken as the error on the contrast 
at each radius. The resulting contrast curves for HATS-26 (blue) and HATS-30 (orange) are shown on Figure \ref{fig:lucky-contrast}, where the grey bands show the uncertainty 
of the contrast at each radius. The corresponding contrast curve for HATS-27 is shown in Figure \ref{fig:lucky-contrast2}. Code to model the PSFs of images as explained here 
and to generate these contrast curves can be found at \url{https://github.com/nespinoza/luckyimg-reduction}.

\clearpage

%
%
\ifthenelse{\boolean{emulateapj}}{
    \begin{deluxetable*}{llrrrrl}
}{
    \begin{deluxetable}{llrrrrl}
}
\tablewidth{0pc}
\tablecaption{
    Light curve data for \hatcur{25}--\hatcur{30}\label{tab:phfu}.
}
\tablehead{
    \colhead{Object\tablenotemark{a}} &
    \colhead{BJD\tablenotemark{b}} & 
    \colhead{Mag\tablenotemark{c}} & 
    \colhead{\ensuremath{\sigma_{\rm Mag}}} &
    \colhead{Mag(orig)\tablenotemark{d}} & 
    \colhead{Filter} &
    \colhead{Instrument} \\
    \colhead{} &
    \colhead{\hbox{~~~~(2,400,000$+$)~~~~}} & 
    \colhead{} & 
    \colhead{} &
    \colhead{} & 
    \colhead{} &
    \colhead{}
}
\startdata
   HATS-27 & $ 56076.42690 $ & $   0.00302 $ & $   0.00469 $ & $ \cdots $ & $ r$ &         HS\\
   HATS-27 & $ 56090.33807 $ & $   0.00531 $ & $   0.00449 $ & $ \cdots $ & $ r$ &         HS\\
   HATS-27 & $ 55955.86458 $ & $  -0.00442 $ & $   0.00406 $ & $ \cdots $ & $ r$ &         HS\\
   HATS-27 & $ 56113.52396 $ & $   0.00956 $ & $   0.00462 $ & $ \cdots $ & $ r$ &         HS\\
   HATS-27 & $ 56016.14672 $ & $   0.00478 $ & $   0.00406 $ & $ \cdots $ & $ r$ &         HS\\
   HATS-27 & $ 56062.51729 $ & $  -0.00009 $ & $   0.00864 $ & $ \cdots $ & $ r$ &         HS\\
   HATS-27 & $ 56020.78399 $ & $   0.00065 $ & $   0.00445 $ & $ \cdots $ & $ r$ &         HS\\
   HATS-27 & $ 56006.87363 $ & $  -0.01152 $ & $   0.00436 $ & $ \cdots $ & $ r$ &         HS\\
   HATS-27 & $ 56030.05986 $ & $  -0.00319 $ & $   0.00425 $ & $ \cdots $ & $ r$ &         HS\\
   HATS-27 & $ 56076.43037 $ & $   0.00830 $ & $   0.00472 $ & $ \cdots $ & $ r$ &         HS\\
\enddata
\tablenotetext{a}{
    Either \hatcur{25}, \hatcur{26}, \hatcur{27}, \hatcur{28}, \hatcur{29} or \hatcur{30}.
}
\tablenotetext{b}{
    Barycentric Julian Date is computed directly from the UTC time
    without correction for leap seconds.
}
\tablenotetext{c}{
    The out-of-transit level has been subtracted. For observations
    made with the HATSouth instruments (identified by ``HS'' in the
    ``Instrument'' column) these magnitudes have been corrected for
    trends using the EPD and TFA procedures applied {\em prior} to
    fitting the transit model. This procedure may lead to an
    artificial dilution in the transit depths. The blend factors for
    the HATSouth light curves are listed in
    Tables~\ref{tab:planetparam}~and~\ref{tab:planetparamtwo}. For
    observations made with follow-up instruments (anything other than
    ``HS'' in the ``Instrument'' column), the magnitudes have been
    corrected for a quadratic trend in time, and for variations
    correlated with three PSF shape parameters, fit simultaneously
    with the transit.
}
\tablenotetext{d}{
    Raw magnitude values without correction for the quadratic trend in
    time, or for trends correlated with the shape of the PSF. These are only
    reported for the follow-up observations.
}
\tablecomments{
    This table is available in a machine-readable form in the online
    journal.  A portion is shown here for guidance regarding its form
    and content.
}
\ifthenelse{\boolean{emulateapj}}{
    \end{deluxetable*}
}{
    \end{deluxetable}
}

\section{Analysis}
\label{sec:analysis}
\subsection{Properties of the parent stars}
\label{sec:stelparam}

We determine the properties of the host stars using the Zonal Atmospherical Stellar Parameter Estimator 
(\texttt{ZASPE}, Brahm et al., in preparation) on median combined FEROS spectra for all our 
systems except for HATS-25, where only one FEROS spectrum was used. With the effective temperature 
($T_{\textnormal{eff}*}$), log-gravity (\loggstar) metallicity ([Fe/H]) and the projected stellar rotational velocity of the star 
($v\sin i$) calculated for each of our systems, the Yonsei-Yale \citep[Y$^2$, ][]{yi:2001} 
isochrones were used to obtain the physical parameters of the host stars. However, instead of using $\loggstar$ 
to search for the best-fit isochrone, we follow \citet{sozzetti:2007} in using the 
stellar density ($\rho_*$), which is well constrained parameter by our transit fits. 
Once this was done and physical parameters were found, a second ZASPE iteration was 
done for all systems except for HATS-27, for which a second iteration did not improve the results. In this second iteration, the revised value of $\loggstar$ 
was used as input in order to derive the final properties of the stars. In order to calculate the distances to these stars, 
we compared their measured broad-band photometry  to the predicted magnitudes in each
filter from the isochrones, assuming an extinction law from \citet{cardelli:1989} with $R_{V} = 3.1$.
The resulting parameters for \hatcur{25}, \hatcur{26} and \hatcur{27} are given in \reftabl{stellar}, and 
for \hatcur{28}, \hatcur{29} and \hatcur{30} in \reftabl{stellartwo}. The locations of each star on an $\teffstar$--$\rhostar$ 
diagram (similar to a Hertzsprung-Russell diagram) are shown in \reffigl{iso}.

It is interesting to note that while HATS-25, HATS-28, HATS-29 and HATS-30 are typical G dwarfs, 
HATS-26 and HATS-27 stand out as slightly evolved F stars which are just after and in the 
turn-off points, respectively. Consequently, they have radii of $2.04^{+0.15}_{-0.11}R_\Sun$ and 
$1.74^{+0.17}_{-0.10}R_\Sun$ which (combined with their effective 
temperatures of $6071\pm 81$ K and $6438 \pm 64$ K, respectively) implies relatively 
large luminosities of $5.06^{+0.90}_{-0.64}L_\Sun$ and $4.67^{+0.92}_{-0.58}L_\Sun$. Because of this, their 
planets receive larger insolation levels than typical hot Jupiters with the same periods.

\ifthenelse{\boolean{emulateapj}}{
    \begin{figure*}[!ht]
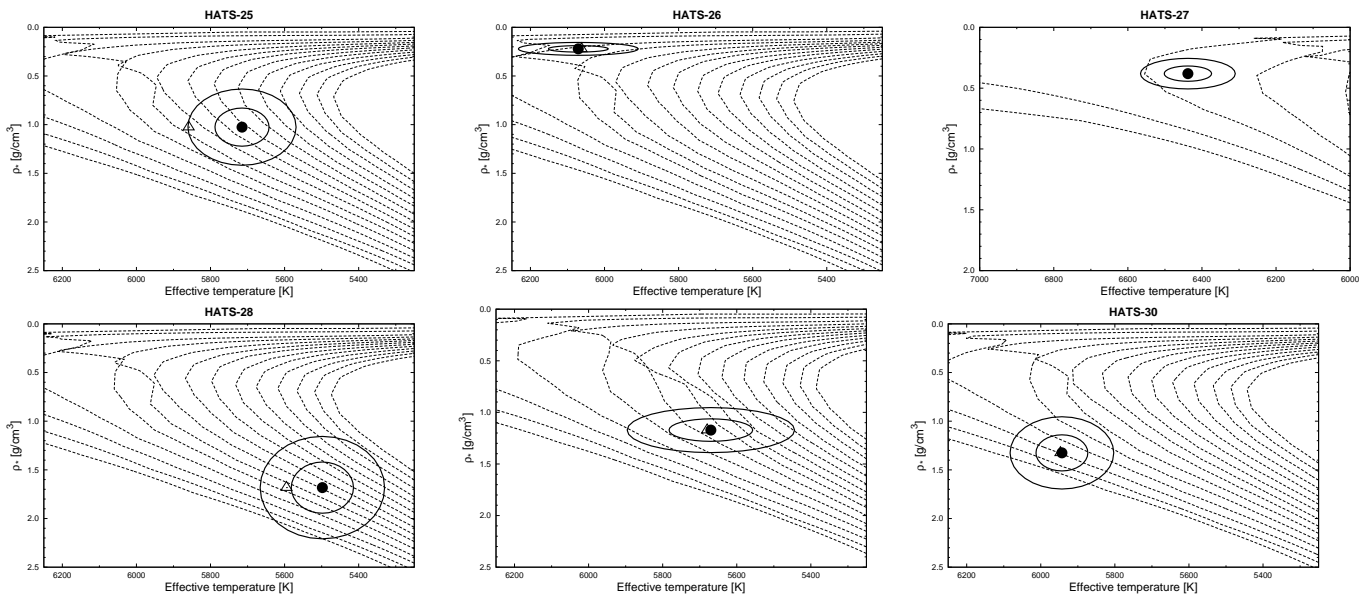

}{
    \begin{figure}[!ht]
}
{
\centering
\setlength{\plotwidthtwo}{0.31\linewidth}
\includegraphics[width={\plotwidthtwo}]{\hatcurhtr{25}-iso-rho.eps}
\hfil
\includegraphics[width={\plotwidthtwo}]{\hatcurhtr{26}-iso-rho.eps}
\hfil
\includegraphics[width={\plotwidthtwo}]{\hatcurhtr{27}-iso-rho.eps}
}
{
\centering
\setlength{\plotwidthtwo}{0.31\linewidth}
\includegraphics[width={\plotwidthtwo}]{\hatcurhtr{28}-iso-rho.eps}
\hfil
\includegraphics[width={\plotwidthtwo}]{\hatcurhtr{29}-iso-rho.eps}
\hfil
\includegraphics[width={\plotwidthtwo}]{\hatcurhtr{30}-iso-rho.eps}
}
\caption{
    Model isochrones from \cite{\hatcurisocite{25}} for the measured
    metallicities of each of the six new transiting planet host stars. We show models for ages of 0.2\,Gyr and 1.0 to 14.0\,Gyr in 1.0\,Gyr increments (ages increasing from left to right). The
    adopted values of $\teffstar$ and \rhostar\ are shown together with
    their 1$\sigma$ and 2$\sigma$ confidence ellipsoids.  The initial
    values of \teffstar\ and \rhostar\ from the first ZASPE and \lc\
    analyses are represented with a triangle.
}
\label{fig:iso}
\ifthenelse{\boolean{emulateapj}}{
    \end{figure*}
}{
    \end{figure}
}

%
%
\ifthenelse{\boolean{emulateapj}}{
    \begin{deluxetable*}{lcccl}
}{
    \begin{deluxetable}{lcccl}
}
\tablewidth{0pc}
\tabletypesize{\footnotesize}
\tablecaption{
    Stellar parameters for \hatcur{25}, \hatcur{26} and \hatcur{27}
    \label{tab:stellar}
}
\tablehead{
    \multicolumn{1}{c}{} &
    \multicolumn{1}{c}{\bf HATS-25} &
    \multicolumn{1}{c}{\bf HATS-26} &
    \multicolumn{1}{c}{\bf HATS-27} &
    \multicolumn{1}{c}{} \\
    \multicolumn{1}{c}{~~~~~~~~Parameter~~~~~~~~} &
    \multicolumn{1}{c}{Value}                     &
    \multicolumn{1}{c}{Value}                     &
    \multicolumn{1}{c}{Value}                     &
    \multicolumn{1}{c}{Source}
}
\startdata
\noalign{\vskip -3pt}
\sidehead{Astrometric properties and cross-identifications}
~~~~2MASS-ID\dotfill               & \hatcurCCtwomass{25}  & \hatcurCCtwomass{26} & \hatcurCCtwomass{27} & \\
~~~~GSC-ID\dotfill                 & \hatcurCCgsc{25}      & \hatcurCCgsc{26}     & \hatcurCCgsc{27}     & \\
~~~~R.A. (J2000)\dotfill            & \hatcurCCra{25}       & \hatcurCCra{26}    & \hatcurCCra{27}    & 2MASS\\
~~~~Dec. (J2000)\dotfill            & \hatcurCCdec{25}      & \hatcurCCdec{26}   & \hatcurCCdec{27}   & 2MASS\\
~~~~$\mu_{\rm R.A.}$ (\masy)              & \hatcurCCpmra{25}     & \hatcurCCpmra{26} & \hatcurCCpmra{27} & UCAC4\\
~~~~$\mu_{\rm Dec.}$ (\masy)              & \hatcurCCpmdec{25}    & \hatcurCCpmdec{26} & \hatcurCCpmdec{27} & UCAC4\\
\sidehead{Spectroscopic properties}
~~~~$\teffstar$ (K)\dotfill         &  \hatcurSMEteff{25}   & \hatcurSMEteff{26} & \hatcurSMEteff{27} & ZASPE\tablenotemark{a}\\
~~~~$\feh$\dotfill                  &  \hatcurSMEzfeh{25}   & \hatcurSMEzfeh{26} & \hatcurSMEzfeh{27} & ZASPE               \\
~~~~$\vsini$ (\kms)\dotfill         &  \hatcurSMEvsin{25}   & \hatcurSMEvsin{26} & \hatcurSMEvsin{27} & ZASPE                \\
~~~~$\vmac$ (\kms)\dotfill          &  $3.90$   & $4.44$ & $5.01$ & Assumed              \\
~~~~$\vmic$ (\kms)\dotfill          &  $1.04$   & $1.29$ & $1.67$ & Assumed              \\
~~~~$\gamma_{\rm RV}$ (\ms)\dotfill&  \hatcurRVgammaabs{25}  & \hatcurRVgammaabs{26} & \hatcurRVgammaabs{27} & FEROS or HARPS\tablenotemark{b}  \\
\sidehead{Photometric properties}
~~~~$B$ (mag)\dotfill               &  \hatcurCCtassmB{25}  & \hatcurCCtassmB{26} & \hatcurCCtassmB{27} & APASS\tablenotemark{c} \\
~~~~$V$ (mag)\dotfill               &  \hatcurCCtassmv{25}  & \hatcurCCtassmv{26} & \hatcurCCtassmv{27} & APASS\tablenotemark{c} \\
~~~~$g$ (mag)\dotfill               &  \hatcurCCtassmg{25}  & \hatcurCCtassmg{26} & \hatcurCCtassmg{27} & APASS\tablenotemark{c} \\
~~~~$r$ (mag)\dotfill               &  \hatcurCCtassmr{25}  & \hatcurCCtassmr{26} & \hatcurCCtassmr{27} & APASS\tablenotemark{c} \\
~~~~$i$ (mag)\dotfill               &  \hatcurCCtassmi{25}  & \hatcurCCtassmi{26} & \hatcurCCtassmi{27} & APASS\tablenotemark{c} \\
~~~~$J$ (mag)\dotfill               &  \hatcurCCtwomassJmag{25} & \hatcurCCtwomassJmag{26} & \hatcurCCtwomassJmag{27} & 2MASS           \\
~~~~$H$ (mag)\dotfill               &  \hatcurCCtwomassHmag{25} & \hatcurCCtwomassHmag{26} & \hatcurCCtwomassHmag{27} & 2MASS           \\
~~~~$K_s$ (mag)\dotfill             &  \hatcurCCtwomassKmag{25} & \hatcurCCtwomassKmag{26} & \hatcurCCtwomassKmag{27} & 2MASS           \\
\sidehead{Derived properties}
~~~~$\mstar$ ($\msun$)\dotfill      &  \hatcurISOmlong{25}   & \hatcurISOmlong{26} & \hatcurISOmlong{27} & YY+$\rhostar$+ZASPE \tablenotemark{d}\\
~~~~$\rstar$ ($\rsun$)\dotfill      &  \hatcurISOrlong{25}   & \hatcurISOrlong{26} & \hatcurISOrlong{27} & YY+$\rhostar$+ZASPE         \\
~~~~$\loggstar$ (cgs)\dotfill       &  \hatcurISOlogg{25}    & \hatcurISOlogg{26} & \hatcurISOlogg{27} & YY+$\rhostar$+ZASPE         \\
~~~~$\rhostar$ (\gcmc)\dotfill       &  \hatcurLCrho{25}    & \hatcurLCrho{26} & \hatcurLCrho{27} & Light curves         \\
~~~~$\rhostar$ (\gcmc) \tablenotemark{e}\dotfill       &  \hatcurISOrho{25}    & \hatcurISOrho{26} & \hatcurISOrho{27} & YY+Light curves+ZASPE          \\
~~~~$\lstar$ ($\lsun$)\dotfill      &  \hatcurISOlum{25}     & \hatcurISOlum{26} & \hatcurISOlum{27} & YY+$\rhostar$+ZASPE         \\
~~~~$M_V$ (mag)\dotfill             &  \hatcurISOmv{25}      & \hatcurISOmv{26} & \hatcurISOmv{27} & YY+$\rhostar$+ZASPE         \\
~~~~$M_K$ (mag,\hatcurjhkfilset{25})\dotfill &  \hatcurISOMK{25} & \hatcurISOMK{26} & \hatcurISOMK{27} & YY+$\rhostar$+ZASPE         \\
~~~~Age (Gyr)\dotfill               &  \hatcurISOage{25}     & \hatcurISOage{26} & \hatcurISOage{27} & YY+$\rhostar$+ZASPE         \\
~~~~$A_{V}$ (mag)\dotfill               &  \hatcurXAv{25}     & \hatcurXAv{26} & \hatcurXAv{27} & YY+$\rhostar$+ZASPE         \\
~~~~Distance (pc)\dotfill           &  \hatcurXdistred{25}\phn  & \hatcurXdistred{26} & \hatcurXdistred{27} & YY+$\rhostar$+ZASPE\\ [-1.5ex]
\enddata
\tablecomments{
For \hatcur{25} and \hatcur{26} the fixed-circular-orbit model has a higher Bayesian evidence than the eccentric-orbit model (it is 10 and 8 times greater for these two systems respectively). We therefore assume a fixed circular orbit in generating the parameters listed for both of these systems. For \hatcur{27} the free-eccentricity model has an indistinguishable Bayesian evidence from the fixed-circular model, but in this case the eccentricity is poorly constrained with implausibly high values permitted by the low S/N RV measurements. For this system we also adopt the fixed-circular model parameters.
}
\tablenotetext{a}{
    ZASPE = Zonal Atmospherical Stellar Parameter Estimator routine
    for the analysis of high-resolution spectra (Brahm et al.~2016, in
    preparation), applied to the FEROS spectra of \hatcur{25} and \hatcur{26}. These
    parameters rely primarily on ZASPE, but have a small dependence
    also on the iterative analysis incorporating the isochrone search
    and global modeling of the data.
}
\tablenotetext{b}{
    From FEROS for \hatcur{26} and from HARPS for \hatcur{25} and
    \hatcur{27}. The error on $\gamma_{\rm RV}$ is determined from the
    orbital fit to the RV measurements, and does not include the
    systematic uncertainty in transforming the velocities to the IAU
    standard system. The velocities have \textbf{not} been corrected for gravitational redshifts.
} \tablenotetext{c}{
    From APASS DR6 for as
    listed in the UCAC 4 catalog \citep{zacharias:2012:ucac4}.  
}
\tablenotetext{d}{
    \hatcurisoshort{25}+\rhostar+ZASPE = Based on the \hatcurisoshort{25}
    isochrones \citep{\hatcurisocite{25}}, \rhostar\ as a luminosity
    indicator, and the ZASPE results.
}
\tablenotetext{e}{
    In the case of $\rhostar$ we list two values. The first value is
    determined from the global fit to the light curves and RV data,
    without imposing a constraint that the parameters match the
    stellar evolution models. The second value results from
    restricting the posterior distribution to combinations of
    $\rhostar$+$\teffstar$+$\feh$ that match to a \hatcurisoshort{25}
    stellar model.
}
\ifthenelse{\boolean{emulateapj}}{
    \end{deluxetable*}
}{
    \end{deluxetable}
}

%
%
\ifthenelse{\boolean{emulateapj}}{
    \begin{deluxetable*}{lcccl}
}{
    \begin{deluxetable}{lcccl}
}
\tablewidth{0pc}
\tabletypesize{\footnotesize}
\tablecaption{
    Stellar parameters for \hatcur{28}, \hatcur{29} and \hatcur{30}
    \label{tab:stellartwo}
}
\tablehead{
    \multicolumn{1}{c}{} &
    \multicolumn{1}{c}{\bf HATS-28} &
    \multicolumn{1}{c}{\bf HATS-29} &
    \multicolumn{1}{c}{\bf HATS-30} &
    \multicolumn{1}{c}{} \\
    \multicolumn{1}{c}{~~~~~~~~Parameter~~~~~~~~} &
    \multicolumn{1}{c}{Value}                     &
    \multicolumn{1}{c}{Value}                     &
    \multicolumn{1}{c}{Value}                     &
    \multicolumn{1}{c}{Source}
}
\startdata
\noalign{\vskip -3pt}
\sidehead{Astrometric properties and cross-identifications}
~~~~2MASS-ID\dotfill               & \hatcurCCtwomass{28}  & \hatcurCCtwomass{29} & \hatcurCCtwomass{30} & \\
~~~~GSC-ID\dotfill                 & \hatcurCCgsc{28}      & \hatcurCCgsc{29}     & \hatcurCCgsc{30}     & \\
~~~~R.A. (J2000)\dotfill            & \hatcurCCra{28}       & \hatcurCCra{29}    & \hatcurCCra{30}    & 2MASS\\
~~~~Dec. (J2000)\dotfill            & \hatcurCCdec{28}      & \hatcurCCdec{29}   & \hatcurCCdec{30}   & 2MASS\\
~~~~$\mu_{\rm R.A.}$ (\masy)              & \hatcurCCpmra{28}     & \hatcurCCpmra{29} & \hatcurCCpmra{30} & UCAC4\\
~~~~$\mu_{\rm Dec.}$ (\masy)              & \hatcurCCpmdec{28}    & \hatcurCCpmdec{29} & \hatcurCCpmdec{30} & UCAC4\\
\sidehead{Spectroscopic properties}
~~~~$\teffstar$ (K)\dotfill         &  \hatcurSMEteff{28}   & \hatcurSMEteff{29} & \hatcurSMEteff{30} & ZASPE\tablenotemark{a}\\
~~~~$\feh$\dotfill                  &  \hatcurSMEzfeh{28}   & \hatcurSMEzfeh{29} & \hatcurSMEzfeh{30} & ZASPE               \\
~~~~$\vsini$ (\kms)\dotfill         &  \hatcurSMEvsin{28}   & \hatcurSMEvsin{29} & \hatcurSMEvsin{30} & ZASPE                \\
~~~~$\vmac$ (\kms)\dotfill          &  $3.56$   & $3.83$ & $4.25$ & Assumed              \\
~~~~$\vmic$ (\kms)\dotfill          &  $0.93$   & $1.02$ & $1.19$ & Assumed              \\
~~~~$\gamma_{\rm RV}$ (\ms)\dotfill&  \hatcurRVgammaabs{28}  & \hatcurRVgammaabs{29} & \hatcurRVgammaabs{30} & FEROS or HARPS\tablenotemark{b}  \\
\sidehead{Photometric properties}
~~~~$B$ (mag)\dotfill               &  \hatcurCCtassmB{28}  & \hatcurCCtassmB{29} & \hatcurCCtassmB{30} & APASS\tablenotemark{c} \\
~~~~$V$ (mag)\dotfill               &  \hatcurCCtassmv{28}  & \hatcurCCtassmv{29} & \hatcurCCtassmv{30} & APASS\tablenotemark{c} \\
~~~~$g$ (mag)\dotfill               &  \hatcurCCtassmg{28}  & \hatcurCCtassmg{29} & \hatcurCCtassmg{30} & APASS\tablenotemark{c} \\
~~~~$r$ (mag)\dotfill               &  \hatcurCCtassmr{28}  & \hatcurCCtassmr{29} & \hatcurCCtassmr{30} & APASS\tablenotemark{c} \\
~~~~$i$ (mag)\dotfill               &  \hatcurCCtassmi{28}  & \hatcurCCtassmi{29} & \hatcurCCtassmi{30} & APASS\tablenotemark{c} \\
~~~~$J$ (mag)\dotfill               &  \hatcurCCtwomassJmag{28} & \hatcurCCtwomassJmag{29} & \hatcurCCtwomassJmag{30} & 2MASS           \\
~~~~$H$ (mag)\dotfill               &  \hatcurCCtwomassHmag{28} & \hatcurCCtwomassHmag{29} & \hatcurCCtwomassHmag{30} & 2MASS           \\
~~~~$K_s$ (mag)\dotfill             &  \hatcurCCtwomassKmag{28} & \hatcurCCtwomassKmag{29} & \hatcurCCtwomassKmag{30} & 2MASS           \\
\sidehead{Derived properties}
~~~~$\mstar$ ($\msun$)\dotfill      &  \hatcurISOmlong{28}   & \hatcurISOmlong{29} & \hatcurISOmlong{30} & YY+$\rhostar$+ZASPE \tablenotemark{d}\\
~~~~$\rstar$ ($\rsun$)\dotfill      &  \hatcurISOrlong{28}   & \hatcurISOrlong{29} & \hatcurISOrlong{30} & YY+$\rhostar$+ZASPE         \\
~~~~$\loggstar$ (cgs)\dotfill       &  \hatcurISOlogg{28}    & \hatcurISOlogg{29} & \hatcurISOlogg{30} & YY+$\rhostar$+ZASPE         \\
~~~~$\rhostar$ (\gcmc)\dotfill       &  \hatcurLCrho{28}    & \hatcurLCrho{29} & \hatcurLCrho{30} & Light curves         \\
~~~~$\rhostar$ (\gcmc) \tablenotemark{e}\dotfill       &  \hatcurISOrho{28}    & \hatcurISOrho{29} & \hatcurISOrho{30} & YY+Light Curves+ZASPE         \\
~~~~$\lstar$ ($\lsun$)\dotfill      &  \hatcurISOlum{28}     & \hatcurISOlum{29} & \hatcurISOlum{30} & YY+$\rhostar$+ZASPE         \\
~~~~$M_V$ (mag)\dotfill             &  \hatcurISOmv{28}      & \hatcurISOmv{29} & \hatcurISOmv{30} & YY+$\rhostar$+ZASPE         \\
~~~~$M_K$ (mag,\hatcurjhkfilset{28})\dotfill &  \hatcurISOMK{28} & \hatcurISOMK{29} & \hatcurISOMK{30} & YY+$\rhostar$+ZASPE         \\
~~~~Age (Gyr)\dotfill               &  \hatcurISOage{28}     & \hatcurISOage{29} & \hatcurISOage{30} & YY+$\rhostar$+ZASPE         \\
~~~~$A_{V}$ (mag)\dotfill               &  \hatcurXAv{28}     & \hatcurXAv{29} & \hatcurXAv{30} & YY+$\rhostar$+ZASPE         \\
~~~~Distance (pc)\dotfill           &  \hatcurXdistred{28}\phn  & \hatcurXdistred{29} & \hatcurXdistred{30} & YY+$\rhostar$+ZASPE\\ [-1.5ex]
\enddata
\tablecomments{
For all three systems the fixed-circular-orbit model has a higher Bayesian evidence than the eccentric-orbit model (it is 5, 660, and 3 times greater for \hatcur{28}, \hatcur{29} and \hatcur{30}, respectively). We therefore assume a fixed circular orbit in generating the parameters listed for these systems.
}
\tablenotetext{a}{
    ZASPE = Zonal Atmospherical Stellar Parameter Estimator routine
    for the analysis of high-resolution spectra (Brahm et al.~2016, in
    preparation), applied to the FEROS spectra of \hatcur{28} and \hatcur{26}. These
    parameters rely primarily on ZASPE, but have a small dependence
    also on the iterative analysis incorporating the isochrone search
    and global modeling of the data.
}
\tablenotetext{b}{
    From FEROS for \hatcur{28} and \hatcur{30}, and from HARPS for
    \hatcur{29}. The error on $\gamma_{\rm RV}$ is determined from the
    orbital fit to the RV measurements, and does not include the
    systematic uncertainty in transforming the velocities to the IAU
    standard system. The velocities have \textbf{not} been corrected for gravitational redshifts.
} \tablenotetext{c}{
    From APASS DR6 for as
    listed in the UCAC 4 catalog \citep{zacharias:2012:ucac4}.  
}
\tablenotetext{d}{
    \hatcurisoshort{25}+\rhostar+ZASPE = Based on the \hatcurisoshort{25}
    isochrones \citep{\hatcurisocite{25}}, \rhostar\ as a luminosity
    indicator, and the ZASPE results.
}
\tablenotetext{e}{
    In the case of $\rhostar$ we list two values. The first value is
    determined from the global fit to the light curves and RV data,
    without imposing a constraint that the parameters match the
    stellar evolution models. The second value results from
    restricting the posterior distribution to combinations of
    $\rhostar$+$\teffstar$+$\feh$ that match to a \hatcurisoshort{25}
    stellar model.
}
\ifthenelse{\boolean{emulateapj}}{
    \end{deluxetable*}
}{
    \end{deluxetable}
}

\subsection{Excluding blend scenarios}
\label{sec:blend}


In order to exclude blend scenarios we carried out an analysis following
\citet{hartman:2012:hat39hat41}. We attempt to model the available
photometric data (including light curves and catalog broad-band
photometric measurements) for each object as a blend between an
eclipsing binary star system and a third star along the line of
sight. The physical properties of the stars are constrained using the
Padova isochrones \citep{girardi:2000}, while we also require that the
brightest of the three stars in the blend have atmospheric parameters
consistent with those measured with ZASPE. We also simulate composite
cross-correlation functions (CCFs) and use them to predict RVs and BSs
for each blend scenario considered.

Based on this analysis we rule out blended stellar eclipsing binary
scenarios for all six systems. However, in general we cannot rule out
the possibility that one or more of these objects may be an unresolved
binary star system with one component hosting a transiting planet, although 
limits can be placed on those scenarios for HATS-26, HATS-27 and HATS-30 based on 
our lucky imaging observations shown on Section 2.4. The results for 
each object are as follows:
\begin{itemize}
\item {\em \hatcur{25}}: All blend models tested give higher $\chi^2$ than a model of single star with a planet. Those blend models which cannot be rejected with greater than $5\sigma$ confidence predict either RV or BS variations greater than 1\,\kms, which are excluded by the observations.
\item {\em \hatcur{26}}: All blend models tested can be rejected with greater than $5\sigma$ confidence based on the photometry alone. In particular, the blend models predict a large out-of-transit variation due to the tidal distortion of the binary star components. Such a variation is ruled out by the HATSouth photometry.
\item {\em \hatcur{27}}: Same conclusion as for \hatcur{25}.
\item {\em \hatcur{28}}: All blend models tested can be rejected with greater than $4\sigma$ confidence based on the photometry alone.
\item {\em \hatcur{29}}: Blend models which cannot be rejected with greater than $5\sigma$ confidence based on the photometry alone generally predict large RV and BS variations exceeding 1\,\kms. There is a narrow region of parameter space where the blend models are rejected at $4\sigma$ confidence based on the photometry, and the simulated RVs and BSs have scatters of a few 100\,\ms, which is not much greater than the measured values. However, the simulated RVs do not phase with the photometric ephemeris.
\item {\em \hatcur{30}}: All blend models tested can be rejected with greater than $4\sigma$ confidence based on the photometry alone.
\end{itemize}

\subsection{Global modeling of the data}
\label{sec:globmod}

We modeled the HATSouth photometry, the follow-up photometry, and the
high-precision RV measurements following
\citet{pal:2008:hat7,bakos:2010:hat11,hartman:2012:hat39hat41}. We fit
\citet{mandel:2002} transit models to the light curves, allowing for a
dilution of the HATSouth transit depth as a result of blending from
neighboring stars and over-correction by the trend-filtering
method. For the follow-up light curves we include a quadratic trend in
time, and linear trends with up to three parameters describing the
shape of the PSF, in our model for each event to correct for
systematic errors in the photometry. We fit Keplerian orbits to the RV
curves allowing the zero-point for each instrument to vary
independently in the fit, and allowing for RV jitter which we we also
vary as a free parameter for each instrument. We used a Differential
Evolution Markov Chain Monte Carlo procedure to explore the fitness
landscape and to determine the posterior distributions of the
parameters. Note that we tried fitting both fixed-circular-orbits and
free-eccentricity models to the data, and for all six systems find
that the data are consistent with a circular orbit. We estimate the
Bayesian evidence for the fixed-circular and free-eccentricity models
for each system, and find that in all six cases the fixed-circular
model has greater evidence. In particular, for the HATS-25, HATS-26, HATS-28, 
HATS-29 and HATS-30 systems, the Bayesian evidence for the fixed-circular-orbit model is 
10, 8, 5, 660 and 3 times greater, respectively, than the eccentric-orbit model, favouring 
the former in these cases. For HATS-27, both models are indistinguishable, but the eccentricity is 
poorly constrained by the data at hand, giving implausibly high values for it. We therefore adopt the 
parameters that come from the fixed-circular-orbit models for all of the systems. The
resulting parameters for \hatcurb{25}, \hatcurb{26} and \hatcurb{27}
are listed in \reftabl{planetparam}, while for \hatcurb{28},
\hatcurb{29} and \hatcurb{30} they are listed in
\reftabl{planetparamtwo}.

As can be observed from the tables, all the presented planets can be 
classified as typical hot Jupiters, with short-periods, similar masses of 
$\sim 0.6M_J$ and larger-than-Jupiter radii. 

%
\ifthenelse{\boolean{emulateapj}}{
    \begin{deluxetable*}{lccc}
}{
    \begin{deluxetable}{lccc}
}
\tabletypesize{\scriptsize}
\tablecaption{Orbital and planetary parameters for \hatcurb{25}, \hatcurb{26} and \hatcurb{27}\label{tab:planetparam}}
\tablehead{
    \multicolumn{1}{c}{} &
    \multicolumn{1}{c}{\bf HATS-25b} &
    \multicolumn{1}{c}{\bf HATS-26b} &
    \multicolumn{1}{c}{\bf HATS-27b} \\ 
    \multicolumn{1}{c}{~~~~~~~~~~~~~~~Parameter~~~~~~~~~~~~~~~} &
    \multicolumn{1}{c}{Value} &
    \multicolumn{1}{c}{Value} &
    \multicolumn{1}{c}{Value}
}
\startdata
\noalign{\vskip -3pt}
\sidehead{\Lc{} parameters}
~~~$P$ (days)             \dotfill    & $\hatcurLCP{25}$ & $\hatcurLCP{26}$ & $\hatcurLCP{27}$ \\
~~~$T_c$ (${\rm BJD}$)    
      \tablenotemark{a}   \dotfill    & $\hatcurLCT{25}$ & $\hatcurLCT{26}$ & $\hatcurLCT{27}$ \\
~~~$T_{14}$ (days)
      \tablenotemark{a}   \dotfill    & $\hatcurLCdur{25}$ & $\hatcurLCdur{26}$ & $\hatcurLCdur{27}$ \\
~~~$T_{12} = T_{34}$ (days)
      \tablenotemark{a}   \dotfill    & $\hatcurLCingdur{25}$ & $\hatcurLCingdur{26}$ & $\hatcurLCingdur{27}$ \\
~~~$\arstar$              \dotfill    & $\hatcurPPar{25}$ & $\hatcurPPar{26}$ & $\hatcurPPar{27}$ \\
~~~$\zrstar$ \tablenotemark{b}             \dotfill    & $\hatcurLCzeta{25}$\phn & $\hatcurLCzeta{26}$\phn & $\hatcurLCzeta{27}$\phn \\
~~~$\rpl/\rstar$          \dotfill    & $\hatcurLCrprstar{25}$ & $\hatcurLCrprstar{26}$ & $\hatcurLCrprstar{27}$ \\
~~~$b^2$                  \dotfill    & $\hatcurLCbsq{25}$ & $\hatcurLCbsq{26}$ & $\hatcurLCbsq{27}$ \\
~~~$b \equiv a \cos i/\rstar$
                          \dotfill    & $\hatcurLCimp{25}$ & $\hatcurLCimp{26}$ & $\hatcurLCimp{27}$ \\
~~~$i$ (deg)              \dotfill    & $\hatcurPPi{25}$\phn & $\hatcurPPi{26}$\phn & $\hatcurPPi{27}$\phn \\

\sidehead{HATSouth blend factors \tablenotemark{c}}
~~~Blend factor \dotfill & $\hatcurLCiblend{25}$ & $\hatcurLCiblend{26}$ & $\hatcurLCiblend{27}$ \\

\sidehead{Limb-darkening coefficients \tablenotemark{d}}
~~~$c_1,R$                  \dotfill    & $\cdots$ & $\cdots$ & $\hatcurLBiR{27}$ \\
~~~$c_2,R$                  \dotfill    & $\cdots$ & $\cdots$ & $\hatcurLBiiR{27}$ \\
~~~$c_1,r$                  \dotfill    & $\hatcurLBir{25}$ & $\hatcurLBir{26}$ & $\hatcurLBir{27}$ \\
~~~$c_2,r$                  \dotfill    & $\hatcurLBiir{25}$ & $\hatcurLBiir{26}$ & $\hatcurLBiir{27}$ \\
~~~$c_1,i$                  \dotfill    & $\hatcurLBii{25}$ & $\hatcurLBii{26}$ & $\hatcurLBii{27}$ \\
~~~$c_2,i$                  \dotfill    & $\hatcurLBiii{25}$ & $\hatcurLBiii{26}$ & $\hatcurLBiii{27}$ \\

\sidehead{RV parameters}
~~~$K$ (\ms)              \dotfill    & $\hatcurRVK{25}$\phn\phn & $\hatcurRVK{26}$\phn\phn & $\hatcurRVK{27}$\phn\phn \\
~~~$e$ \tablenotemark{e}               \dotfill    & $\hatcurRVeccentwosiglimeccen{25}$ & $\hatcurRVeccentwosiglimeccen{26}$ & $\hatcurRVeccentwosiglimeccen{27}$ \\
~~~RV jitter FEROS (\ms) \tablenotemark{f}       \dotfill    & $\cdots$ & \hatcurRVjittertwosiglimA{26} & \hatcurRVjitterA{27} \\
~~~RV jitter HARPS (\ms)        \dotfill    & \hatcurRVjittertwosiglim{25} & \hatcurRVjittertwosiglimC{26} & \hatcurRVjittertwosiglimC{27} \\
~~~RV jitter Coralie (\ms)        \dotfill    & $\cdots$ & \hatcurRVjittertwosiglimB{26} & \hatcurRVjittertwosiglimB{27} \\

\sidehead{Planetary parameters}
~~~$\mpl$ ($\mjup$)       \dotfill    & $\hatcurPPmlong{25}$ & $\hatcurPPmlong{26}$ & $\hatcurPPmlong{27}$ \\
~~~$\rpl$ ($\rjup$)       \dotfill    & $\hatcurPPrlong{25}$ & $\hatcurPPrlong{26}$ & $\hatcurPPrlong{27}$ \\
~~~$C(\mpl,\rpl)$
    \tablenotemark{g}     \dotfill    & $\hatcurPPmrcorr{25}$ & $\hatcurPPmrcorr{26}$ & $\hatcurPPmrcorr{27}$ \\
~~~$\rhopl$ (\gcmc)       \dotfill    & $\hatcurPPrho{25}$ & $\hatcurPPrho{26}$ & $\hatcurPPrho{27}$ \\
~~~$\log g_p$ (cgs)       \dotfill    & $\hatcurPPlogg{25}$ & $\hatcurPPlogg{26}$ & $\hatcurPPlogg{27}$ \\
~~~$a$ (AU)               \dotfill    & $\hatcurPParel{25}$ & $\hatcurPParel{26}$ & $\hatcurPParel{27}$ \\
~~~$T_{\rm eq}$ (K)        \dotfill   & $\hatcurPPteff{25}$ & $\hatcurPPteff{26}$ & $\hatcurPPteff{27}$ \\
~~~$\Theta$ \tablenotemark{h} \dotfill & $\hatcurPPtheta{25}$ & $\hatcurPPtheta{26}$ & $\hatcurPPtheta{27}$ \\
~~~$\log_{10}\langle F \rangle$ (cgs) \tablenotemark{i}
                          \dotfill    & $\hatcurPPfluxavglog{25}$ & $\hatcurPPfluxavglog{26}$ & $\hatcurPPfluxavglog{27}$ \\ [-1.5ex]
\enddata
\tablenotetext{a}{
    Times are in Barycentric Julian Date calculated directly from UTC {\em without} correction for leap seconds.
    \ensuremath{T_c}: Reference epoch of
    mid transit that minimizes the correlation with the orbital
    period.
    \ensuremath{T_{14}}: total transit duration, time
    between first to last contact;
    \ensuremath{T_{12}=T_{34}}: ingress/egress time, time between first
    and second, or third and fourth contact.
}
\tablecomments{
For \hatcur{25} and \hatcur{26} the fixed-circular-orbit model has a higher Bayesian evidence than the eccentric-orbit model (it is 10 and 8 times greater for these two systems respectively). We therefore assume a fixed circular orbit in generating the parameters listed for both of these systems. For \hatcur{27} the free-eccentricity model has an indistinguishable Bayesian evidence from the fixed-circular model, but in this case the eccentricity is poorly constrained with implausibly high values permitted by the low S/N RV measurements. For this system we also adopt the fixed-circular model parameters.
}
\tablenotetext{b}{
   Reciprocal of the half duration of the transit used as a jump parameter in our MCMC analysis in place of $\arstar$. It is related to $\arstar$ by the expression $\zrstar = \arstar(2\pi(1+e\sin\omega))/(P\sqrt{1-b^2}\sqrt{1-e^2})$ \citep{bakos:2010:hat11}.
}
\tablenotetext{c}{
    Scaling factor applied to the model transit that is fit to the HATSouth light curves. This factor accounts for dilution of the transit due to blending from neighboring stars and over-filtering of the light curve.  These factors are varied in the fit, and we allow independent factors for observations obtained with different HATSouth camera and field combinations.
}
\tablenotetext{d}{
    Values for a quadratic law, adopted from the tabulations by
    \cite{claret:2004} according to the spectroscopic (ZASPE) parameters
    listed in \reftabl{stellar}.
}
\tablenotetext{e}{
    For fixed circular orbit models we list
    the 95\% confidence upper limit on the eccentricity determined
    when $\sqrt{e}\cos\omega$ and $\sqrt{e}\sin\omega$ are allowed to
    vary in the fit.
}
\tablenotetext{f}{
    Term added in quadrature to the formal RV uncertainties for each
    instrument. This is treated as a free parameter in the fitting
    routine. In cases where the jitter is consistent with zero we list the 95\% confidence upper limit.
}
\tablenotetext{g}{
    Correlation coefficient between the planetary mass \mpl\ and radius
    \rpl\ estimated from the posterior parameter distribution.
}
\tablenotetext{h}{
    The Safronov number is given by $\Theta = \frac{1}{2}(V_{\rm
    esc}/V_{\rm orb})^2 = (a/\rpl)(\mpl / \mstar )$
    \citep[see][]{hansen:2007}.
}
\tablenotetext{i}{
    Incoming flux per unit surface area, averaged over the orbit.
}
\ifthenelse{\boolean{emulateapj}}{
    \end{deluxetable*}
}{
    \end{deluxetable}
}

%
\ifthenelse{\boolean{emulateapj}}{
    \begin{deluxetable*}{lccc}
}{
    \begin{deluxetable}{lccc}
}
\tabletypesize{\scriptsize}
\tablecaption{Orbital and planetary parameters for \hatcurb{28}, \hatcurb{29} and \hatcurb{30}\label{tab:planetparamtwo}}
\tablehead{
    \multicolumn{1}{c}{} &
    \multicolumn{1}{c}{\bf HATS-28b} &
    \multicolumn{1}{c}{\bf HATS-29b} &
    \multicolumn{1}{c}{\bf HATS-30b} \\ 
    \multicolumn{1}{c}{~~~~~~~~~~~~~~~Parameter~~~~~~~~~~~~~~~} &
    \multicolumn{1}{c}{Value} &
    \multicolumn{1}{c}{Value} &
    \multicolumn{1}{c}{Value}
}
\startdata
\noalign{\vskip -3pt}
\sidehead{\Lc{} parameters}
~~~$P$ (days)             \dotfill    & $\hatcurLCP{28}$ & $\hatcurLCP{29}$ & $\hatcurLCP{30}$ \\
~~~$T_c$ (${\rm BJD}$)    
      \tablenotemark{a}   \dotfill    & $\hatcurLCT{28}$ & $\hatcurLCT{29}$ & $\hatcurLCT{30}$ \\
~~~$T_{14}$ (days)
      \tablenotemark{a}   \dotfill    & $\hatcurLCdur{28}$ & $\hatcurLCdur{29}$ & $\hatcurLCdur{30}$ \\
~~~$T_{12} = T_{34}$ (days)
      \tablenotemark{a}   \dotfill    & $\hatcurLCingdur{28}$ & $\hatcurLCingdur{29}$ & $\hatcurLCingdur{30}$ \\
~~~$\arstar$              \dotfill    & $\hatcurPPar{28}$ & $\hatcurPPar{29}$ & $\hatcurPPar{30}$ \\
~~~$\zrstar$ \tablenotemark{b}             \dotfill    & $\hatcurLCzeta{28}$\phn & $\hatcurLCzeta{29}$\phn & $\hatcurLCzeta{30}$\phn \\
~~~$\rpl/\rstar$          \dotfill    & $\hatcurLCrprstar{28}$ & $\hatcurLCrprstar{29}$ & $\hatcurLCrprstar{30}$ \\
~~~$b^2$                  \dotfill    & $\hatcurLCbsq{28}$ & $\hatcurLCbsq{29}$ & $\hatcurLCbsq{30}$ \\
~~~$b \equiv a \cos i/\rstar$
                          \dotfill    & $\hatcurLCimp{28}$ & $\hatcurLCimp{29}$ & $\hatcurLCimp{30}$ \\
~~~$i$ (deg)              \dotfill    & $\hatcurPPi{28}$\phn & $\hatcurPPi{29}$\phn & $\hatcurPPi{30}$\phn \\

\sidehead{HATSouth blend factors \tablenotemark{c}}
~~~Blend factor 1 \dotfill & $\hatcurLCiblend{28}$ & $\hatcurLCiblend{29}$ & $\hatcurLCiblendA{30}$ \\
~~~Blend factor 2 \dotfill & $\cdots$ & $\cdots$ & $\hatcurLCiblendB{30}$ \\
~~~Blend factor 3 \dotfill & $\cdots$ & $\cdots$ & $\hatcurLCiblendC{30}$ \\

\sidehead{Limb-darkening coefficients \tablenotemark{d}}
~~~$c_1,r$                  \dotfill    & $\hatcurLBir{28}$ & $\hatcurLBir{29}$ & $\hatcurLBir{30}$ \\
~~~$c_2,r$                  \dotfill    & $\hatcurLBiir{28}$ & $\hatcurLBiir{29}$ & $\hatcurLBiir{30}$ \\
~~~$c_1,i$                  \dotfill    & $\hatcurLBii{28}$ & $\hatcurLBii{29}$ & $\hatcurLBii{30}$ \\
~~~$c_2,i$                  \dotfill    & $\hatcurLBiii{28}$ & $\hatcurLBiii{29}$ & $\hatcurLBiii{30}$ \\

\sidehead{RV parameters}
~~~$K$ (\ms)              \dotfill    & $\hatcurRVK{28}$\phn\phn & $\hatcurRVK{29}$\phn\phn & $\hatcurRVK{30}$\phn\phn \\
~~~$e$ \tablenotemark{e}               \dotfill    & $\hatcurRVeccentwosiglimeccen{28}$ & $\hatcurRVeccentwosiglimeccen{29}$ & $\hatcurRVeccentwosiglimeccen{30}$ \\
~~~RV jitter FEROS (\ms) \tablenotemark{f}       \dotfill    & \hatcurRVjitter{28} & $\cdots$ & \hatcurRVjittertwosiglimA{30} \\
~~~RV jitter HARPS (\ms)        \dotfill    & $\cdots$ & \hatcurRVjittertwosiglimB{29} & $\cdots$ \\
~~~RV jitter Coralie (\ms)        \dotfill    & $\cdots$ & \hatcurRVjittertwosiglimA{29} & \hatcurRVjittertwosiglimB{30} \\
~~~RV jitter CYCLOPS (\ms)        \dotfill    & $\cdots$ & \hatcurRVjitterC{29} & $\cdots$ \\

\sidehead{Planetary parameters}
~~~$\mpl$ ($\mjup$)       \dotfill    & $\hatcurPPmlong{28}$ & $\hatcurPPmlong{29}$ & $\hatcurPPmlong{30}$ \\
~~~$\rpl$ ($\rjup$)       \dotfill    & $\hatcurPPrlong{28}$ & $\hatcurPPrlong{29}$ & $\hatcurPPrlong{30}$ \\
~~~$C(\mpl,\rpl)$
    \tablenotemark{g}     \dotfill    & $\hatcurPPmrcorr{28}$ & $\hatcurPPmrcorr{29}$ & $\hatcurPPmrcorr{30}$ \\
~~~$\rhopl$ (\gcmc)       \dotfill    & $\hatcurPPrho{28}$ & $\hatcurPPrho{29}$ & $\hatcurPPrho{30}$ \\
~~~$\log g_p$ (cgs)       \dotfill    & $\hatcurPPlogg{28}$ & $\hatcurPPlogg{29}$ & $\hatcurPPlogg{30}$ \\
~~~$a$ (AU)               \dotfill    & $\hatcurPParel{28}$ & $\hatcurPParel{29}$ & $\hatcurPParel{30}$ \\
~~~$T_{\rm eq}$ (K)        \dotfill   & $\hatcurPPteff{28}$ & $\hatcurPPteff{29}$ & $\hatcurPPteff{30}$ \\
~~~$\Theta$ \tablenotemark{h} \dotfill & $\hatcurPPtheta{28}$ & $\hatcurPPtheta{29}$ & $\hatcurPPtheta{30}$ \\
~~~$\log_{10}\langle F \rangle$ (cgs) \tablenotemark{i}
                          \dotfill    & $\hatcurPPfluxavglog{28}$ & $\hatcurPPfluxavglog{29}$ & $\hatcurPPfluxavglog{30}$ \\ [-1.5ex]
\enddata
\tablenotetext{a}{
    Times are in Barycentric Julian Date calculated directly from UTC {\em without} correction for leap seconds.
    \ensuremath{T_c}: Reference epoch of
    mid transit that minimizes the correlation with the orbital
    period.
    \ensuremath{T_{14}}: total transit duration, time
    between first to last contact;
    \ensuremath{T_{12}=T_{34}}: ingress/egress time, time between first
    and second, or third and fourth contact.
}
\tablecomments{
For all three systems the fixed-circular-orbit model has a higher Bayesian evidence than the eccentric-orbit model (it is 5, 660, and 3 times greater for \hatcur{28}, \hatcur{29} and \hatcur{30}, respectively). We therefore assume a fixed circular orbit in generating the parameters listed for these systems.
}
\tablenotetext{b}{
   Reciprocal of the half duration of the transit used as a jump parameter in our MCMC analysis in place of $\arstar$. It is related to $\arstar$ by the expression $\zrstar = \arstar(2\pi(1+e\sin\omega))/(P\sqrt{1-b^2}\sqrt{1-e^2})$ \citep{bakos:2010:hat11}.
}
\tablenotetext{c}{
    Scaling factor applied to the model transit that is fit to the HATSouth light curves. This factor accounts for dilution of the transit due to blending from neighboring stars and over-filtering of the light curve.  These factors are varied in the fit, and we allow independent factors for observations obtained with different HATSouth camera and field combinations. For \hatcur{30} blend factors 1 through 3 are used for the G754.3, G754.4 and G755.1 observations, respectively.
}
\tablenotetext{d}{
    Values for a quadratic law, adopted from the tabulations by
    \cite{claret:2004} according to the spectroscopic (ZASPE) parameters
    listed in \reftabl{stellar}.
}
\tablenotetext{e}{
    For fixed circular orbit models we list
    the 95\% confidence upper limit on the eccentricity determined
    when $\sqrt{e}\cos\omega$ and $\sqrt{e}\sin\omega$ are allowed to
    vary in the fit.
}
\tablenotetext{f}{
    Term added in quadrature to the formal RV uncertainties for each
    instrument. This is treated as a free parameter in the fitting
    routine. In cases where the jitter is consistent with zero we list the 95\% confidence upper limit.
}
\tablenotetext{g}{
    Correlation coefficient between the planetary mass \mpl\ and radius
    \rpl\ estimated from the posterior parameter distribution.
}
\tablenotetext{h}{
    The Safronov number is given by $\Theta = \frac{1}{2}(V_{\rm
    esc}/V_{\rm orb})^2 = (a/\rpl)(\mpl / \mstar )$
    \citep[see][]{hansen:2007}.
}
\tablenotetext{i}{
    Incoming flux per unit surface area, averaged over the orbit.
}
\ifthenelse{\boolean{emulateapj}}{
    \end{deluxetable*}
}{
    \end{deluxetable}
}



\section{Discussion}
\label{sec:discussion}

\begin{figure*}
\setlength{\plotwidthtwo}{0.31\linewidth}
\epsscale{1.0}
\plottwo{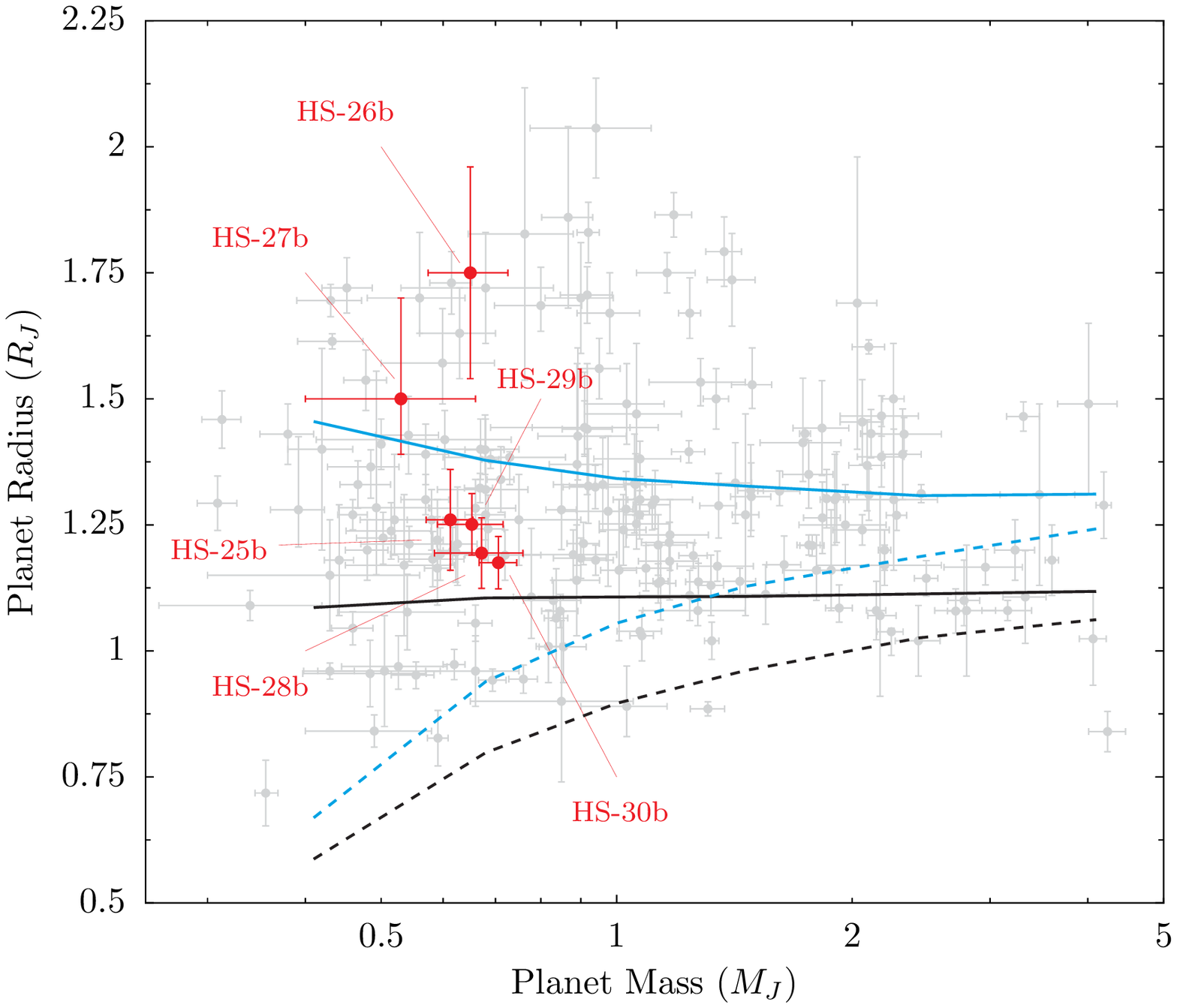}{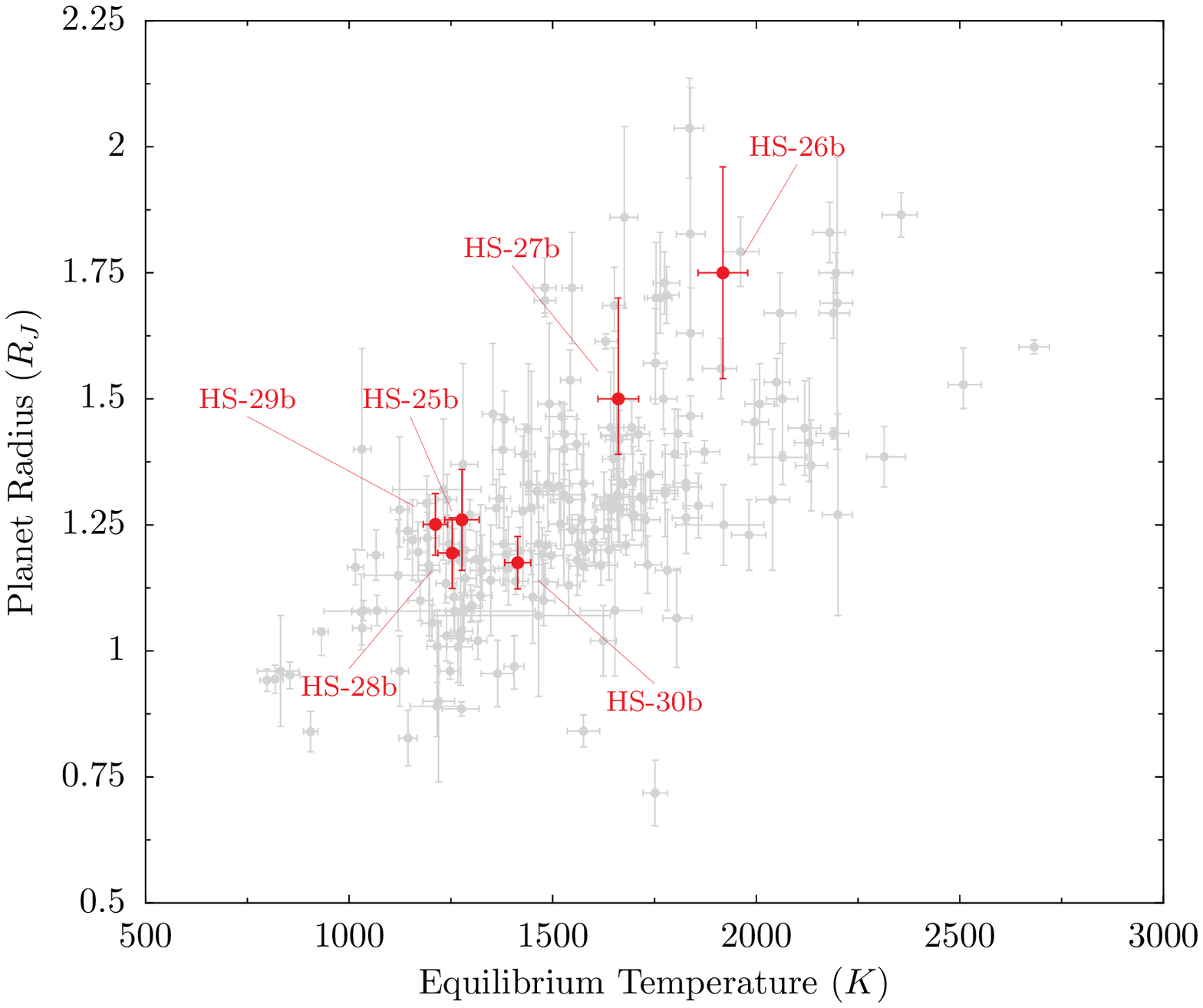}
\caption{(\textit{Left}) Mass-radius diagram for all the transiting hot Jupiters discovered to date (grey points). Red points 
indicate the discovered exoplanets presented in this work. The black lines show the mass-radius relations of $4.5$ Gyr old 
planets at $0.045$ AU from the Sun obtained from \cite{fortney:2007} for core-free giant planets (solid line) and for planets with 
$100M_\Earth$ cores (dashed line), which are appropriate for the insolation levels received by HATS-25b, HATS-28b, HATS-29b 
and HATS-30b. The blue lines show the same relations but for planets at $0.02$ AU, more (but not exactly) appropriate for the 
insolation levels received by HATS-26b and HATS-27b. We note, however, that these relations imply insolation levels around $2500$ times the 
solar insolation level at Earth, while the actual insolation levels for HATS-26b and HATS-27b are closer to $2250$ and $1250$ times 
the solar flux at Earth, respectively. (\textit{Right}) Equilibrium temperature-radius diagram for all the transiting hot Jupiters discovered to date along with 
the discovered exoplanets presented in this work with the same colors as in the left plot.}
\label{fig:mrd}
\end{figure*}

In this paper we present six new transiting planets discovered by the HAT-South survey. 
Figure \ref{fig:mrd} puts the discovered exoplanets in the context of all known transiting hot Jupiters 
(here defined as planets with $0.1M_J<M<5M_J$ and periods $P<10d$) discovered to date\footnote{Data 
taken from exoplanets.eu on 2016/02/01.} with secure masses and radii (i.e., masses and radii inconsistent with 
zero at $3-\sigma$). We can see that the discovered exoplanets all fall in a
heavily populated region of the mass distribution of hot Jupiters near $\sim 0.6M_J$. However, although HATS-30b, 
HATS-29b, HATS-28b and HATS-25b all fall in the peak of the radius distribution, with radii of $\sim 1.2R_J$, 
making them all moderately inflated planets, HATS-26b ($1.75R_J$) and HATS-27b ($1.50R_J$) fall on the 
high-end part of it, making them highly inflated planets.  These two 
hot Jupiters also have the lowest densities of the group: HATS-26b has a density of only $0.153 \pm 0.042$ \gcmc, 
while HATS-27b has a density of $0.180^{+0.083}_{-0.057}$ \gcmc. These densities are quite unusual not only in this 
group of planets, but also among the population of hot Jupiters in general: of the known systems, only $\sim 10$ have 
densities lower than $0.2$ \gcmc. 

The empirical relations in equation (9) of \cite{enoch:2012} predict the radii of these six new exoplanets to within the uncertainties. Therefore, these exoplanets 
appear to follow the trends followed by other close-in exoplanets, namely, that both increasing their semi-major axes and the effective 
temperatures leads to an increase in planetary radii. To further illustrate 
this, the right panel of Figure~\ref{fig:mrd} shows the equilibrium temperature-radius diagram for the same exoplanets as on the 
left plot. We can clearly see that the correlation followed by most of the discovered transiting hot Jupiters to date is also 
followed by our newly discovered exoplanets.

\begin{figure}
\setlength{\plotwidthtwo}{0.31\linewidth}
\epsscale{1.0}
\plotone{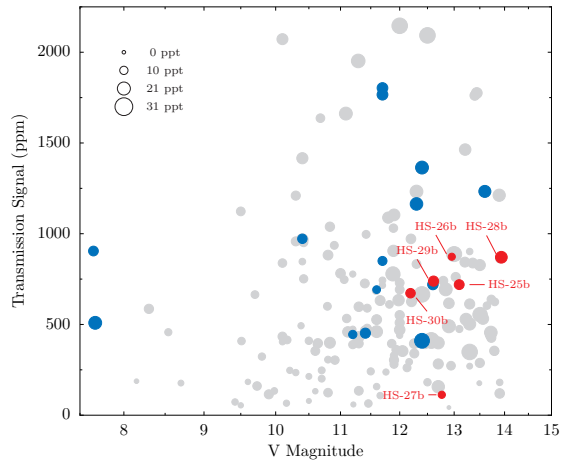}
\caption{Visual magnitude versus expected transmission signal for all the hot Jupiters discovered to date (grey points). Blue points indicate 
systems that have already been characterised via transmission spectroscopy, while red points indicate the exoplanets presented in this work. 
The size of the points indicate the transit depth with larger points indicating larger transit depths; the legend in the upper left corner indicates the 
corresponding depths in parts per thousand (ppt).}
\label{fig:char}
\end{figure}

In terms of future characterization, all the presented planets (except HATS-27b) have expected transmission signals between $\sim 700-900$ ppm 
and all (except HATS-28) have magnitudes between $V\sim12-13$, making them interesting targets for future atmospheric studies. Figure~\ref{fig:char} 
illustrates $V$ band magnitude versus the expected transmission signals for our newly discovered planets along with planets discovered to date, where 
the formula used to calculate the signal assumes an atmosphere that is five scale-heights thick, and is given by
\begin{eqnarray*}
\delta_\textnormal{transpec} = \frac{10R_p H}{R_*^2},
\end{eqnarray*}
where $R_p$ is the planetary radius, $R_*$ is the stellar radius and $H = k_B T_p/mg_p$ is the planetary scale-height, calculated 
using Boltzmann's constant, $k_B$, the planetary equilibrium temperature, $T_p$, the mean mass of the constituents that make up the 
atmosphere of the planet (assumed to be $\textnormal{H}_2$), $m$, and the acceleration due to gravity on the planetary surface, $g_p$. 
Systems already characterized by transmission spectroscopy are indicated in blue. As can be seen, the discovered exoplanets 
add to the increasing fraction of planets that have expected transmission signals on the same order as those already characterized. 
The most interesting systems in this respect are HATS-26b ($V = 12.9$), which has an expected transmission signal of $\sim 900$ ppm and a long transit duration 
of $5.2$ hours, and HATS-29b ($V=12.6$), which has an expected transmission signal of $\sim 700$ ppm, a transit depth two times that of HATS-26b 
and a transit duration of $3.2$ hours. 

Although not a good target for transmission, HATS-27b ($V = 12.8$) is an attractive system if one is interested in estimating the projected spin-orbit alignment of  the 
system: despite its modest planet-to-star ratio of  ($R_p/R_* = 0.0895\pm0.0041$), the host star rotates at a moderately high rate ($v\sin(i)$ of $9.32 \pm 0.5$ km/s) which, 
coupled with the long transit duration of $4.8$ hours, makes this inflated hot Jupiter a good target for follow-up Rossiter-McLaughlin (RM) observations. In particular, 
using equation (6) of \cite{gaudi:2007}, the amplitude of the RM effect, $K_R$, should be $\approx 75$ m/s. We 
obtained a precision of $\sim 30$ m/s in 10 minute exposures with HARPS for this star, making the RM effect readily detectable. In addition, given that the temperature of the host star is  $6428 \pm 64$ K, the system 
lies in a very interesting regime at which it has been claimed that planetary orbits of hot Jupiters shift from aligned to misaligned \citep{albrecht:2012,addison:2016}.

\acknowledgements 

Development of the HATSouth project was funded by NSF MRI grant
NSF/AST-0723074, operations have been supported by NASA grants NNX09AB29G and NNX12AH91H, and
follow-up observations receive partial support from grant
NSF/AST-1108686.
N.E. is supported by CONICYT-PCHA/Doctorado Nacional.
A.J.\ acknowledges support from FONDECYT project 1130857, BASAL CATA PFB-06, and project IC120009 ``Millennium Institute of Astrophysics (MAS)'' of the Millenium Science Initiative, Chilean Ministry of Economy. R.B.\ and N.E.\ acknowledge  support from project IC120009 ``Millenium Institute of Astrophysics  (MAS)'' of the Millennium Science Initiative, Chilean Ministry of Economy.  V.S.\ acknowledges support form BASAL CATA PFB-06. 
This work is based on observations made with ESO Telescopes at the La
Silla Observatory.
This paper also uses observations obtained with facilities of the Las
Cumbres Observatory Global Telescope.
Work at the Australian National University is supported by ARC Laureate
Fellowship Grant FL0992131.
Work at UNSW is supported by ARC Discovery Project DP130102695.
We acknowledge the use of the AAVSO Photometric All-Sky Survey (APASS),
funded by the Robert Martin Ayers Sciences Fund, and the SIMBAD
database, operated at CDS, Strasbourg, France.
Operations at the MPG~2.2\,m Telescope are jointly performed by the
Max Planck Gesellschaft and the European Southern Observatory.  The
imaging system GROND has been built by the high-energy group of MPE in
collaboration with the LSW Tautenburg and ESO\@.  We thank the
MPG~2.2\,m telescope support team for their technical assistance during observations. 
We thank Helmut Steinle and Jochen Greiner for
supporting the GROND observations presented in this manuscript.
Observing time were obtained through proposals CN2013-B55, CN2014A-104, CN2014B-57, 
CN2015A-51 and ESO 096.C-0544.
We are grateful to P.Sackett for her help in the early phase of the
HATSouth project.



\clearpage
\LongTables

%
%
\tabletypesize{\scriptsize}
\ifthenelse{\boolean{emulateapj}}{
    \begin{deluxetable*}{llrrrrrl}
}{
    \begin{deluxetable}{llrrrrrl}
}
\tablewidth{0pc}
\tablecaption{
    Relative radial velocities and bisector spans for \hatcur{25}--\hatcur{30}.
    \label{tab:rvs}
}
\tablehead{
    \colhead{Star} &
    \colhead{BJD} &
    \colhead{RV\tablenotemark{a}} &
    \colhead{\ensuremath{\sigma_{\rm RV}}\tablenotemark{b}} &
    \colhead{BS} &
    \colhead{\ensuremath{\sigma_{\rm BS}}} &
    \colhead{Phase} &
    \colhead{Instrument}\\
    \colhead{} &
    \colhead{\hbox{(2,450,000$+$)}} &
    \colhead{(\ms)} &
    \colhead{(\ms)} &
    \colhead{(\ms)} &
    \colhead{(\ms)} &
    \colhead{} &
    \colhead{}
}
\startdata
\multicolumn{8}{c}{\bf HATS-25} \\
\hline\\
HATS-25 & $ 7067.85231 $ & $    34.25 $ & $    10.00 $ & $  -19.0 $ & $   40.0 $ & $   0.941 $ & HARPS \\
HATS-25 & $ 7067.87380 $ & $     4.25 $ & $    11.00 $ & $   -2.0 $ & $   40.0 $ & $   0.946 $ & HARPS \\
HATS-25 & $ 7068.86477 $ & $   -70.75 $ & $     7.00 $ & $  -38.0 $ & $   28.0 $ & $   0.176 $ & HARPS \\
HATS-25 & $ 7070.85785 $ & $    57.25 $ & $     9.00 $ & $  -27.0 $ & $   32.0 $ & $   0.640 $ & HARPS \\
HATS-25 & $ 7071.87936 $ & $    54.25 $ & $    21.00 $ & $   15.0 $ & $   70.0 $ & $   0.878 $ & HARPS \\
HATS-25 & $ 7072.88746 $ & $   -48.75 $ & $     9.00 $ & $  -23.0 $ & $   36.0 $ & $   0.112 $ & HARPS \\
HATS-25 & $ 7118.73269 $ & $    80.25 $ & $    16.00 $ & $  -36.0 $ & $   50.0 $ & $   0.777 $ & HARPS \\
HATS-25 & $ 7120.73472 $ & $   -77.75 $ & $    11.00 $ & $  -14.0 $ & $   40.0 $ & $   0.243 $ & HARPS \\
\cutinhead{\bf HATS-26}
HATS-26 & $ 6828.49681 $ & $   -65.10 $ & $    28.00 $ & $  150.0 $ & $   24.0 $ & $   0.213 $ & Coralie \\
HATS-26 & $ 6829.52934 $ & $     2.90 $ & $    30.00 $ & $   80.0 $ & $   26.0 $ & $   0.526 $ & Coralie \\
HATS-26 & $ 7031.72967 $ & $    58.90 $ & $    15.00 $ & $   30.0 $ & $   12.0 $ & $   0.754 $ & FEROS \\
HATS-26 & $ 7035.82606 $ & $    34.90 $ & $    16.00 $ & $   68.0 $ & $   13.0 $ & $   0.995 $ & FEROS \\
HATS-26 & $ 7037.84686 $ & $    22.90 $ & $    15.00 $ & $   34.0 $ & $   12.0 $ & $   0.607 $ & FEROS \\
HATS-26 & $ 7049.79153 $ & $   -56.10 $ & $    16.00 $ & $   71.0 $ & $   12.0 $ & $   0.224 $ & FEROS \\
HATS-26 & $ 7050.84666 $ & $    -1.10 $ & $    15.00 $ & $   64.0 $ & $   12.0 $ & $   0.543 $ & FEROS \\
HATS-26 & $ 7053.88112 $ & $    -3.10 $ & $    17.00 $ & $   59.0 $ & $   13.0 $ & $   0.462 $ & FEROS \\
HATS-26 & $ 7054.81498 $ & $    97.90 $ & $    14.00 $ & $   32.0 $ & $   11.0 $ & $   0.745 $ & FEROS \\
HATS-26 & $ 7056.81639 $ & $   -65.10 $ & $    17.00 $ & $   69.0 $ & $   13.0 $ & $   0.351 $ & FEROS \\
HATS-26 & $ 7067.70058 $ & $    68.91 $ & $    18.00 $ & $   67.0 $ & $   38.0 $ & $   0.647 $ & HARPS \\
HATS-26 & $ 7069.77078 $ & $   -92.09 $ & $    20.00 $ & $    8.0 $ & $   42.0 $ & $   0.273 $ & HARPS \\
HATS-26 & $ 7070.73942 $ & $    41.91 $ & $    18.00 $ & $   92.0 $ & $   38.0 $ & $   0.567 $ & HARPS \\
HATS-26 & $ 7072.71165 $ & $   -72.09 $ & $    16.00 $ & $   28.0 $ & $   34.0 $ & $   0.164 $ & HARPS \\
\cutinhead{\bf HATS-27}
HATS-27 & $ 6828.57385 $ & $    29.59 $ & $    40.00 $ & $   24.0 $ & $   29.0 $ & $   0.704 $ & Coralie \\
HATS-27 & $ 6828.62287 $ & $   154.59 $ & $    41.00 $ & $  107.0 $ & $   29.0 $ & $   0.715 $ & Coralie \\
HATS-27 & $ 6829.58090 $ & $    16.59 $ & $    37.00 $ & $   79.0 $ & $   27.0 $ & $   0.922 $ & Coralie \\
HATS-27 & $ 6841.56122 $ & $    50.37 $ & $    21.00 $ & $   14.0 $ & $   14.0 $ & $   0.505 $ & FEROS \\
HATS-27 & $ 6842.51976 $ & $    22.37 $ & $    17.00 $ & $   71.0 $ & $   12.0 $ & $   0.712 $ & FEROS \\
HATS-27 & $ 6845.58436 $ & $   -47.63 $ & $    17.00 $ & $   48.0 $ & $   12.0 $ & $   0.373 $ & FEROS \\
HATS-27 & $ 6846.47434 $ & $    30.37 $ & $    26.00 $ & $  -69.0 $ & $   15.0 $ & $   0.565 $ & FEROS \\
HATS-27 & $ 6847.47811 $ & $    28.37 $ & $    14.00 $ & $   24.0 $ & $   10.0 $ & $   0.781 $ & FEROS \\
HATS-27 & $ 6850.59743 $ & $    54.37 $ & $    23.00 $ & $    8.0 $ & $   14.0 $ & $   0.454 $ & FEROS \\
HATS-27 & $ 6851.54418 $ & $   -33.63 $ & $    18.00 $ & $   11.0 $ & $   12.0 $ & $   0.658 $ & FEROS \\
HATS-27 & $ 6852.48123 $ & $   -16.63 $ & $    18.00 $ & $   74.0 $ & $   12.0 $ & $   0.860 $ & FEROS \\
HATS-27 & $ 6852.58575 $ & $    63.37 $ & $    22.00 $ & $   56.0 $ & $   14.0 $ & $   0.883 $ & FEROS \\
HATS-27 & $ 6854.49043 $ & $   -92.63 $ & $    20.00 $ & $   48.0 $ & $   13.0 $ & $   0.293 $ & FEROS \\
HATS-27 & $ 6855.47772 $ & $   -37.63 $ & $    15.00 $ & $   99.0 $ & $   11.0 $ & $   0.506 $ & FEROS \\
HATS-27 & $ 6856.49742 $ & $    29.37 $ & $    13.00 $ & $   32.0 $ & $   10.0 $ & $   0.726 $ & FEROS \\
HATS-27 & $ 7067.80560 $ & $   -38.64 $ & $    25.00 $ & $   73.0 $ & $   48.0 $ & $   0.296 $ & HARPS \\
HATS-27 & $ 7068.84310 $ & $   -24.64 $ & $    15.00 $ & $   18.0 $ & $   30.0 $ & $   0.520 $ & HARPS \\
HATS-27 & $ 7069.87176 $ & $    74.36 $ & $    24.00 $ & $   -6.0 $ & $   48.0 $ & $   0.741 $ & HARPS \\
HATS-27 & $ 7070.83755 $ & $   -18.64 $ & $    20.00 $ & $   30.0 $ & $   38.0 $ & $   0.950 $ & HARPS \\
HATS-27 & $ 7071.86649 $ & $  -100.64 $ & $    40.00 $ & $   58.0 $ & $   72.0 $ & $   0.172 $ & HARPS \\
HATS-27 & $ 7072.87384 $ & $    -1.64 $ & $    25.00 $ & $  -29.0 $ & $   48.0 $ & $   0.389 $ & HARPS \\
HATS-27 & $ 7118.60461 $ & $   -45.64 $ & $    23.00 $ & $  247.0 $ & $   44.0 $ & $   0.251 $ & HARPS \\
HATS-27 & $ 7119.69411 $ & $   224.37 $ & $    26.00 $ & $  468.0 $ & $   16.0 $ & $   0.486 $ & FEROS \\
HATS-27 & $ 7119.76445 $ & $    68.37 $ & $    16.00 $ & $  147.0 $ & $   11.0 $ & $   0.501 $ & FEROS \\
HATS-27 & $ 7120.70644 $ & $   105.36 $ & $    27.00 $ & $   55.0 $ & $   48.0 $ & $   0.704 $ & HARPS \\
HATS-27 & $ 7121.56784 $ & $    33.37 $ & $    15.00 $ & $  100.0 $ & $   11.0 $ & $   0.890 $ & FEROS \\
HATS-27 & $ 7466.59338 $ & $    -6.64 $ & $    33.00 $ & $   27.0 $ & $   48.0 $ & $   0.296 $ & HARPS \\
HATS-27 & $ 7467.57122 $ & $    25.36 $ & $    27.00 $ & $   37.0 $ & $   38.0 $ & $   0.507 $ & HARPS \\
HATS-27 & $ 7468.57169 $ & $    62.36 $ & $    27.00 $ & $   47.0 $ & $   38.0 $ & $   0.723 $ & HARPS \\
\cutinhead{\bf HATS-28}
HATS-28 & $ 7181.60951 $ & $  -138.64 $ & $    18.00 $ & $ -149.0 $ & $   25.0 $ & $   0.313 $ & FEROS \\
HATS-28 & $ 7182.79396 $ & $   124.36 $ & $    22.00 $ & $   25.0 $ & $   30.0 $ & $   0.686 $ & FEROS \\
HATS-28 & $ 7183.58139 $ & $    54.36 $ & $    22.00 $ & $  -35.0 $ & $   30.0 $ & $   0.933 $ & FEROS \\
HATS-28 & $ 7184.65002 $ & $   -21.64 $ & $    22.00 $ & $   72.0 $ & $   30.0 $ & $   0.269 $ & FEROS \\
HATS-28 & $ 7187.76095 $ & $   -58.64 $ & $    15.00 $ & $  -36.0 $ & $   20.0 $ & $   0.247 $ & FEROS \\
HATS-28 & $ 7188.73064 $ & $    58.36 $ & $    14.00 $ & $  -94.0 $ & $   18.0 $ & $   0.552 $ & FEROS \\
HATS-28 & $ 7189.88679 $ & $    86.36 $ & $    13.00 $ & $   -5.0 $ & $   18.0 $ & $   0.915 $ & FEROS \\
HATS-28 & $ 7191.66177 $ & $   -53.64 $ & $    12.00 $ & $  -33.0 $ & $   16.0 $ & $   0.473 $ & FEROS \\
HATS-28 & $ 7192.67821 $ & $    88.36 $ & $    19.00 $ & $    2.0 $ & $   25.0 $ & $   0.793 $ & FEROS \\
HATS-28 & $ 7193.76250 $ & $   -67.64 $ & $    11.00 $ & $    1.0 $ & $   15.0 $ & $   0.134 $ & FEROS \\
HATS-28 & $ 7194.62609 $ & $   -48.64 $ & $    12.00 $ & $   45.0 $ & $   16.0 $ & $   0.405 $ & FEROS \\
HATS-28 & $ 7196.85775 $ & $   -96.64 $ & $    15.00 $ & $  -38.0 $ & $   20.0 $ & $   0.107 $ & FEROS \\
HATS-28 & $ 7218.71803 $ & $    17.36 $ & $    25.00 $ & $  -13.0 $ & $   34.0 $ & $   0.979 $ & FEROS \\
HATS-28 & $ 7220.77528 $ & $    55.36 $ & $    20.00 $ & $    4.0 $ & $   26.0 $ & $   0.625 $ & FEROS \\
HATS-28 & $ 7223.54993 $ & $   -89.64 $ & $    24.00 $ & $  -25.0 $ & $   32.0 $ & $   0.498 $ & FEROS \\
HATS-28 & $ 7224.55108 $ & $    73.36 $ & $    14.00 $ & $  -46.0 $ & $   19.0 $ & $   0.812 $ & FEROS \\
HATS-28 & $ 7227.52090 $ & $   114.36 $ & $    10.00 $ & $  -46.0 $ & $   14.0 $ & $   0.746 $ & FEROS \\
HATS-28 & $ 7230.76298 $ & $    79.36 $ & $    11.00 $ & $  -35.0 $ & $   16.0 $ & $   0.765 $ & FEROS \\
\cutinhead{\bf HATS-29}
HATS-29 & $ 7118.83135 $ & $    44.32 $ & $    19.00 $ & $  -66.0 $ & $   60.0 $ & $   0.862 $ & HARPS \\
HATS-29 & $ 7119.83810 $ & $   -26.68 $ & $    12.00 $ & $  -42.0 $ & $   44.0 $ & $   0.080 $ & HARPS \\
HATS-29 & $ 7120.82811 $ & $   -75.68 $ & $     7.00 $ & $  -41.0 $ & $   28.0 $ & $   0.295 $ & HARPS \\
HATS-29 & $ 7149.24168 $ & $   -72.08 $ & $    16.80 $ & \nodata      & \nodata      & $   0.464 $ & CYCLOPS \\
HATS-29 & $ 7149.25763 $ & $   -22.98 $ & $    17.70 $ & \nodata      & \nodata      & $   0.468 $ & CYCLOPS \\
HATS-29 & $ 7149.27360 $ & $   -11.78 $ & $     7.20 $ & \nodata      & \nodata      & $   0.471 $ & CYCLOPS \\
HATS-29 & $ 7150.27180 $ & $    96.82 $ & $    16.30 $ & \nodata      & \nodata      & $   0.688 $ & CYCLOPS \\
HATS-29 & $ 7150.28776 $ & $    11.02 $ & $    17.20 $ & \nodata      & \nodata      & $   0.691 $ & CYCLOPS \\
HATS-29 & $ 7150.30372 $ & $   127.52 $ & $    12.80 $ & \nodata      & \nodata      & $   0.695 $ & CYCLOPS \\
HATS-29 & $ 7152.13921 $ & $    -1.98 $ & $     9.70 $ & \nodata      & \nodata      & $   0.093 $ & CYCLOPS \\
HATS-29 & $ 7152.15453 $ & $   -36.58 $ & $     8.00 $ & \nodata      & \nodata      & $   0.097 $ & CYCLOPS \\
HATS-29 & $ 7152.16986 $ & $   -15.58 $ & $    11.00 $ & \nodata      & \nodata      & $   0.100 $ & CYCLOPS \\
HATS-29 & $ 7179.75812 $ & $   -28.10 $ & $    15.00 $ & $   12.0 $ & $   26.0 $ & $   0.090 $ & Coralie \\
HATS-29 & $ 7180.75107 $ & $   -88.10 $ & $    16.00 $ & $   92.0 $ & $   29.0 $ & $   0.305 $ & Coralie \\
HATS-29 & $ 7181.76069 $ & $    13.90 $ & $    13.00 $ & $  -15.0 $ & $   24.0 $ & $   0.525 $ & Coralie \\
HATS-29 & $ 7182.73909 $ & $    87.90 $ & $    14.00 $ & $  -73.0 $ & $   26.0 $ & $   0.737 $ & Coralie \\
\cutinhead{\bf HATS-30}
HATS-30 & $ 6932.62878 $ & $   -59.90 $ & $    10.00 $ & $    4.0 $ & $   11.0 $ & $   0.411 $ & FEROS \\
HATS-30 & $ 6939.66985 $ & $    86.61 $ & $    15.00 $ & $  -10.0 $ & $   19.0 $ & $   0.629 $ & Coralie \\
HATS-30 & $ 6940.55468 $ & $    31.61 $ & $    14.00 $ & $   -5.0 $ & $   18.0 $ & $   0.908 $ & Coralie \\
HATS-30 & $ 6941.71844 $ & $   -64.39 $ & $    15.00 $ & $   23.0 $ & $   21.0 $ & $   0.274 $ & Coralie \\
HATS-30 & $ 6968.73018 $ & $    70.61 $ & $    12.00 $ & $  -43.0 $ & $   15.0 $ & $   0.784 $ & Coralie \\
HATS-30 & $ 6970.67182 $ & $   -57.39 $ & $    15.00 $ & $  -77.0 $ & $   19.0 $ & $   0.395 $ & Coralie \\
HATS-30 & $ 6972.60936 $ & $   -27.39 $ & $    14.00 $ & $   -1.0 $ & $   19.0 $ & $   0.006 $ & Coralie \\
HATS-30 & $ 6982.70613 $ & $   -83.90 $ & $    10.00 $ & $   19.0 $ & $   11.0 $ & $   0.186 $ & FEROS \\
HATS-30 & $ 6984.64881 $ & $    89.10 $ & $    10.00 $ & $   34.0 $ & $   11.0 $ & $   0.798 $ & FEROS \\
HATS-30 & $ 6985.58892 $ & $   -49.90 $ & $    10.00 $ & $   29.0 $ & $   11.0 $ & $   0.095 $ & FEROS \\
HATS-30 & $ 6997.56102 $ & $    67.10 $ & $    11.00 $ & $  -12.0 $ & $   13.0 $ & $   0.866 $ & FEROS \\
HATS-30 & $ 6998.61894 $ & $   -97.90 $ & $    10.00 $ & $   13.0 $ & $   10.0 $ & $   0.199 $ & FEROS \\
HATS-30 & $ 6999.66282 $ & $    31.10 $ & $    10.00 $ & $   -8.0 $ & $   11.0 $ & $   0.528 $ & FEROS \\
\enddata
\tablenotetext{a}{
    The zero-point of these velocities is arbitrary. An overall offset
    $\gamma_{\rm rel}$ fitted independently to the velocities from
    each instrument has been subtracted.
}
\tablenotetext{b}{
    Internal errors excluding the component of astrophysical jitter
    considered in \refsecl{globmod}.
}
\ifthenelse{\boolean{rvtablelong}}{
}{
} 
\ifthenelse{\boolean{emulateapj}}{
    \end{deluxetable*}
}{
    \end{deluxetable}
}

\end{document}